\documentclass[longauth]{aa}  
\usepackage{graphicx}
\usepackage{txfonts}
\usepackage{natbib}
\usepackage{longtable}
\usepackage{multirow}
\usepackage{color}
\usepackage{url}
\usepackage{mathtools}
\usepackage{amsmath}
\usepackage{soul}
\usepackage{adjustbox}
\sethlcolor{yellow}

\usepackage[pdfpagemode={UseOutlines},bookmarks=true,bookmarksopen=true,
   bookmarksopenlevel=0,bookmarksnumbered=true,hypertexnames=false,
   colorlinks,linkcolor={blue},citecolor={blue},urlcolor={red},
   pdfstartview={FitV},unicode,breaklinks=true]{hyperref}

\usepackage[switch]{lineno}

\def\fermi{{\it Fermi}-LAT }
\def\deg{$^{\circ}$}
\def\ergs{\ifmmode \mathrm{erg\hspace{1mm}s}^{-1} \else erg s$^{-1}$\fi}
\def\ergscm{erg s$^{-1}$ cm$^{-2}$}
\def\phcms{photons cm$^{-2}$ s$^{-1}$}

\newcommand{\myemail}{roberto.angioni@ssdc.asi.it}

\begin{document} 

   \title{Gamma-ray emission in radio galaxies under the VLBI scope}
   \subtitle{II. The relationship between gamma-ray emission 
   and parsec-scale jets in radio galaxies}

      \author{R.~Angioni
          \inst{1,2,3,4}
	  \and
          E.~Ros
          \inst{1}
          \and
          M.~Kadler
          \inst{2}
          \and
          R.~Ojha
          \inst{5,6,7}
          \and
          C.~M\"uller
          \inst{8,1}
          \and
          P.~G.~Edwards
          \inst{9}
          \and
          P.~R.~Burd
          \inst{2}
          \and
          B.~Carpenter
          \inst{5,6,7}
          \and
          M.~S.~Dutka
          \inst{5,10}
          \and
          S.~Gulyaev
          \inst{11}
          \and
          H.~Hase
          \inst{12}
          \and
          S.~Horiuchi
          \inst{13}
          \and
          F.~Krau\ss
          \inst{14}
          \and
          J.~E.~J.~Lovell
          \inst{15}
          \and
          T.~Natusch
          \inst{11}
          \and
          C.~Phillips
          \inst{9}
          \and
          C.~Pl\"otz
          \inst{12}
          \and
          J.~F.~H.~Quick
          \inst{16}
          \and
          F.~R\"osch
          \inst{2}
          \and
          R.~Schulz
          \inst{17,2,18}
          \and
          J.~Stevens
          \inst{9}
          \and
          A.~K.~Tzioumis
          \inst{9}
          \and
          S.~Weston
          \inst{11}
          \and
          J.~Wilms
          \inst{18}
          \and
          J.~A.~Zensus
          \inst{1}
                      }

   \institute{Max-Planck-Institut f\"ur Radioastronomie, Auf dem H\"ugel 69, 53121 Bonn, Germany, email: \myemail
   \and
   Institut f\"ur Theoretische Physik und Astrophysik, Universit\"at
   W\"urzburg, Emil-Fischer-Str. 31, 97074 W\"urzburg, Germany
  \and
  ASI Space Science Data Center, Via del Politecnico snc, 00133 Rome, Italy
  \and
  INFN - Roma Tor Vergata, Via della Ricerca Scientifica 1, 00133 Rome, Italy
  \and
   NASA Goddard Space Flight Center, Greenbelt, MD 20771, USA
   \and
   Catholic University of America, Washington, DC 20064, USA
   \and
   University of Maryland, Baltimore County, 1000 Hilltop Cir, Baltimore, MD 21250, USA
   \and
   Department of Astrophysics/IMAPP, Radboud University Nijmegen, PO Box 9010, 6500 GL Nijmegen, the Netherlands
   \and
   CSIRO Astronomy and Space Science, PO Box 76, Epping, NSW 1710, Australia
   \and
   Wyle  Science,  Technology  and  Engineering  Group,  Greenbelt, MD 20771, USA
   \and
   Institute for Radio Astronomy \& Space Research, AUT University, 1010 Auckland, New Zealand
   \and
   Bundesamt  f\"ur  Kartographie  und  Geod\"asie,  93444  Bad  K\"otzting, Germany
   \and
   CSIRO  Astronomy  and  Space  Science,  Canberra  Deep  Space Communications Complex, PO Box 1035, Tuggeranong, ACT 2901, Australia
   \and
   Department of Astronomy \& Astrophysics, Pennsylvania State University, University Park, PA 16802, USA
   \and
   School of Mathematics \& Physics, University of Tasmania, Private Bag 37, Hobart, 7001 Tasmania, Australia
   \and
   Hartebeesthoek Radio Astronomy Observatory, PO Box 443, 1740 Krugersdorp, South Africa
   \and
   ASTRON,  the  Netherlands  Institute  for  Radio  Astronomy,  Post-bus 2, 7990 AA Dwingeloo, The Netherlands
   \and
   Dr.~Remeis-Sternwarte  \&  ECAP,  Universit\"at  Erlangen-N\"urnberg, Sternwartstra\ss e 7, 96049 Bamberg, Germany
   }

   \date{Received 23 April 2020; accepted 20 July 2020}

 
  \abstract
   {}
   {This is
the second paper in our series studying the evolution of parsec-scale
radio emission in radio galaxies in the southern hemisphere. Following
our study of the radio and high-energy properties of $\gamma$-ray-emitting sources, here we investigate the kinematic and spectral
properties of the parsec-scale jets of radio galaxies that have not
yet been detected by the \textit{Fermi} Large Area Telescope (\textit{Fermi}-LAT) instrument
on board NASA's \textit{Fermi Gamma-ray Space Telescope}. For many sources, these results represent
the first milliarcsecond resolution information in the literature.
These studies were conducted within the framework of the Tracking Active Nuclei with Austral Milliarcsecond Interferometry (TANAMI)
monitoring program and in the context of high-energy $\gamma$-ray
observations from \textit{Fermi}-LAT.}
   {We take advantage of the regular 8.4 GHz and 22.3 GHz Very Long Baseline Interferometry (VLBI) observations provided by the TANAMI monitoring program, and explore the kinematic properties of six $\gamma$-ray-faint radio galaxies. We complement this with $\sim8.5$ years of \textit{Fermi}-LAT data, deriving updated upper limits on the $\gamma$-ray emission from this subsample of TANAMI radio galaxies. We include publicly available VLBI
kinematics of $\gamma$-ray-quiet radio galaxies monitored by the MOJAVE
program and perform a consistent \textit{Fermi}-LAT analysis. We combine these results with those from our previous paper to construct the largest sample of radio galaxies with combined VLBI and $\gamma$-ray measurements to date. The connection between parsec-scale jet emission and high-energy
properties in the misaligned jets of radio galaxies is explored.}
   {We report for the first time evidence of superluminal motion up to $\beta_\mathrm{app}=3.6$ in the jet of the $\gamma$-ray-faint radio galaxy PKS\,2153$-$69. We find a clear trend of higher apparent speed as a function of distance from the jet core, which indicates that the jet is still being accelerated on scales of tens of parsecs, or $\sim10^5\,R_s$, corresponding to the end of the collimation and acceleration zone in nearby radio galaxies. We find evidence of subluminal apparent motion in the jets of PKS\,1258$-$321 and IC\,4296, and no measurable apparent motion for PKS\,1549$-$79, PKS\,1733$-$565 and PKS\,2027$-$308. For all these sources, TANAMI provides the first multi-epoch kinematic analysis on parsec scales. We then compare the VLBI properties of $\gamma$-ray-detected and undetected radio galaxies, and find that the two populations show a significantly different distribution of median core flux density, and, possibly, of median core brightness temperature. In terms of correlation between VLBI and $\gamma$-ray properties, we find a significant correlation between median core flux density and $\gamma$-ray flux, but no correlation with typical Doppler boosting indicators such as median core brightness temperature and core dominance.}
   {Our study suggests that high-energy emission from radio galaxies is related to parsec-scale radio emission from the inner jet, but is not driven by Doppler boosting effects, in contrast to the situation in their blazar counterparts. This implies that $\gamma$-ray loudness does not necessarily reflect a higher prevalence of boosting effects.}


   \maketitle
%

\section{Introduction}
Gamma-ray-detected AGN typically belong to the class of blazars that show a spectral energy distribution (SED) comprising two peaks. The low-energy peak is located between the radio and the
soft-X-ray band, while the high-energy peak can range from hard-X-rays
to TeV $\gamma$-rays (see e.g., \citealt{2017A&ARv..25....2P} for a
general review). The low-energy peak is associated with synchrotron
emission from ultra-relativistic electrons in the jet magnetic field (e.g., \citealt{2013EPJWC..6105001G}), while the
high-energy peak could be produced by the same electron population
through inverse Compton (IC) processes, with the seed photon field
being the synchrotron emission itself (Synchrotron Self Compton, SSC,
\citealt{1992ApJ...397L...5M}), or emission from the Broad Line Region
(BLR) or AGN torus (external Compton, EC, \citealt{1994ApJ...421..153S}).
These models alone cannot reproduce the entirety of observed AGN
phenomena, especially at very high energies (VHE, E$>$100 GeV).
Therefore, more sophisticated models involving contributions from
relativistic protons, in addition to the electrons, have been
developed (e.g., \citealt{1993A&A...269...67M, 2013ApJ...768...54B}).

In all of these models, the same electrons that produce the
low-energy emission make at least a contribution to the observed
high-energy component. This implies that we expect to observe a
connection between the two SED peaks. Indeed, several studies have
found such a connection between the radio-mm band and the $\gamma$-ray
band, in the form of correlations of fluxes (e.g.
\citealt{2009ApJ...696L..17K,2011ApJ...741...30A,2017A&A...606A.138L}), $\gamma$-ray activity episodes
concurrent with morphological or kinematical changes in the pc-scale
radio jet (e.g., \citealt{2012A&A...537A..70S,2013ApJ...773..147J,2015ApJ...808..162C}), or correlated variability
(e.g., \citealt{2014MNRAS.441.1899F,2016A&A...585A..91K,2016A&A...588A.146S}). Simply the fact that most of the
sources observed by \textit{Fermi}-LAT are radio-loud AGN
\citep{2011ApJ...741...30A,2015ApJS..218...23A} is also a strong indication of the bond
between the two bands. However, all the aforementioned studies deal with large AGN samples, which are typically heavily dominated by blazars. In these sources, the observed emission is strongly affected by orientation-dependent Doppler boosting effects, which are difficult to disentangle from the intrinsic emission. On the other hand, radio galaxies, i.e., the misaligned parent population of blazars \citep[see e.g.][]{1995PASP..107..803U}, show significantly smaller Doppler factors, providing a more direct tool to study high-energy emission in AGN jets. 

In \citet{2019A&A...627A.148A} (hereafter Paper I) we reported on the interplay between the radio VLBI properties and high-energy emission in $\gamma$-ray-detected radio galaxies, taking advantage of multi-epoch VLBI measurements from the TANAMI and MOJAVE monitoring programs, complemented by \fermi data. In this paper, we complement this study by presenting the kinematic analysis and \fermi upper limits of $\gamma$-ray-undetected TANAMI radio galaxies, and discussing the radio--$\gamma$-ray relation in radio galaxies in a broader context and using a larger sample. We compare the parsec-scale jet properties of $\gamma$-ray-detected and undetected radio galaxies, and investigate the presence of correlations with \fermi properties.

The paper is organized as follows. Sect.~\ref{sec:data} introduces the source sample and data sets and analysis methods used in this paper. Sect.~\ref{sec:res} reports the results obtained from the radio VLBI data (Sect.~\ref{sec:res_im},~\ref{sec:res_kin}) and \textit{Fermi}-LAT data (Sect.~\ref{sec:res_lat}). We discuss the implications of our results in Sect.~\ref{sec:disc}, and summarize our findings in Sect.~\ref{sec:conc}. All the results are presented on a source-by-source basis. Throughout the paper we assume a cosmology with $H_0 =  73$ km s$^{-1}$ Mpc$^{-1}$, $\Omega_{\mathrm{m}}$ =   0.27, $\Omega_{\mathrm{\Lambda}}$ =   0.73~\citep{2011ApJS..192...18K}, the radio spectral indices refer to the convention $S\propto\nu^{+\alpha}$, while the $\gamma$-ray photon indices are in the form $F\propto E^{-\Gamma}$.


\section{Sample, observations and data reduction}
\label{sec:data}
\subsection{The TANAMI radio galaxy sample}
A detailed introduction of the TANAMI radio galaxy sample was presented in Paper I (see Section 2). The full TANAMI sample was constructed with sources south of  $\delta = -30^\circ$ from a radio-selected subsample (all sources in the \citet{1994A&AS..105..211S} catalogue, with $S_\mathrm{5\,GHz}>2$ Jy and spectral index $\alpha^\mathrm{2.7\,GHz}_\mathrm{5\,GHz} > -0.5$) and a $\gamma$-ray-selected subsample consisting of known $\gamma$-ray sources. This selection method is naturally biased towards blazars, since they are easier to detect in $\gamma$-rays and have flatter spectral indices in the radio. Therefore, out of $\sim100$ sources, the TANAMI sample includes a total of just fifteen radio galaxies (see Table~\ref{tab:rgs}). Of these, seven have been associated with a $\gamma$-ray source based on \textit{Fermi}-LAT data, while eight have not yet been detected. In this paper, we focus on the LAT-undetected subsample. The individual TANAMI sources studied in this paper therefore include PKS\,1258$-$321 (1258$-$321), IC\,4296 (PKS\,1333$-$33, 1333$-$337), PKS\,1549$-$79 (1549$-$790), PKS\,1733$-$56 (1733$-$565), PKS\,1814$-$63 (1814$-$638), PKS\,2027$-$308 (2027$-$308) and PKS\,2153$-$69 (2152$-$699). 

\subsection{Radio data}
We refer the reader to Section 3.1 of Paper I for a detailed description of the VLBI data analysis. Briefly, we imaged all available epochs
of 8.4\,GHz VLBI observations of the $\gamma$-ray-undetected radio
galaxies (see Table~\ref{tab:rgs}). The time range covered by these observations goes from 2008 to 2013 (the exact range varies for each source). The resulting full-resolution maps and the corresponding image parameters are presented in the appendix. We then modeled the resulting images using circular Gaussian components and tracked those across time to measure apparent speeds, using simple linear regression fits. We combined the measured apparent speed with the jet-to-counterjet flux ratio in order to constrain the intrinsic jet speed and viewing angle. Additionally, we have used selected contemporaneous observations at 22.3~GHz to produce spectral index maps.


\begin{table*}[h!tbp]
\caption{Radio galaxies in the TANAMI sample.}
\label{tab:rgs}
\begin{center}
\begin{tabular}{llllllcc}
\hline
\hline
B1950 name & Other name & Class$^a$ & Redshift & RA(J2000) & Dec(J2000) & 4FGL$^b$ & Ref$^c$\\
\hline
0518$-$458 & Pictor~A & FR II & 0.035 & 79.957 & $-$45.779 & J0519.6$-$4544 & [1]\\
0521$-$365 & PKS\,0521$-$36	& RG/SSRQ & 0.057 & 80.742 & $-$36.459 & J0522.9$-$3628 & [2]\\
0625$-$354 & PKS\,0625$-$35 & FR I/BLL & 0.055 & 96.778 & $-$35.487 & J0627.0$-$3529 & [3]\\
1258$-$321 & PKS\,1258$-$321 &  FR I & 0.017 & 195.253 & $-$32.441 & - & [4]\\
1322$-$428 & Centaurus~A & FR I & 0.0018 & 201.365 & $-$43.019 & J1325.5$-$4300 & [5]\\
1333$-$337 & IC\,4296 &  FR I & 0.013 & 204.162 &	$-$33.966 & - & [4]\\
1343$-$601 & Centaurus~B &  FR I & 0.013 & 206.704 & $-$60.408 & J1346.3$-$6026 & [6]\\
1549$-$790 & PKS\,1549$-$79	&  RG/CFS & 0.15 & 239.245 &	$-$79.234 & - & [7]\\
1600$-$489 & PMN~J1603$-$4904 & GPS$^d$ & 0.23 & 240.961 & $-$49.068 & J1603.8$-$4903 & [8]\\
1718$-$649 & PKS\,1718$-$649 &  GPS/CSO & 0.014 & 260.921 & $-$65.010 & J1724.2$-$6501 & [9]\\
1733$-$565 & PKS\,1733$-$56 &  FR II & 0.098 & 264.399 & $-$56.567 & - & [2]\\
1814$-$637 & PKS\,1814$-$63 &  CSS/CSO & 0.065 & 274.896 & $-$63.763 & - & [7]\\
2027$-$308 & PKS\,2027$-$308 &  RG & 0.539 & 307.741 & $-$30.657 & - & [10]\\
2152$-$699 & PKS\,2153$-$69 &  FR II & 0.028 & 329.275 & $-$69.690 & J2156.0$-$6942$^e$ & [11]\\
\hline
\hline
\end{tabular}
\end{center}
$^a$ FR I: Fanaroff-Riley type 1; FR II: Fanaroff-Riley type 2; BLL:
BL Lac; RG: Radio galaxy; SSRQ: Steep Spectrum Radio Quasar; CFS: Compact Flat Spectrum; GPS: Gigahertz Peaked Spectrum; CSO: Compact
Symmetric Object; CSS: Compact Steep Spectrum (see \citealt{1998PASP..110..493O} for a review on the GPS, CSO and CSS classes).\\
$^b$ $\gamma$-ray counterpart source name from the 4FGL~\citep{4fgl}, if applicable.\\
$^c$ Redshift reference: [1] \citet{1989spce.book.....L}, [2] \citet{2009MNRAS.399..683J}, [3] \citet{1995ApJS...96..343Q}, [4] \citet{2000MNRAS.313..469S}, [5] \citet{1978PASP...90..237G}, [6] \citet{1989A&A...223...61W}, [7] \citet{2008MNRAS.387..639H}, [8] \citet{2016A&A...586L...2G}, [9] \citet{2004MNRAS.350.1195M}, [10] \citet{2008ApJS..175...97H}, [11] \citet{1991ApJS...75..935D}
\\$^d$ Originally misclassified as BL Lac, this source has been classified as a young radio galaxy based on multi-wavelength
studies \citep{2014A&A...562A...4M,2015A&A...574A.117M,2016A&A...593L..19M}.
\\$^e$ see Section~\ref{sec:res_lat}.
\end{table*}


\subsection{\textit{Fermi}-LAT data}
We refer the reader to Section 3.2 of Paper I for a detailed
description of our \textit{Fermi}-LAT data analysis method. In short,
we use 103 months of data in the energy range 0.1-100 GeV, using a Region of Interest (ROI)
of 15$^{\circ}$ and the latest Pass8 analysis software, through the
python package \texttt{Fermipy}~\citep{2017ICRC...35..824W}. We apply a zenith angle cut at
100$^{\circ}$, and model the four PSF types\,\footnote{LAT data can be selected into Point Spread Function (PSF) quartiles, based on the
quality of the direction reconstruction, from the worst quartile
(PSF0) to the best (PSF3)} of LAT data separately,
using the summed-likelihood method. To compensate for the decrease in angular resolution with the worse PSF quartiles, we increase the low-energy selection cut to 400 MeV, 500 MeV and 800 MeV for PSF2, PSF1 and PSF0, respectively.

We fit the ROI with the initial model including all sources from the Third \textit{Fermi}-LAT source catalog \citep{2015ApJS..218...23A} in the region, and iteratively add new
sources based on peaks in the excess Test Statistic (TS\,\footnote{The TS is defined as $2\log(L/L_0)$ where
$L$ is the likelihood of the model with a point source at the target
position, and $L_0$ is the likelihood without the source. A value of TS=25
corresponds to a significance of
4.2$\sigma$~\citep{1996ApJ...461..396M}}) maps, until the ROI is
properly modeled. We intentionally use the 3FGL as a starting model rather than the more up-to-date 4FGL catalog~\citep{4fgl} to keep our analysis consistent with Paper I.

Since the sources studied in this paper are not present in the 3FGL, we center the ROI on the
radio position of the target. If a new source consistent with the
target position is not found by the
source-finding iterative procedure, we place a test source at the ROI
center assuming a typical photon index value $\Gamma=2.2$~\citep[see e.g.][]{4lac}, and derive a 95\% confidence upper limit.

\section{Results}
\label{sec:res}

We
first present the high-resolution VLBI images of our seven $\gamma$-ray-faint radio galaxies:
PKS\,1258$-$321, IC\,4296, PKS\,1549$-$79, PKS\,1733$-$56, PKS\,1814$-$63, PKS\,2027$-$308 and PKS\,2153$-$69. Then we discuss the parsec-scale jet kinematics of these sources.
Finally we present the results of our analysis of the \textit{Fermi}-LAT
observations of these misaligned jet sources.

\subsection{Radio imaging results}
\label{sec:res_im}
\begin{figure*}
    \centering
    \includegraphics[width=0.32\linewidth]{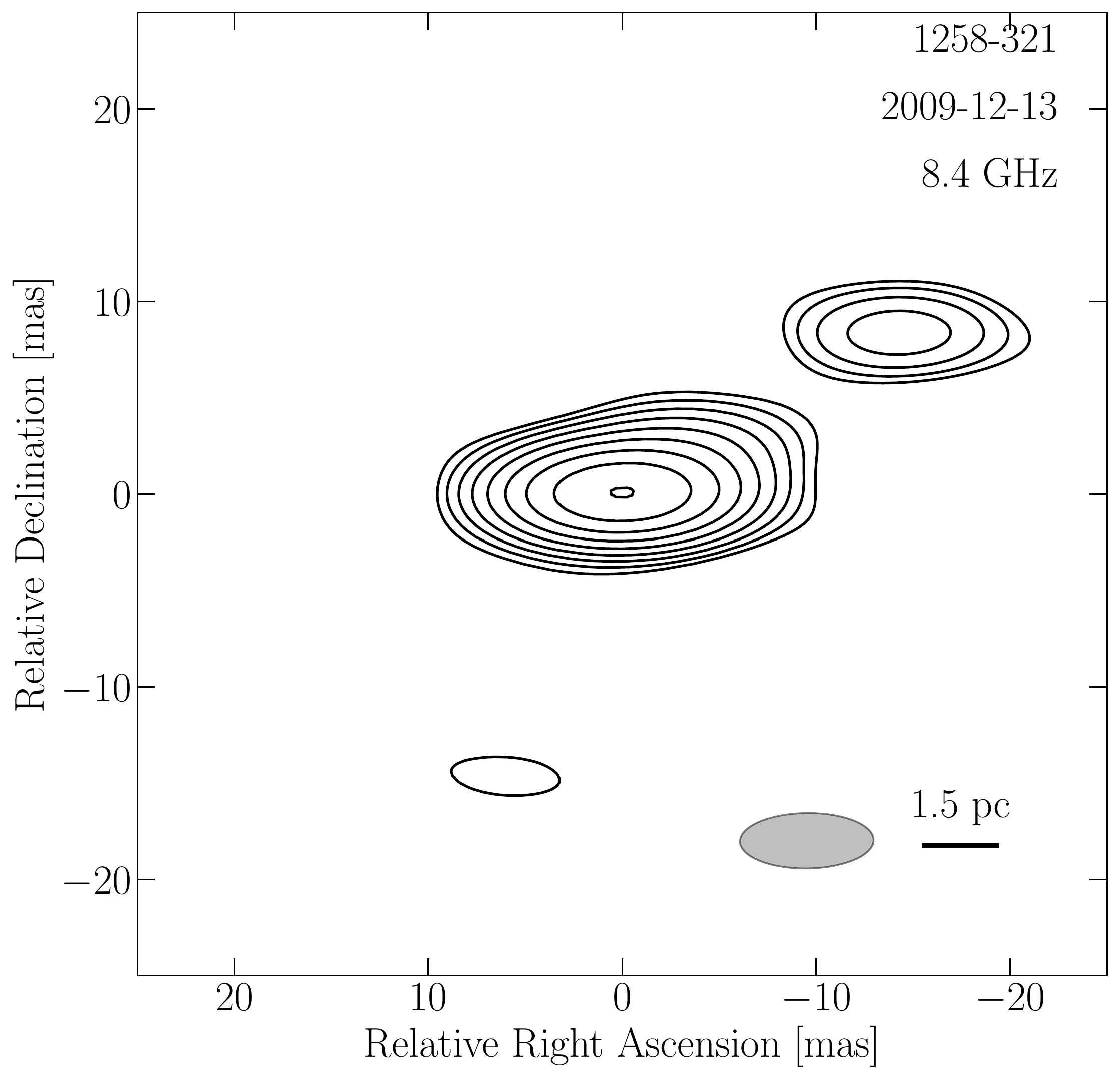}
    \includegraphics[width=0.32\linewidth]{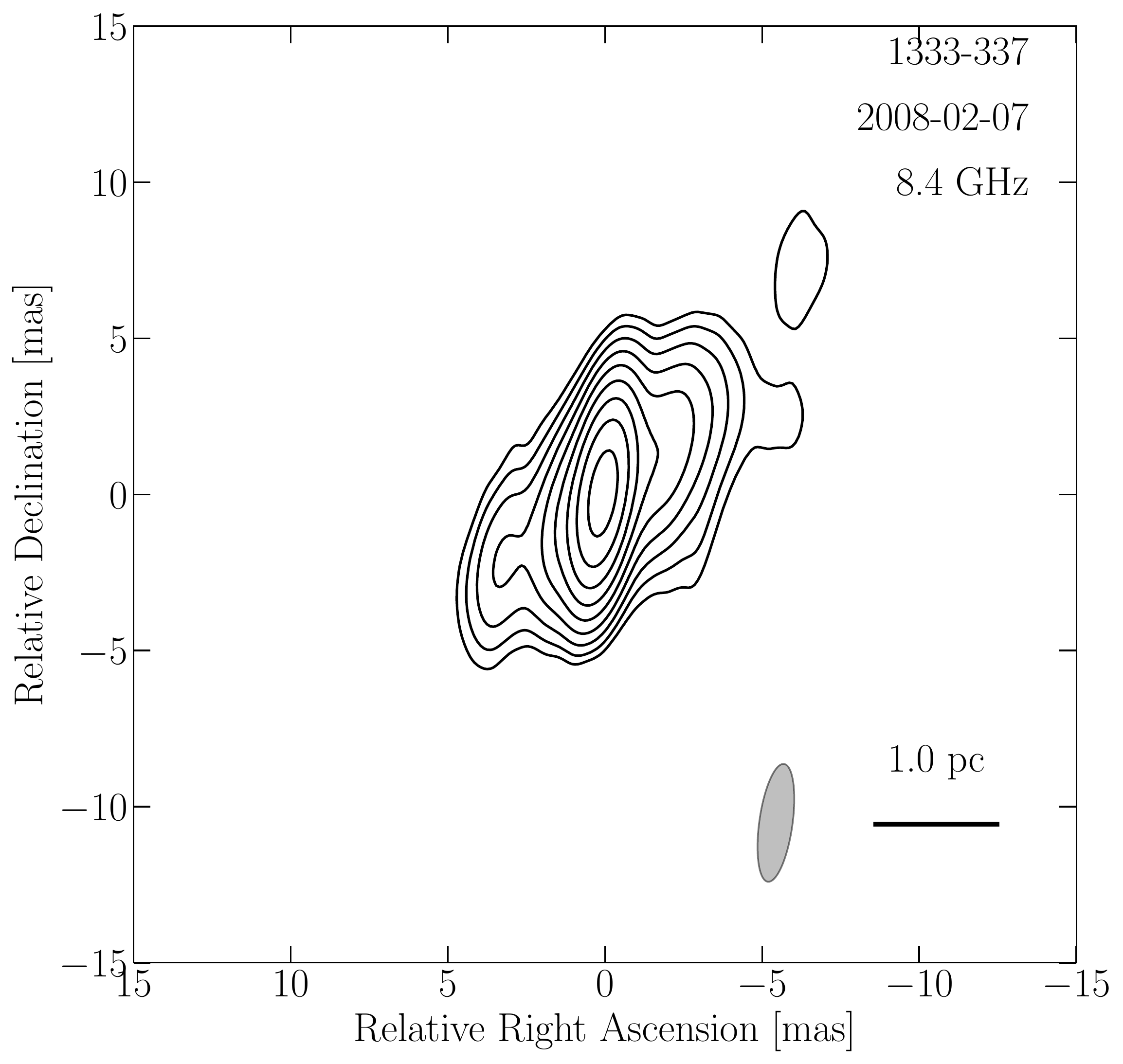}
    \includegraphics[width=0.32\linewidth]{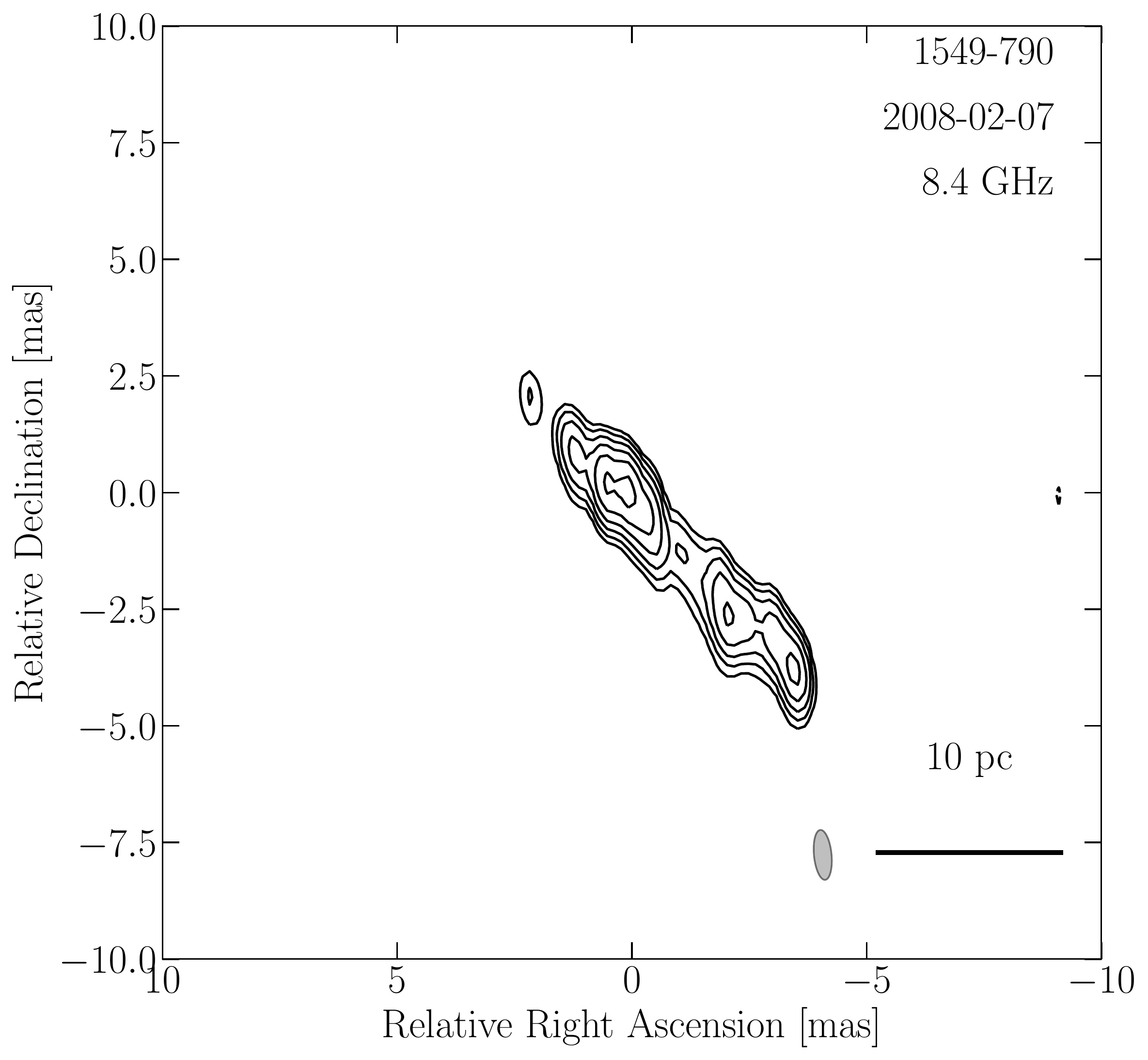}
    \includegraphics[width=0.32\linewidth]{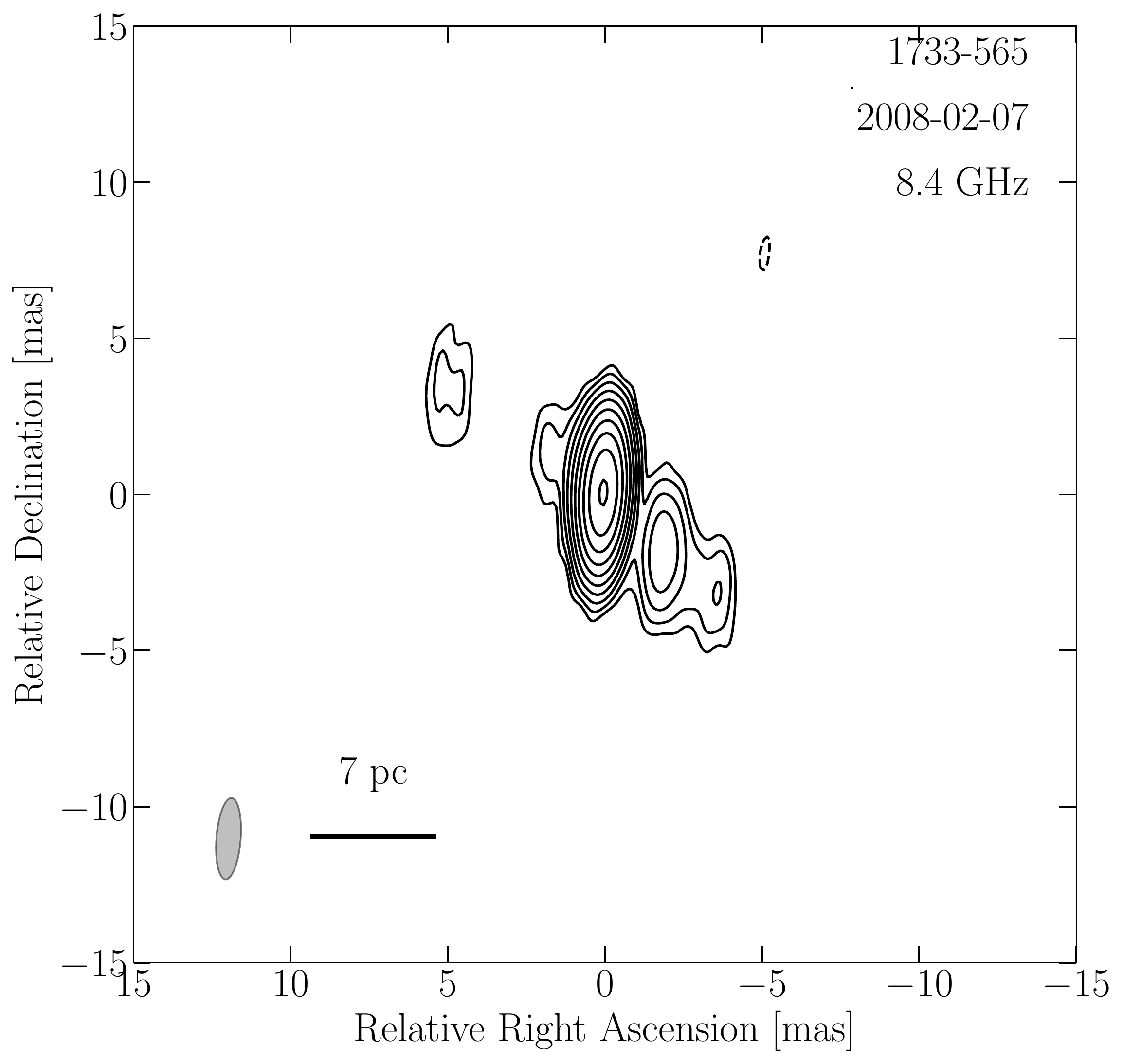}
    \includegraphics[width=0.32\linewidth]{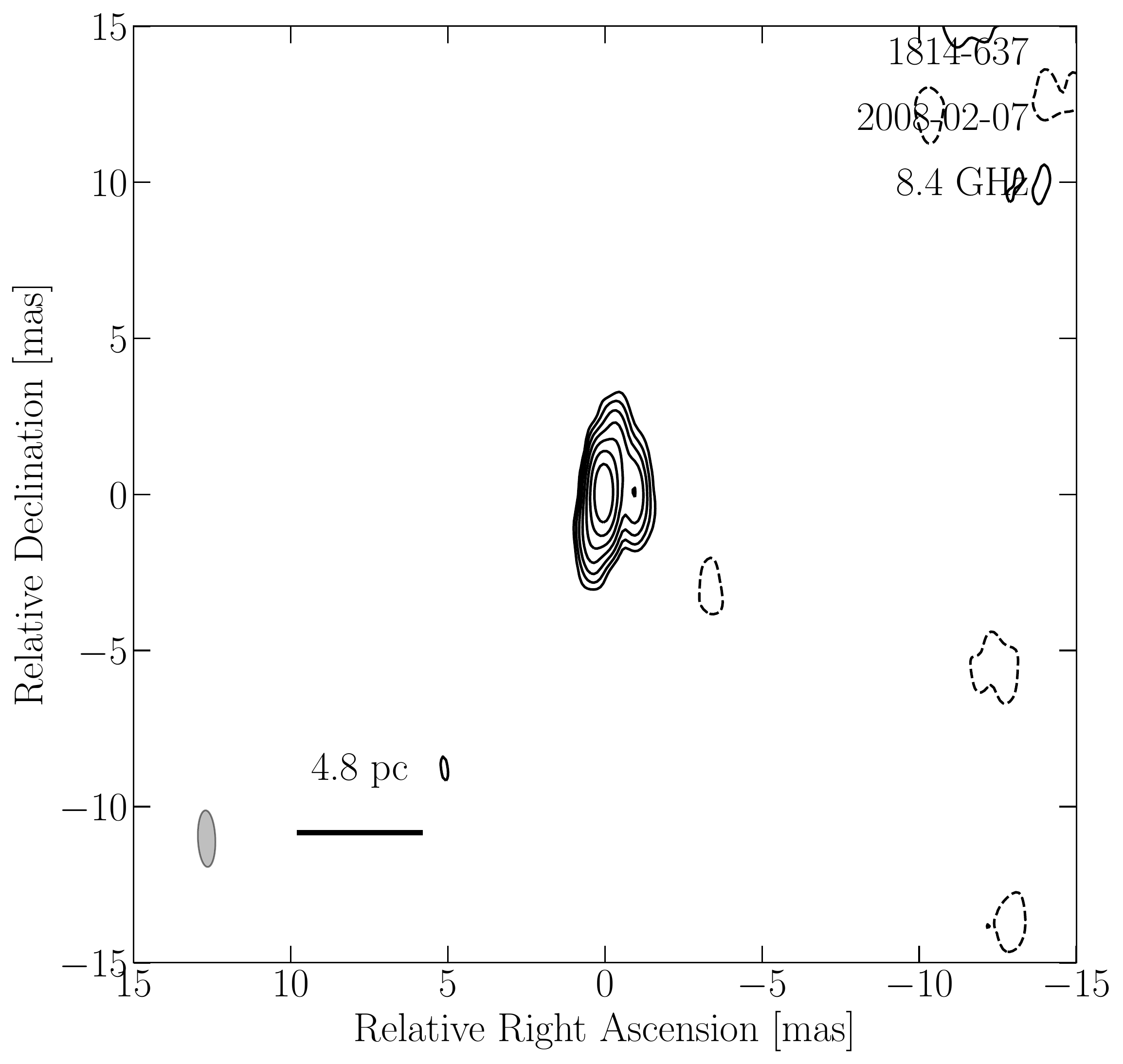}
    \includegraphics[width=0.32\linewidth]{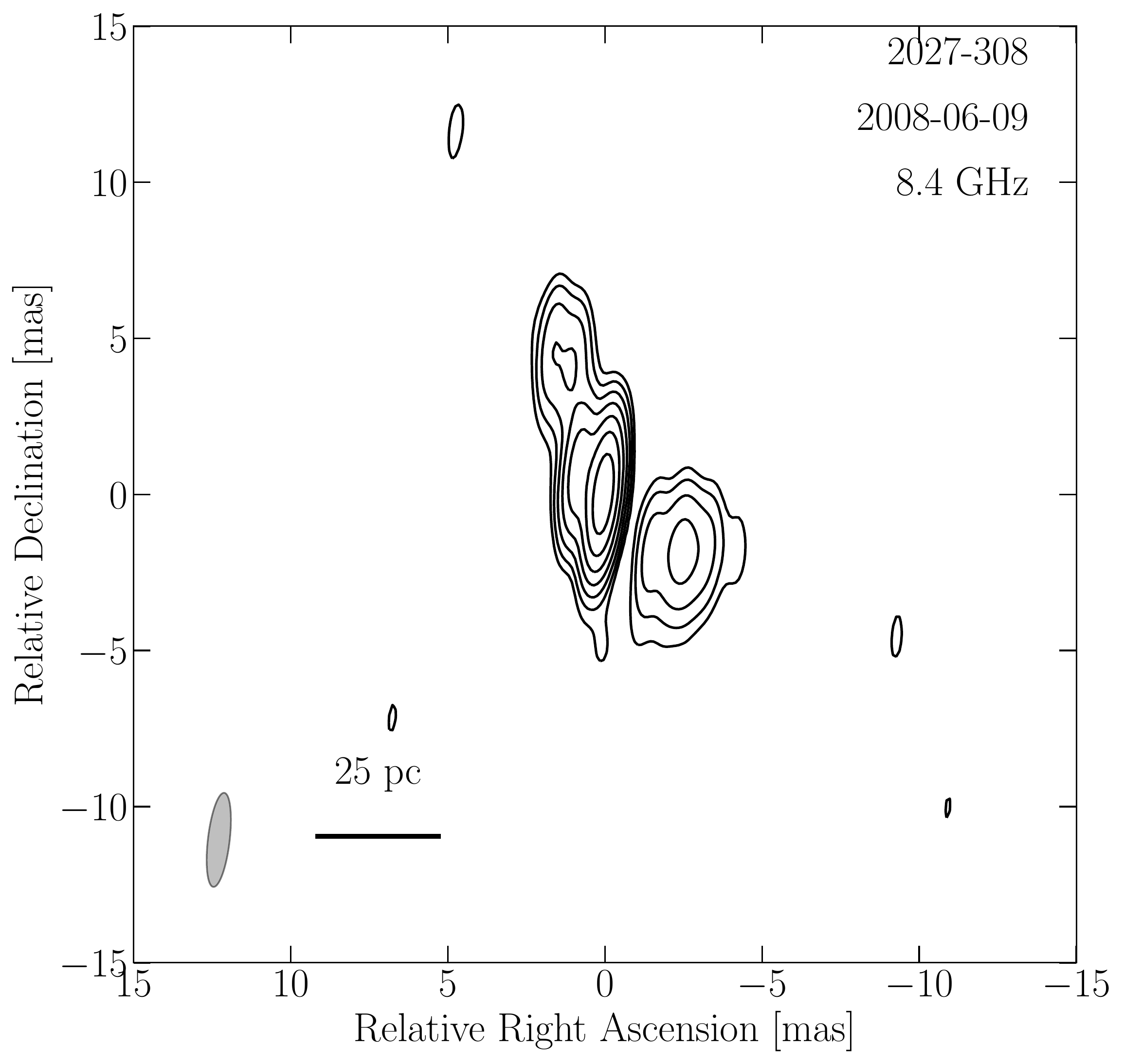}
    \includegraphics[width=0.32\linewidth]{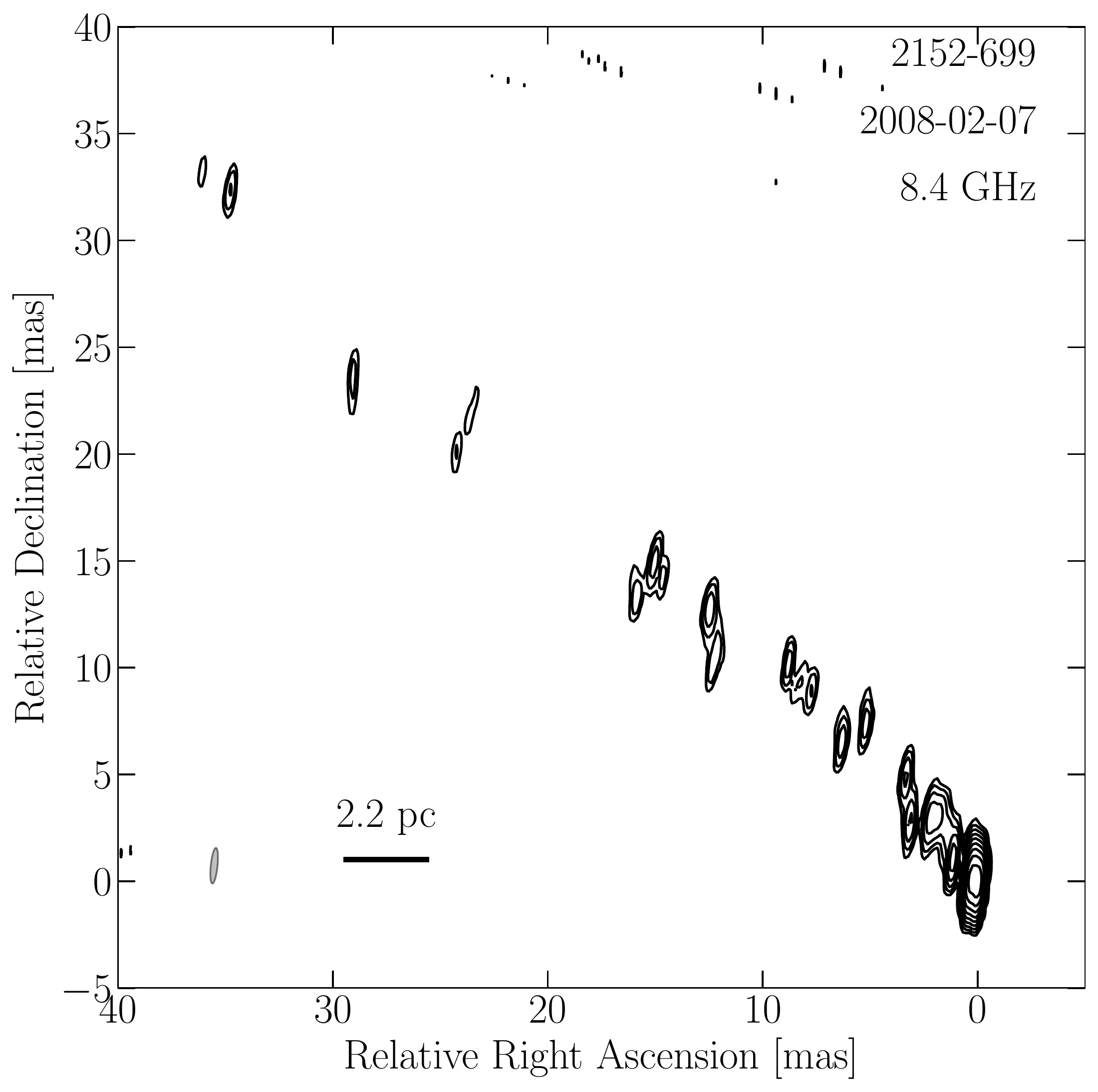}
    \caption{First-epoch 8.4\,GHz contour maps of $\gamma$-ray-faint TANAMI radio galaxies. The grey ellipse in the bottom left of each panel indicates the convolving beam, while the black bar shows the linear scale. The full set of multi-epoch images and the associated map parameter tables are presented in Appendix~\ref{app:maps}.}
    \label{fig:wallpaper}
\end{figure*}

In Fig.~\ref{fig:wallpaper} we present first-epoch VLBI images of the $\gamma$-ray-faint TANAMI radio galaxies, while the full set of multi-epoch images and related map parameters is presented in Appendix~\ref{app:maps}. The imaging results for each individual source are discussed in this subsection.

\paragraph{1258$-$321}
The source shows a faint one-sided jet extending to the north-west (see Fig.~\ref{fig:wallpaper}), aligned with the kpc-scale structure~\citep{2005ApJS..156...13M}. The first-epoch image was presented for the first time in \cite{2018A&A...610A...1M}. 

\begin{figure}[!htbp]
\begin{center}
\includegraphics[width=0.9\linewidth]{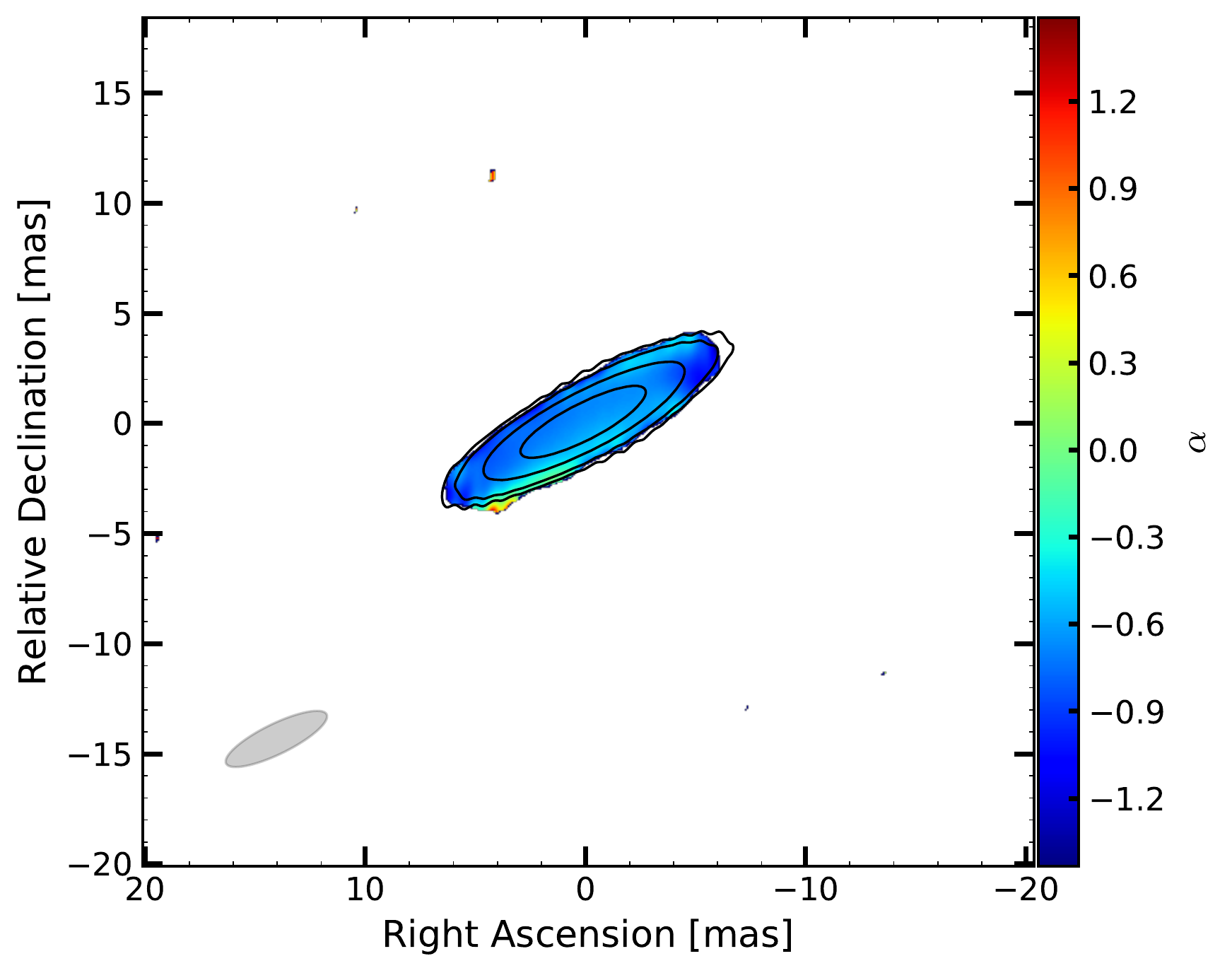}
\end{center}
\caption{Spectral index map of IC\,4296 between 8.4 GHz and 22.3 GHz for epoch 2008-02-07. The black contours are from the 8.4 GHz image. Both
maps were convolved with a common beam that is represented in grey in
the lower-left corner.}
\label{1333_spix}
\end{figure}

\paragraph{1333$-$337 (IC\,4296)}
This FR~I source shows a symmetric double-sided morphology on parsec
scales (see Fig.~\ref{fig:wallpaper}), extending in a direction consistent with its large scale structure \citep{1986ApJ...302..306K}. The spectral index map, presented in Fig.~\ref{1333_spix}, shows an unresolved feature
with a spectral index $\alpha\sim-0.8$.

\begin{figure}[!htbp]
\begin{center}
\includegraphics[width=0.9\linewidth]{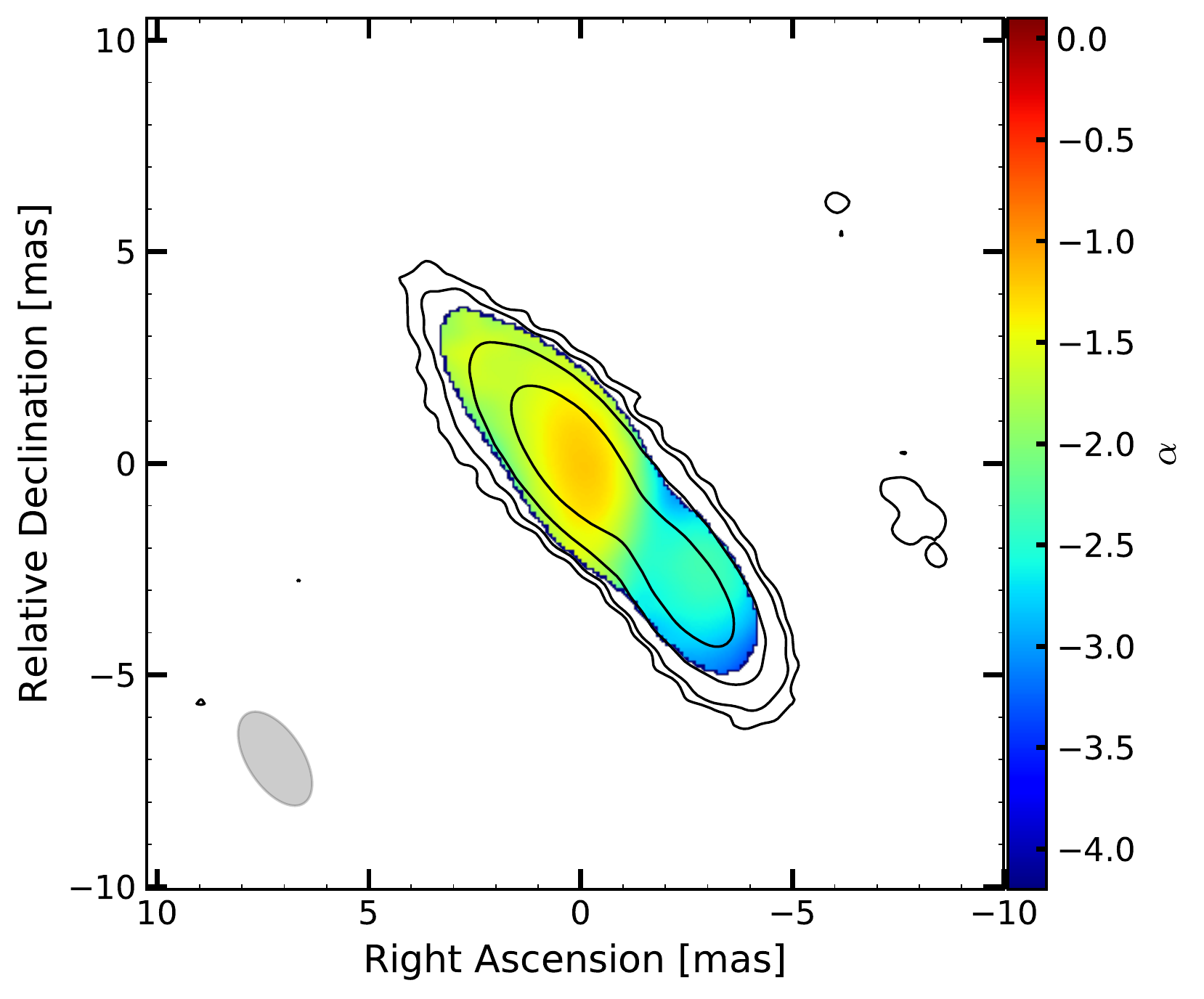}
\end{center}
\caption{Spectral index map of PKS\,1549$-$79 between 8.4 GHz and 22.3 GHz for epoch 2008-11-27. The black contours are from the 8.4 GHz image. Both
maps were convolved with a common beam that is represented in grey in
the lower-left corner.}
\label{1549_spix}
\end{figure}

\begin{figure}[!htbp]
\begin{center}
\includegraphics[width=0.9\linewidth]{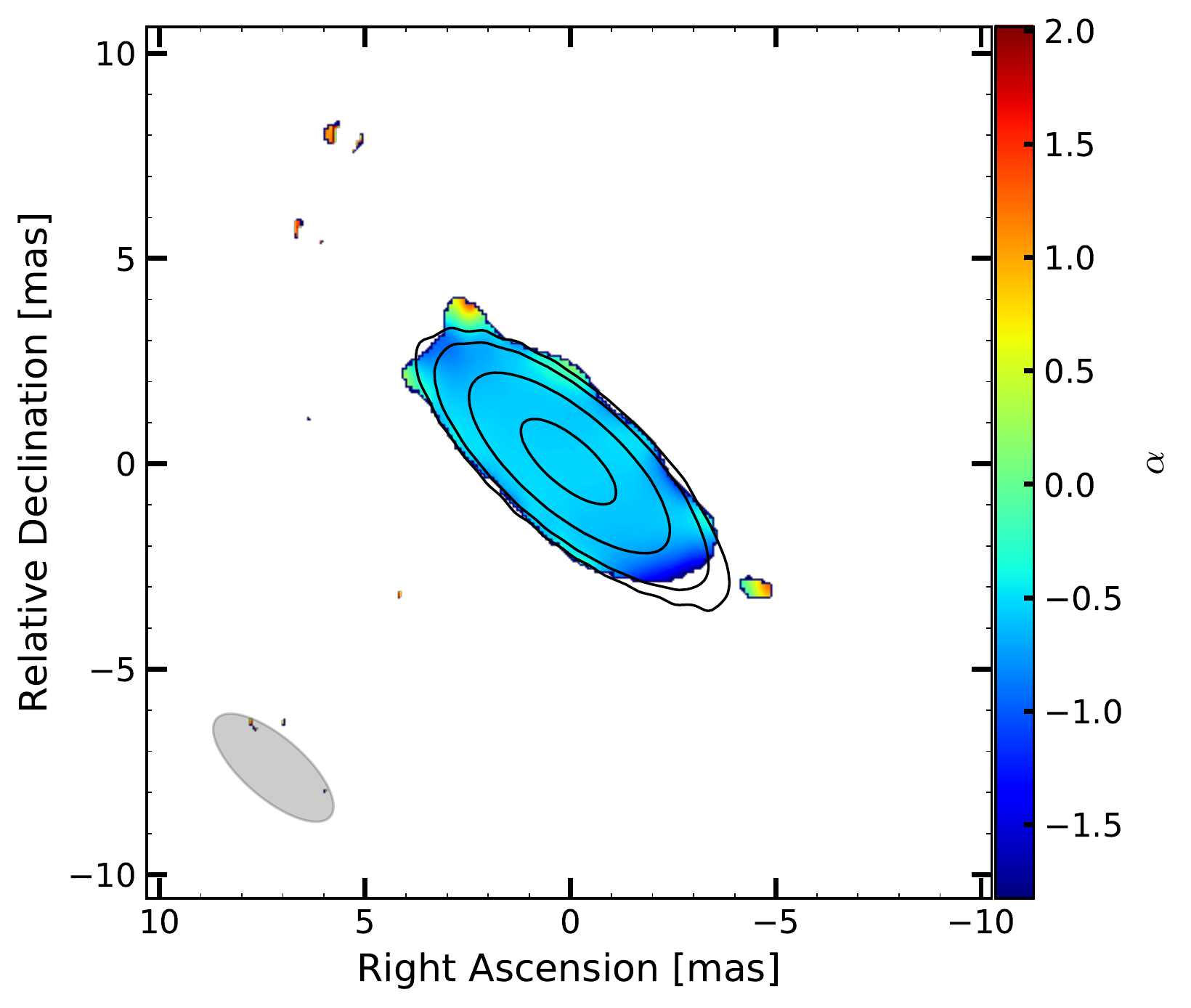}

\end{center}
\caption{Spectral index map of PKS\,1733$-$565 between 8.4 GHz and 22.3 GHz for epoch 2008-02-07. The black contours are from the 8.4 GHz image. Both
maps were convolved with a common beam that is represented in grey in
the lower-left corner.}
\label{1733_spix}
\end{figure}

\paragraph{1549$-$790}
This compact flat-spectrum (CFS) source exhibits a classic
double-sided structure at milliarcsecond resolution (see Fig.~\ref{fig:wallpaper}), suggesting that
its viewing angle is large. \cite{2006MNRAS.370.1633H} presented VLBI maps of PKS\,1549$-$79 at 2.3 and 8.4~GHz, showing a one-sided jet structure, which was interpreted as evidence of a moderately aligned viewing angle. As discussed in \cite{2010A&A...519A..45O}, the difference between this previous map and our image could be due to the higher sensitivity of TANAMI observations. A spectral index map is shown in Fig~\ref{1549_spix}.
\paragraph{1733$-$565}
This FR~II radio galaxy shows a double-sided structure, with jets
extending in the northeast-southwest direction, up to $\sim5$ mas from
the core (see Fig.~\ref{fig:wallpaper}). Our TANAMI data therefore allow us to recover additional structure with respect to previous VLBI images, where the source appeared as a compact core \citep{2004AJ....127.3609O}. The jet orientation follows the one of the large scale radio structure \citep{1982PASAu...4..447H}. The jet structure appears
to be consistent across the epochs. A spectral index image was created for epoch 2008-02-07, and is presented in
Fig.~\ref{1733_spix}. No correction for core-shift effects was
required for this source. The source is practically unresolved in the
spectral map, with an overall spectral index in the range
$-0.8\lesssim \alpha \lesssim-0.3$. We therefore do not detect a
flat-spectrum core in this source.

\paragraph{1814$-$637}
This CSS source has been proposed as a putative young radio source~\citep[][and references therein]{2010A&A...519A..45O}. At full resolution, it shows a compact, marginally resolved component (see Fig.~\ref{fig:wallpaper}). Using tapering, i.e., down-weighting the visibilities from the longest baselines, it is possible to recover possible emission on larger scales. In this case, we detect a symmetric structure in the north-south direction, reminiscent of classic CSOs (see Appendix~\ref{app:maps}). No contemporaneous 22~GHz data are available for this source.

\paragraph{2027$-$308}
This is the most distant radio galaxy in the sample, with a redshift
$z=0.539$ (at this scale, 1 mas corresponds to $\sim6$ pc). Interestingly, the source
shows an asymmetric double-sided structure, with one jet extending to the south-west and a counterjet which seems to bend north (see Fig.~\ref{fig:wallpaper}). No contemporaneous 22~GHz data are available for this source.

\begin{figure}[!htbp]
\begin{center}
\includegraphics[width=0.9\linewidth]{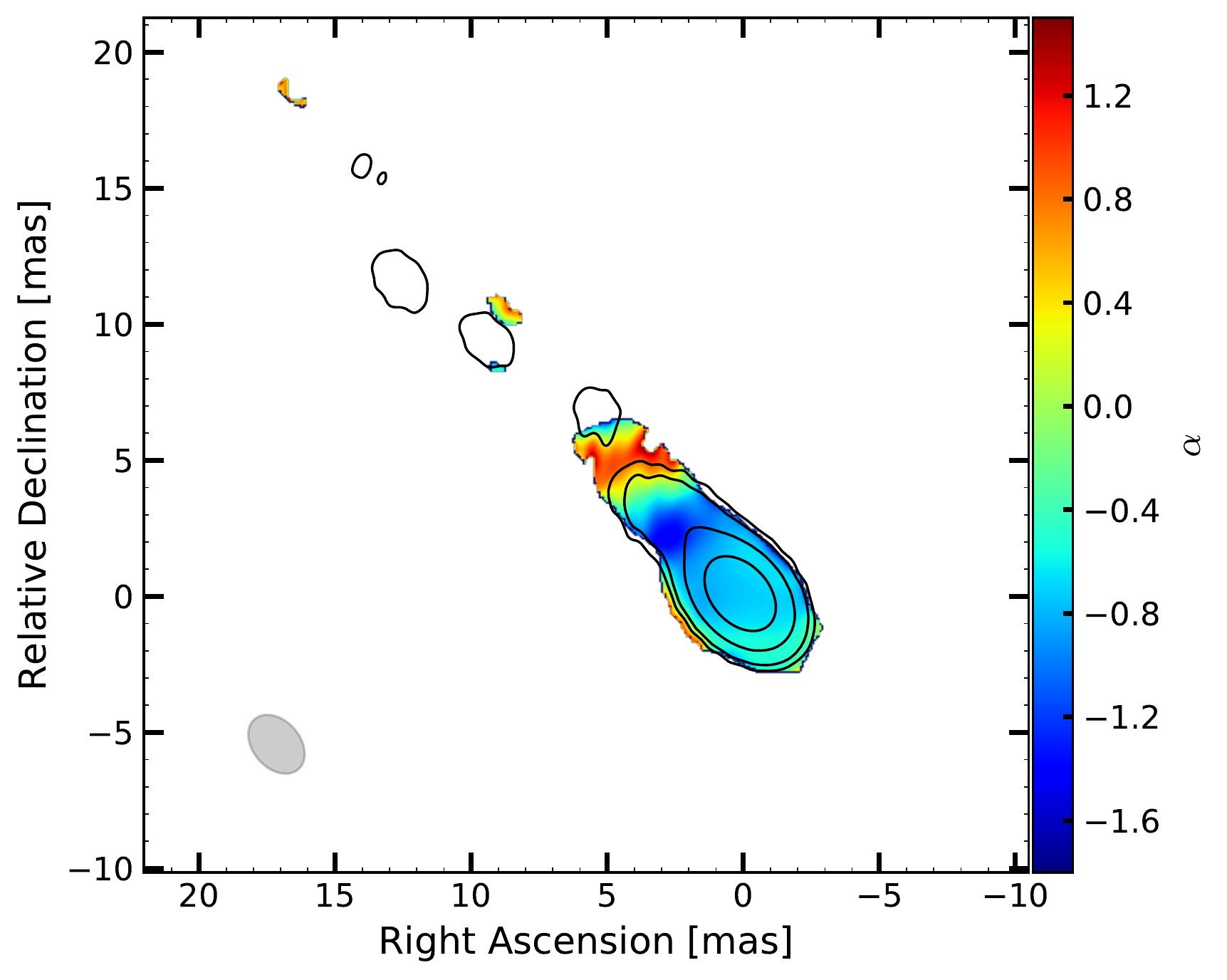}
\end{center}
\caption{Spectral index map of PKS\,2153$-$69 between 8.4 GHz and 22.3 GHz for epoch 2008-11-27. The black contours are from the 8.4 GHz image. Both
maps were convolved with a common beam that is represented in grey in
the lower-left corner.}
\label{2152_spix}
\end{figure}

\paragraph{2152$-$699}
This FR~II source shows a relatively elongated jet, extending $\sim50$ mas
from the core to the north-east. The extended jet is best seen in the first TANAMI epoch (see Fig.~\ref{fig:wallpaper}), due to better $(u,v)$ coverage on short spacings. Previous VLBI observations by \cite{1996AJ....111..718T} showed a one-sided jet extending for $\sim20$ mas to the north-east. Space-VLBI images with VSOP show a narrow, smooth jet extending in the same direction for $\sim6$ mas \citep{2002ApJS..141..311T}. The jet orientation is consistent with the large scale radio structure of the source \citep{1990ESASP.310..513F,1990PASAu...8..252N}. A spectral index map is
shown in Fig.~\ref{2152_spix}, and shows a relatively steep core with a
spectral index around $-0.5<\alpha<-0.7$.
\subsection{Radio kinematic analysis results}
\label{sec:res_kin}
\begin{figure*}[!htbp]
\begin{center}
\includegraphics[width=0.49\linewidth]{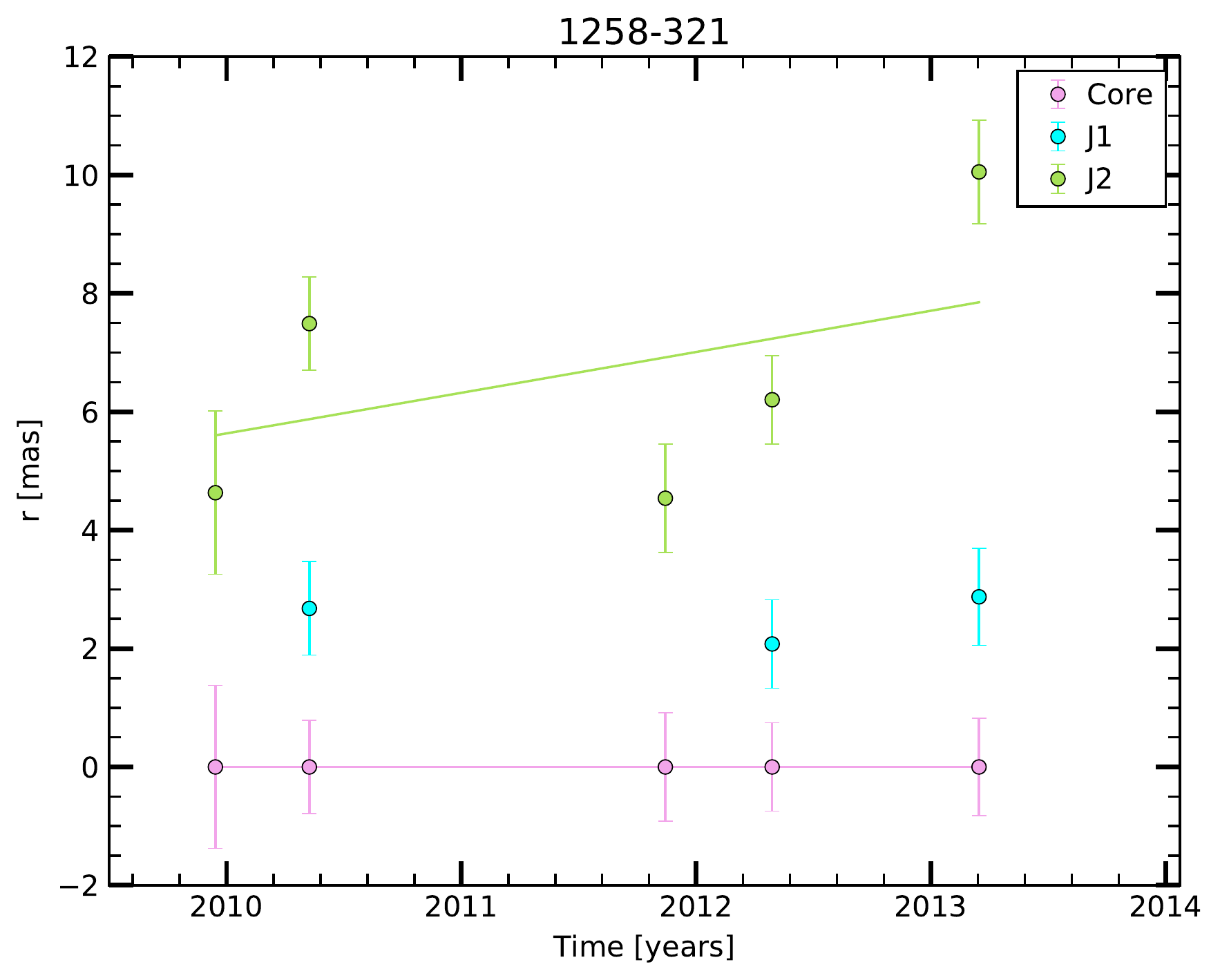}
\includegraphics[width=0.49\linewidth]{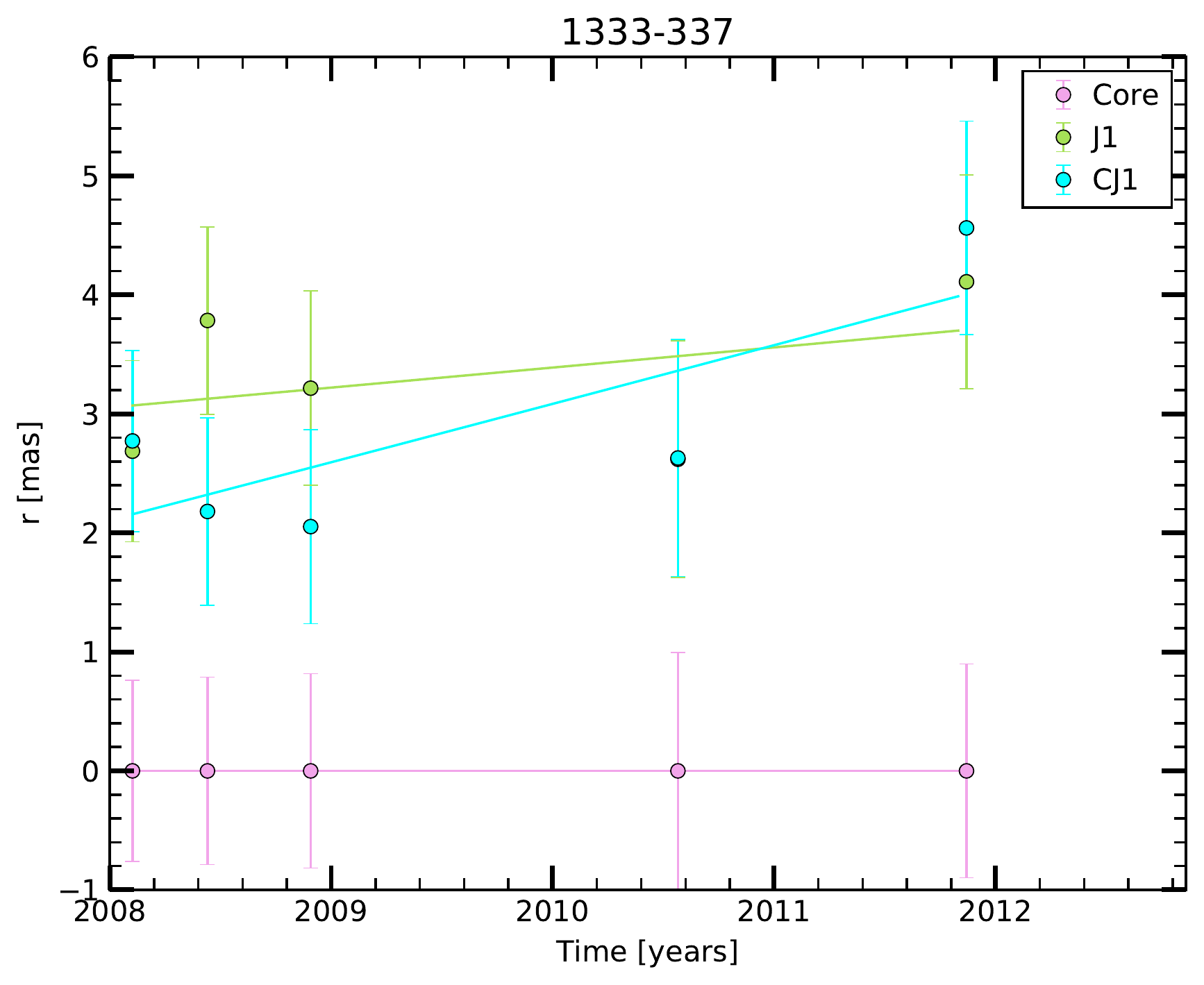}
\includegraphics[width=0.49\linewidth]{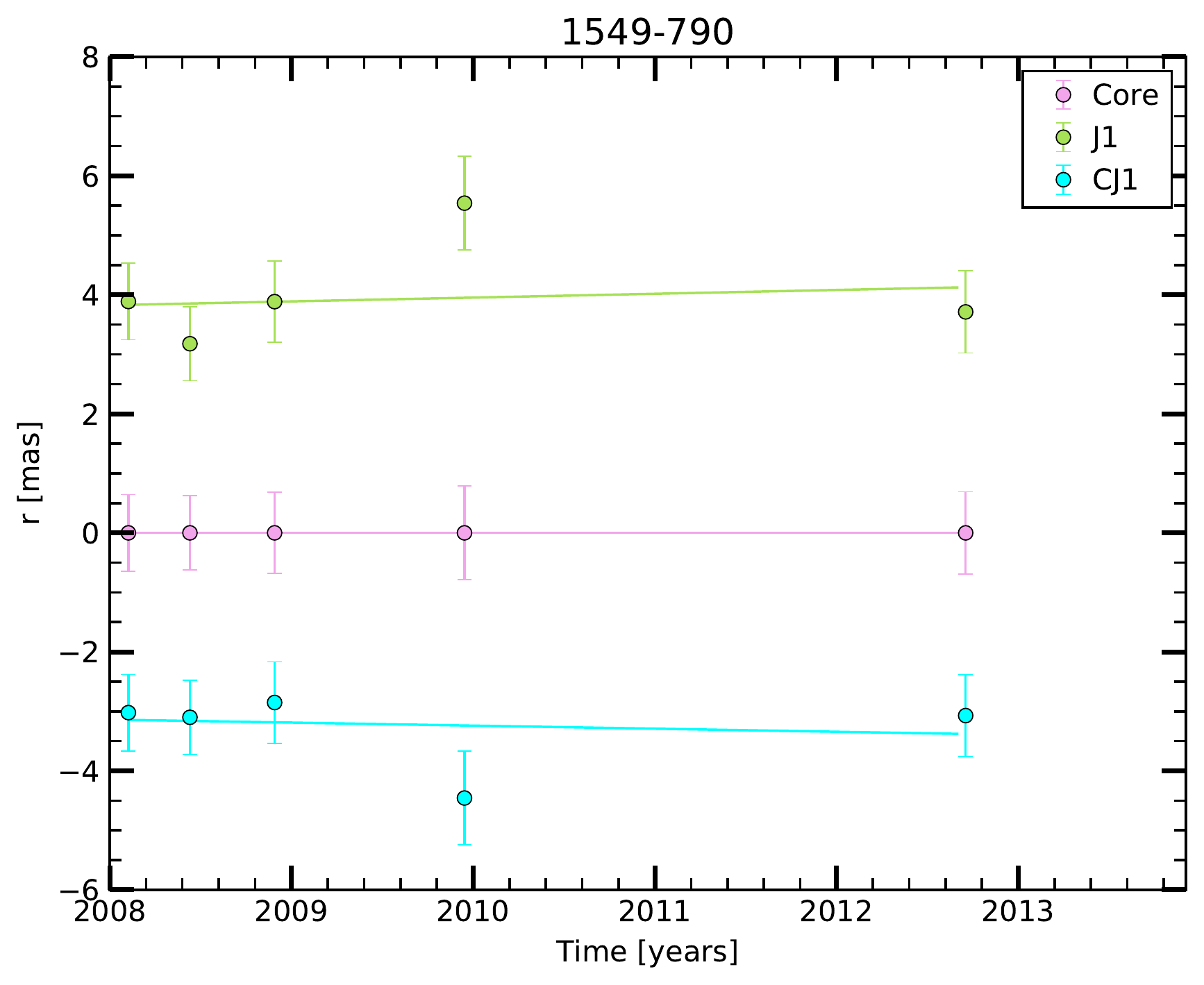}
\includegraphics[width=0.49\linewidth]{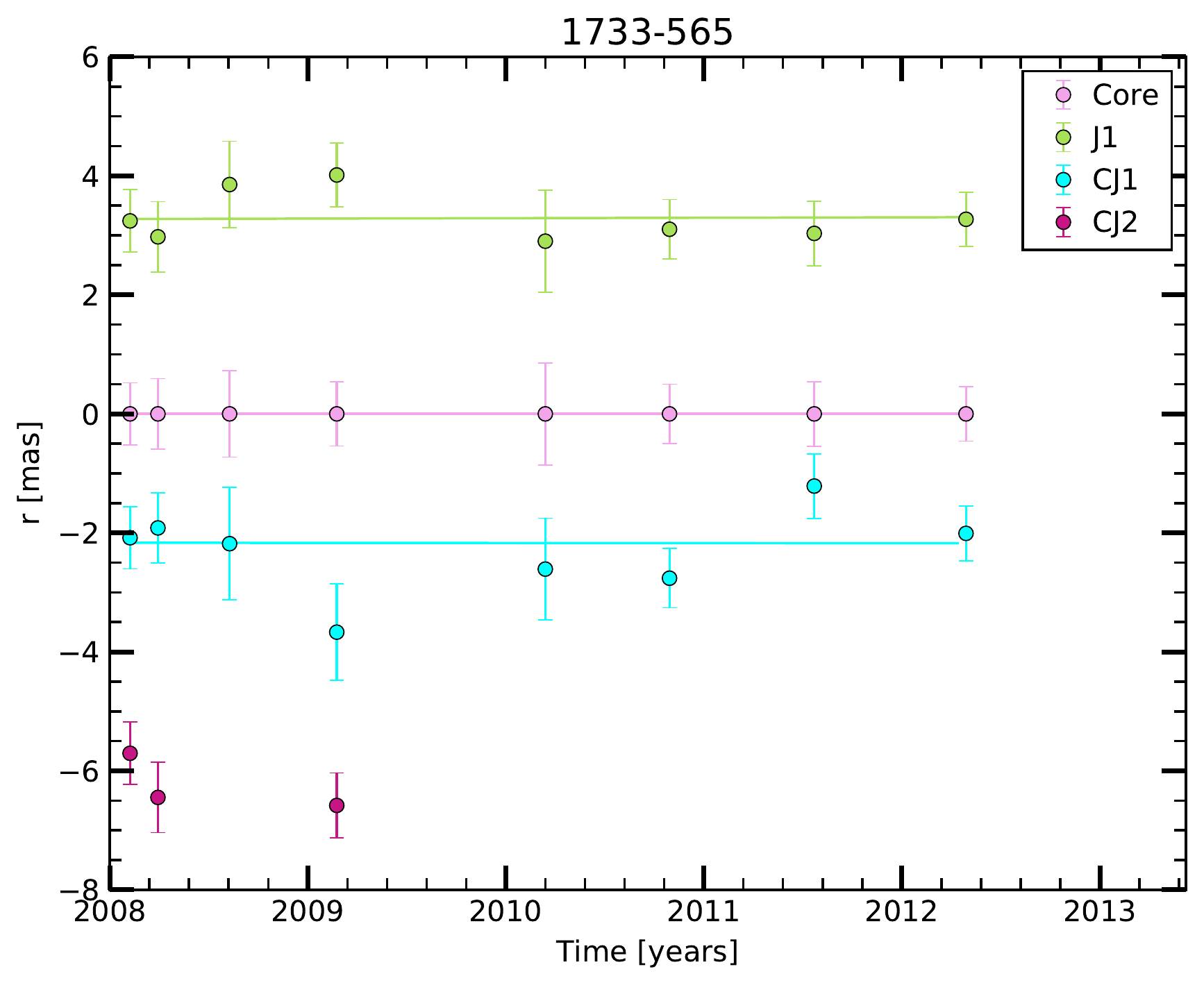}
\includegraphics[width=0.49\linewidth]{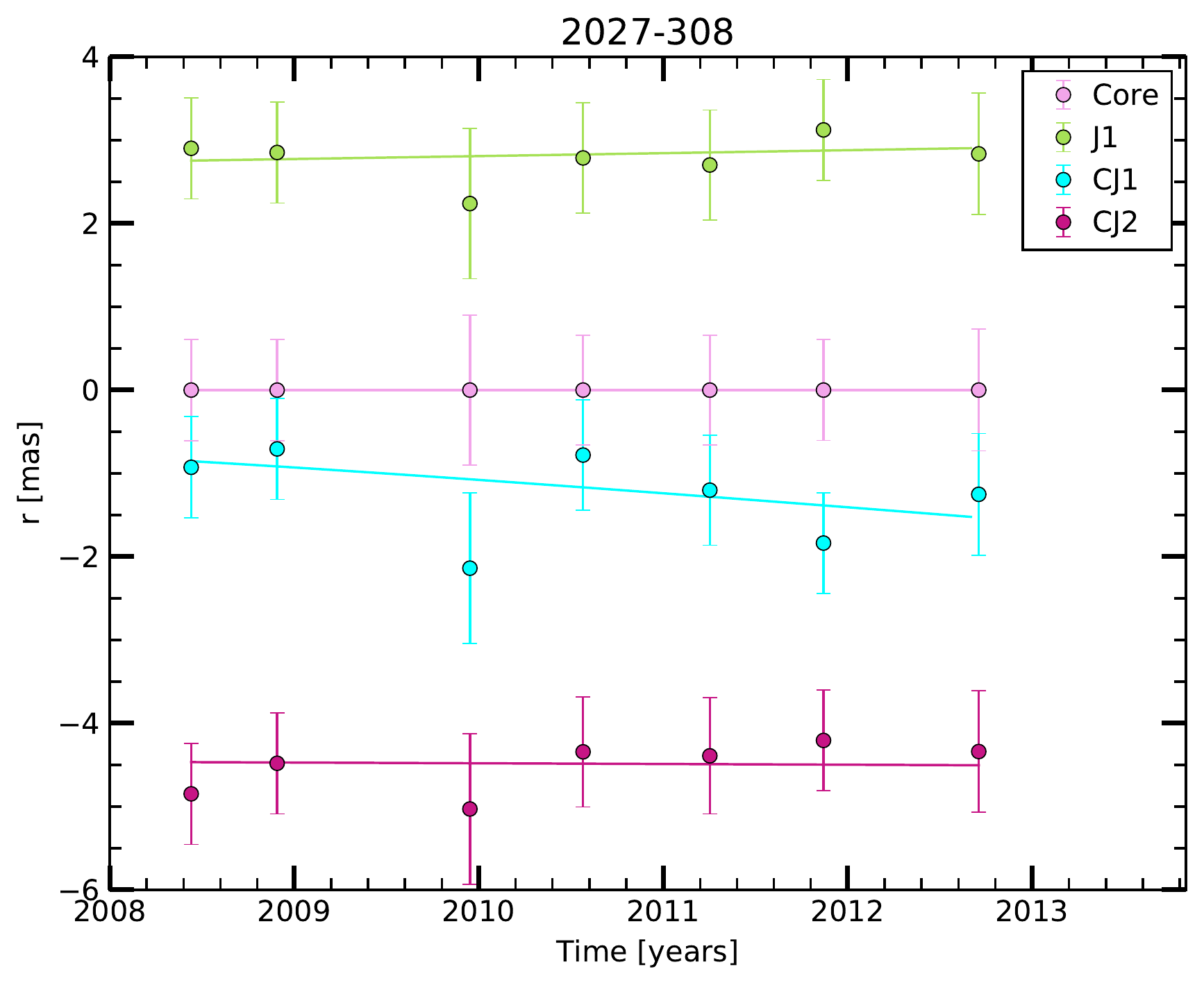}
\includegraphics[width=0.49\linewidth]{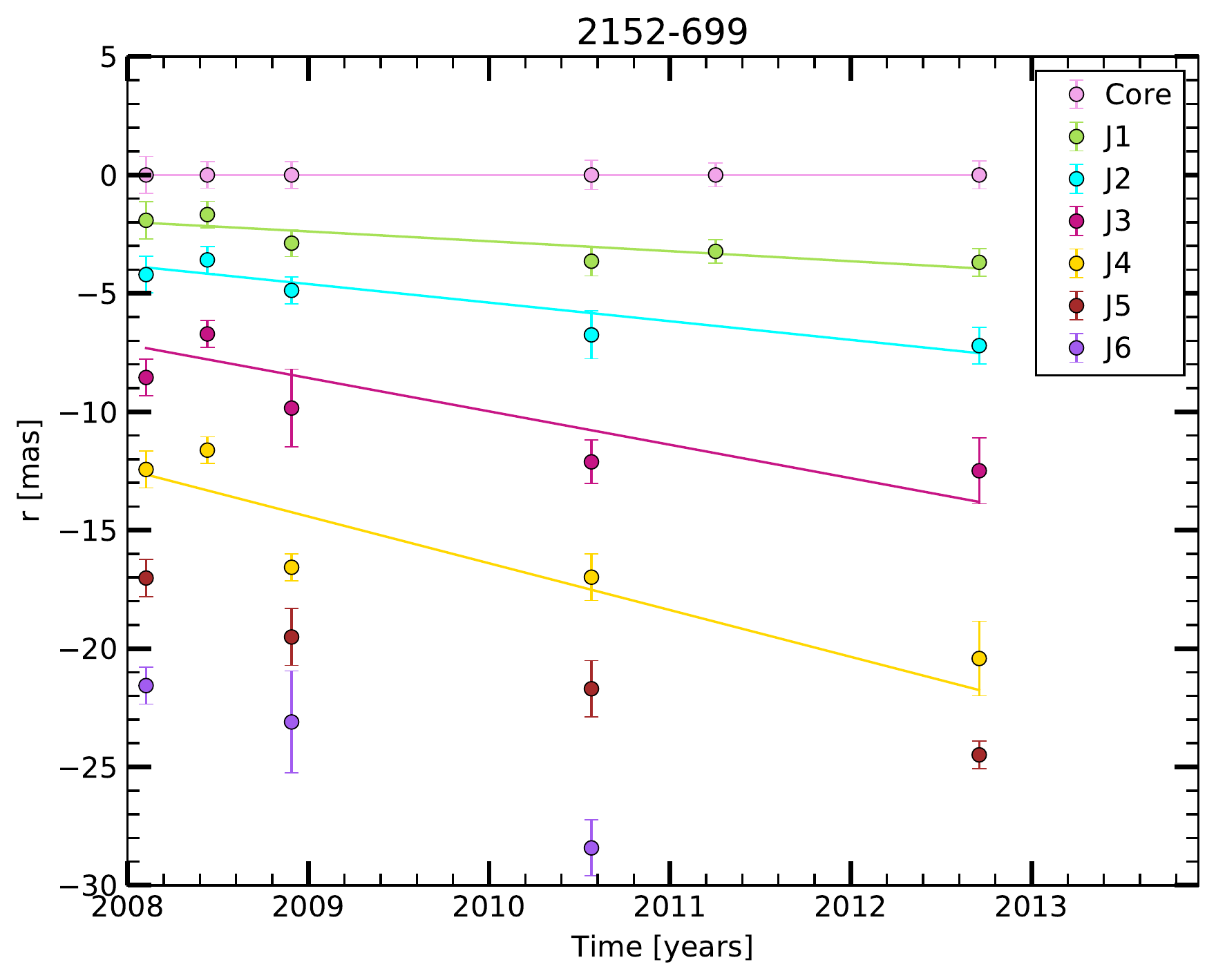}
\end{center}
\caption{Jet kinematics of our radio galaxies: core distance of jet features as a function of time. The solid lines represent a least squares fit to their positions (the slope is the apparent speed).}
\label{kin_dist}
\end{figure*}

Figure~\ref{kin_dist} shows the core separation versus time for each Gaussian component identified and tracked in the jet of each source. We list the corresponding values of the angular speed, apparent speed and estimated ejection date in Table~\ref{kin_tab_all}, where the columns list the source name, component identification, apparent angular and linear speeds, ejection date (when applicable) and number of epochs in which a component is detected. As in Paper I, only components with at least five epochs are listed, and ejection dates are given only if the measured speed is not consistent with zero within the $1\sigma$ errors. Finally, we illustrate the parameter space of intrinsic jet parameters $\beta_\mathrm{app}$ (apparent linear speed) and $\theta$ (viewing angle) allowed by our results in Fig.~\ref{beta_theta}. Additional plots illustrating the kinematic analysis results are provided in Appendix~\ref{app:kin}, namely the multi-epoch component identification and tracking, their flux density evolution with time, and tables reporting the \texttt{Modelfit} component\footnote{\texttt{Modelfit} is a Difmap task which allows to model the clean map with Gaussian components.} parameters.

\paragraph{1258$-$321}
In this source, only one jet component (J1) can be tracked for at least five epochs. The corresponding apparent speed is subluminal.
Figure~\ref{beta_theta} shows the limits on
the intrinsic jet speed and viewing angle for PKS\,1258$-$321 resulting
from our observations. We do not detect a counterjet in this case, therefore we set a lower limit for the jet-to-counterjet flux ratio ($R$).

\paragraph{1333$-$337}
We find that the two symmetric jet components (J1 and CJ1) show almost no motion during our monitoring period. Although we do find a non-zero separation speed for component CJ1, the uncertainty on this value is higher than 50\%. In Fig.~\ref{beta_theta}, we adopt the only non-zero apparent speed value as the central estimate, and use its uncertainty to define the minimum and maximum estimates.

\paragraph{1549$-$790}
We find that the symmetric structure in this source is remarkably stable, with no measured jet motions, and an upper limit on the apparent speed of $\beta_\mathrm{app}<1.7$. We use this value to define the allowed intrinsic jet parameter space in Fig.~\ref{beta_theta}

\paragraph{1733$-$565}
We do not detect any statistically significant motion in this double-sided jet. We use the maximum apparent speed value allowed by our measurements, i.e., $\beta_\mathrm{app}<0.24$, to constrain the intrinsic jet parameter space (see Fig.~\ref{beta_theta}).

\paragraph{1814$-$637}
The source is only marginally resolved in most of the epochs, therefore it is not possible to track any jet component.

\paragraph{2027$-$308}
In this case, we do not measure any significant jet motion. The uncertainty of the component positions is very large, due to their faint flux density. Therefore, we do not attempt to estimate the jet viewing angle and intrinsic speed for PKS\,2027$-$308, as such estimates would be extremely uncertain.

\begin{figure}[!!htbp]
\begin{center}
\includegraphics[width=0.82\linewidth]{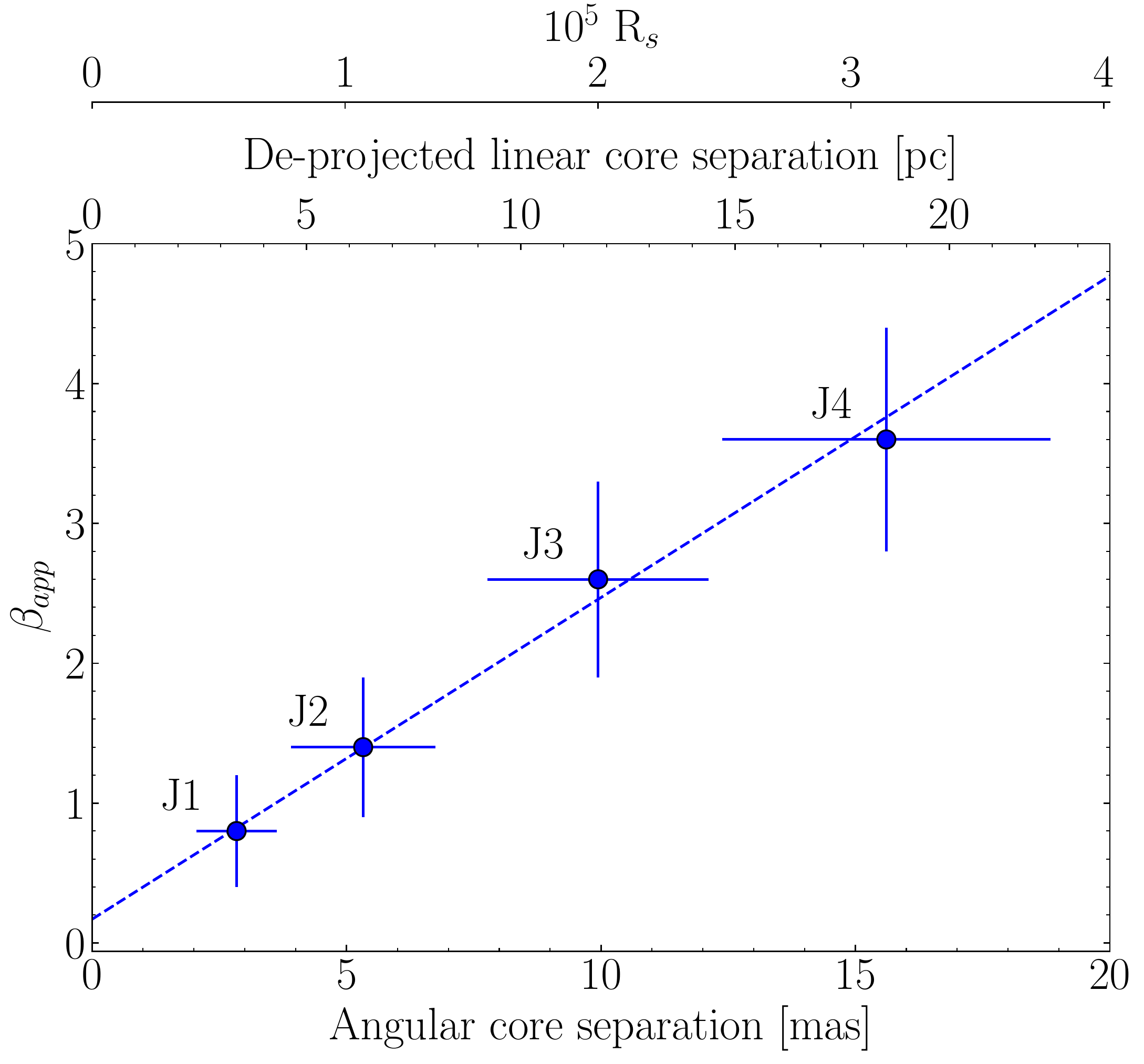}
\end{center}
\caption{Apparent velocity of components in the jet of PKS\,2153$-$69
  as a function of average core distance. A clear linear increasing trend is
  seen, indicating downstream acceleration. The de-projected linear
  core distance has been calculated assuming the maximum possible
  viewing angle of $\sim27^{\circ}$ (see the lower panel of Fig.~\ref{beta_theta}). The topmost x-axis is in units of Schwarzschild radii ($R_s$).}
\label{2152_accel}
\end{figure}

\paragraph{2152$-$699}
This source shows the fastest apparent motions in this sample, up to
$\beta_{app} = 3.6\pm0.8$ for component J4. Interestingly, there is a clear trend of
increasing apparent component speed with increasing core distance.
This can be seen in Fig.~\ref{2152_accel}, where this effect has been
quantified by means of a simple linear fit, which yields $\beta_\mathrm{app} = 0.19d+0.34$ (where $d$ is the angular core separation). This behavior is similar to the one revealed by TANAMI data for the pc-scale
jet of Centaurus~A~\citep{2014A&A...569A.115M}, although both the apparent speed and the distance from the core have much larger values in PKS\,2153$-$69.
The high apparent speed and the absence of a detected counter-jet
allow us to place a stringent constraint of the viewing angle, which
has to be $\theta<27^{\circ}$ (see lower panel of Fig.~\ref{beta_theta}). This value has been used to convert the angular core separation into de-projected linear separation. 

In order to put into context these linear scales within our current understanding of jet collimation and acceleration, we have converted the linear separation into units of Schwarzschild radius $R_s = 2GM_{BH}/c^2$. We did not find any previous measurement of the black hole mass for this source in the literature, however a measurement of the central stellar velocity dispersion is provided by the \href{http://leda.univ-lyon1.fr/}{HyperLeda} database \citep{2014A&A...570A..13M}, with a value $\sigma=(240\pm20)$ km/s. From this, the black hole mass can be estimated using the $M-\sigma$ relation from \citet{2013ApJ...764..184M}, which yields log(M$_{BH}/M_{\odot}$)=8.79. This in turn leads to a Schwarzschild radius $R_s=5.9\times10^{-5}$ pc. The linear scales probed by our kinematic analysis are therefore of the order $\sim10^5\,R_s$. According to recent studies on parsec-scale radio galaxy jets (see, e.g., \citealt{2017A&ARv..25....4B} and references therein), these scales typically correspond to the end of the acceleration and collimation region and the transition between the parabolic and conical jet expansion regime.

\begin{table}[h!tbp]
\caption{Results of the kinematic analysis of LAT-undetected TANAMI radio galaxies. Columns: source name, component identification, apparent angular and linear speeds, ejection date (when applicable) and number of epochs in which a component is detected.}             
\label{kin_tab_all}  
\begin{center}    
\small
\tabcolsep=1.5mm
\begin{tabular}{lccccc}  
\hline\hline 
Source & ID & $\mu$ (mas/yr) & $\beta_{app}$ & Ej. date & \# ep.\\
\hline
1258$-$321 & J1 & 0.7$\pm$0.5 & 0.8$\pm$0.5 & 2003$\pm$3 & 5\\
\hline  
1333$-$337 & J1 & 0.16$\pm$0.35 & 0.14$\pm$0.30 & * & 5\\
 & CJ1 & 0.51$\pm$0.36 & 0.44$\pm$0.32 & 2004.4$\pm$7.9 & 5 \\
\hline  
1549$-$790 & J1 & 0.07$\pm$0.09 & 0.7$\pm$0.8 & * & 5\\
 & CJ1 & 0.06$\pm$0.11 & 0.6$\pm$1.1 & * & 5 \\
\hline  
1733$-$565 & J1 & 0.008$\pm$0.030 & 0.05$\pm$0.19 & * & 8\\
 & CJ1 & 0.002$\pm$0.003 & 0.01$\pm$0.02 & * & 8 \\
\hline  
2027$-$308 & J1 & 0.06$\pm$0.09 & 1.7$\pm$2.7 & * & 6\\
 & CJ1 & 0.2$\pm$0.3 & 5.6$\pm$8.6 & * & 6\\
 & CJ2 & 0.01$\pm$0.25 & 0.3$\pm$7.9 & * & 6 \\
\hline 
2152$-$699 & J1 & 0.4$\pm$0.2 & 0.8$\pm$0.4 & 2005$\pm$2 & 6\\
 & J2 & 0.8$\pm$0.3 & 1.4$\pm$0.5 & 2004$\pm$2 & 5\\
 & J3 & 1.4$\pm$0.4 & 2.6$\pm$0.7 & 2003$\pm$1 & 5\\
 & J4 & 2.0$\pm$0.4 & 3.6$\pm$0.8 & 2002$\pm$1 & 5\\
\hline\hline
\end{tabular}
\end{center}
\end{table}

\clearpage

\begin{figure*}[!htbp]
\begin{center}
\includegraphics[width=0.49\linewidth]{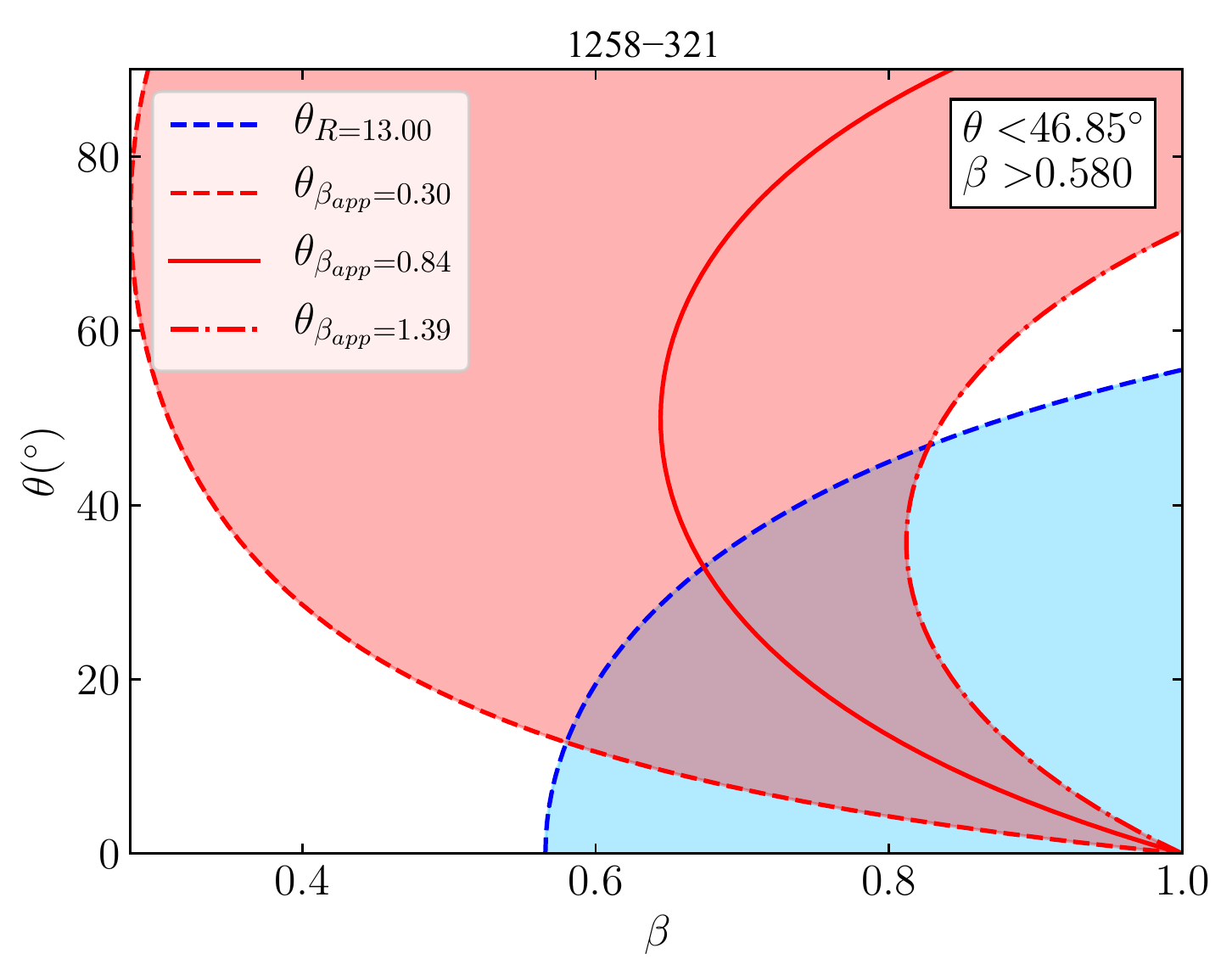}
\includegraphics[width=0.49\linewidth]{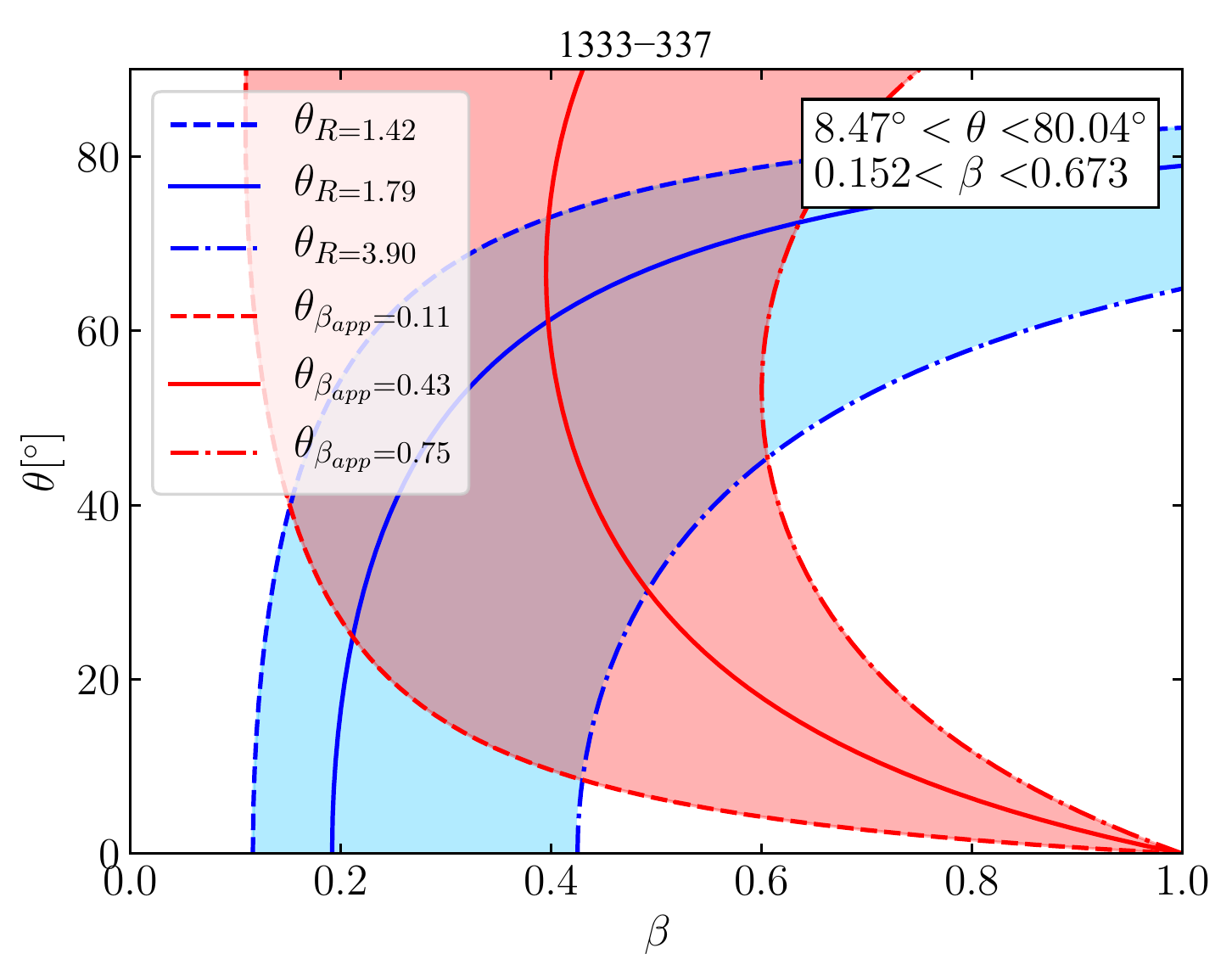}
\includegraphics[width=0.49\linewidth]{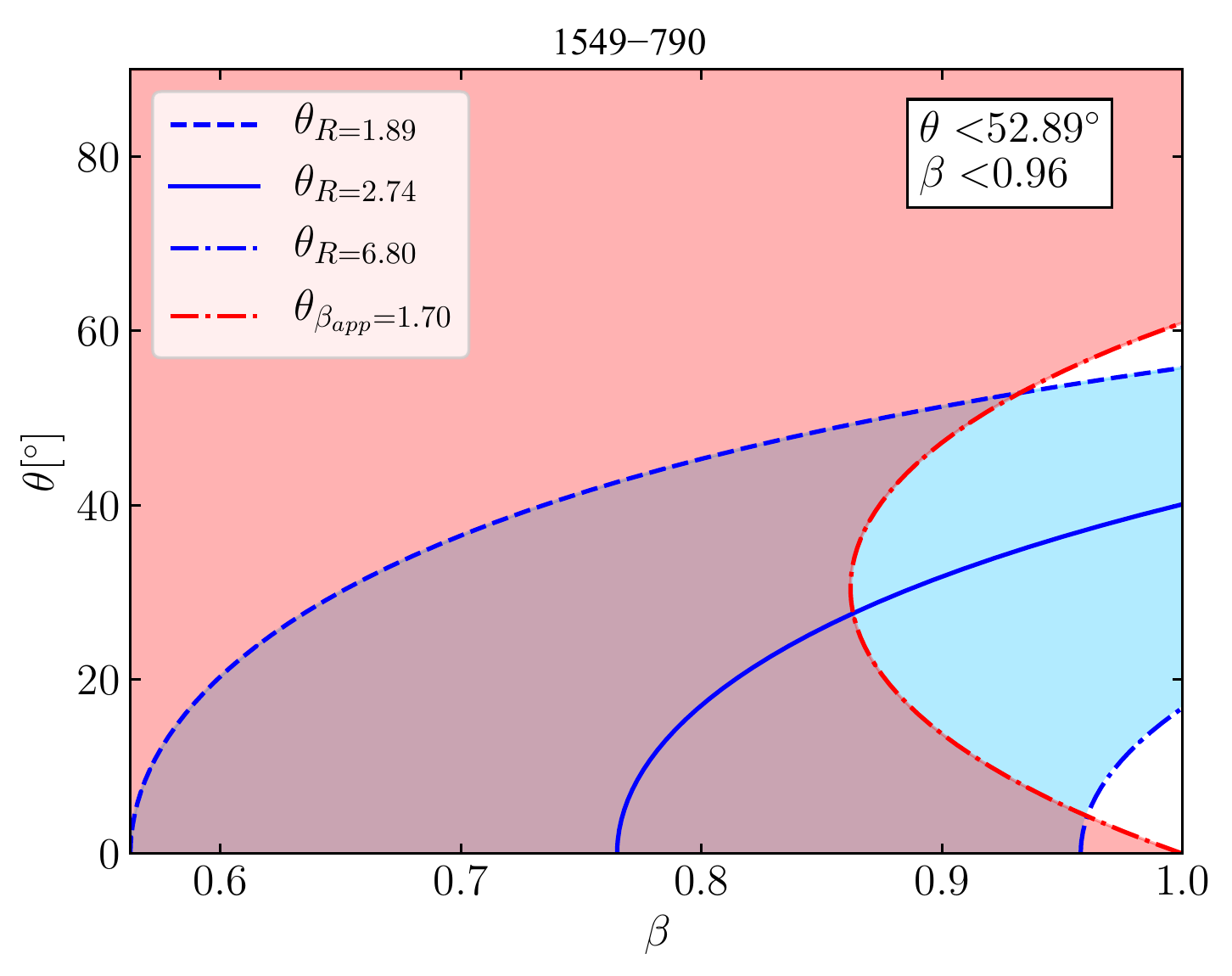}
\includegraphics[width=0.49\linewidth]{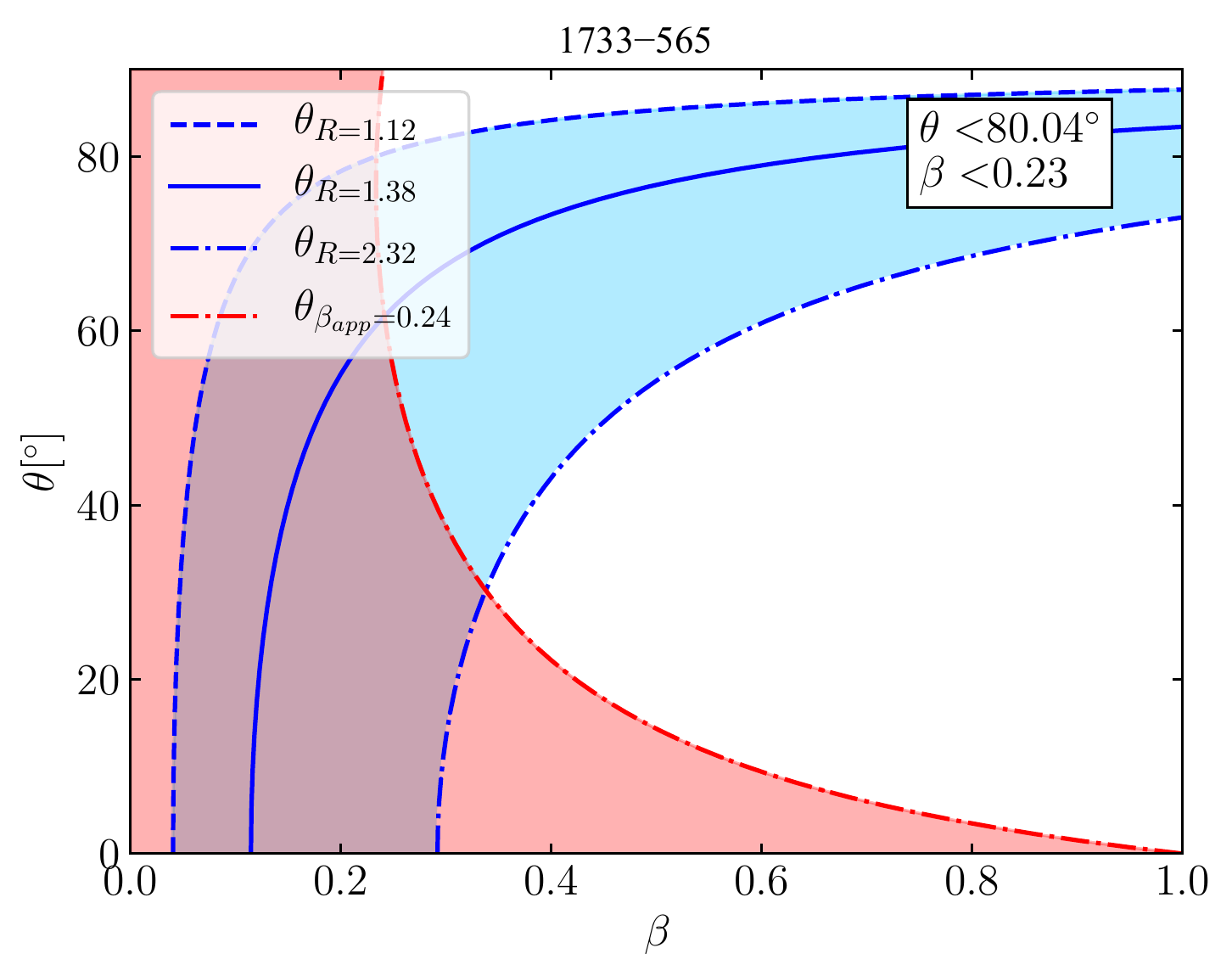}
\includegraphics[width=0.49\linewidth]{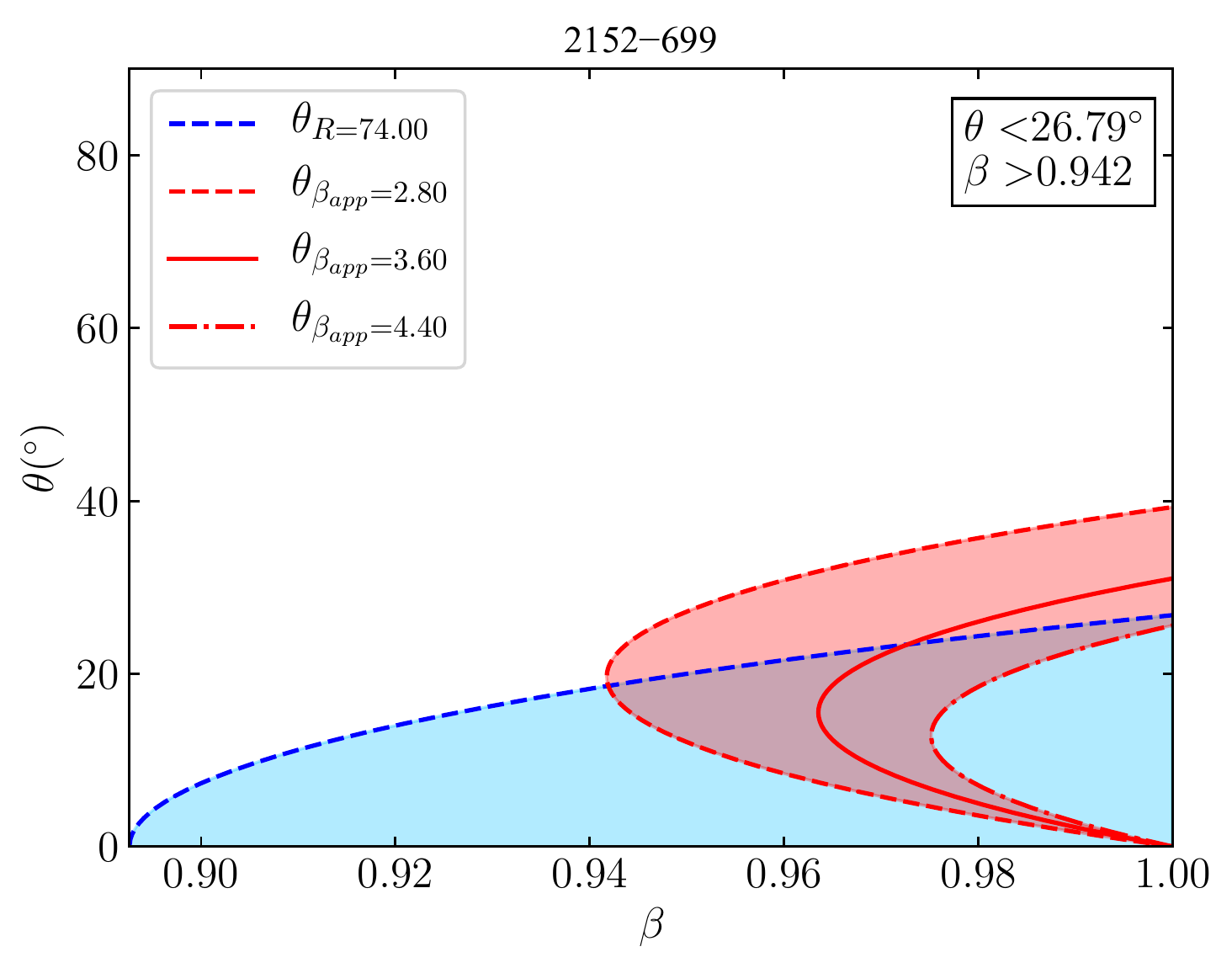}
\end{center}
\caption{Parameter space of intrinsic jet speed $\beta$ and viewing
  angle $\theta$ allowed by our observations. The blue shaded area is
  the one allowed by the measurement of $R$, while the red shaded area
  is the one allowed by the observed $\beta_\mathrm{app}$. When possible, we provide a minimum, maximum and a central estimate of $R$ and $\beta_\mathrm{app}$. The top-right legend reports the resulting limits on $\theta$ and $\beta$.}
\label{beta_theta}
\end{figure*}

\clearpage
\subsection{\textit{Fermi}-LAT results}
\label{sec:res_lat}
In this section we present the results of our analysis of $\gamma$-ray data for so far undetected TANAMI radio galaxies, which are summarized in Table~\ref{rgs-lat-ul}. Here we list the source name, class, photon flux and energy flux upper limits, and TS. We discuss individually the cases of PKS\,1258$-$321 and PKS\,2153$-$69 as they lie close to unidentified $\gamma$-ray sources from the 3FGL catalog.


\begin{table}[h!tbp]
\caption{\fermi upper limits on TANAMI radio galaxies.}
\begin{center}
\tabcolsep=0.1cm
\begin{tabular}{llccc}
\hline
\hline
Source & Class & Flux UL$^a$ & Energy Flux UL$^b$& TS\\
\hline
1258$-$321 & FR~I & 4.68$\times10^{-9}$ & 2.22$\times10^{-12}$ & 4.58\\
1333$-$337 & FR~I & 1.15$\times10^{-8}$ & 2.46$\times10^{-12}$ & 14.18\\
1549$-$790 & RG/CFS & 5.07$\times10^{-9}$ & 1.93$\times10^{-12}$ & 2.98\\
1733$-$565 & FR~II & 1.37$\times10^{-8}$ & 2.91$\times10^{-12}$ & 14.16\\
1814$-$637 & CSS/CSO & 1.89$\times10^{-9}$ & 1.07$\times10^{-12}$ & 0.90\\
2027$-$308 & RG & 6.51$\times10^{-10}$ & 1.20$\times10^{-12}$ & 2.62\\
2152$-$699 & FR~II & 3.44$\times10^{-9}$ & 1.68$\times10^{-12}$ & 7.72\\
\hline
\hline
\end{tabular}
\end{center}
$^a$ 95\% confidence flux upper limit in units of \phcms.\\
$^b$ 95\% confidence energy flux upper limits in units of \ergscm\\

\label{rgs-lat-ul}
\end{table}

\begin{figure}[!htbp]
\begin{center}
\includegraphics[width=\linewidth]{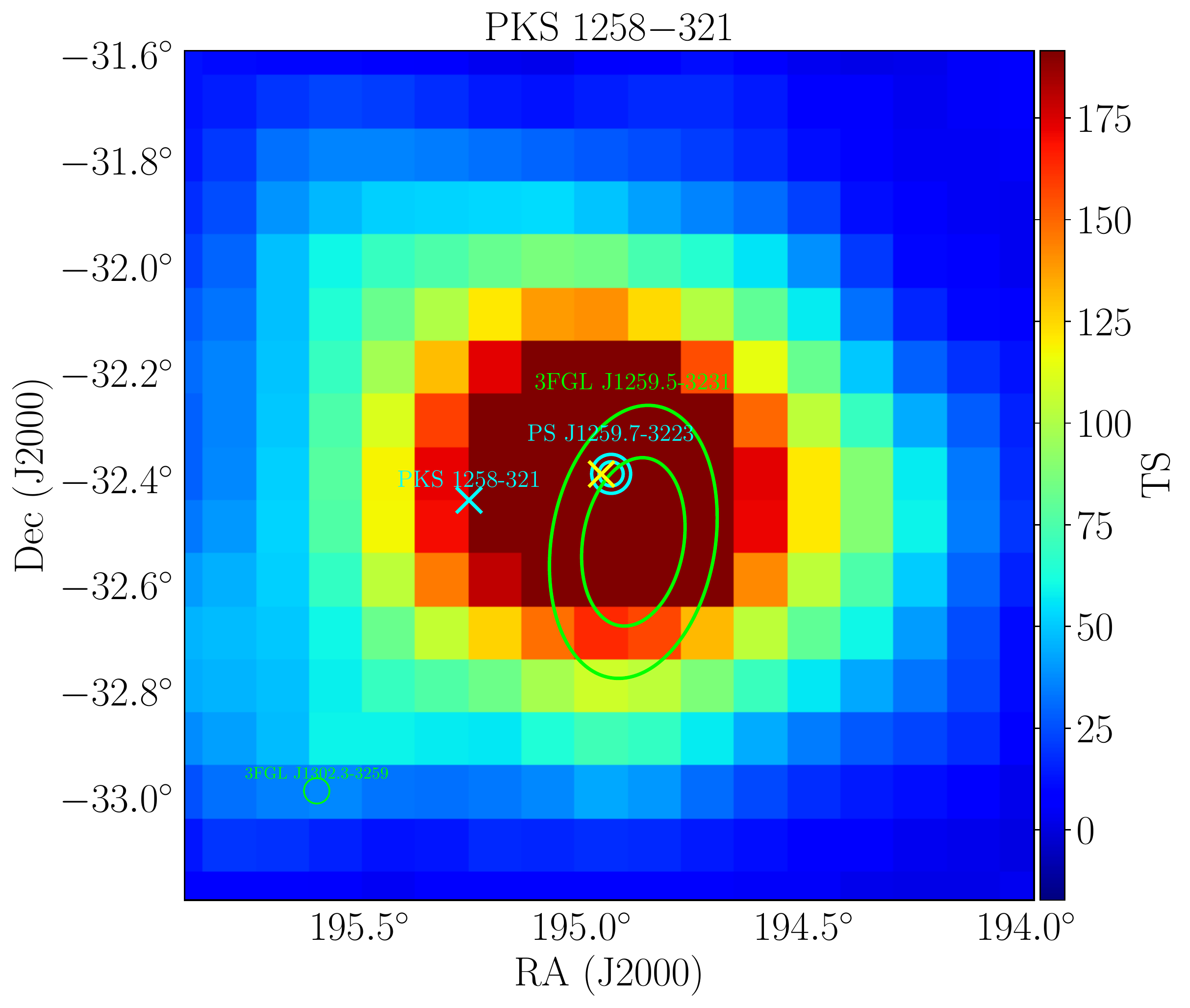}
\includegraphics[width=\linewidth]{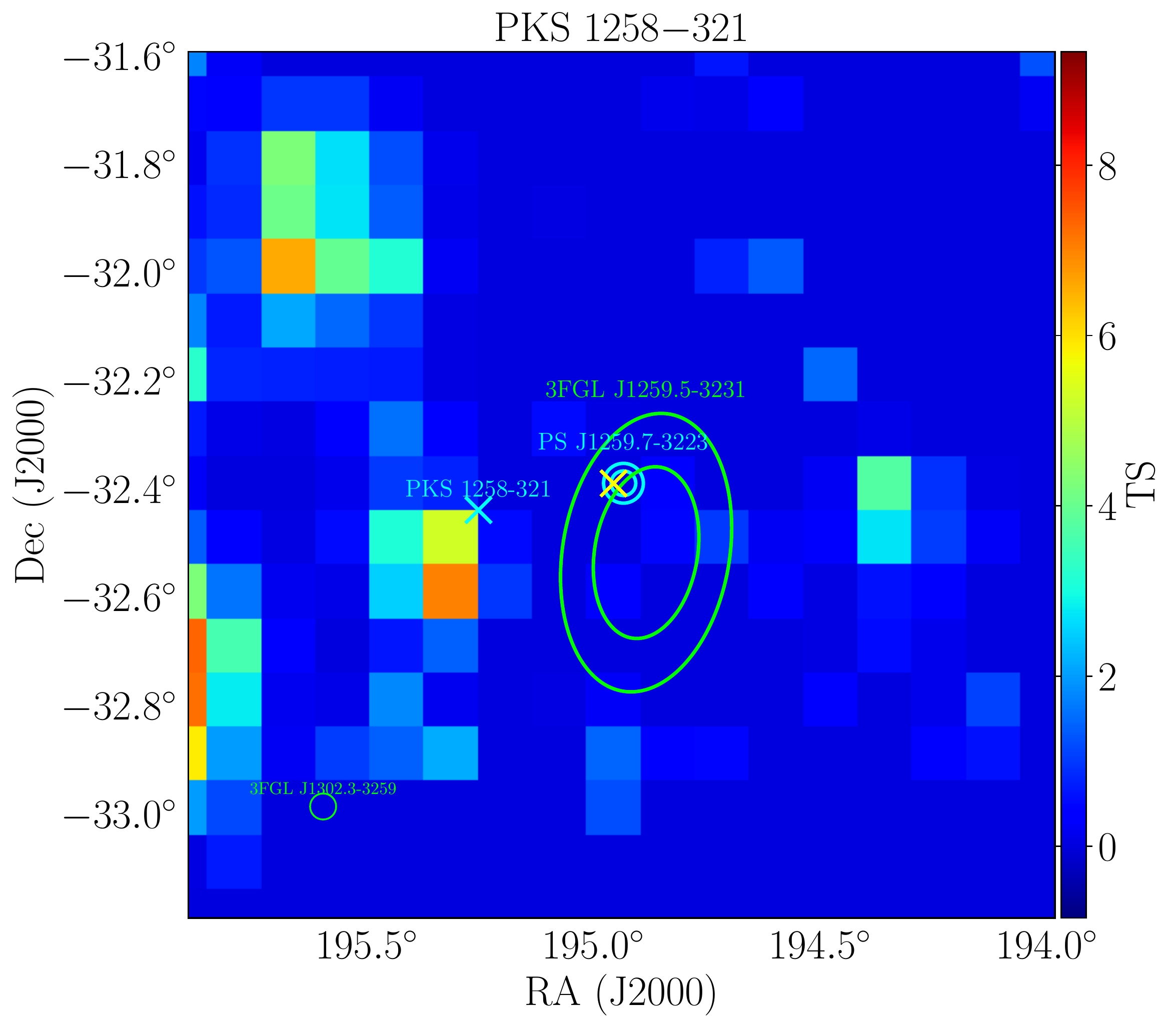}
\end{center}
\caption{\textit{Top panel}: Map of excess TS in the inner region of the ROI centered on
  PKS\,1258$-$321, after removing the unidentified catalog source
  3FGL~J1259.5$-$3231 from the model. \textit{Bottom panel}:
  residual excess TS map after modeling and localizing the new source
  PS~J1259.7$-$3223. \textit{Both panels}:
  The cyan cross represents the radio position of PKS\,1258$-$321. The
  yellow cross represents the position of NVSS~J125949$-$322329. The yellow
  ellipses represent the 68\% and 95\% positional uncertainties from
  the catalog for 3FGL~J1259.5$-$3231. The cyan circles represent the 68\%
  and 95\% positional uncertainties for the new source
  PS~J1259.7-3223. The map radius is 0.8\deg. Each pixel corresponds to 0.1\deg.}
\label{1258_ts}
\end{figure}

   \paragraph{1258$-$321}
This FR~I radio galaxy is not listed as a $\gamma$-ray source in any
\fermi catalog, but it lies close to the unidentified source 3FGL~J1259.5-3231.
To test a possible association, we analyzed 103 months of LAT data
in a 10\deg  ROI around the radio position of PKS\,1258$-$321. We
removed 3FGL~J1259.5$-$3231 from the model, and produced a map of the
excess significance (TS) in the region (see Fig.~\ref{1258_ts}, top panel). We iteratively added new sources to the model
starting from TS$>$25 peaks in the TS map, until no significant excess remained. This procedure resulted in the detection of a new source, dubbed
PS~J1259.8$-$3224, which after localization appears to be consistent
with the catalog position of 3FGL~J1259.5$-$3231, and does not include
the radio position of PKS\,1258$-$321. 

A possible association for the $\gamma$-ray source is the radio source
NVSS~J125949$-$322329 (RA: 194.957542 deg, Dec: $-$32.391361 deg, $z=0.013750$, \citealt{2005ASPC..329...11J}), which lies at
an angular distance of $\sim$54" from the $\gamma$-ray source's best
fit position. This source has been indicated as a
candidate $\gamma$-ray emitter based on its \textit{WISE} colors by
\citet{2014ApJS..215...14D}. Indeed, a $\gamma$-ray source in the Fourth LAT Source Catalog, 4FGL~J1259.7$-$3223, is listed as being associated to this radio source~\citep[4FGL][]{4fgl}.

There is no significant residual excess after modeling this source
(see Fig~\ref{1258_ts}, bottom panel). By placing a test source at the radio position of the PKS\,1258$-$321, we
derive an upper limit, which is listed in Table~\ref{rgs-lat-ul}.

In order to check the consistency of our analysis with the 4FGL catalog, we have repeated the data analysis pipeline using the 4FGL as the starting model, the updated IRF and diffuse models (\texttt{P8R3\_SOURCE\_V2, gll\_iem\_v07.fits, iso\_P8R3\_SOURCE\_V2\_v1.txt}, respectively). The goal was to check that our analysis procedure is able to recover consistent results with respect to the latest catalog, therefore we did not remove any point source from the model in this case. This analysis yields similar results to the 3FGL-based one, i.e. PKS~1258$-$321 is not detected (TS=2.36, $F_{UL}=2.71\times10^{-9}$\phcms).

\begin{figure}[!!htbp]
\begin{center}
\includegraphics[width=\linewidth]{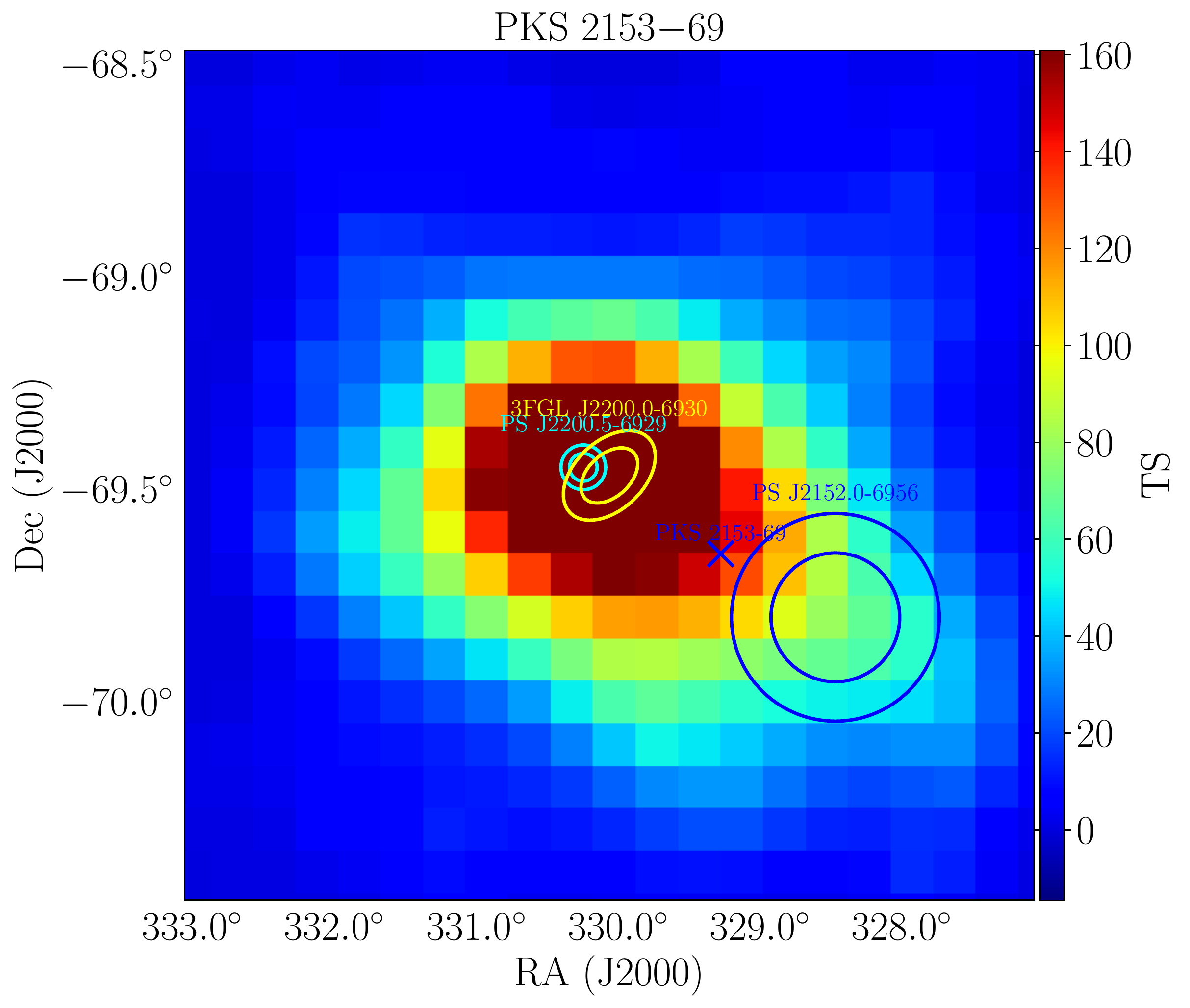}
\includegraphics[width=\linewidth]{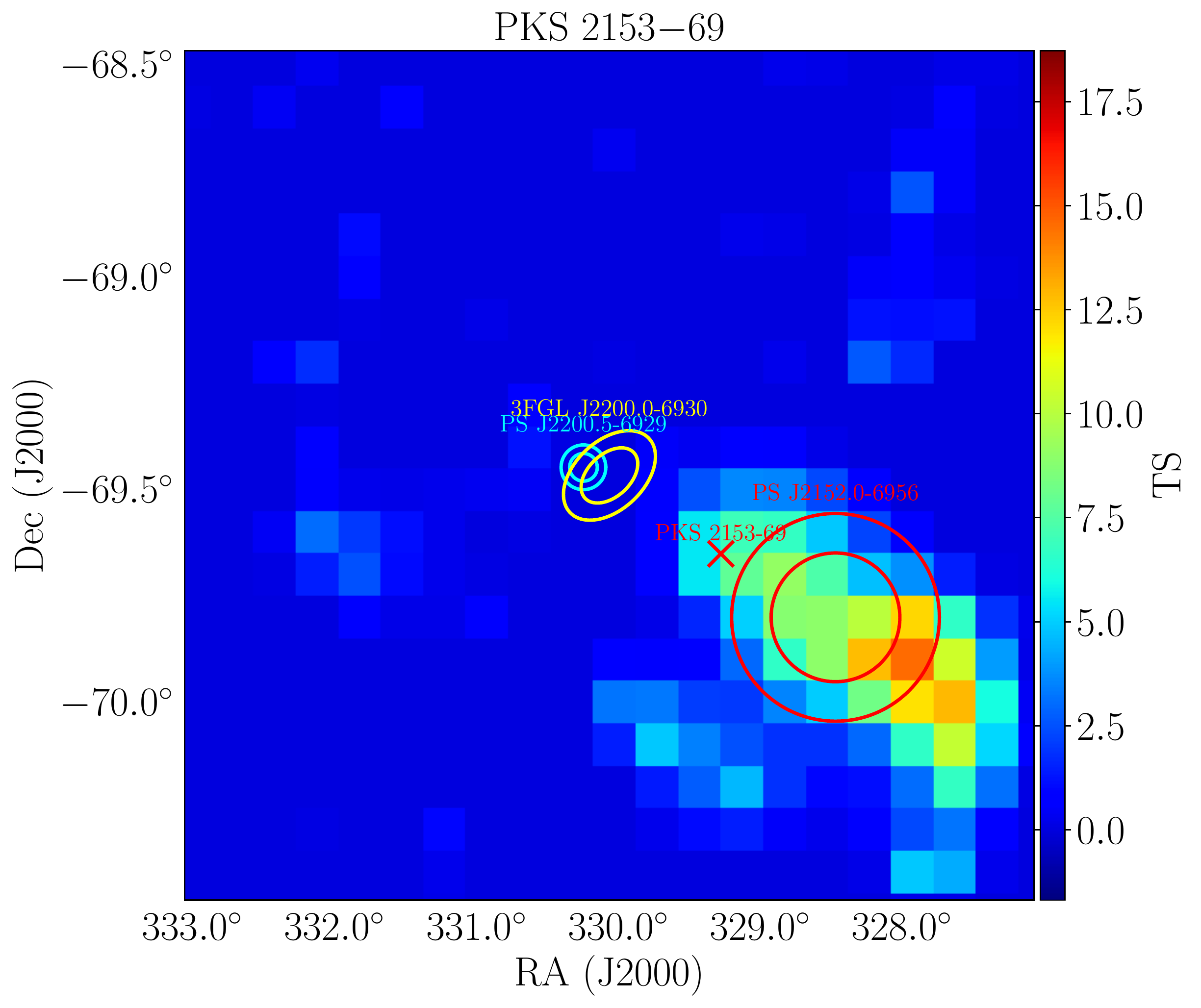}
\end{center}
\caption{\textit{Top panel}: Map of excess TS in the inner region of the ROI centered on
  PKS\,2153$-$69, after removing the unidentified catalog source 3FGL~J2200.0$-$6930 from the model. \textit{Bottom panel}:
  residual excess TS map after modeling and localizing the new source
  PS~J2200.5$-$6930. \textit{Both panels}:
  The cross represents the radio position of PKS\,2153$-$69. The yellow
  ellipses represent the 68\% and 95\% positional uncertainties from
  the catalog for 3FGL~J2200.0$-$6930. The cyan circles represent the 68\%
  and 95\% positional uncertainties for the new source
  PS~J2200.5$-$6930. The blue (top) and red (bottom) circles represent the 68\%
  and 95\% positional uncertainties for the new source PS~J2152.0$-$6956. The map radius is 1\deg. Each pixel corresponds to 0.1\deg.}
\label{2152_ts}
\end{figure}

   \paragraph{2152$-$699}
This source is close to the unidentified catalog source
3FGL~J2200.0$-$6930. We analyzed the region with the same procedure used for PKS\,1258$-$321 and PKS\,1718$-$649 (see Paper I), and found that the newly modeled
$\gamma$-ray source is consistent with the position of
3FGL~J2200.0$-$6930, and not with PKS\,2153$-$69. There is no likely
counterpart to the new $\gamma$-ray source PS~J2200.5$-$6929 within the
95\% error circle (see Fig.~\ref{2152_ts}, top panel). Lowering the threshold of
the source-finding algorithm to TS$>$9, we find a significant source (PS~J2152.0$-$6956) that lies
$\sim0.5^{\circ}$ from the target position. Upon localization, we find
that the 95\% confidence uncertainty region for this source does not
include PKS\,2153$-$69. There is no significant ($>3\sigma$) residual excess after
modeling the latter source (see Fig.~\ref{2152_ts}, bottom panel), therefore we derive an upper limit at the
target position (see Table~\ref{rgs-lat-ul}). 

In the latest \textit{Fermi}-LAT catalog, the 4FGL~\citep{4fgl}, the $\gamma$-ray source 4FGL~J2156.0$-$6942 is associated with PKS~2153$-$69. We performed the same consistency check mentioned in the previous paragraph, using the 4FGL as a starting model, without deleting any point source. 4FGL~J2156.0$-$6942 is detected with TS=87.55, it is spatially distinct from the nearby source 4FGL~J2201.0$-$6928, and it is positionally consistent with PKS~2153$-$69 within the 99\% error region obtained by localization. Therefore, our analysis pipeline successfully reproduces the 4FGL results when using an updated analysis setup. The difference between the 3FGL and 4FGL-based analyses can be ascribed to the different analysis setup (updated catalog, IRFs and diffuse models). Finally, while it may be that PKS~2153$-$69 is indeed a real $\gamma$-ray emitter, one should note that 4FGL~J2156.0$-$6942 is flagged as ``confused'', due to the presence of a nearby brighter source.

\section{Discussion} 
\label{sec:disc}
In this section, we investigate the radio and $\gamma$-ray properties of the radio galaxies in the TANAMI program as a full sample. As in Paper I, in order to increase the sample size we have included all MOJAVE radio galaxies with measured apparent speed and performed a consistent LAT analysis. The results for LAT-detected sources have already been presented in Paper I. Here we list the upper limits obtained for LAT-undetected MOJAVE radio galaxies in Table~\ref{mojave_lat}. We can now combine the TANAMI+MOJAVE $\gamma$-ray-detected and undetected radio galaxy samples. With a total of 35 objects, this is the largest radio galaxy sample with combined $\gamma$-ray and VLBI measurements studied so far. 

\begin{figure*}[htbp]
\begin{center}
\includegraphics[width=0.4\linewidth]{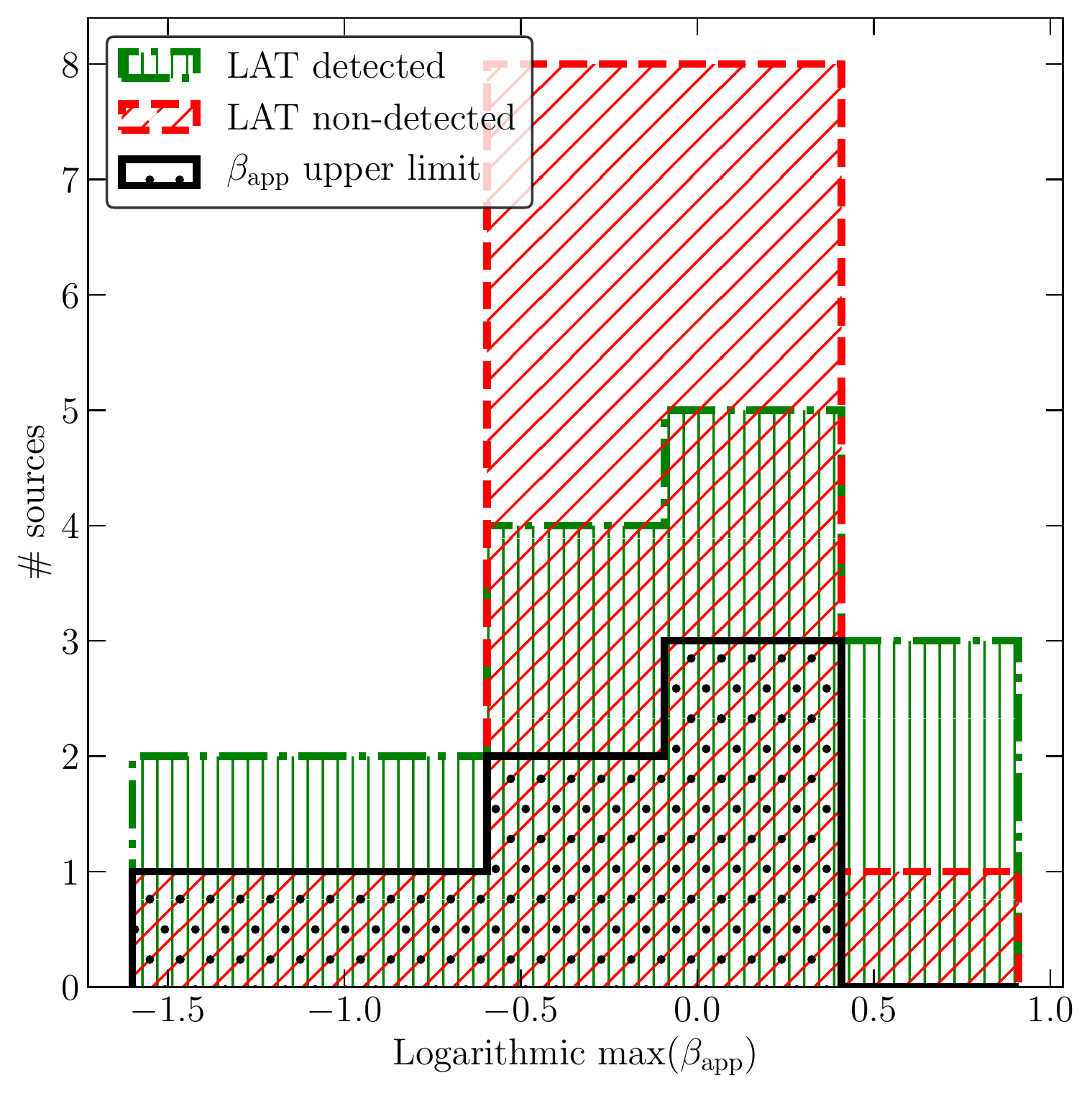}
\includegraphics[width=0.4\linewidth]{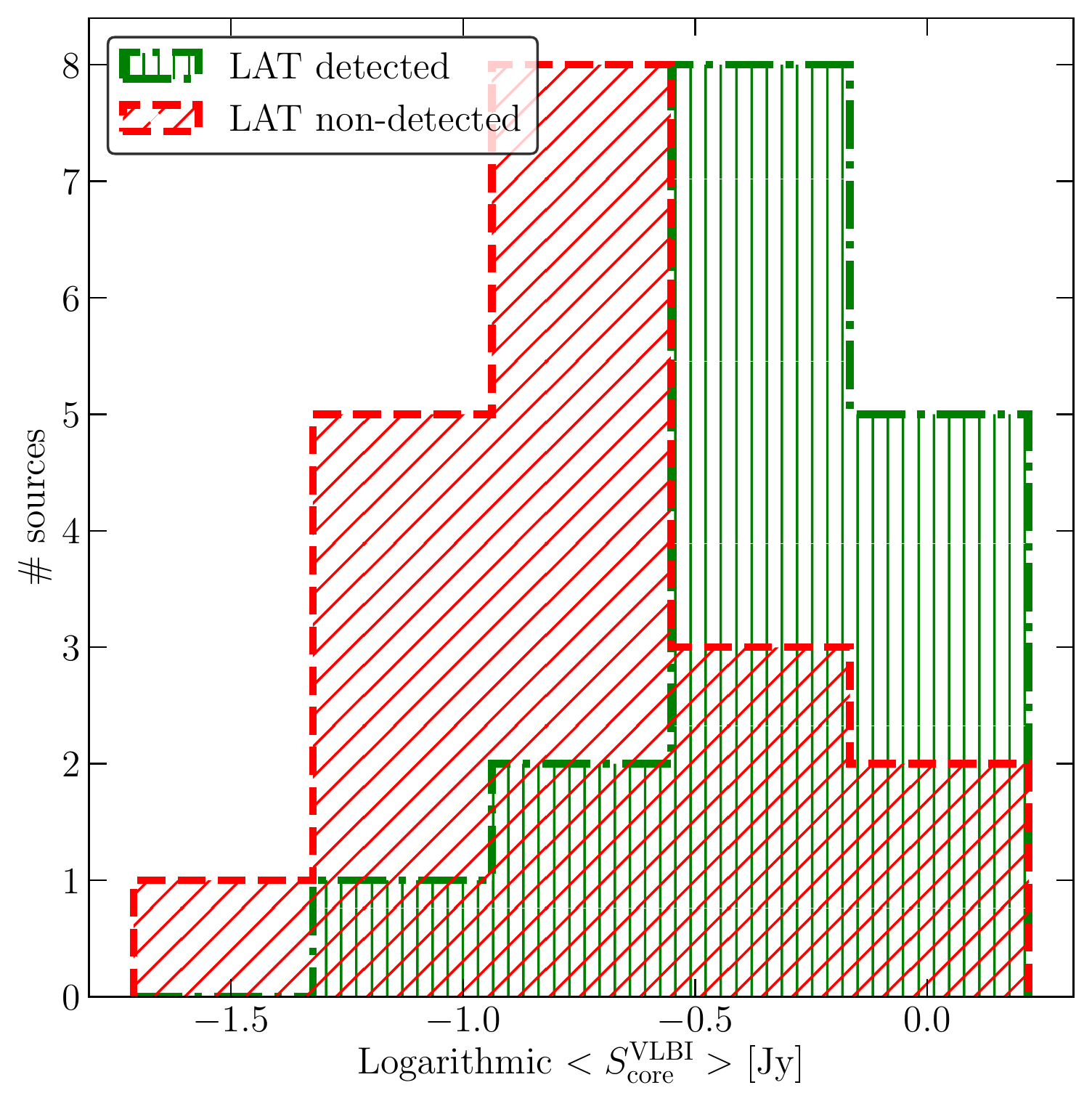}
\includegraphics[width=0.4\linewidth]{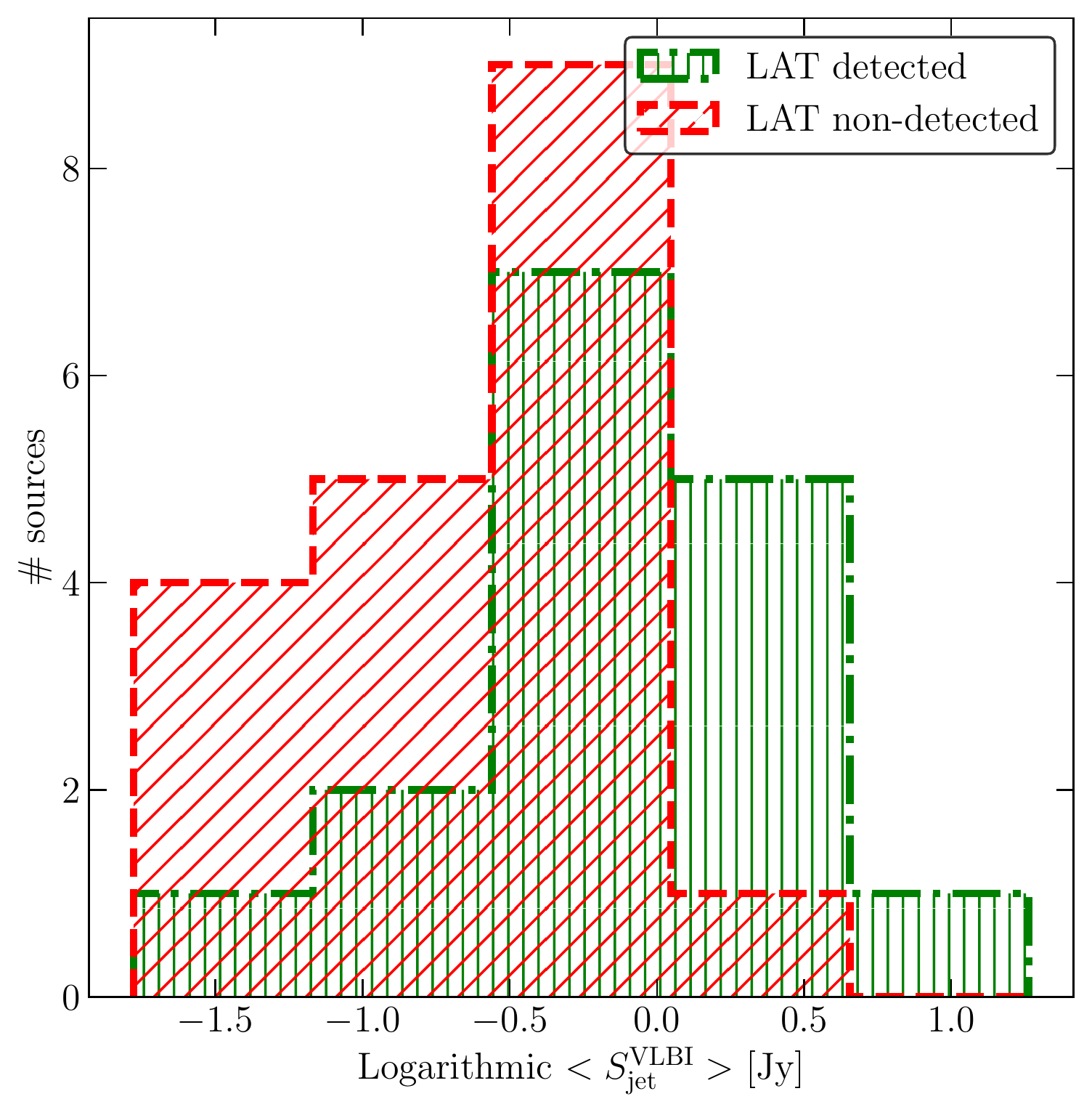}
\includegraphics[width=0.4\linewidth]{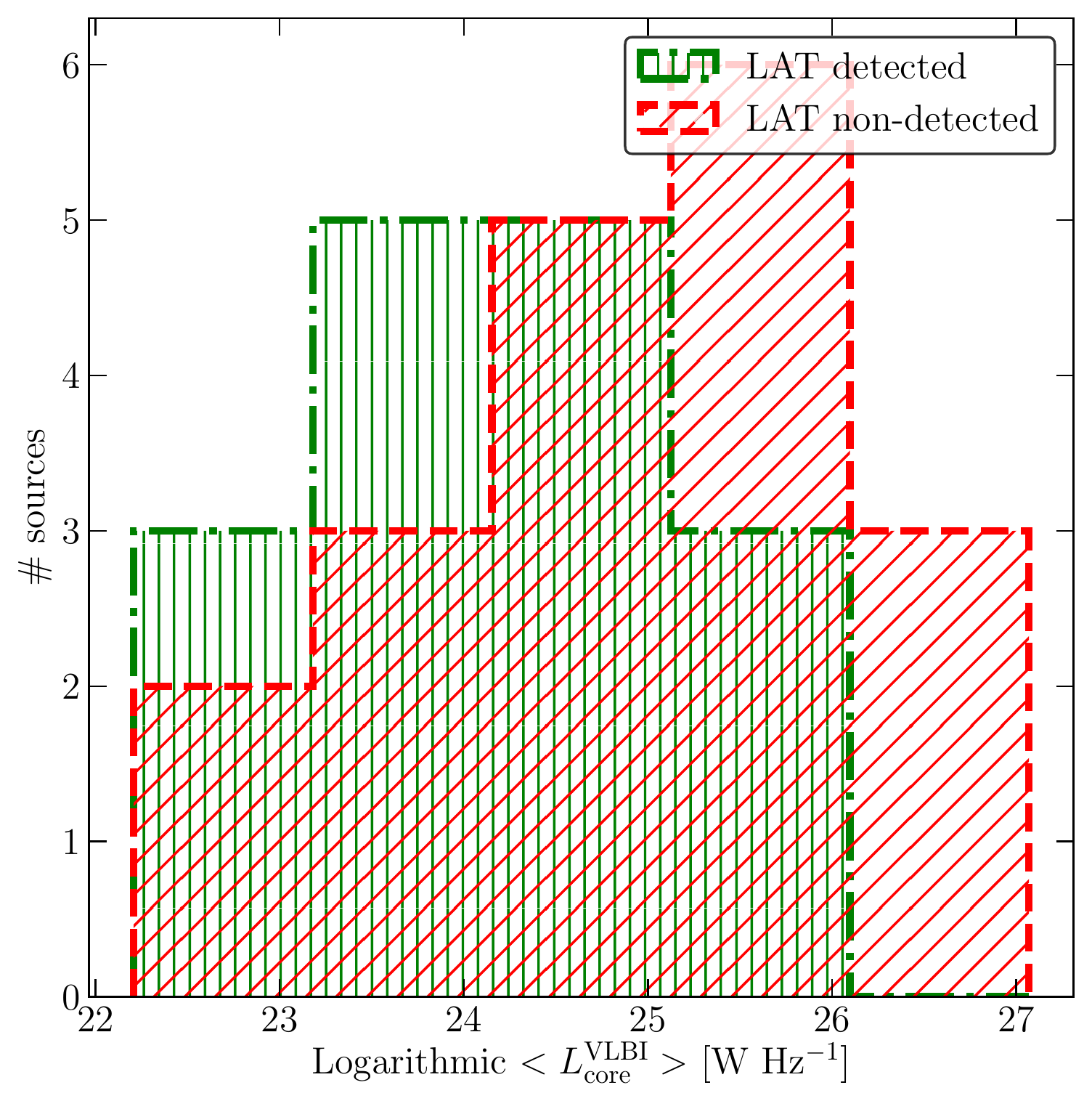}
\includegraphics[width=0.4\linewidth]{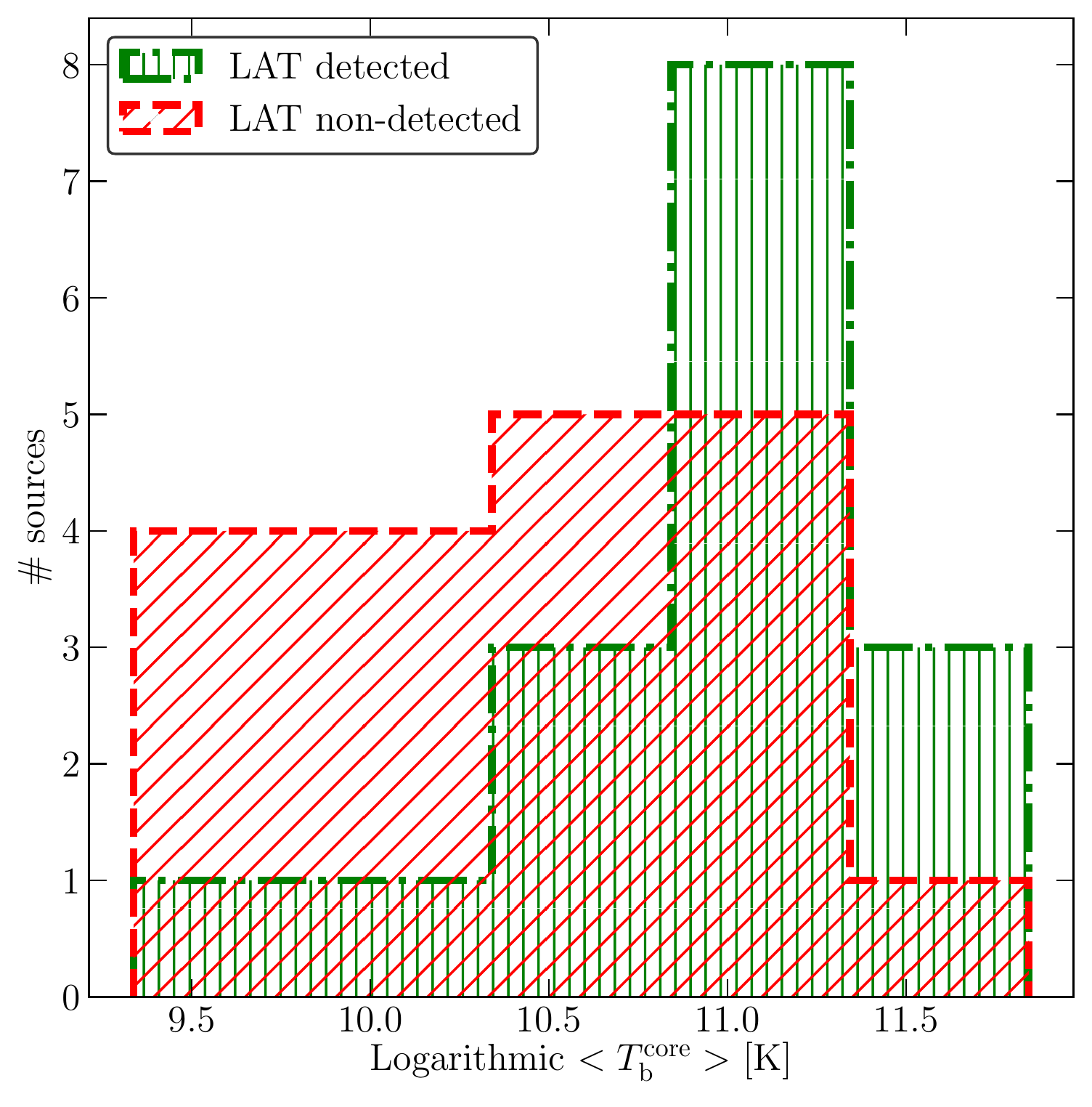}
\includegraphics[width=0.4\linewidth]{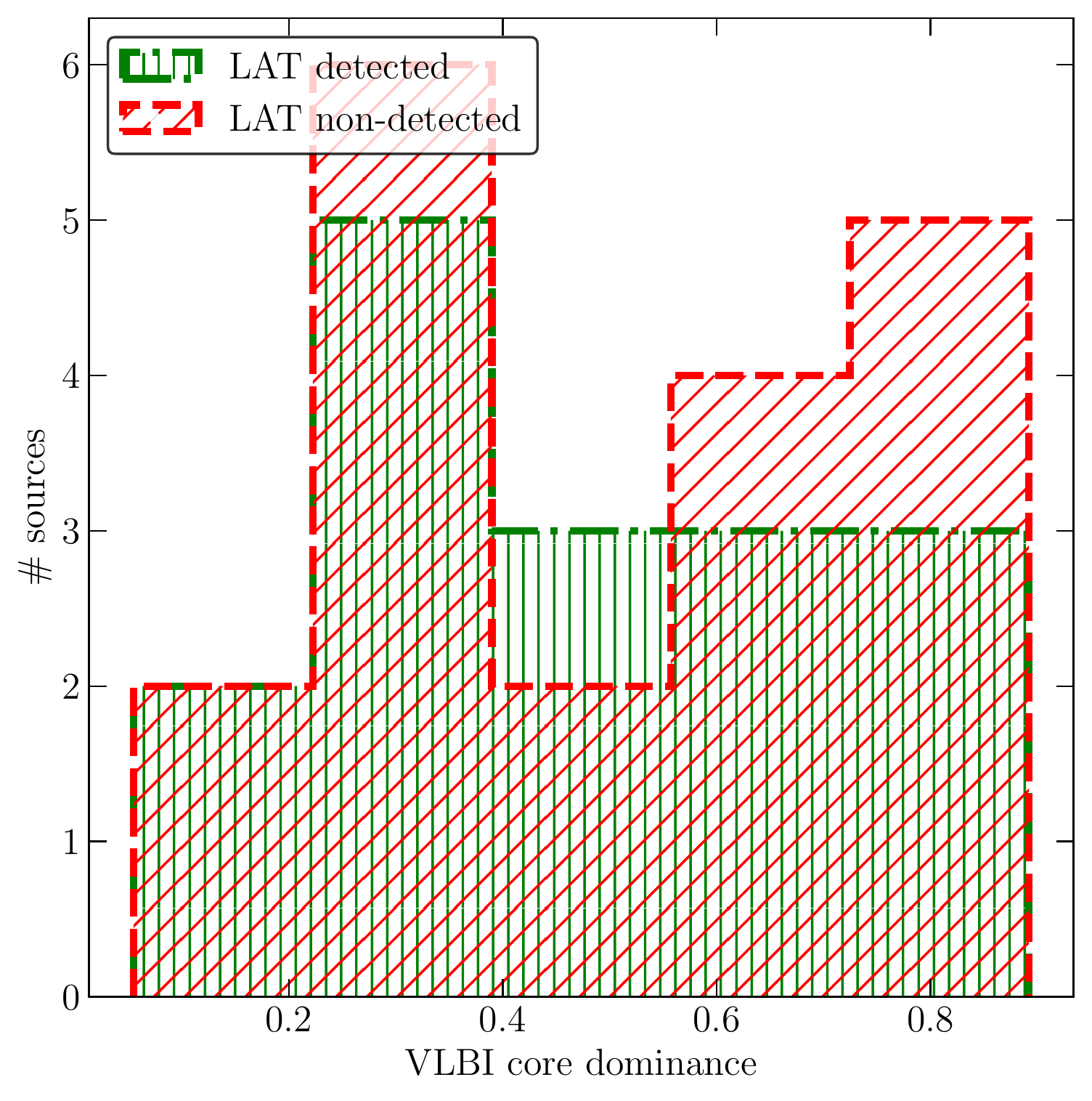}
\end{center}
\caption{Histograms of radio VLBI properties for LAT detected (green vertical hatch) and non-detected radio galaxies (red slashed hatch) from the MOJAVE and TANAMI monitoring programs.\textit{Top left}: maximum measured apparent speed. \textit{Top right}: median VLBI core flux density. \textit{Center left}: median logarithmic VLBI jet flux density. \textit{Center right}: median logarithmic VLBI core luminosity. \textit{Bottom left}: median logarithmic core brightness temperature. \textit{Bottom right}: median VLBI core dominance.}
\label{fig:radio_hist}
\end{figure*}

\begin{table}[h!tbp]
\small
\caption{0.1-100 GeV \textit{Fermi}-LAT upper limits on MOJAVE radio galaxies.}

\begin{center}
\tabcolsep=0.1cm
\begin{tabular}{llccc}
\hline
\hline
B1950 name & Common name & Redshift & Flux$^a$ & TS\\
\hline
  0007$+$106 & Mrk 1501 & 0.0893 & $<4\times10^{-9}$ & 1.87\\
  0026+346 & B2 0026+34 & 0.517 & $<3\times10^{-9}$ & 6.76\\
  0108+388 & GB6 J0111+3906 & 0.668 & $<5\times10^{-9}$ & 2.95\\
  0710+439 & B3 0710+439 & 0.518 & $<6\times10^{-10}$ & 0.0\\
  1345+125 & 4C +12.50 & 0.121 & $<1\times10^{-9}$ & 0.97\\
  1509+054 & PMN J1511+0518 & 0.084 & $<2\times10^{-9}$ & 0.35\\
  1607+268 & CTD 93 & 0.473 & $<7\times10^{-9}$ & 5.88\\
  1845+797 & 3C 390.3 & 0.0555 & $<2\times10^{-9}$ & 5.35\\
  1957+405 & Cygnus A & 0.0561 & $<4\times10^{-9}$ & 2.76\\
  2021+614 & OW 637 & 0.227 & $<1\times10^{-8}$ & 18.6\\
  2128+048 & PKS 2127+04 & 0.99 & $<2\times10^{-9}$ & 0.2\\
\hline
\hline
\end{tabular}
\end{center}
$^a$ \textit{Fermi}-LAT flux upper limit between 0.1-100 GeV in \phcms.\\

\label{mojave_lat}
\end{table}
In order to compare the radio properties of $\gamma$-ray-detected and undetected subsamples, we visualize their distribution with histograms, shown in Fig.~\ref{fig:radio_hist}. We test whether there is a statistically significant difference in the distribution of LAT detected and undetected radio galaxies using the Kolmogorov-Smirnov statistic (KS). The results of this test are listed in Table~\ref{tab:ks}. 
\begin{table}[h!tbp]
\caption{Results of a KS test to assess whether the LAT detected and undetected subsample are drawn from the same parent population, according to their average radio properties as illustrated in Fig.~\ref{fig:radio_hist}.}
\begin{center}
\begin{tabular}{lccc}
\hline
\hline
Variable & KS statistic & $p$-value & Sign. ($\sigma$)\\
\hline
$<S_\mathrm{core}^\mathrm{VLBI}>$ & 0.53 & 0.009 & 2.6\\
$<T_b^\mathrm{core}>$ & 0.50 & 0.017 & 2.4\\
$<L_\mathrm{core}^\mathrm{VLBI}>$ & 0.41 & 0.073 & 1.8\\
$<S_\mathrm{jet}^\mathrm{VLBI}>$ & 0.38 & 0.13 & 1.5\\
max($\beta_\mathrm{app}$) & 0.23 & 0.70 & 0.4\\
$<CD_\mathrm{VLBI}>$ & 0.18 & 0.91 & 0.1\\

\hline
\hline
\end{tabular}
\end{center}
\label{tab:ks}
\end{table}

The most statistically significant result is obtained for the median VLBI core flux density $<S_\mathrm{core}^\mathrm{VLBI}>$, which shows a quite clear dichotomy between the two subsamples, with the LAT-undetected sources occupying the lower end, and the LAT-detected sources dominating the upper end of the distribution (see top right panel of Fig.~\ref{fig:radio_hist}). A similar significance is found for the median core brightness temperature ($<T_b^\mathrm{core}>$, see bottom left panel of Fig.~\ref{fig:radio_hist}). The other VLBI parameters for which we tested for significant differences in the distribution of LAT detected and undetected radio galaxies, i.e., maximum apparent speed (max($\beta_\mathrm{app}$), top left panel of Fig.~\ref{fig:radio_hist}), median jet flux density ($<S_\mathrm{jet}^\mathrm{VLBI}>$, i.e., total minus core, see middle left panel of Fig.~\ref{fig:radio_hist}), median core luminosity ($<L_\mathrm{core}^\mathrm{VLBI}>$, middle right panel of Fig.~\ref{fig:radio_hist}), and median VLBI core dominance ($<CD_\mathrm{VLBI}>$, i.e., core flux over total flux, see bottom right panel of Fig.~\ref{fig:radio_hist}) yield less significant results ($p$-value$>0.05$).

\begin{figure*}[!htbp]
 
\begin{center}
\includegraphics[width=0.42\linewidth]{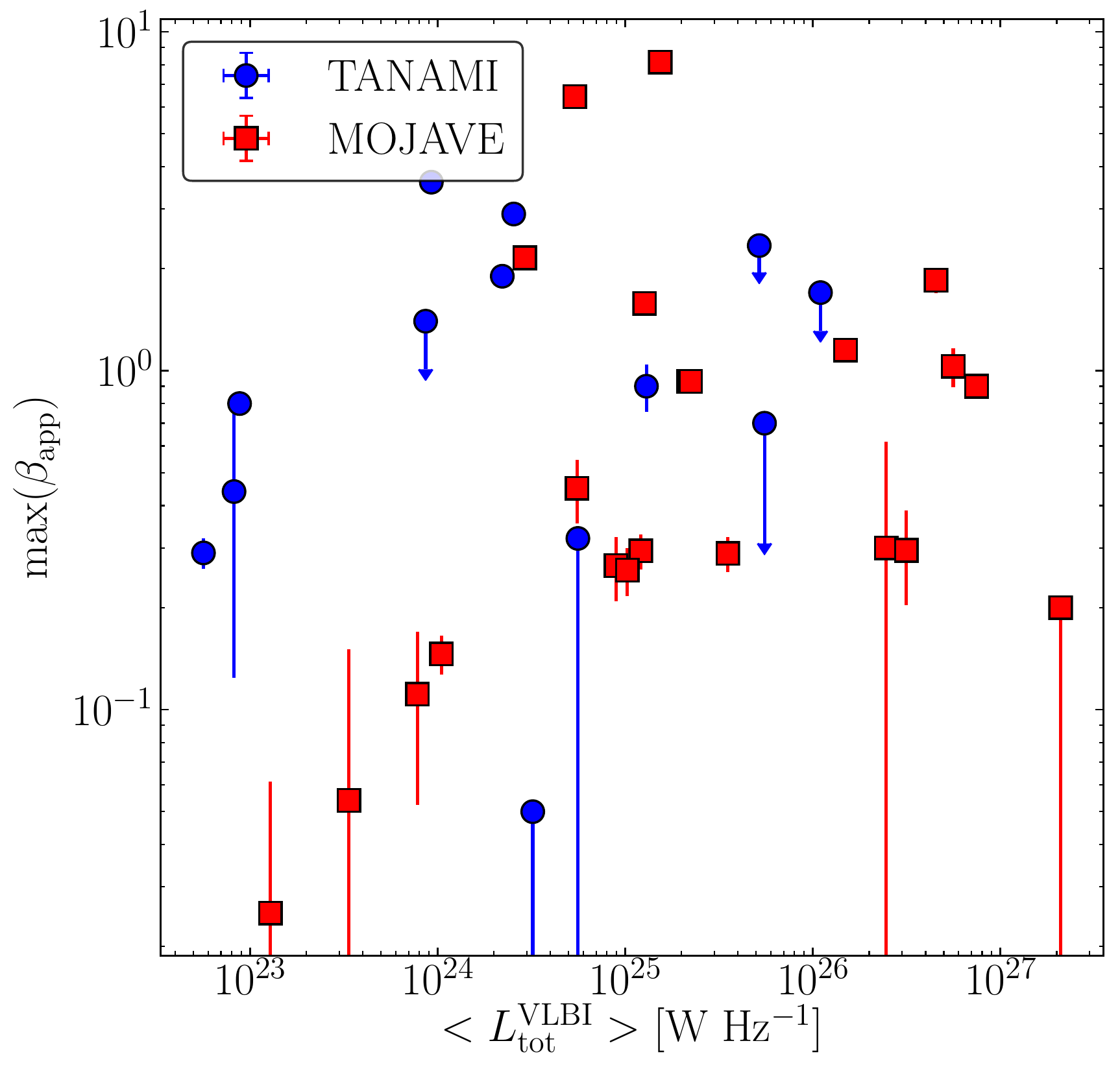}
\includegraphics[width=0.42\linewidth]{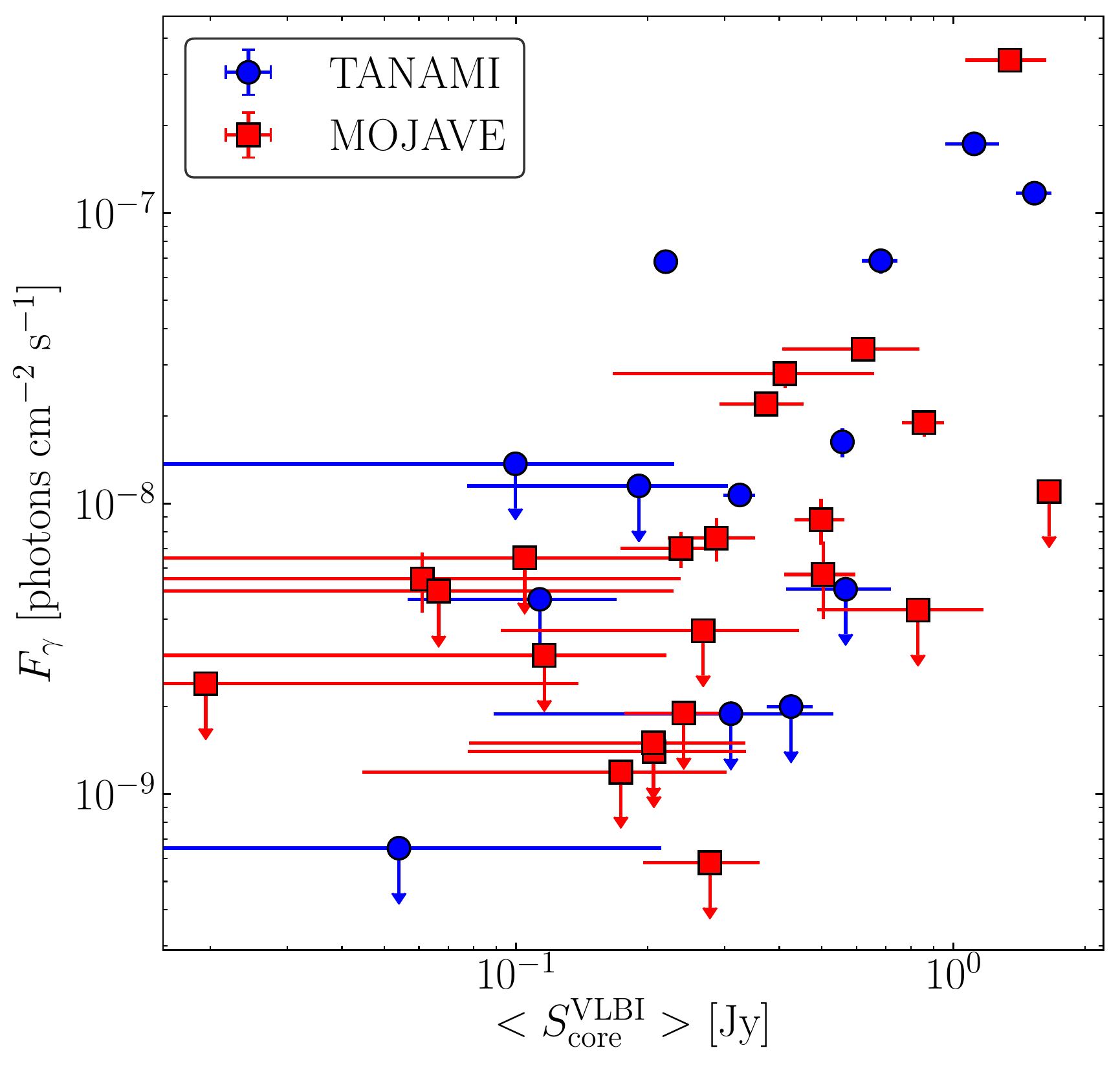}
\includegraphics[width=0.42\linewidth]{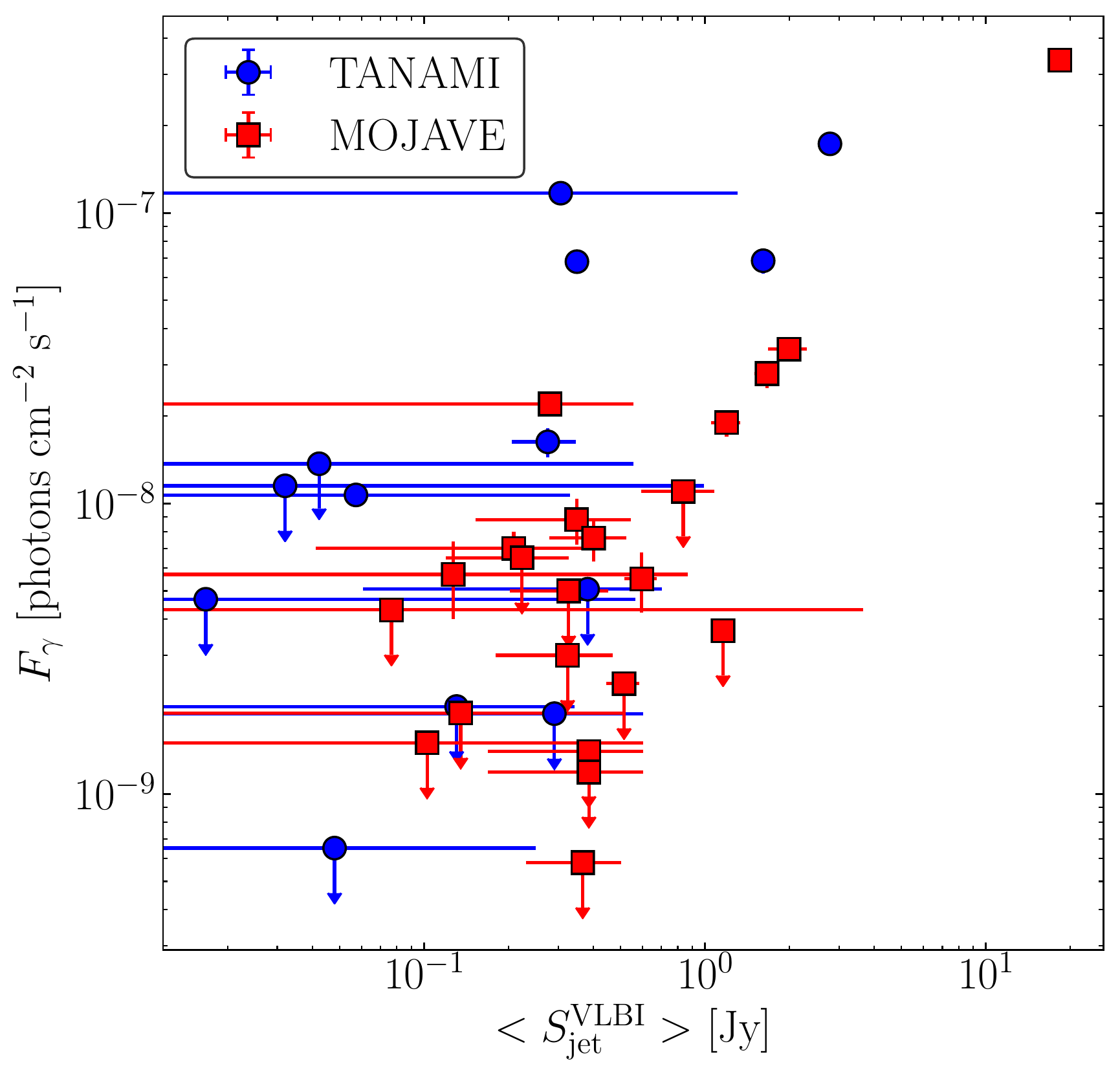}
\includegraphics[width=0.42\linewidth]{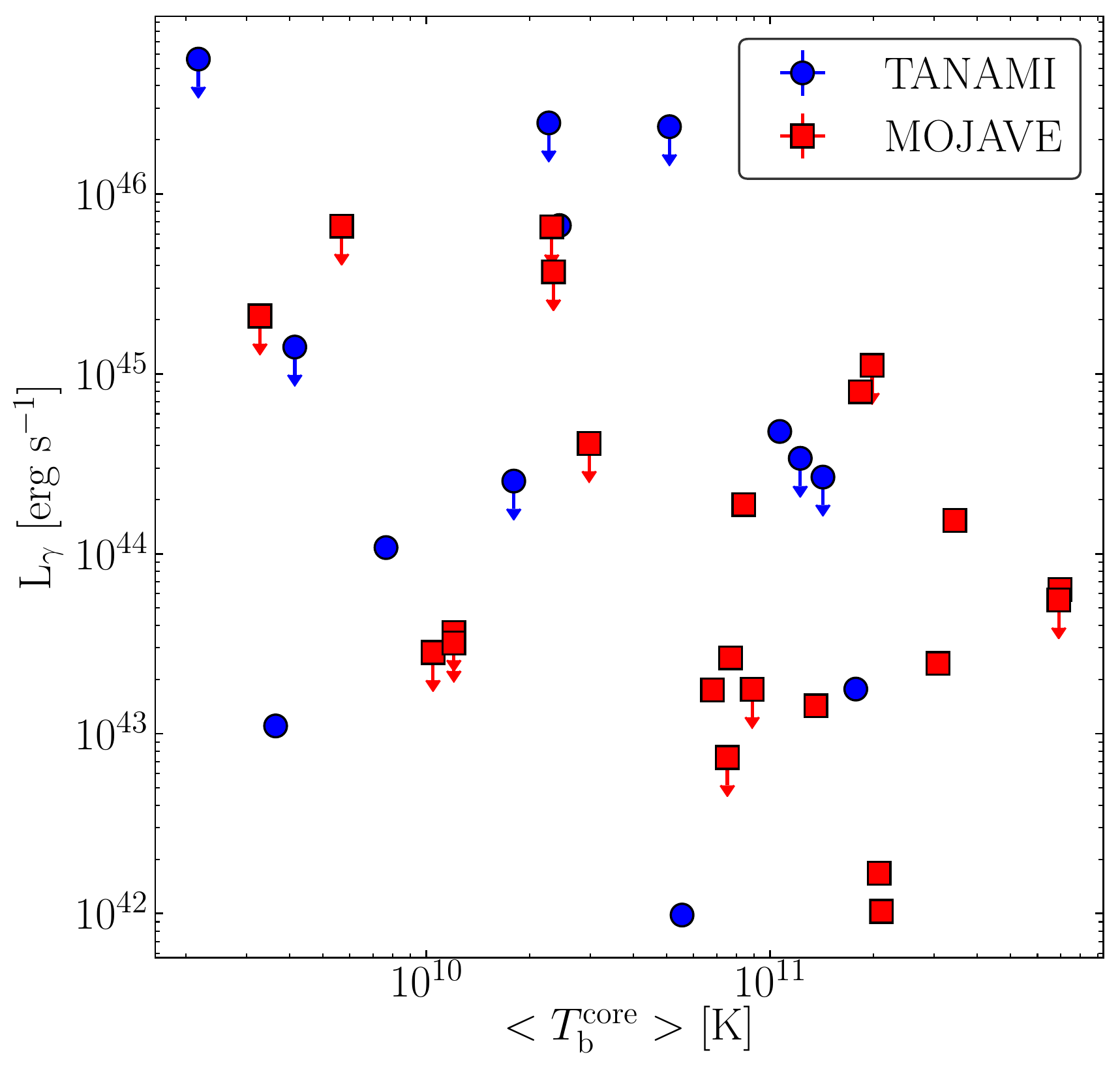}
\includegraphics[width=0.42\linewidth]{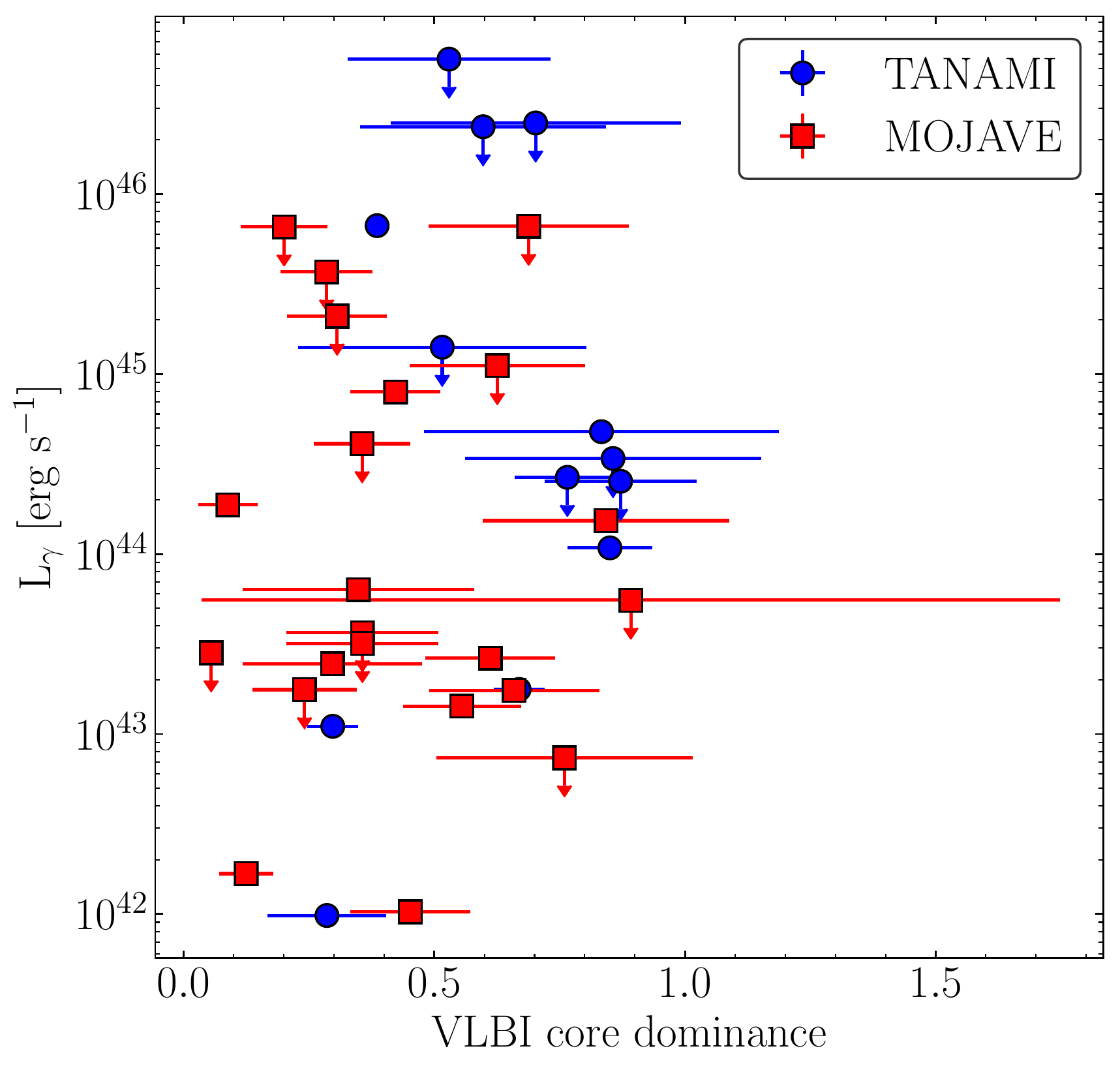}
\end{center}
\caption{\textit{Top left}: Maximum observed apparent speed as a function of total VLBI luminosity. \textit{Top right}: \textit{Fermi}-LAT flux as a function of average VLBI core flux. \textit{Center left}: \textit{Fermi}-LAT flux as a function of average VLBI jet flux. \textit{Center right}: Average VLBI core brightness temperature as a function of \textit{Fermi}-LAT luminosity. \textit{Center bottom}: \textit{Fermi}-LAT luminosity as a function of average VLBI core dominance.}
\label{fig:corrs}

\end{figure*}

In order to investigate possible correlations between the radio and $\gamma$-ray properties of the sources in this sample, we have used the Kendall's correlation coefficient ($\tau$) adapted to take into account the presence of upper limits, following~\cite{1996MNRAS.278..919A}. The correlation coefficient is equal to zero in the case of uncorrelated data, one in case of maximum correlation, and minus one in case of maximum anti-correlation. The resulting correlation coefficients with errors and the relative $p$-values are listed in Table~\ref{tab:corrs}. The scatter plots of radio vs. $\gamma$-ray properties are presented in Fig.~\ref{fig:corrs}.
\begin{table}[h!tbp]
\caption{Correlation coefficients between radio and $\gamma$-ray properties of our radio galaxy sample, and corresponding significance.}
\begin{center}
\begin{tabular}{lccc}
\hline
\hline
Variables & Kendall's $\tau$ & $p$-value & Sign. ($\sigma$)\\
\hline
  $<S_\mathrm{core}^\mathrm{VLBI}>$ vs. $F_\gamma$ & 0.55 & 0.003 & 2.97\\
  $<S_\mathrm{jet}^\mathrm{VLBI}>$ vs. $F_\gamma$ & 0.19 & 0.1 & 1.65\\
  $<CD_\mathrm{VLBI}>$ vs. $L_\gamma$ & 0.16 & 0.17 & 1.37\\
  $L_\gamma$ vs. $<T_\mathrm{b}^\mathrm{core}>$  & 0.08 & 0.5 & 0.67\\
\hline
\hline
\end{tabular}
\end{center}
\label{tab:corrs}
\end{table}

The resulting correlation properties corroborate the ones presented in Paper I. The only radio and $\gamma$-ray properties showing a statistically significant correlation are the VLBI core flux density and LAT flux. There is no correlation between $\gamma$-ray flux ($F_{\gamma}$) and pc-scale jet flux. A simple linear correlation between the luminosity in the two bands ($L_{\gamma}$ and $<L_\mathrm{core}^\mathrm{VLBI}>$) yields $\tau=0.78$ and $p=2\times10^{-5}$. However, when computing the partial correlation coefficient accounting for redshift \citep[see][]{1996MNRAS.278..919A} this is reduced to $\tau=0.39$ and $p=0.05$, which hints at a physical correlation, but with reduced statistical significance. Finally, the $\gamma$-ray luminosity appears to be entirely uncorrelated with common VLBI Doppler boosting indicators such as core brightness temperature and core dominance.


\section{Summary and conclusions}
\label{sec:conc}
In this paper, the second of a series, we continue a sample study on the connection between radio emission from the parsec-scale jet and high-energy $\gamma$-ray properties of radio galaxies. While such a connection has been firmly established for large AGN samples, these were dominated by blazars, i.e., aligned jets whose emission is strongly affected by Doppler boosting effects. In this work, completing the findings of Paper I, we have explored the pc-scale radio - $\gamma$-ray connection in a large sample of radio galaxies, i.e., AGN with misaligned jets.

Here, we report the kinematics of the VLBI jet in LAT-undetected radio galaxies from the TANAMI monitoring program. We found that most of the sources show slowly moving jet components with subluminal or zero apparent speed. The only $\gamma$-ray-faint radio galaxy showing fast apparent motion, up to $\beta_\mathrm{app}=3.6$, is the FR~II source PKS\,2153$-$69. We found a clear linear trend of increasing apparent speed with average core separation in the jet of the latter source, which indicates that it is still undergoing acceleration on scales of tens of parsecs or $\sim10^5\,R_s$, corresponding to the termination of the acceleration and collimation region in radio galaxies.

We present updated LAT upper limits on $\gamma$-ray-faint TANAMI radio galaxies. Specifically, we have investigated whether PKS\,1258$-$321 and PKS\,2153$-$69 might be the counterparts to the nearby unassociated $\gamma$-ray sources 3FGL~J1259.5-3231 and 3FGL~J2200.0$-$6930, respectively. We found that based on $\sim8.5$ years of LAT data, the two radio galaxies are not included within the positional uncertainty of the $\gamma$-ray sources.

Combining our VLBI and \textit{Fermi}-LAT results with those of Paper I, and merging the TANAMI and MOJAVE radio galaxy samples, we assembled the largest sample of radio galaxies with measured pc-scale kinematics and $\gamma$-ray properties to date, counting a total of 35 sources. We have investigated whether the LAT-detected and undetected sub-populations show differences in their radio VLBI properties. Based on a Kolmogorov-Smirnov two-sample test, we found differences in the distribution of median VLBI core flux at the level of 2.6$\sigma$, and indications of a separation in the distribution of median core brightness temperature at the level of 2.4$\sigma$. In terms of correlations between radio VLBI and $\gamma$-ray properties, the results on the full sample corroborate the ones reported in Paper I on the LAT-detected subsample, with a significant correlation between median radio core flux and $\gamma$-ray flux, and no correlation between $\gamma$-ray luminosity and common parsec-scale Doppler boosting indicators such as core brightness temperature and core dominance. Overall, by comparing several VLBI and \textit{Fermi}-LAT properties of radio galaxies, we are able to quantitatively show that high-energy emission in misaligned jets is related to the brightness of their innermost parsec-scale regions, as is observed for blazars. While there is general agreement that beaming effects in radio galaxies are less pronounced than in blazars (cf. the AGN unification model; e.g., \citealt{1995PASP..107..803U}), it is clear that significant relativistic beaming is present in misaligned jet sources as well. This is demonstrated, e.g., by their usually one-sided or apparently asymmetric parsec-scale jets and their high core brightness temperatures. Until now, however, the impact of the remaining Doppler boosting still has on the $\gamma$-ray brightness of radio galaxies has not been investigated. Our results indicate quantitatively, for the first time, that the existing effect of boosting observed in radio galaxies does not dominate the observed $\gamma$-ray properties from their misaligned jets.

\begin{acknowledgements}
We thank Laura Vega Garc\'ia for the development of the Python
GUI-based code that was used for the kinematic analysis and the
spectral index maps. We thank Frank Schinzel as the \textit{Fermi}-LAT collaboration internal referee. We also thank the journal referee for a constructive report.

This research has made use of the NASA/IPAC Extragalactic Database (NED),
which is operated by the Jet Propulsion Laboratory, California Institute of Technology, under contract with the National Aeronautics and Space Administration; data from the MOJAVE database that is maintained by the MOJAVE team \citep{2018ApJS..234...12L}; APLpy, an open-source plotting package for Python \citep{aplpy}; Astropy,\footnote{http://www.astropy.org} a community-developed core Python package for Astronomy \citep{astropy:2013, astropy:2018}

C.M. acknowledges support from the ERC
Synergy  Grant  “BlackHoleCam:  Imaging  the  Event  Horizon  of  Black  Holes”
(Grant 610058).

R.S.
gratefully acknowledges support from the European Research Council under
the European Union's Seventh Framework Programme (FP/2007-2013)/ERC
Advanced Grant RADIOLIFE-320745.

F.K. was supported as an Eberly Research Fellow by the Eberly College of Science at the Pennsylvania State University.

The \textit{Fermi}-LAT Collaboration acknowledges generous ongoing support from a
number of agencies and institutes that have supported both the
development and the operation of the LAT, as well as scientific data
analysis. These include the National Aeronautics and Space
Administration and the Department of Energy in the United States; the
Commissariat \'a l’Energie Atomique and the Centre National de la
Recherche Scientifique/Institut National de Physique Nucl\'eaire et de
Physique des Particules in France; the Agenzia Spaziale Italiana and
the Istituto Nazionale di Fisica Nucleare in Italy; the Ministry of
Education, Culture, Sports, Science and Technology (MEXT), High Energy
Accelerator Research Organization (KEK), and Japan Aerospace
Exploration Agency (JAXA) in Japan; and the K. A. Wallenberg
Foundation, the Swedish Research  Council,
and  the  Swedish  National  Space  Board in Sweden.

Additional support for science analysis during the operations
phase is gratefully acknowledged from the Istituto Nazionale di
Astrofisica in Italy and the Centre National d'Etudes Spatiales in
France. This work was
performed in part under DOE Contract DE-AC02-76SF00515.

The Long Baseline Array is part of the Australia Telescope National
Facility which is funded by the Commonwealth of Australia for
operation as a National Facility managed by CSIRO. This study made use
of data collected through the AuScope initiative. AuScope Ltd is
funded under the National Collaborative Research Infrastructure
Strategy (NCRIS), an Australian Commonwealth Government Programme.
This work made use of the Swinburne University of Technology software
correlator, developed as part of the Australian Major National
Research Facilities Programme. This work was supported by resources
provided by the Pawsey Supercomputing Centre with funding from the
Australian Government and the Government of Western Australia.
Hartebeesthoek Radio Astronomy Observatory (HartRAO) is a facility of
the National Research Foundation (NRF) of South Africa. This research
was funded in part by NASA through Fermi Guest Investigator grants
NNH10ZDA001N, NNH12ZDA001N, and NNH13ZDA001N-FERMI (proposal numbers
41213, 61089, and 71326, respectively). This research was supported by
an appointment to the NASA Postdoctoral Program at the Goddard Space
Flight Center, administered by Universities Space Research Association
through a contract with NASA.
\end{acknowledgements}

\clearpage

\bibliographystyle{aa}
\bibliography{aa}
\clearpage
\begin{appendix}
\section{Full resolution VLBI maps and image parameters}
\label{app:maps}
Here we present the full-resolution VLBI images for the sources studied in this paper, along with the tables including the details of each observation, for each source.

\begin{figure*}[!htbp]
\begin{center}
\includegraphics[width=0.43\linewidth]{Figures/Maps_new/1258-321_2009-12-13.pdf}
\includegraphics[width=0.43\linewidth]{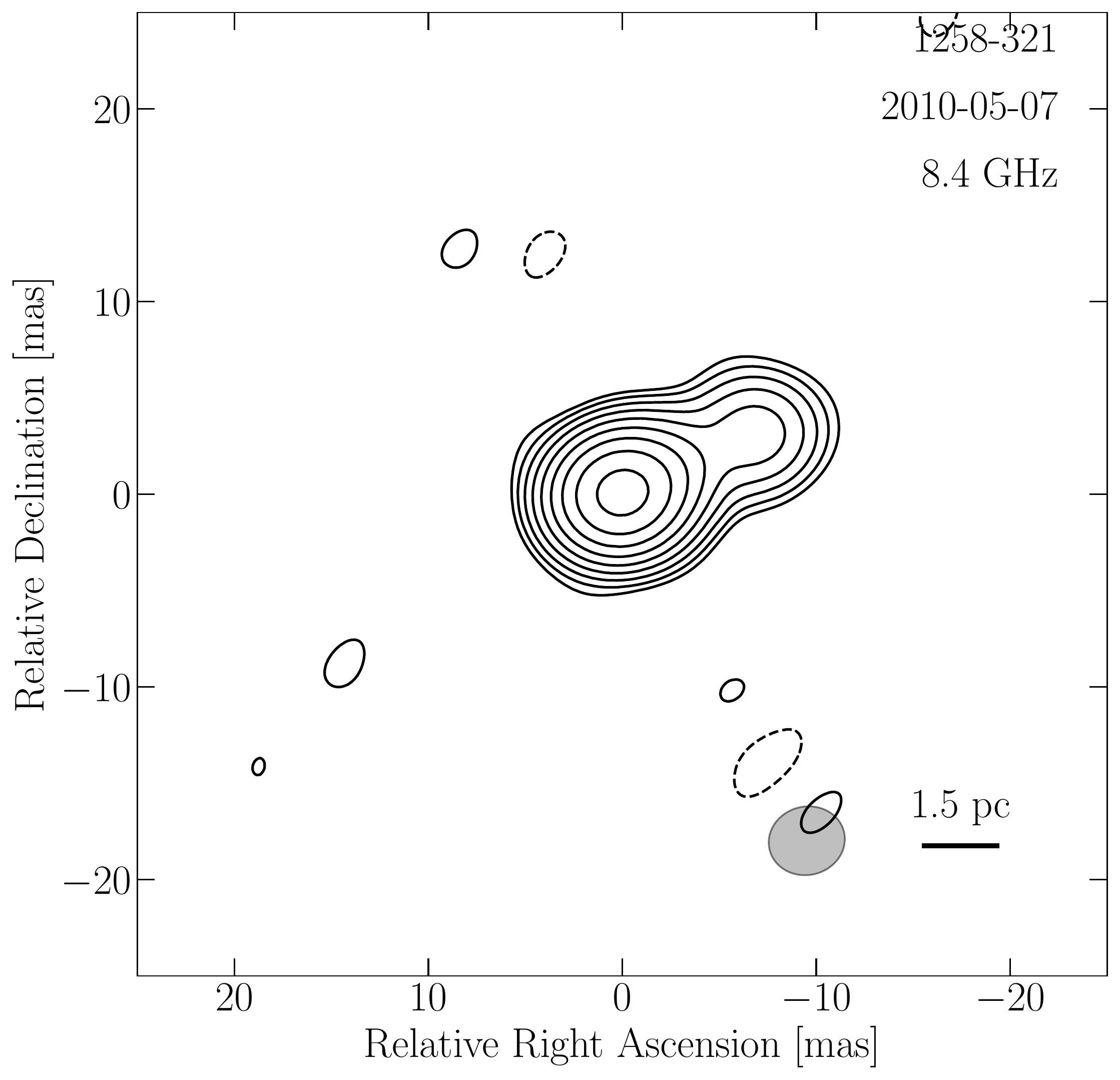}
\includegraphics[width=0.43\linewidth]{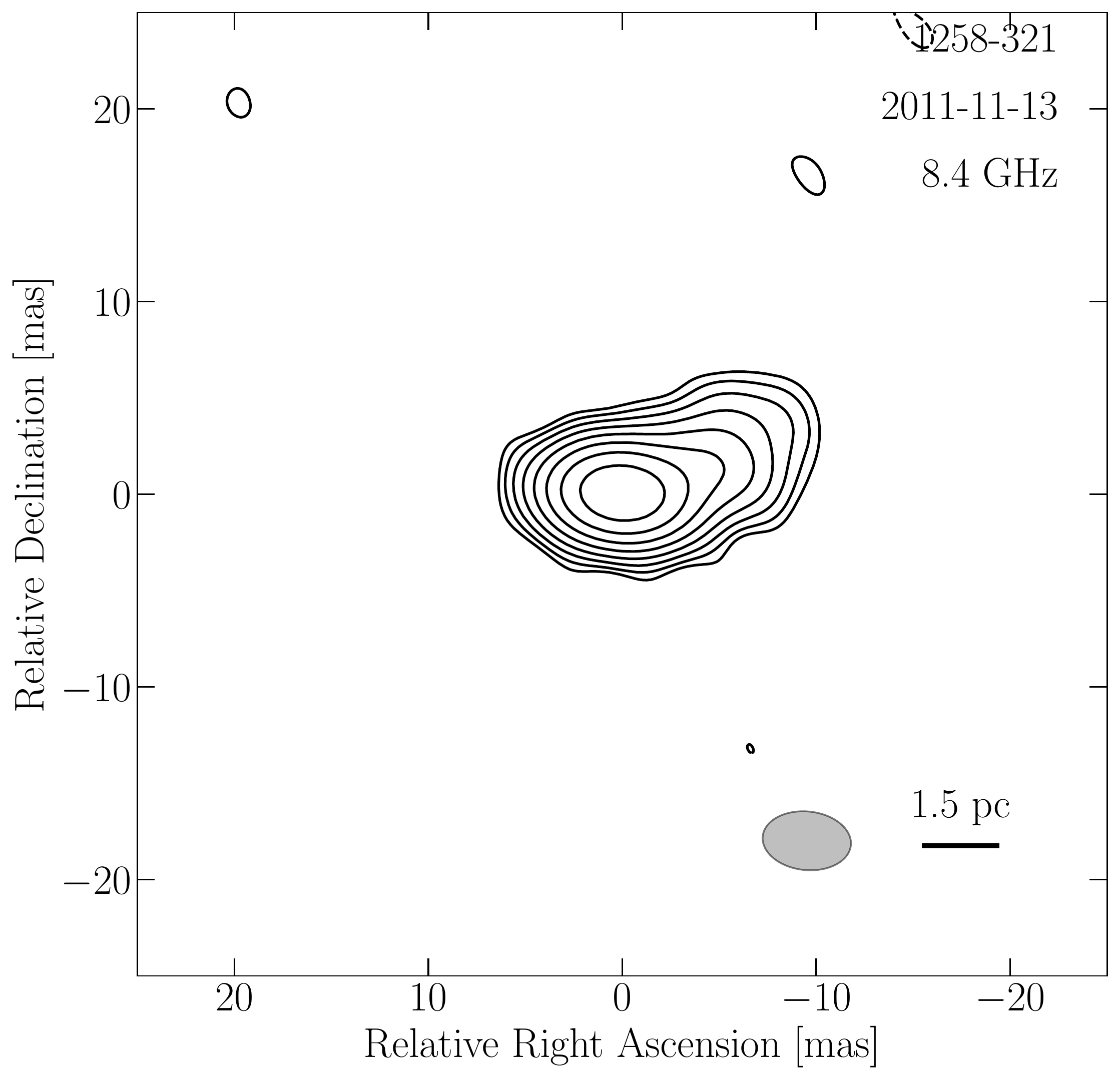}
\includegraphics[width=0.43\linewidth]{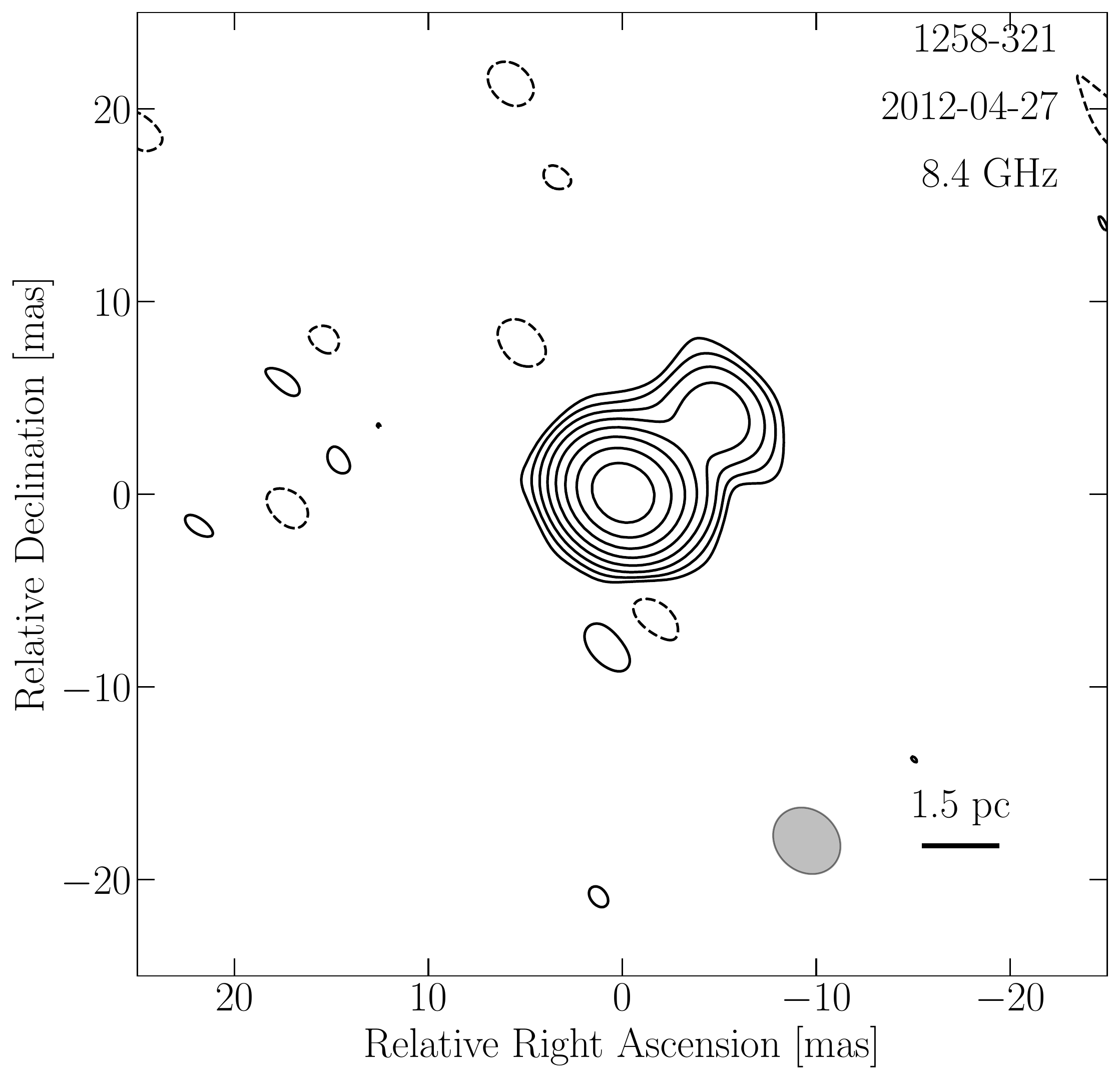}
\includegraphics[width=0.43\linewidth]{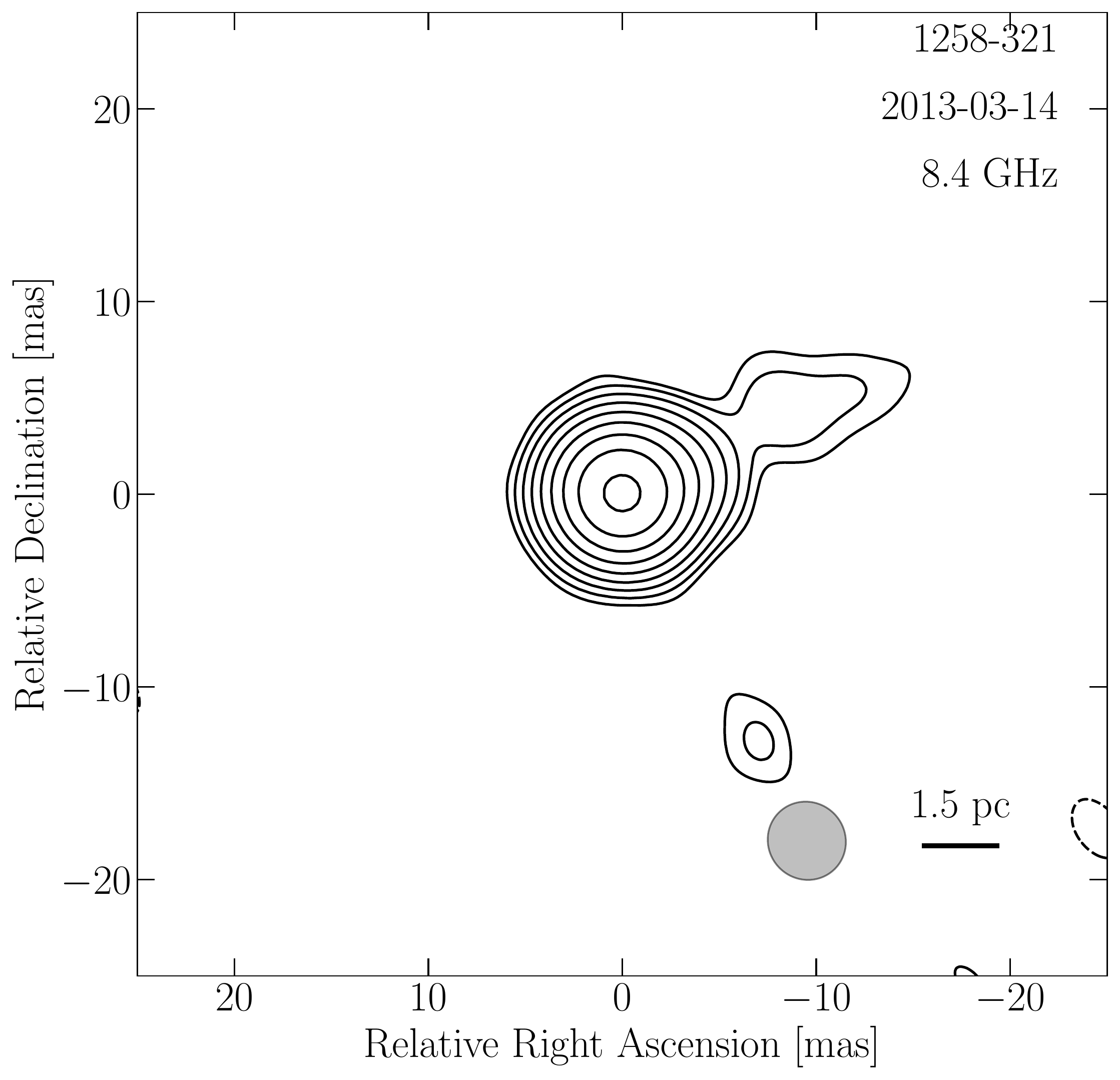}
\end{center}
\caption{Full-resolution images of PKS\,1258$-$321. The map parameters for
  each epoch can be found in Table~\ref{1258_tab}. The grey ellipse
  represents the beam size, while the black line indicates the linear
  scale at the source's redshift. Contours increase in steps of two starting from 0.5, 0.8, 0.8, 1, 0.8 times the noise level in each map, from top left to bottom, respectively.}
\label{1258_full}
\end{figure*}
\begin{figure*}[!htbp]
\begin{center}
\includegraphics[width=0.43\linewidth]{Figures/Maps_new/1333-337_2008-02-07.pdf}
\includegraphics[width=0.43\linewidth]{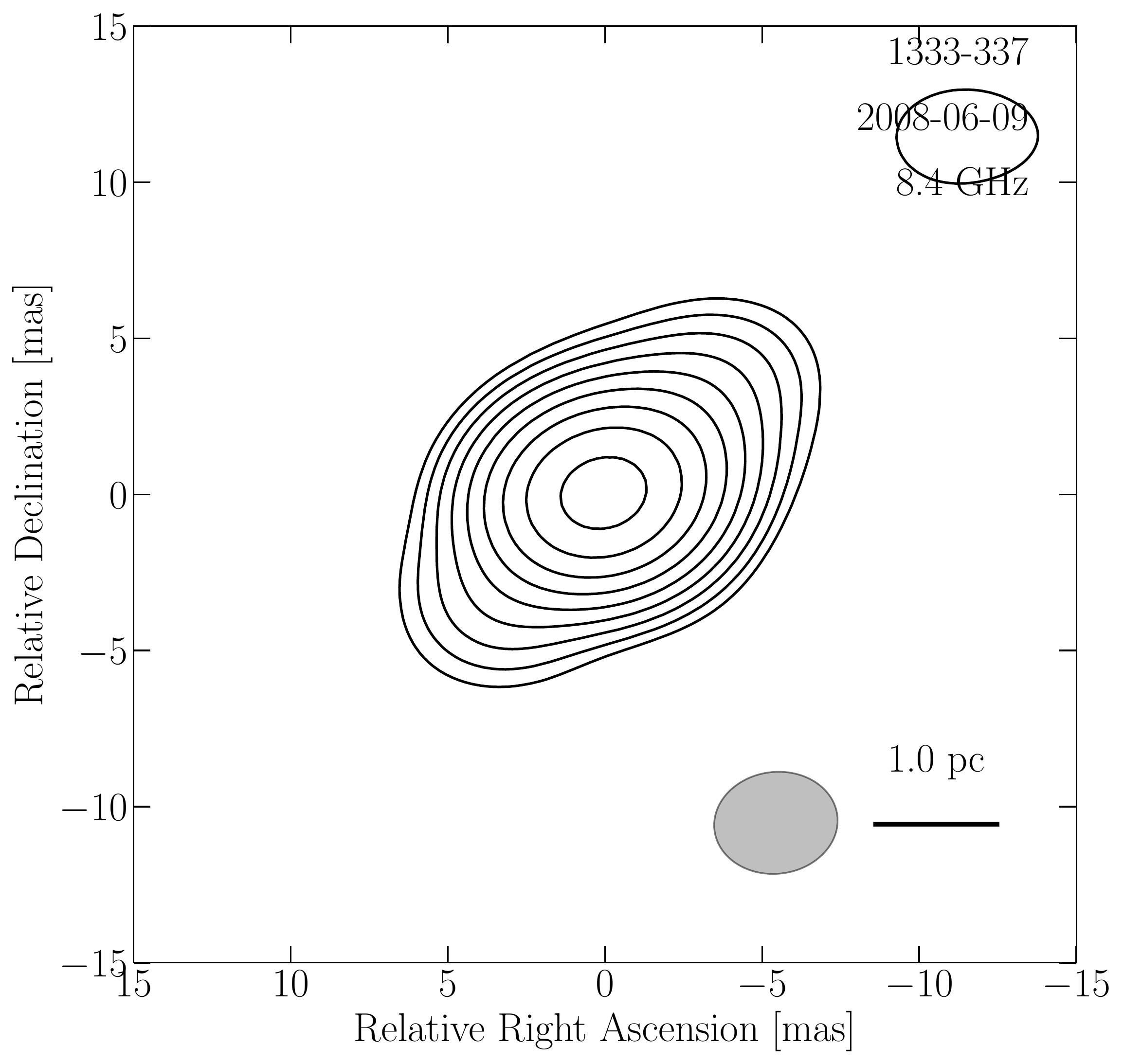}
\includegraphics[width=0.43\linewidth]{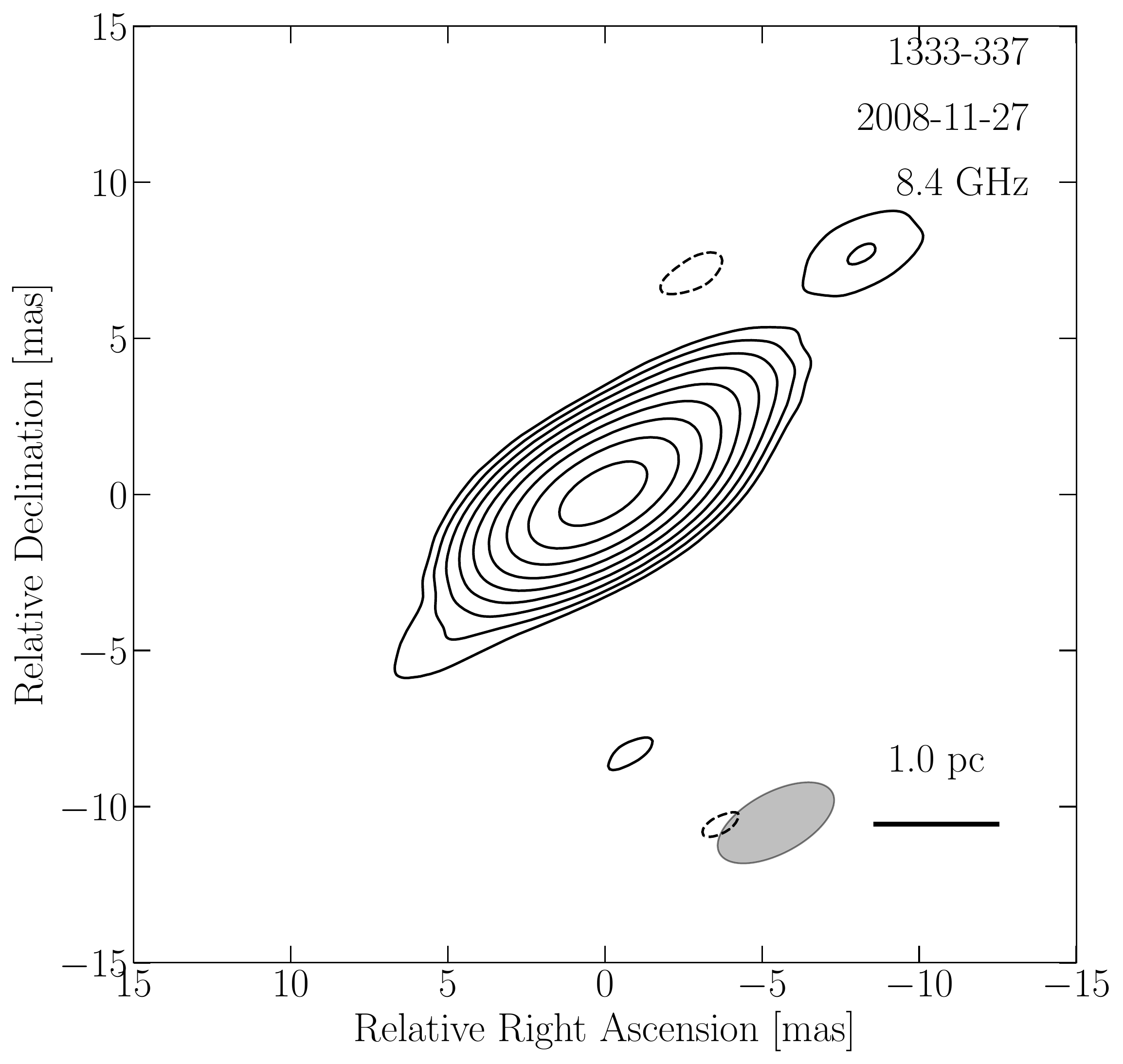}
\includegraphics[width=0.43\linewidth]{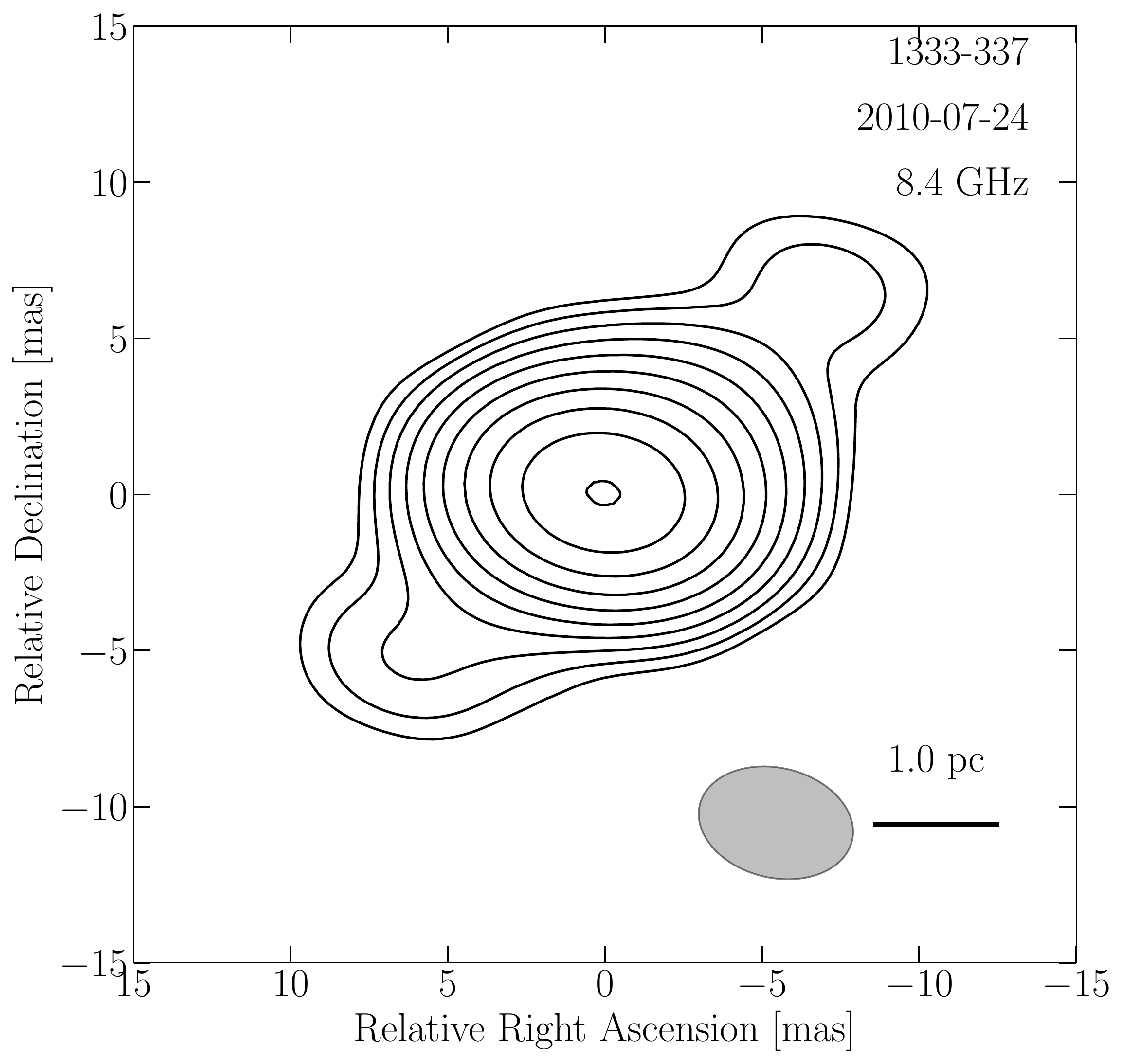}
\includegraphics[width=0.43\linewidth]{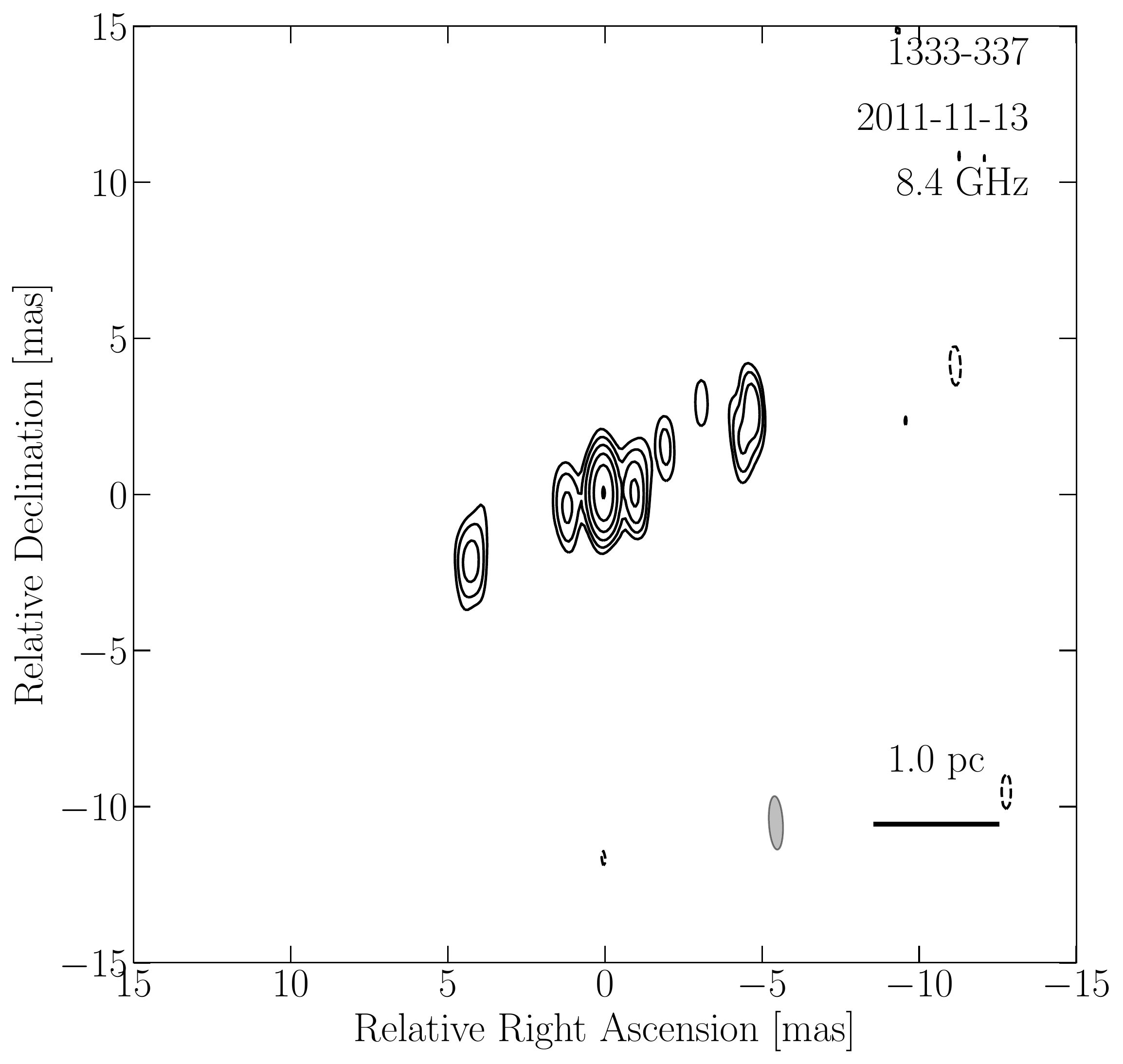}
\end{center}
\caption{Full-resolution images of IC\,4296. The grey ellipse
  represents the beam size, while the black line indicates the linear
  scale at the source's redshift. Contours increase in steps of two starting from 2, 2, 3, 3, 3 times the noise level in each map, from top left to bottom, respectively.}
\label{1333_fulla}
\end{figure*}
\begin{figure*}[!htbp]
\begin{center}
\includegraphics[width=0.43\linewidth]{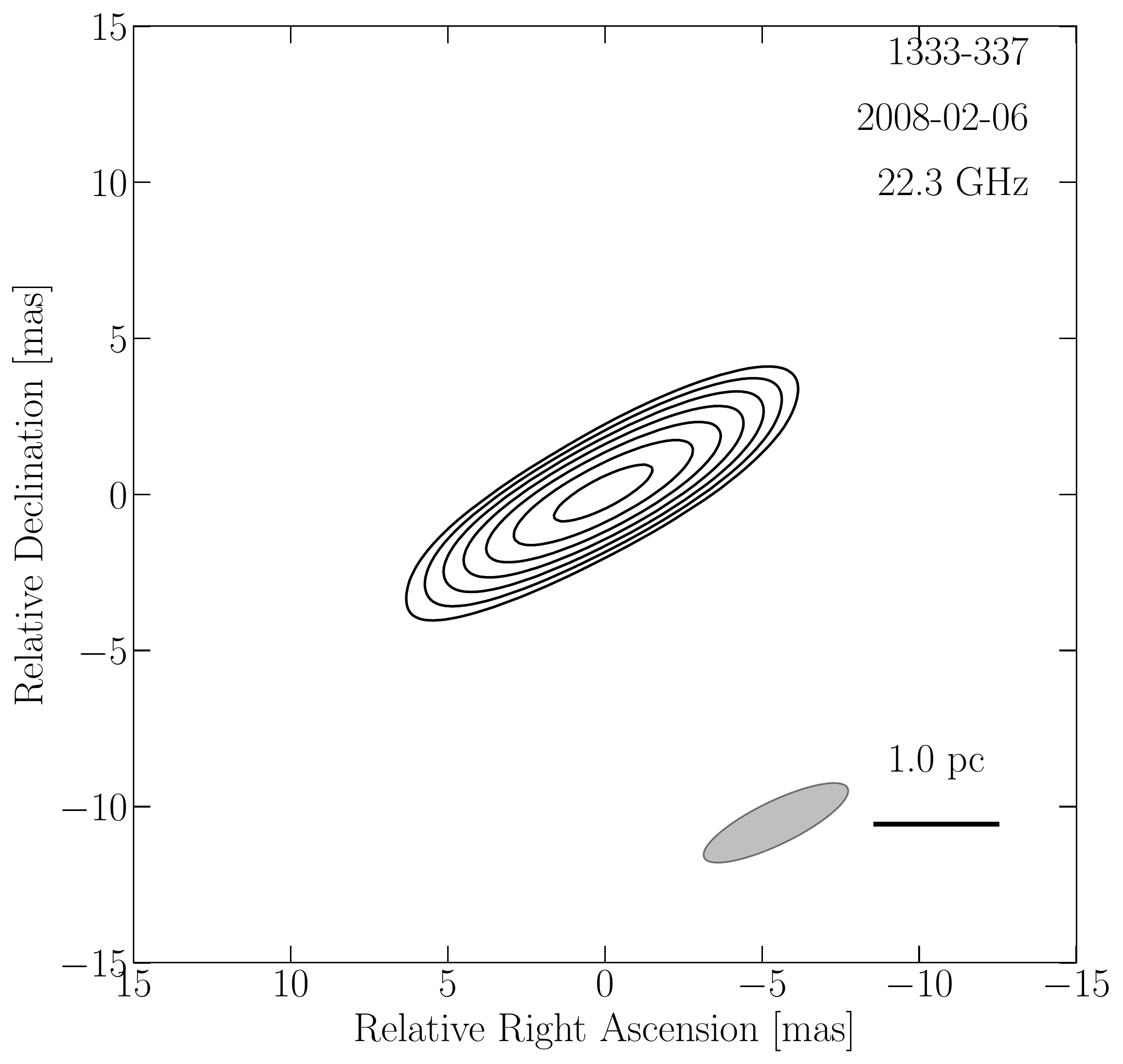}
\includegraphics[width=0.43\linewidth]{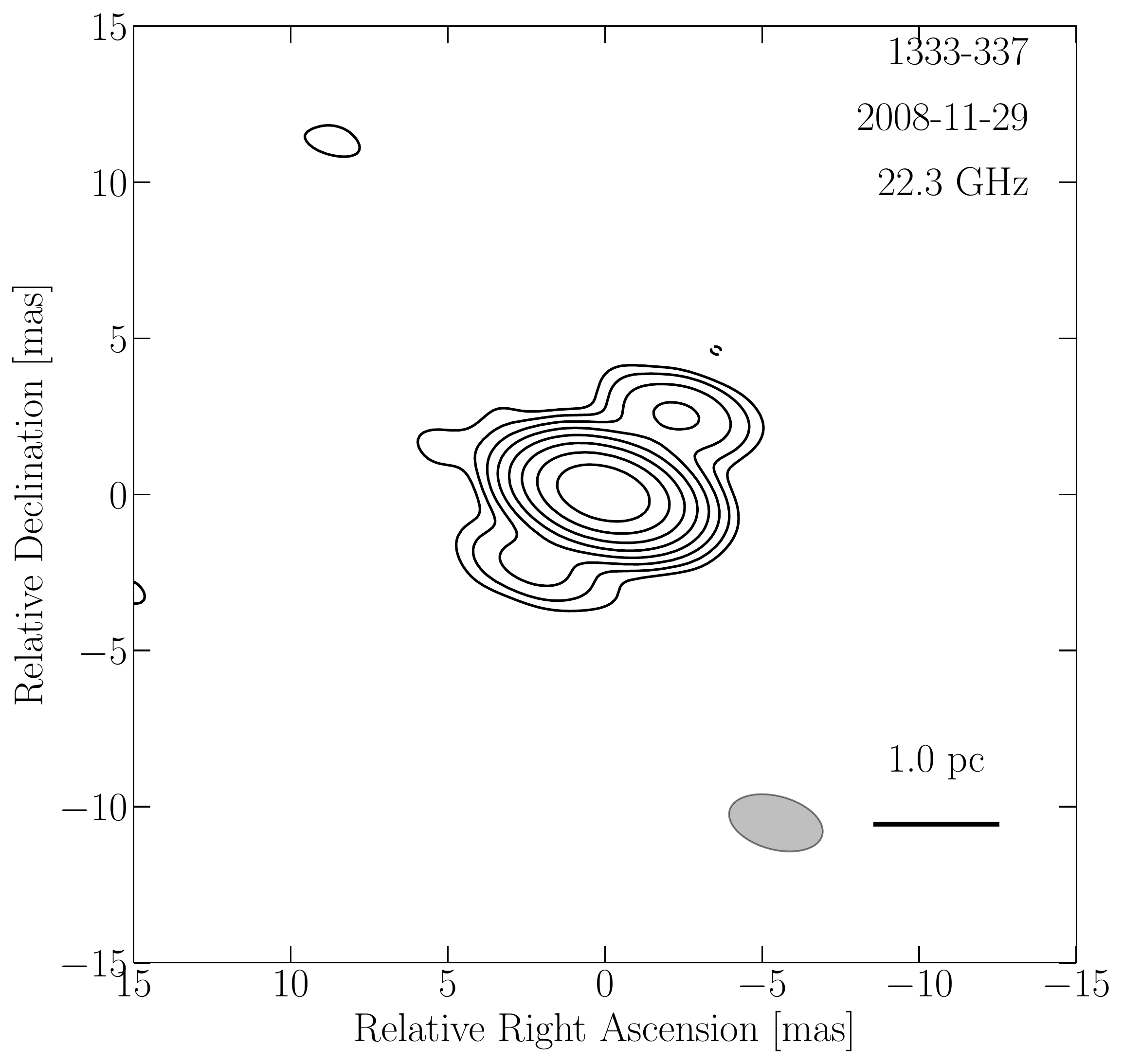}
\end{center}
\caption{Full-resolution images of IC\,4296. The grey ellipse
  represents the beam size, while the black line indicates the linear
  scale at the source's redshift. Contours increase in steps of two starting from 3, 1 times the noise level in each map, from left to right, respectively.}
\label{1333_fullb}
\end{figure*}
\begin{figure*}[!htbp]
\begin{center}
\includegraphics[width=0.43\linewidth]{Figures/Maps_new/1549-790_2008-02-07.pdf}
\includegraphics[width=0.43\linewidth]{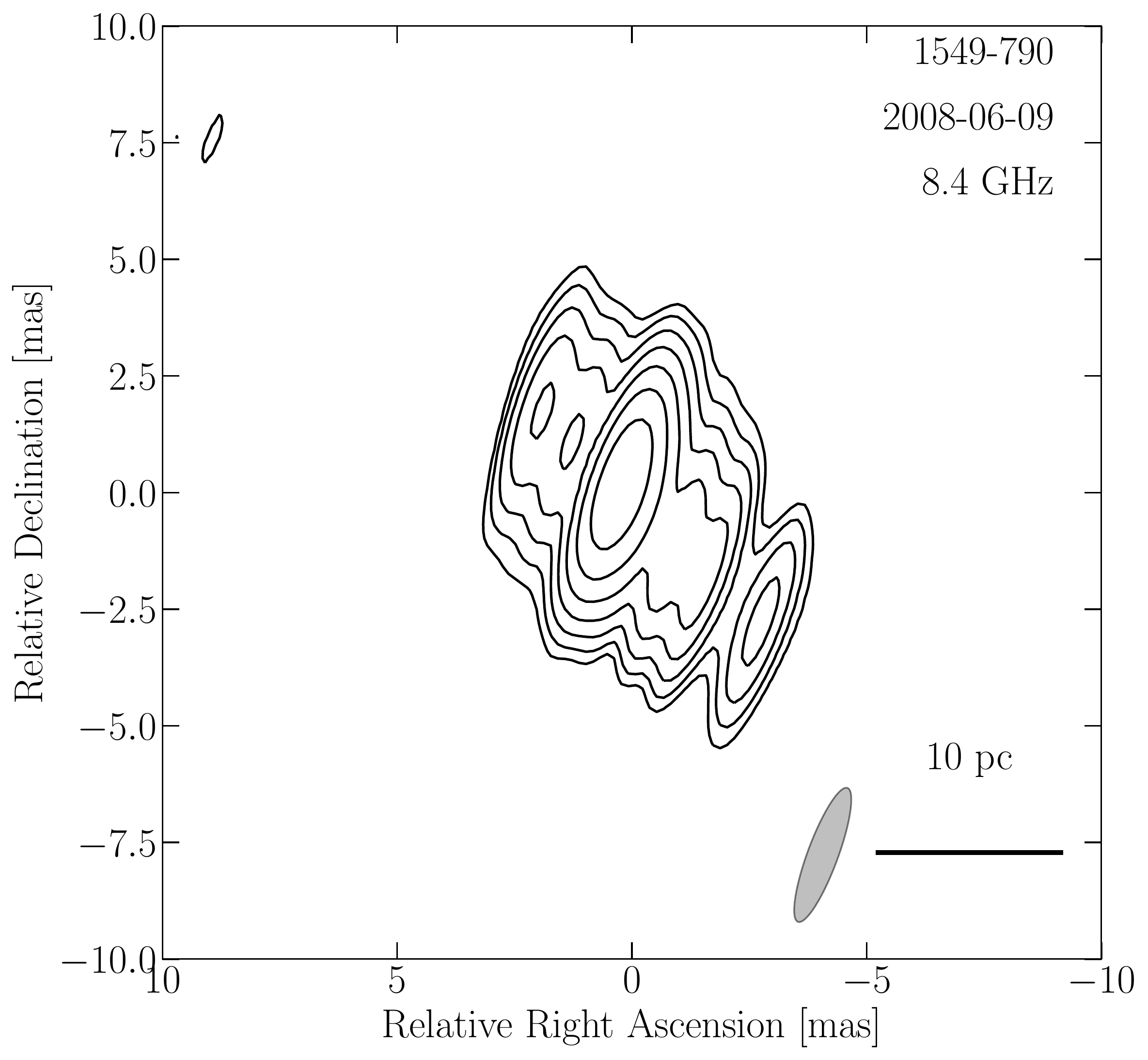}
\includegraphics[width=0.43\linewidth]{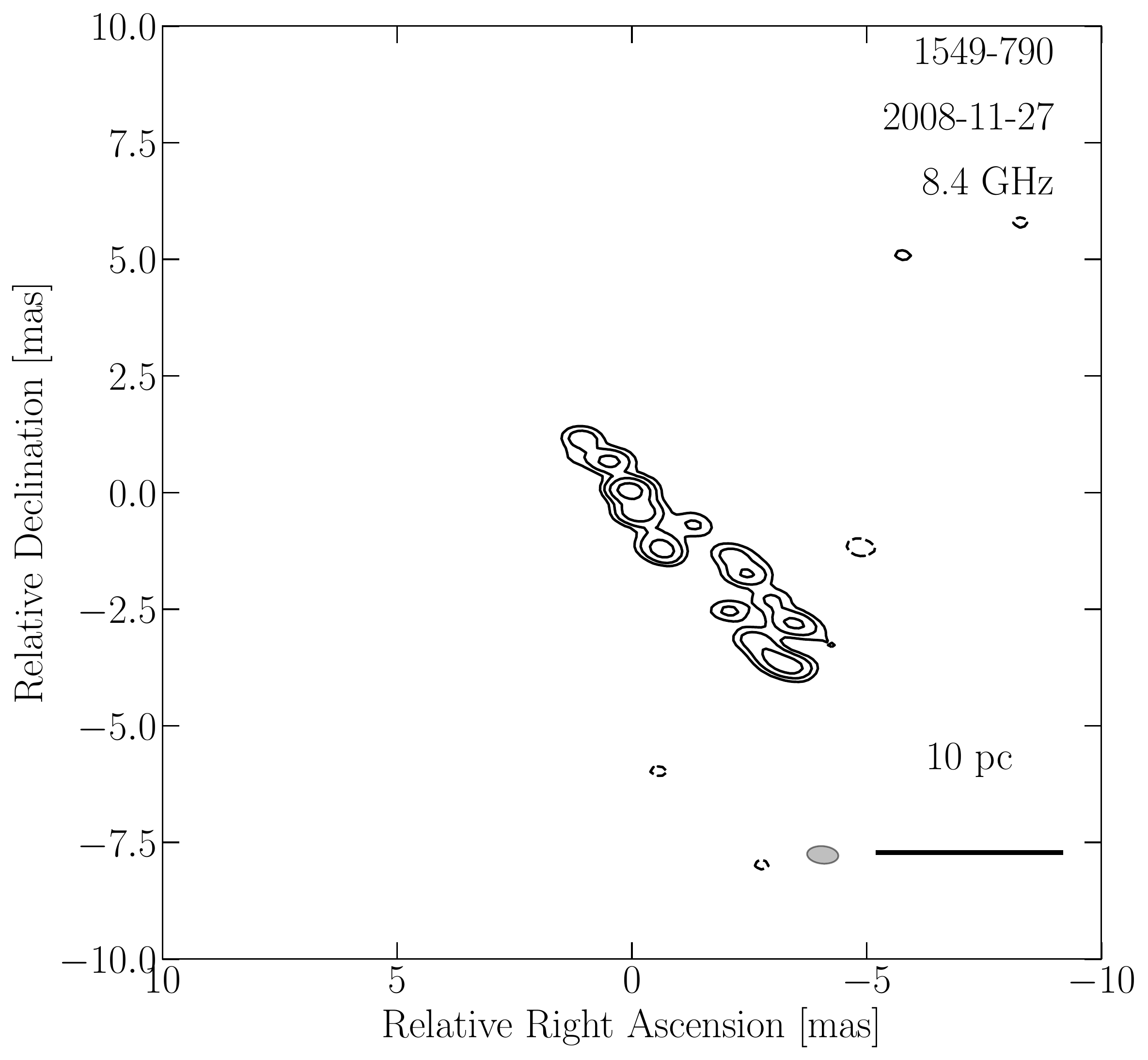}
\includegraphics[width=0.43\linewidth]{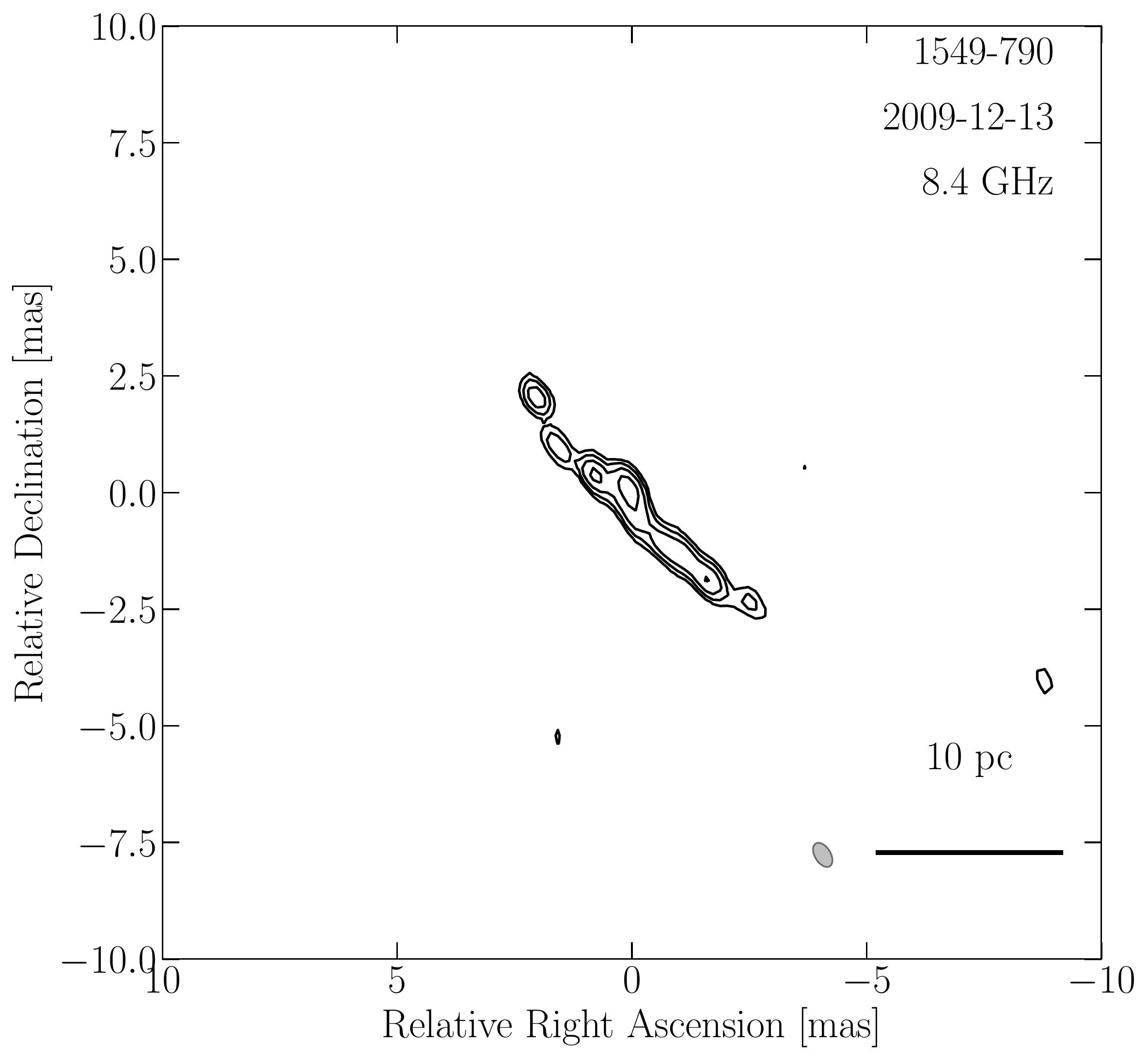}
\includegraphics[width=0.43\linewidth]{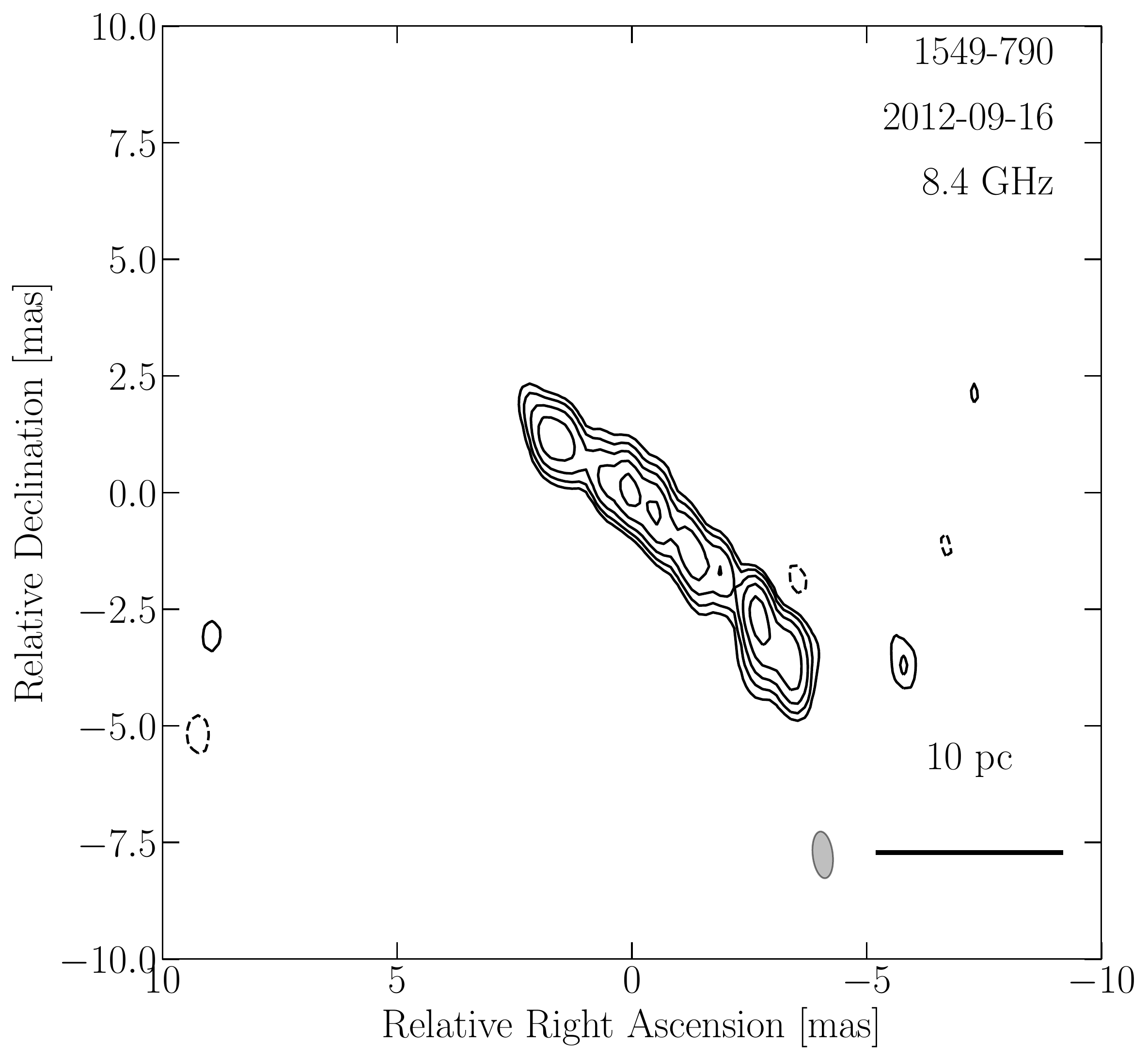}
\end{center}
\caption{Full-resolution images of PKS\,1549$-$79. The grey ellipse
  represents the beam size, while the black line indicates the linear
  scale at the source's redshift. Contours increase in steps of two starting from 1.2, 4, 0.6, 1.5, 0.5 times the noise level in each map, from top left to bottom, respectively.}
\label{1549_fulla}
\end{figure*}
\begin{figure*}[!htbp]
\begin{center}
\includegraphics[width=0.43\linewidth]{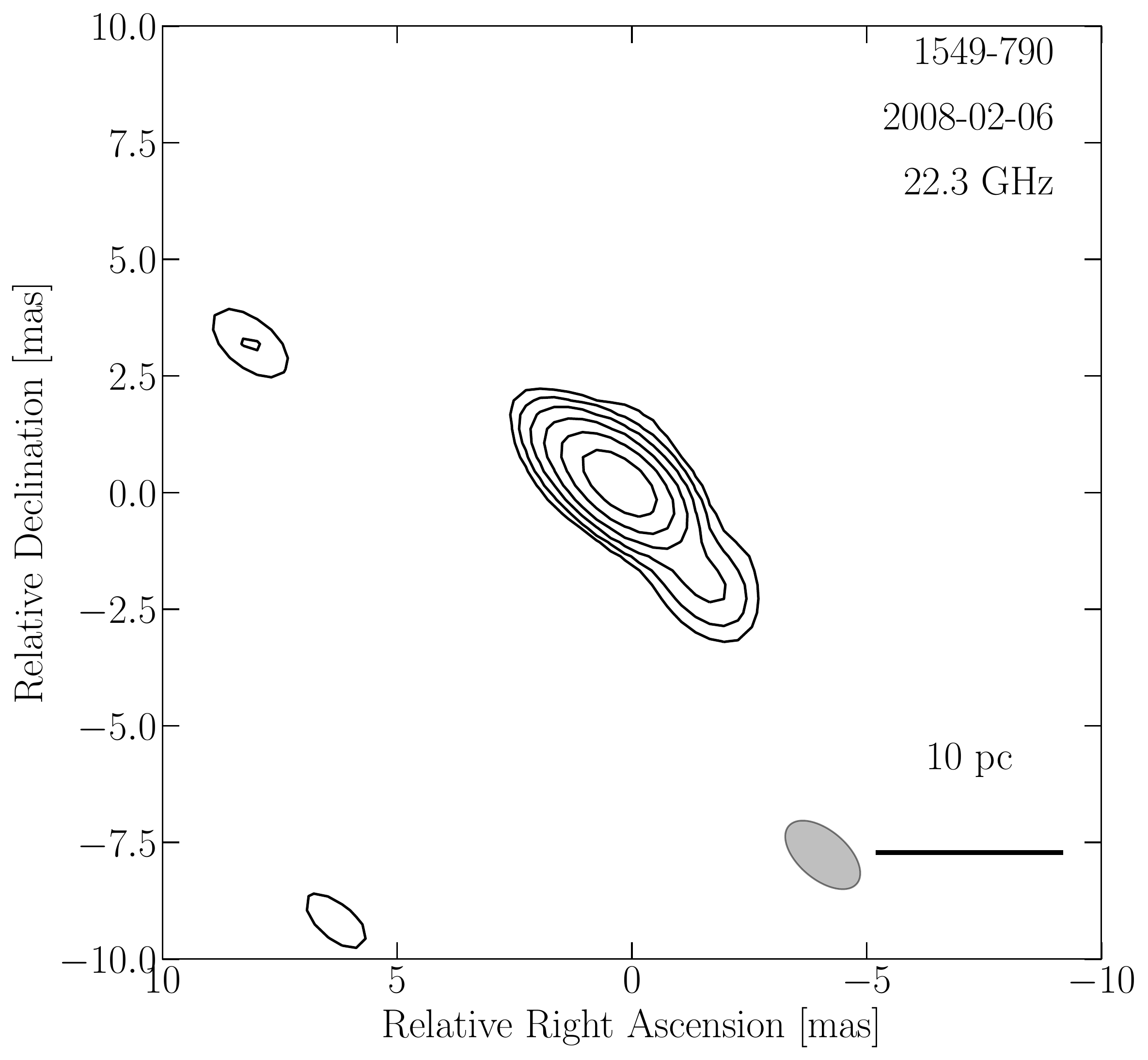}
\includegraphics[width=0.43\linewidth]{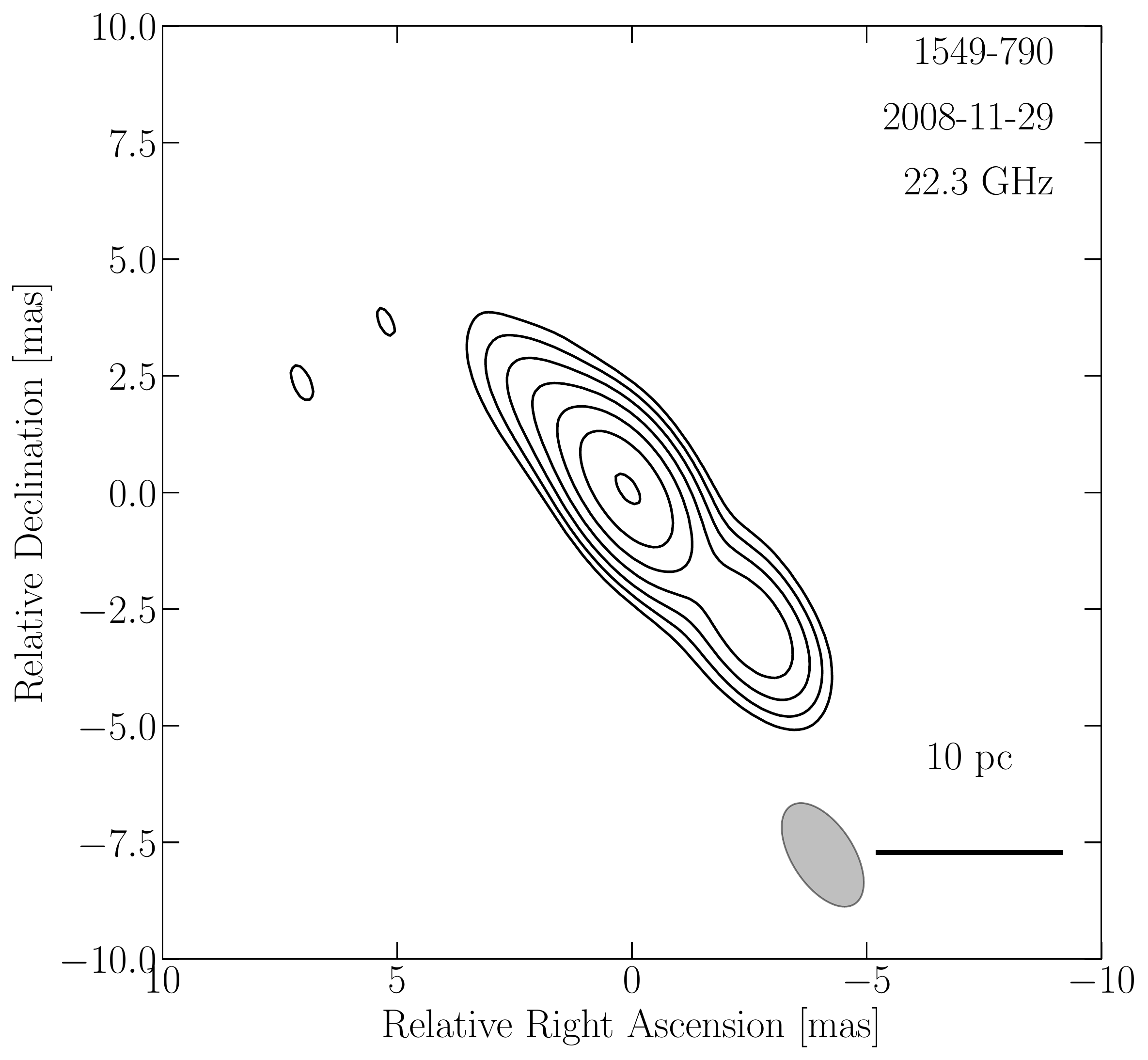}
\end{center}
\caption{Full-resolution images of PKS\,1549$-$79. The grey ellipse
  represents the beam size, while the black line indicates the linear
  scale at the source's redshift. Contours increase in steps of two starting from 1.2, 3 times the noise level in each map, from left to right, respectively.}
\label{1549_fullb}
\end{figure*}
\begin{figure*}[!htbp]
\begin{center}
\includegraphics[width=0.43\linewidth]{Figures/Maps_new/1733-565_2008-02-07.pdf}
\includegraphics[width=0.43\linewidth]{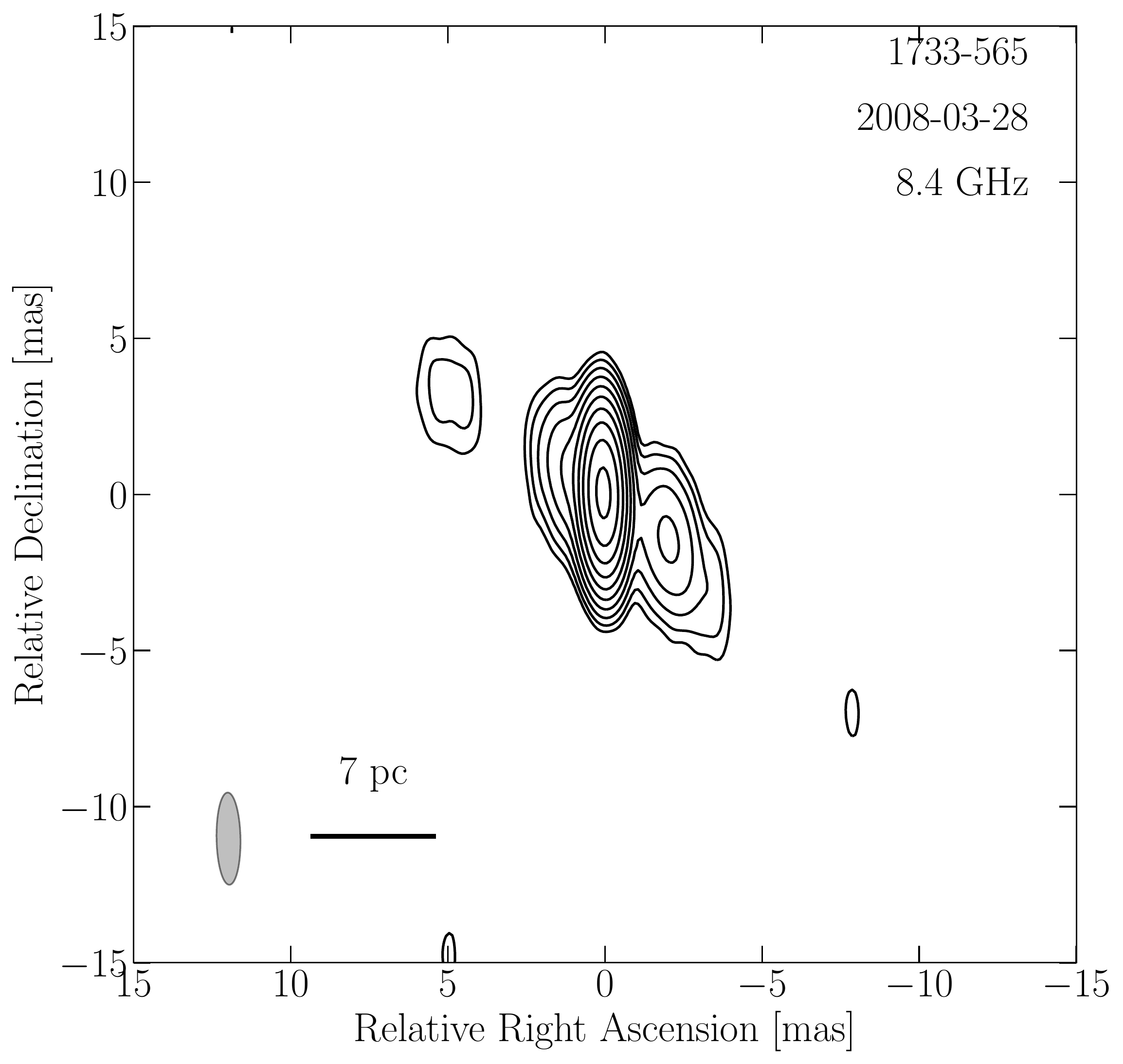}
\includegraphics[width=0.43\linewidth]{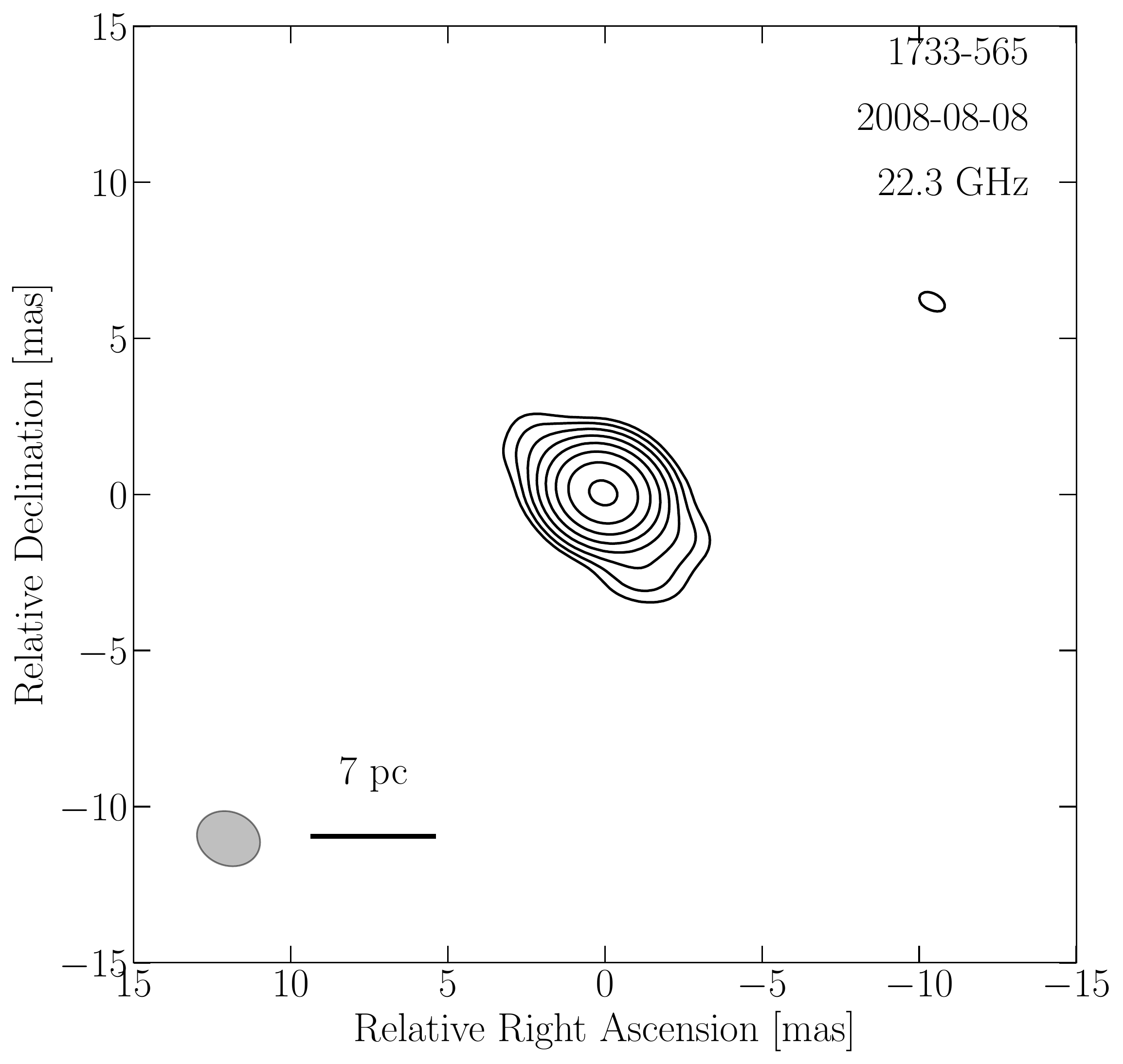}
\includegraphics[width=0.43\linewidth]{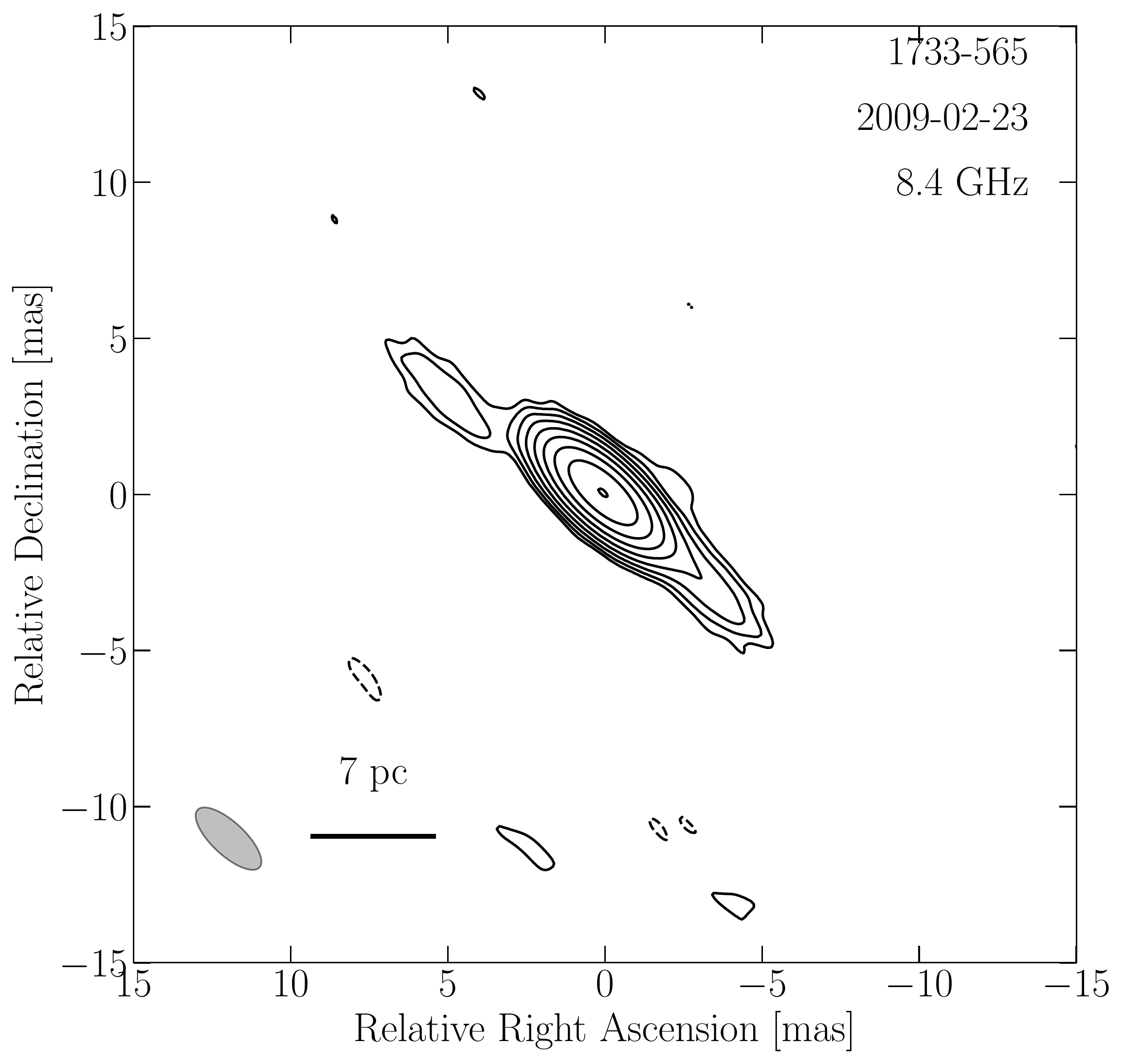}
\includegraphics[width=0.43\linewidth]{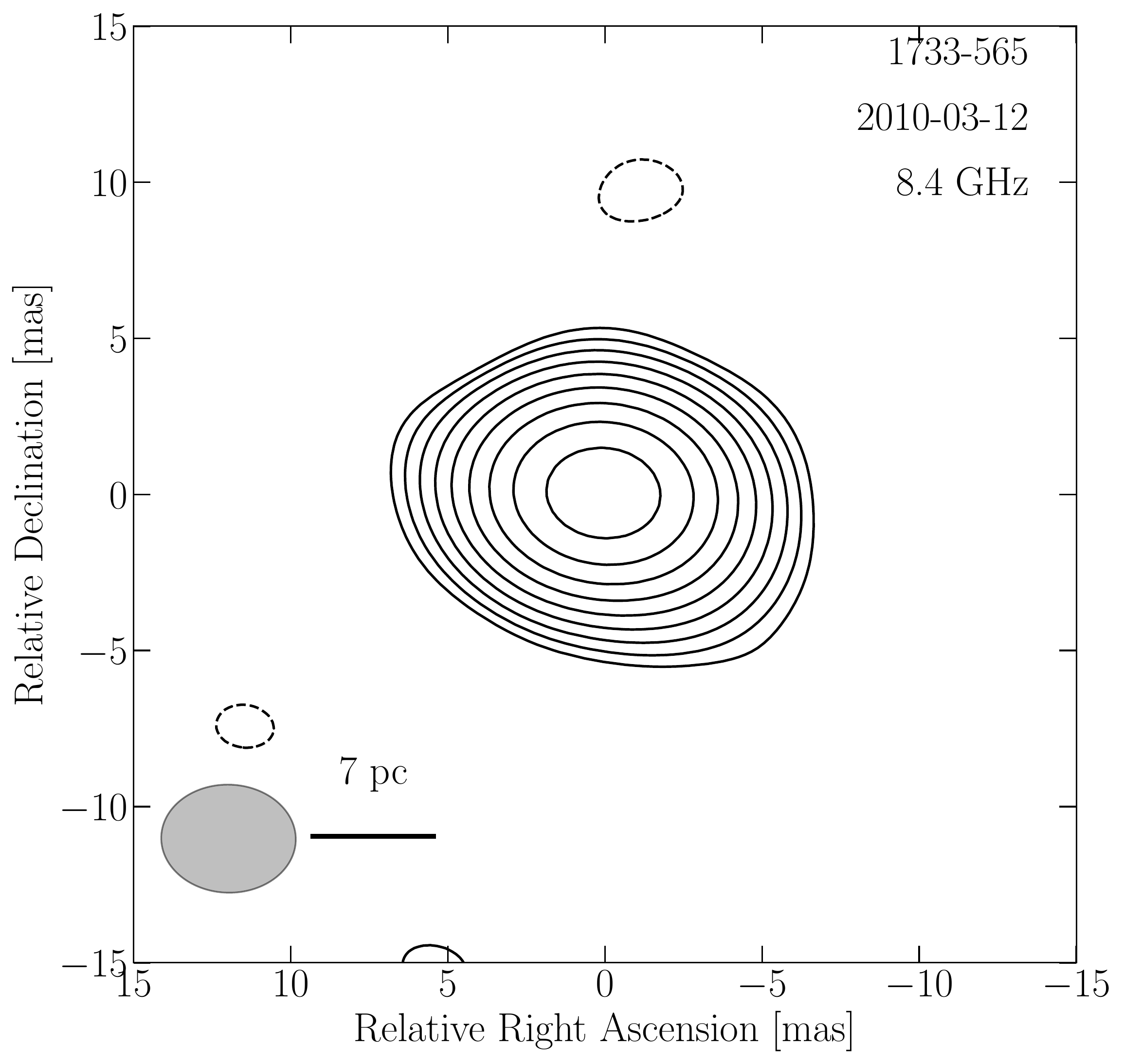}
\includegraphics[width=0.43\linewidth]{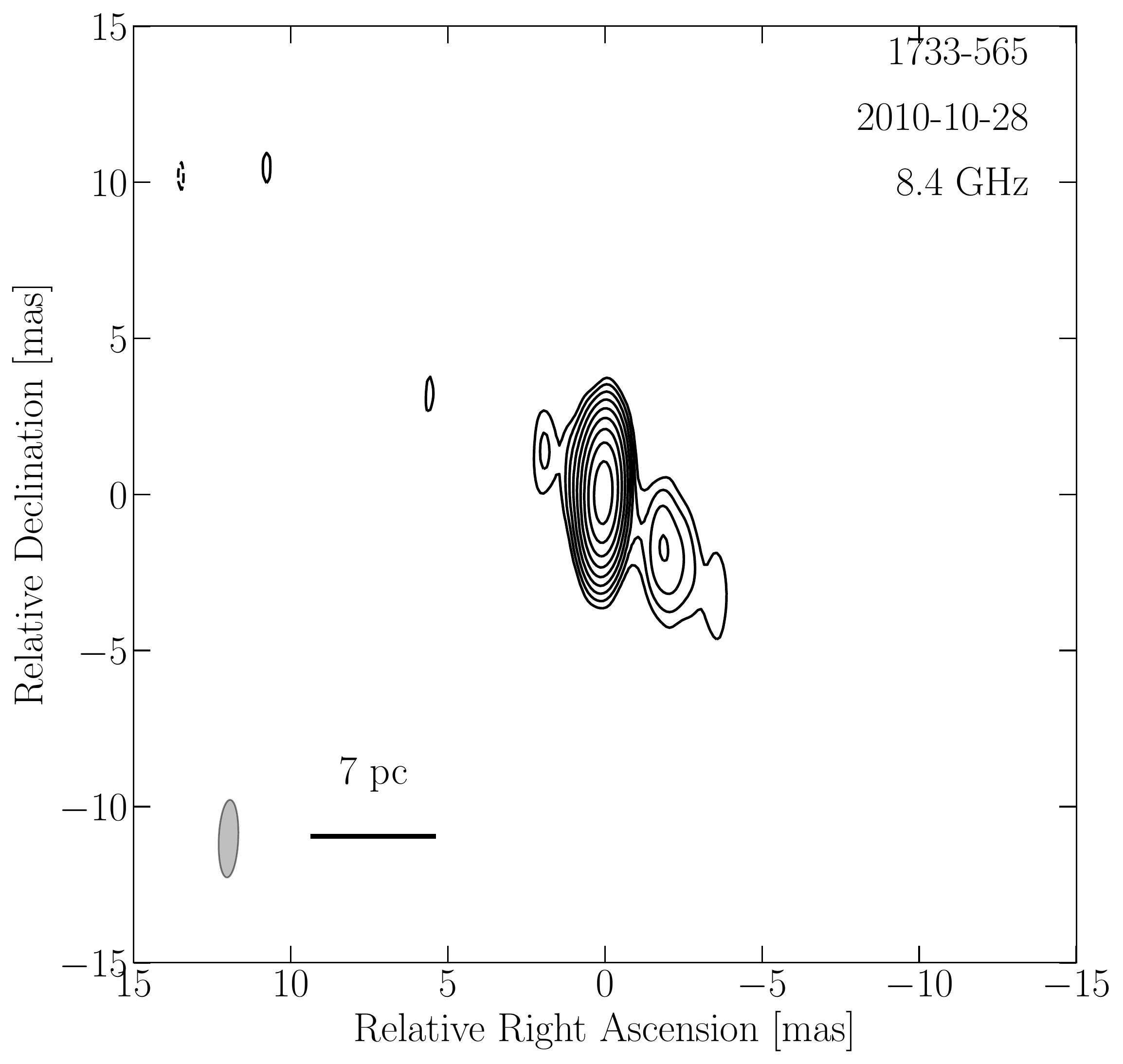}
\end{center}
\caption{Full-resolution images of PKS\,1733$-$56. The grey ellipse
  represents the beam size, while the black line indicates the linear
  scale at the source's redshift. Contours increase in steps of two starting from 1.8 times the noise level in each map.}
\label{1733_fulla}
\end{figure*}
\begin{figure*}[!htbp]
\begin{center}
\includegraphics[width=0.43\linewidth]{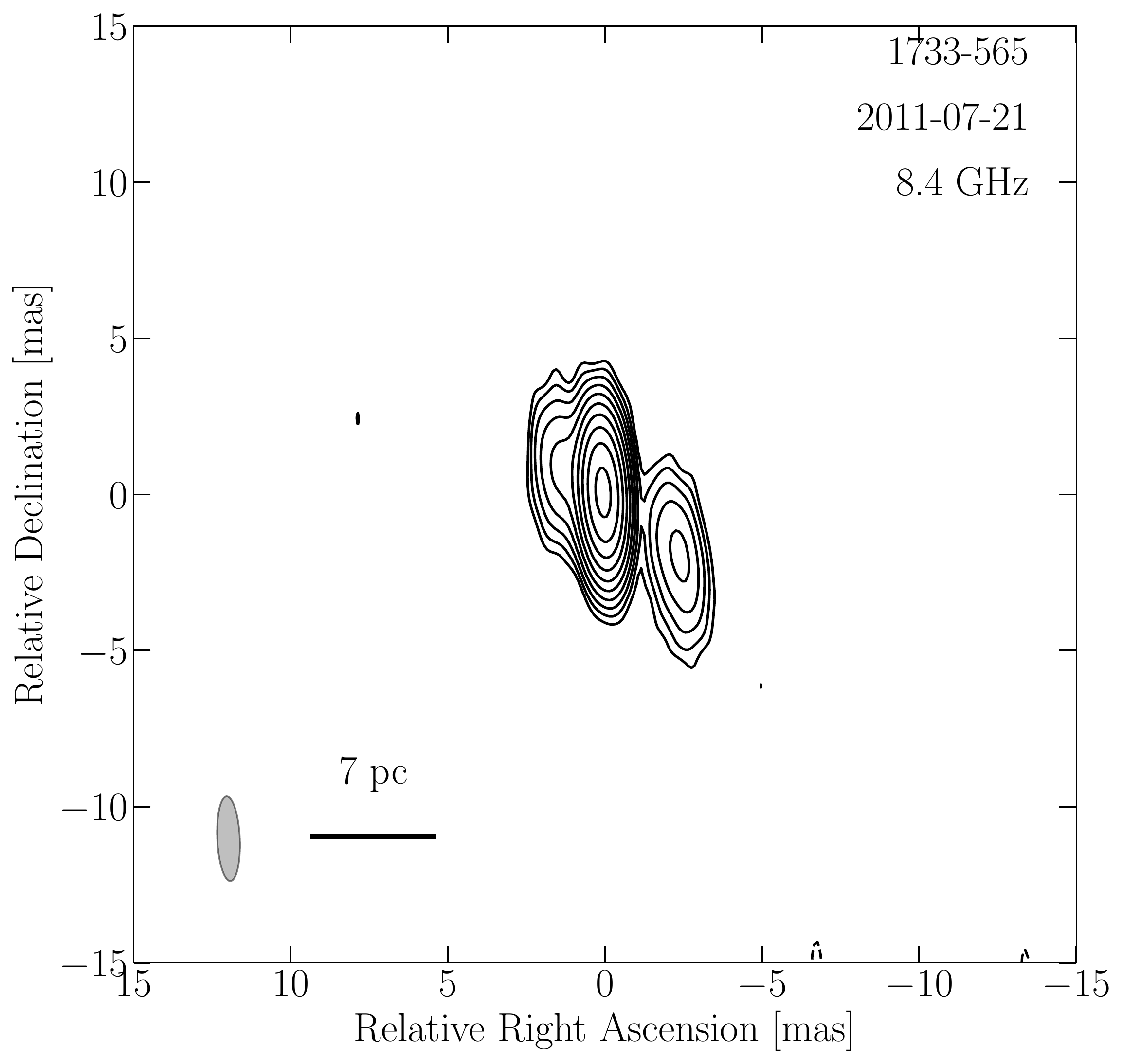}
\includegraphics[width=0.43\linewidth]{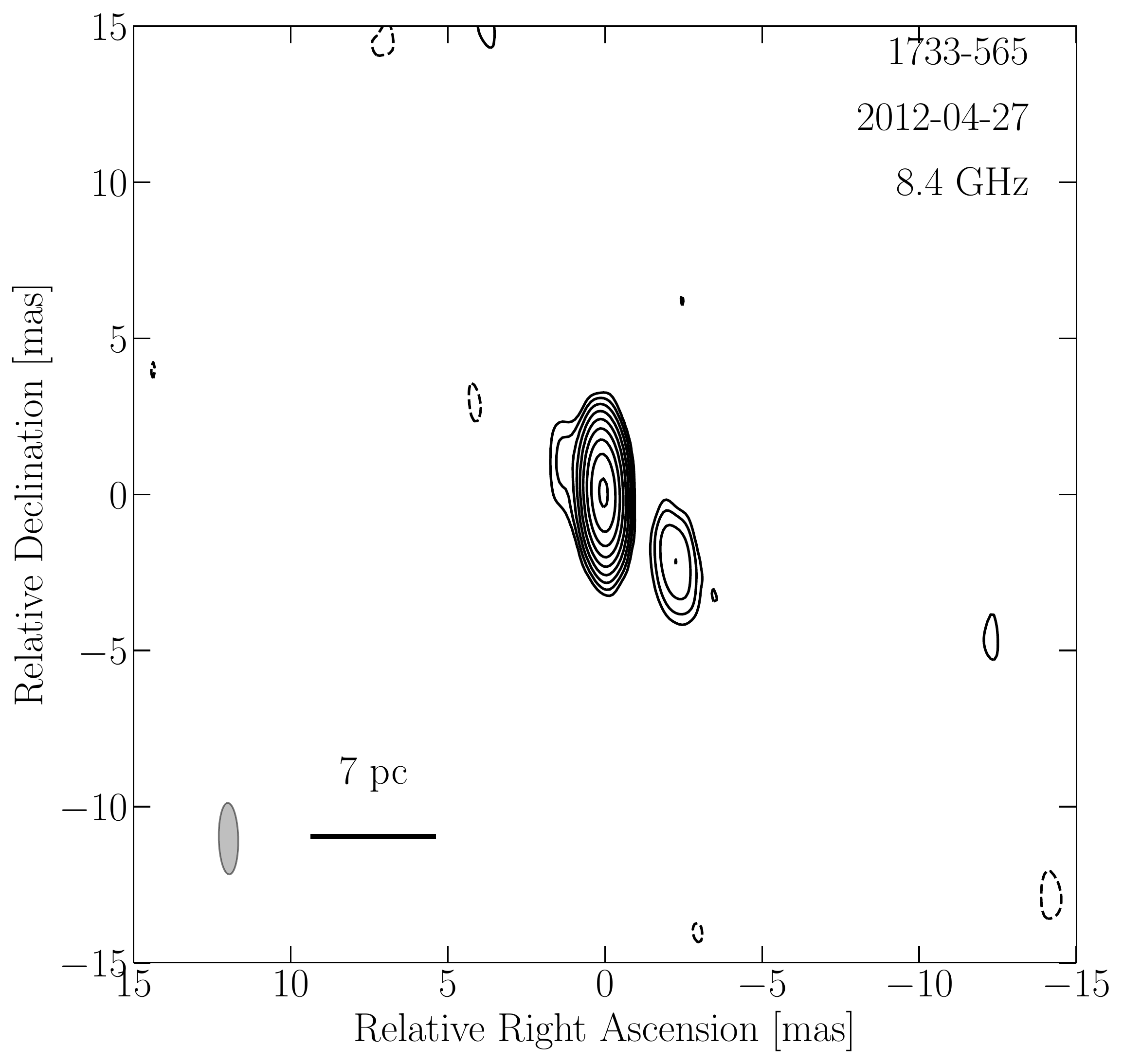}
\includegraphics[width=0.43\linewidth]{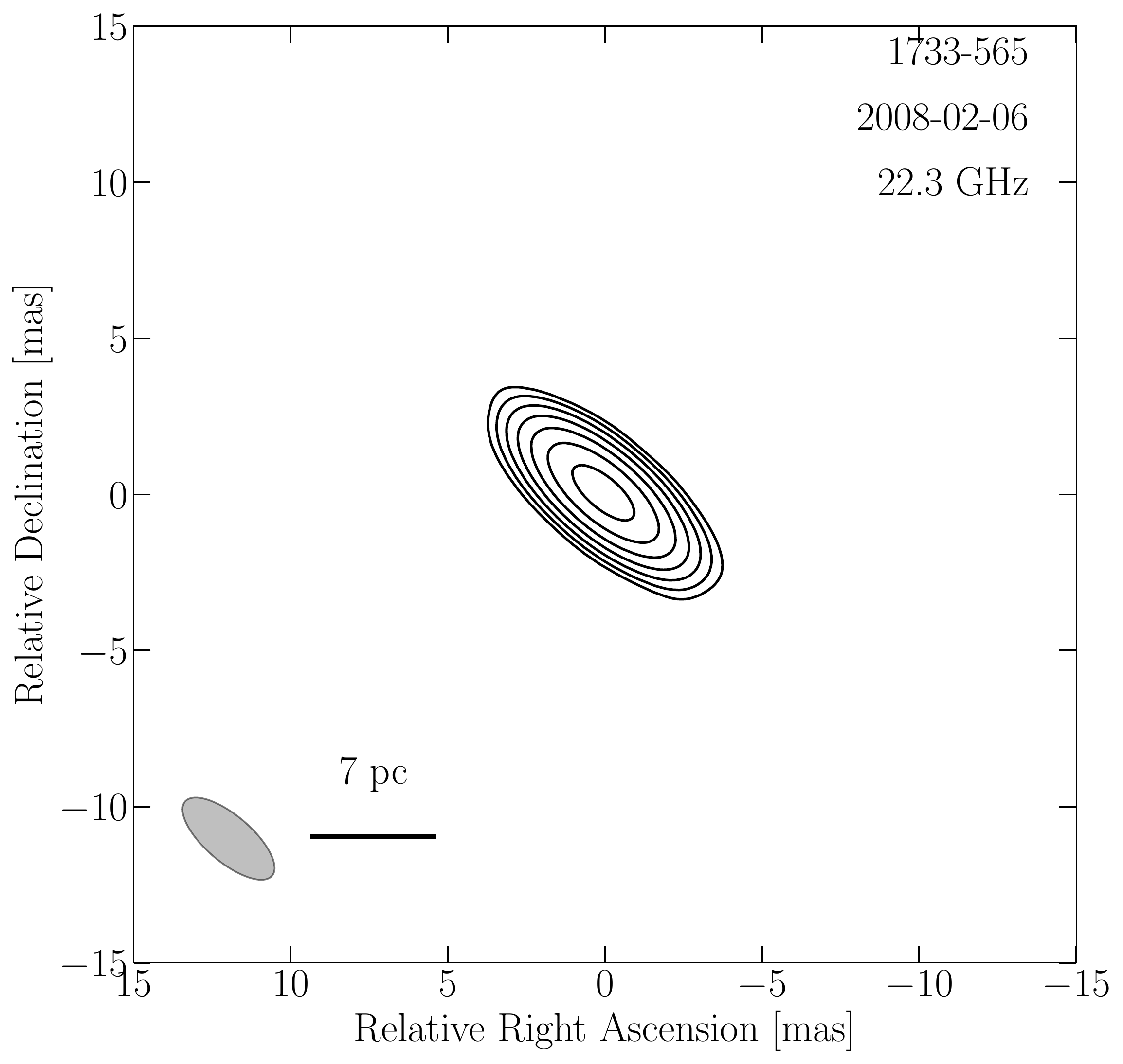}
\includegraphics[width=0.43\linewidth]{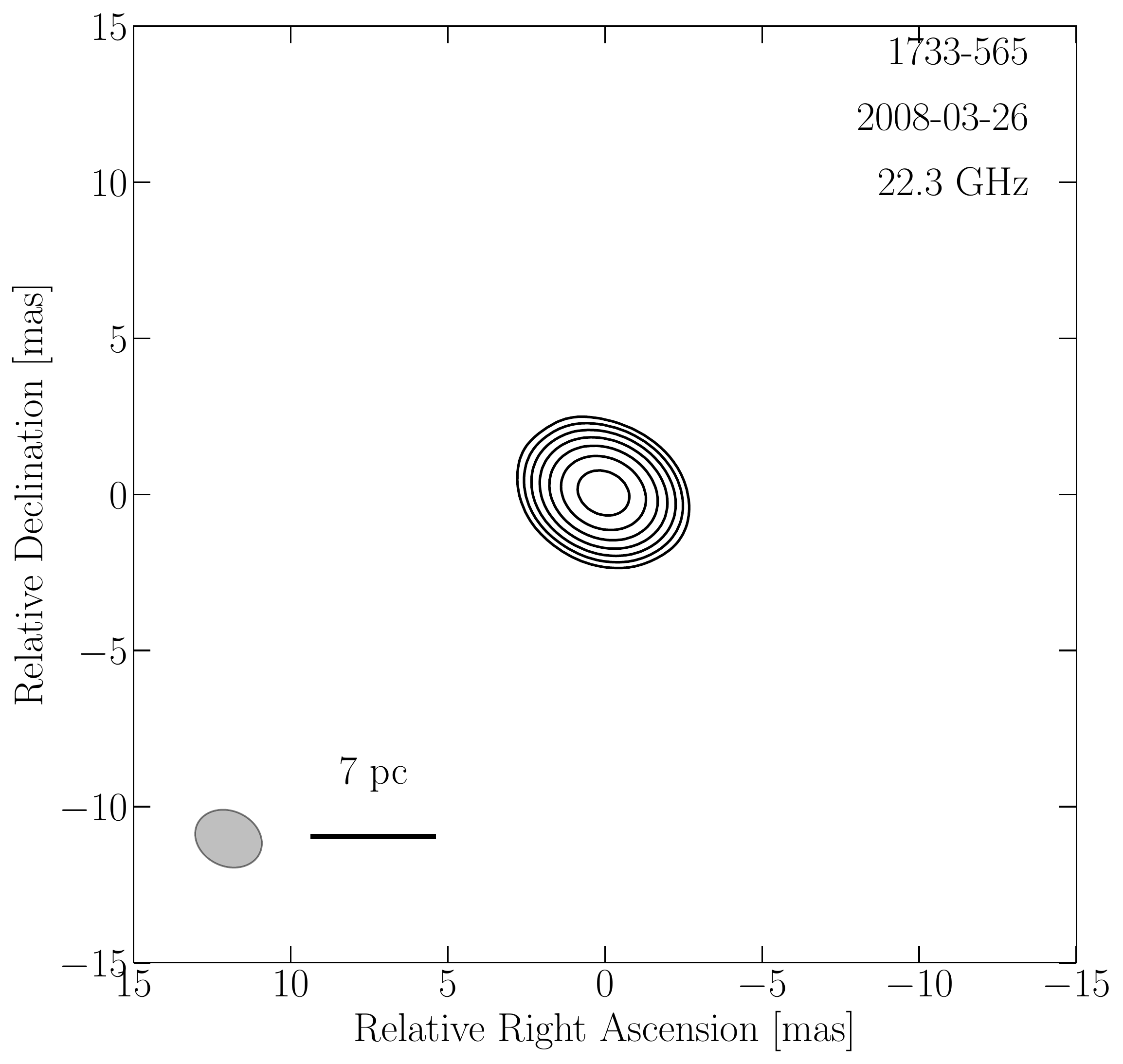}
\includegraphics[width=0.43\linewidth]{Figures/Maps_new/1733-565_2008-08-08.pdf}
\end{center}
\caption{Full-resolution images of PKS\,1733$-$56 (continued). The grey ellipse
  represents the beam size, while the black line indicates the linear
  scale at the source's redshift. Contours increase in steps of two starting from 1.8 times the noise level in each map.}
\label{1733_fullb}
\end{figure*}
\begin{figure*}[!htbp]
\begin{center}
\includegraphics[width=0.43\linewidth]{Figures/Maps_new/1814-637_2008-02-07.pdf}
\includegraphics[width=0.43\linewidth]{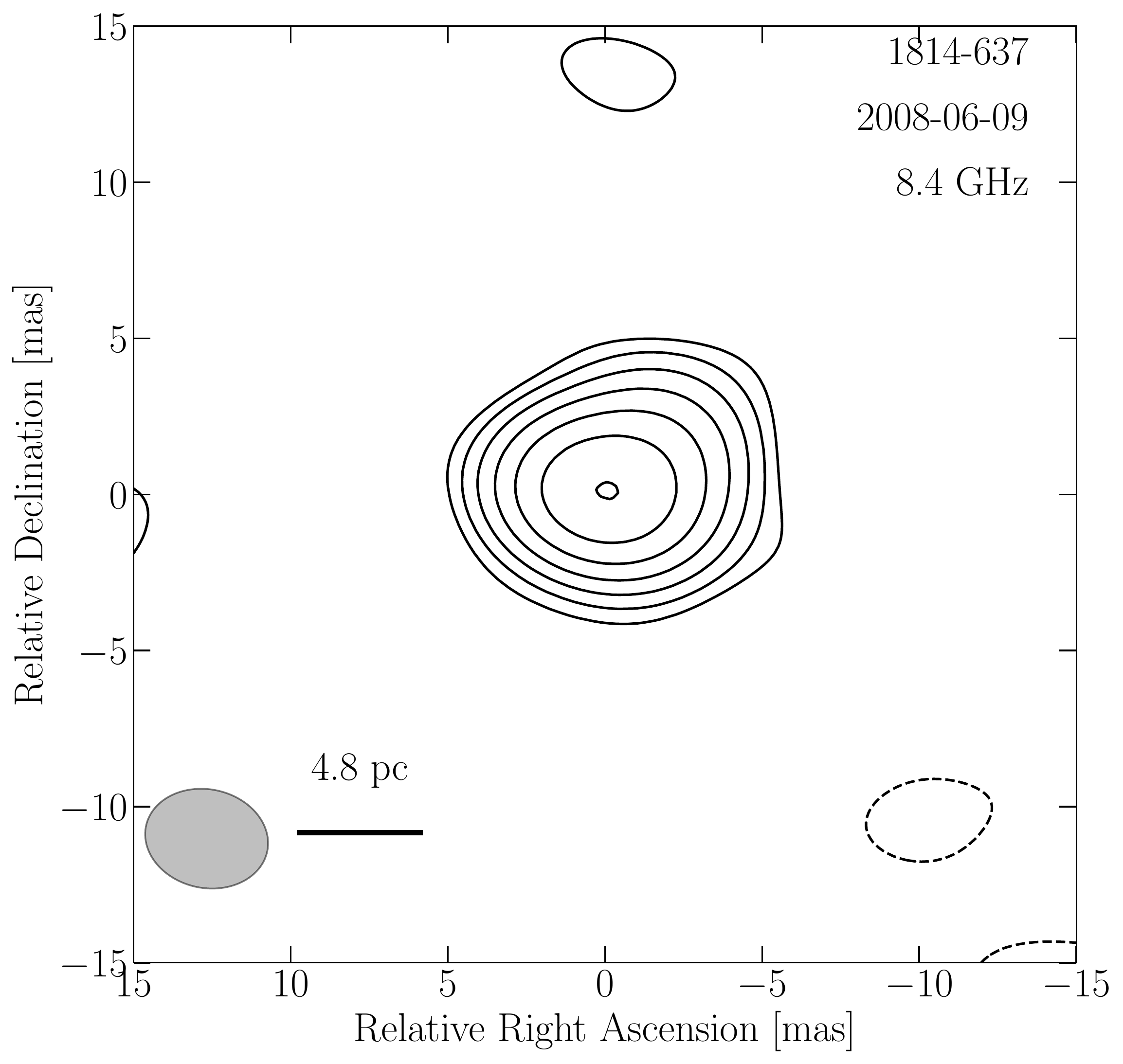}
\includegraphics[width=0.43\linewidth]{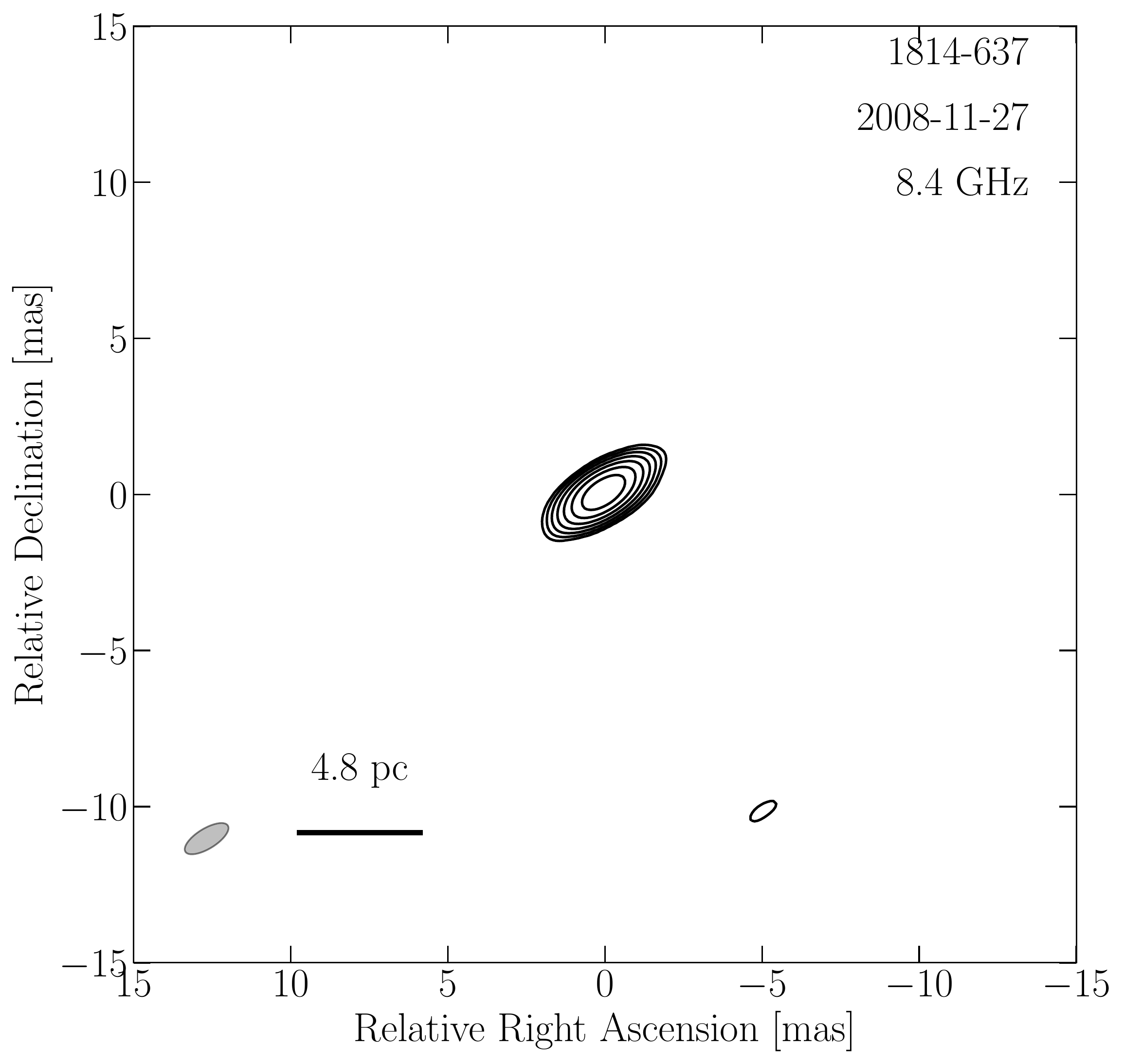}
\includegraphics[width=0.43\linewidth]{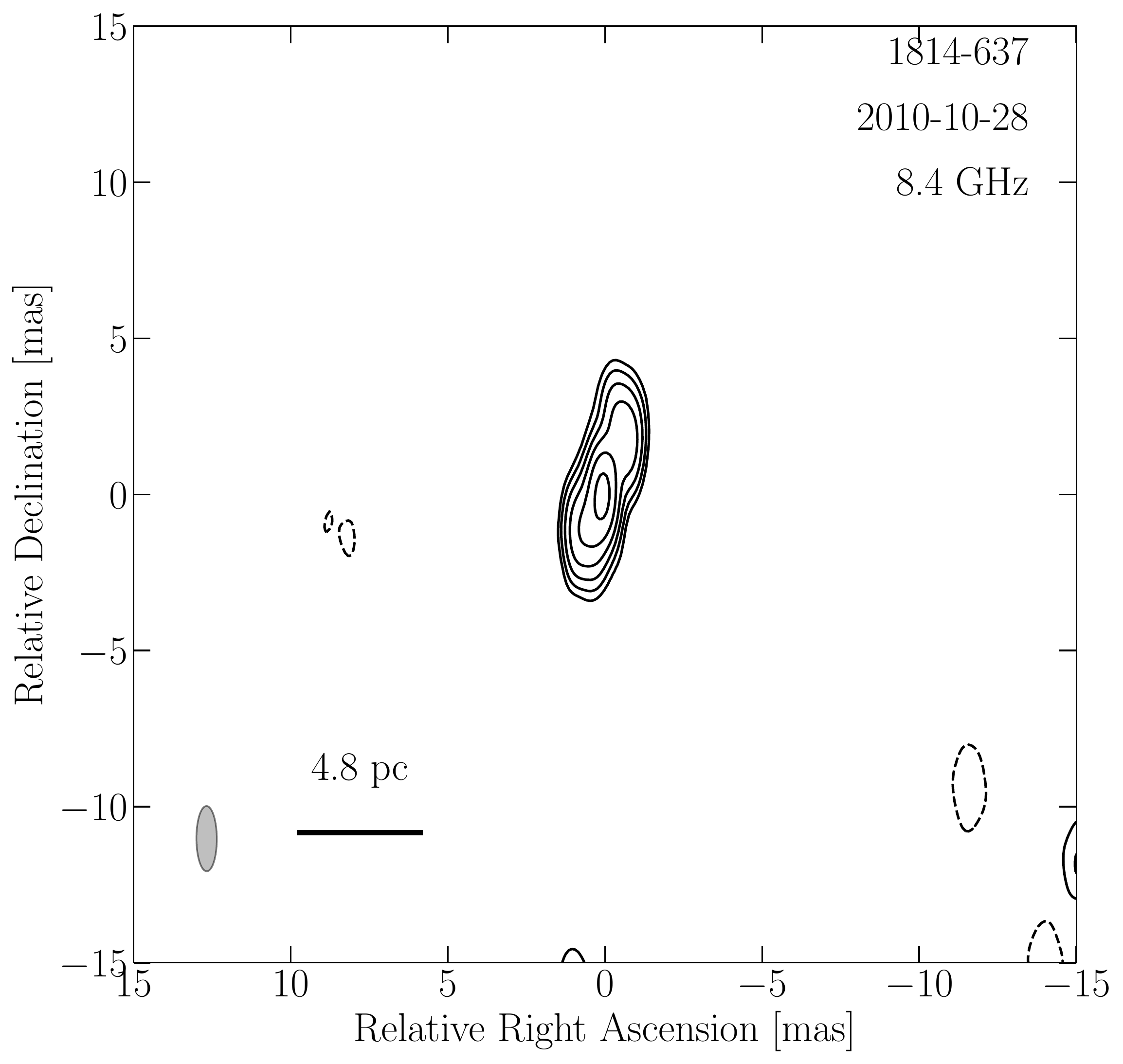}
\includegraphics[width=0.43\linewidth]{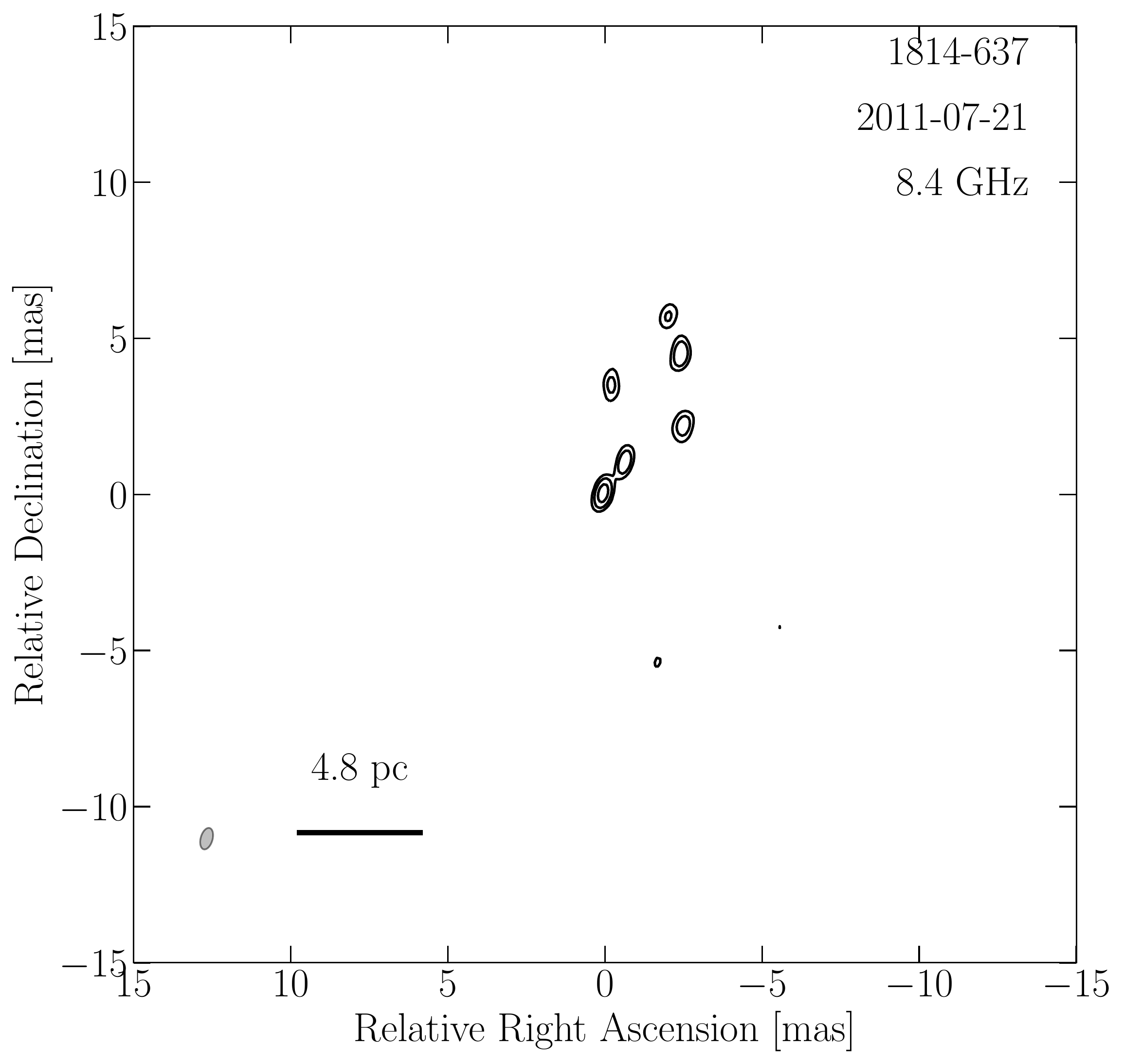}
\end{center}
\caption{Full-resolution images of PKS\,1814$-$63. The grey ellipse
  represents the beam size, while the black line indicates the linear
  scale at the source's redshift. Contours increase in steps of two starting from 9, 22, 5, 12, 3 times the noise level in each map, from top left to bottom, respectively.}
\label{1814_full}
\end{figure*}
\begin{figure*}[!htbp]
\begin{center}
\includegraphics[width=0.43\linewidth]{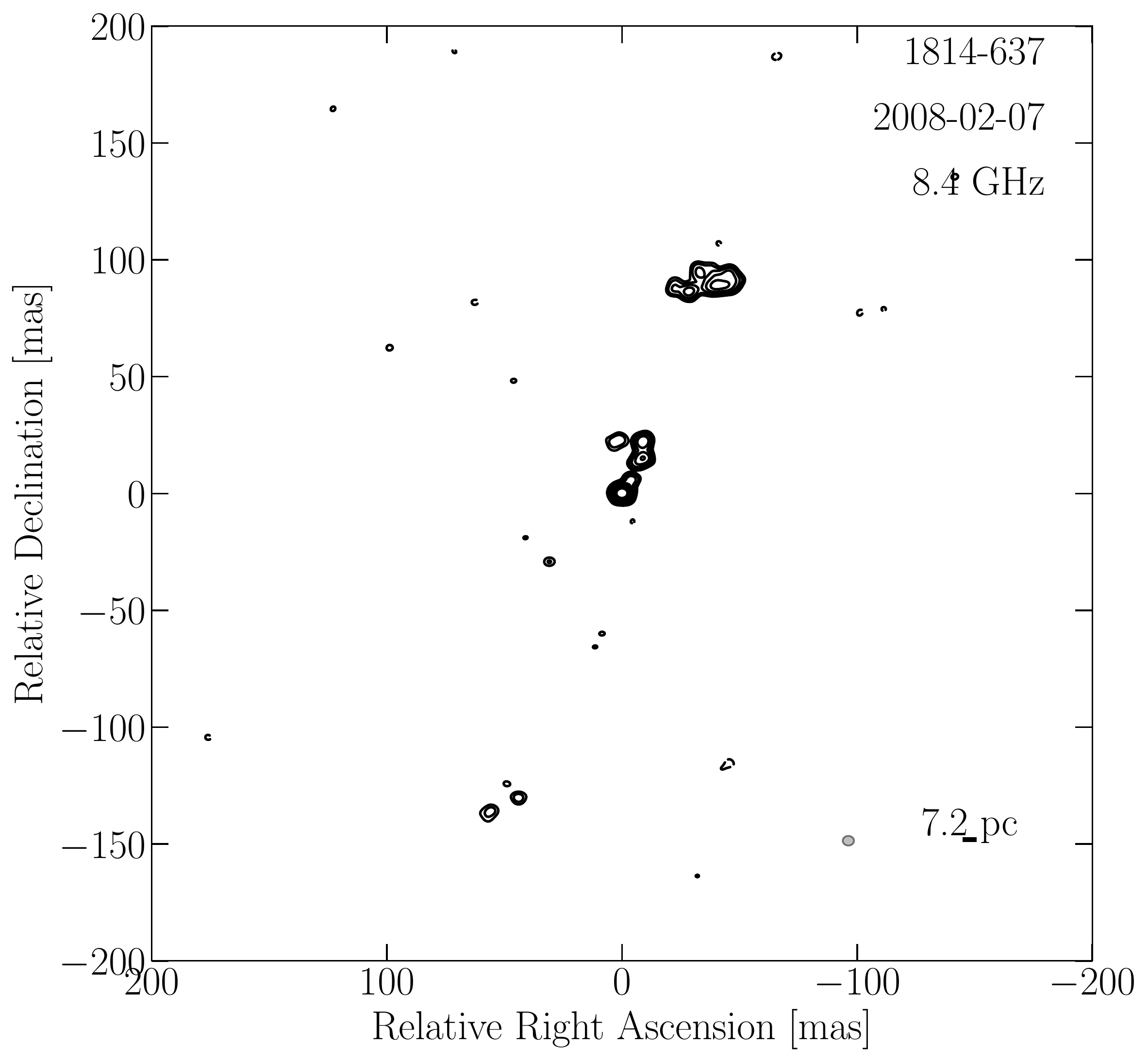}
\includegraphics[width=0.43\linewidth]{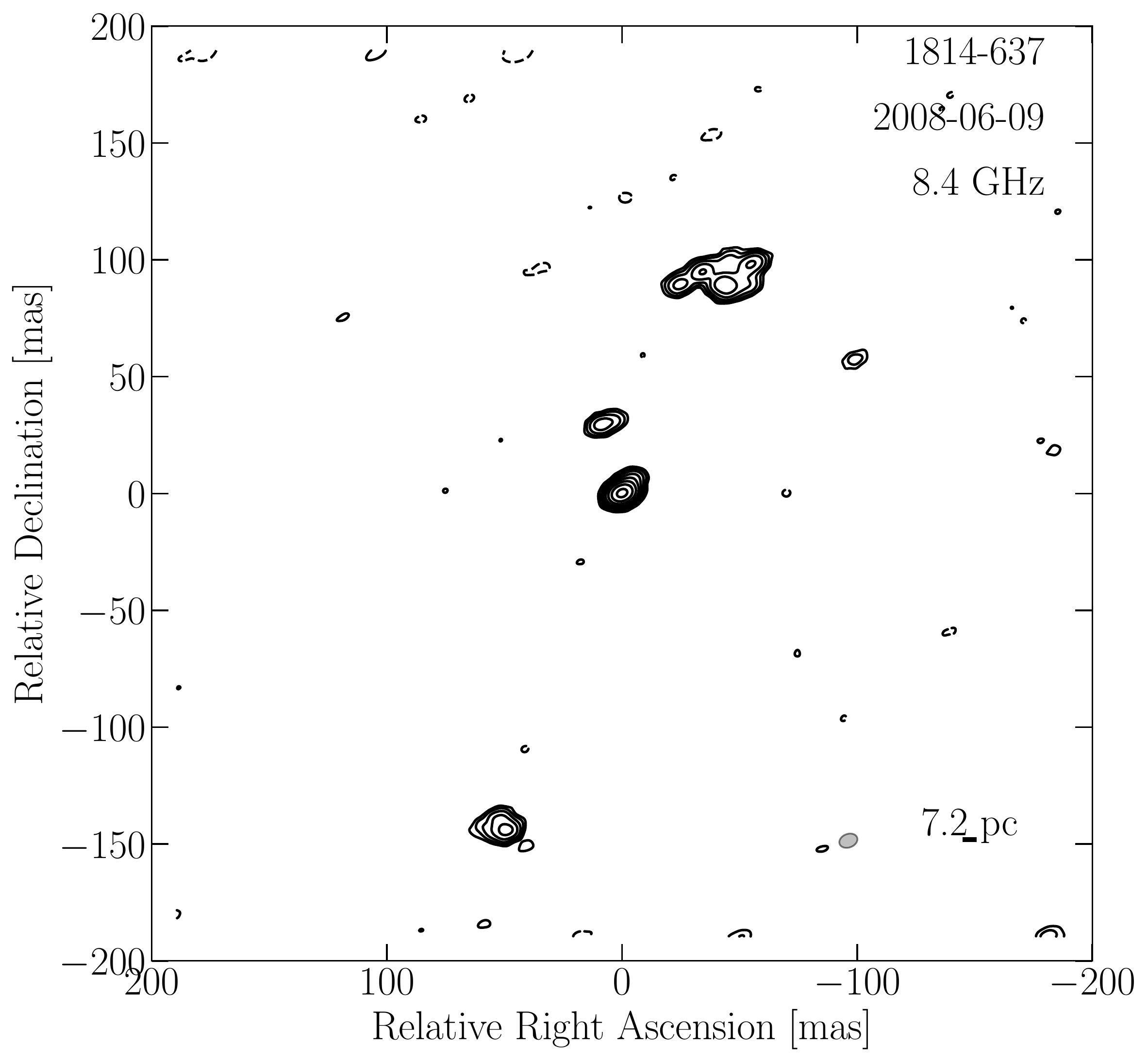}
\includegraphics[width=0.43\linewidth]{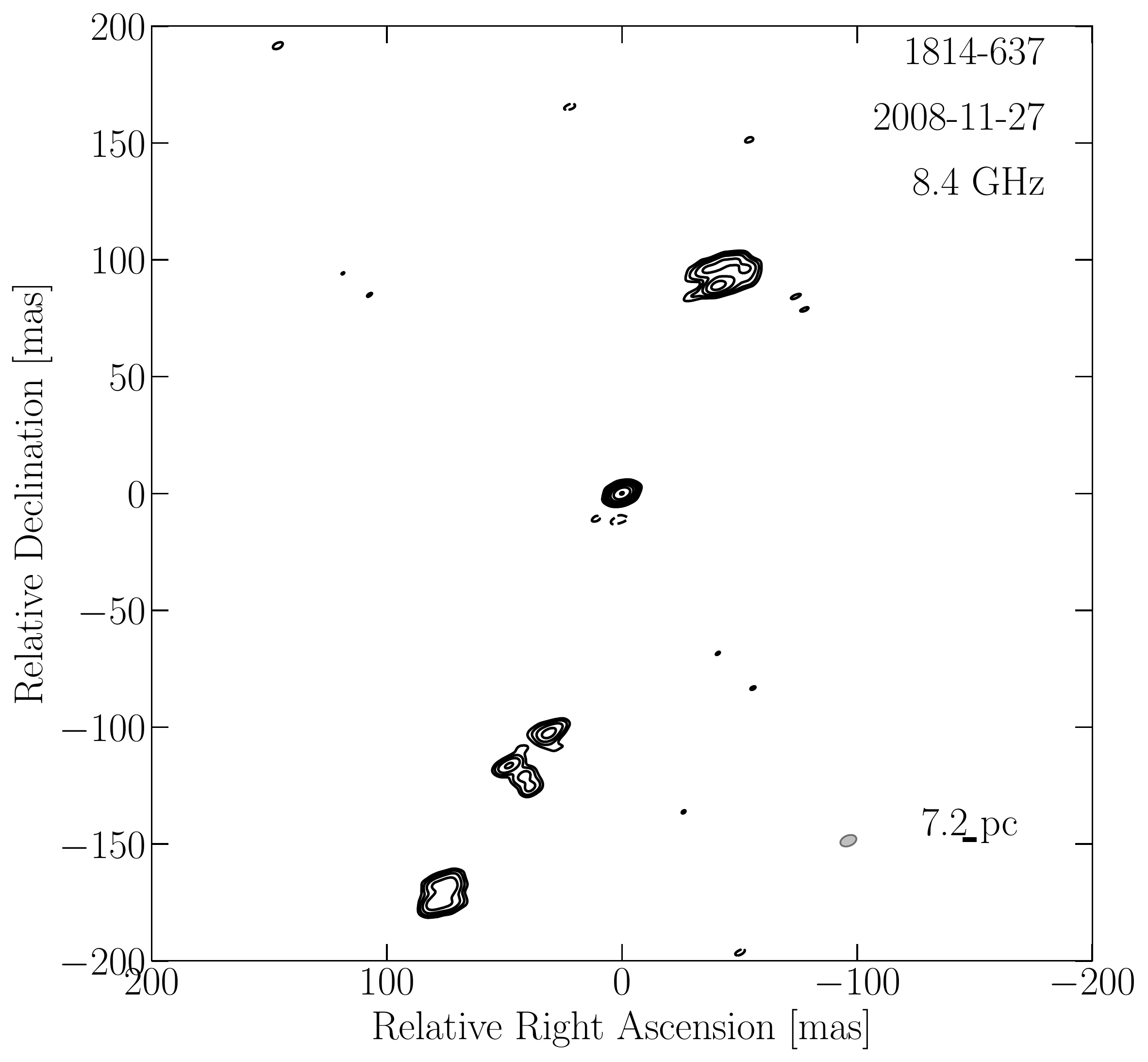}
\includegraphics[width=0.43\linewidth]{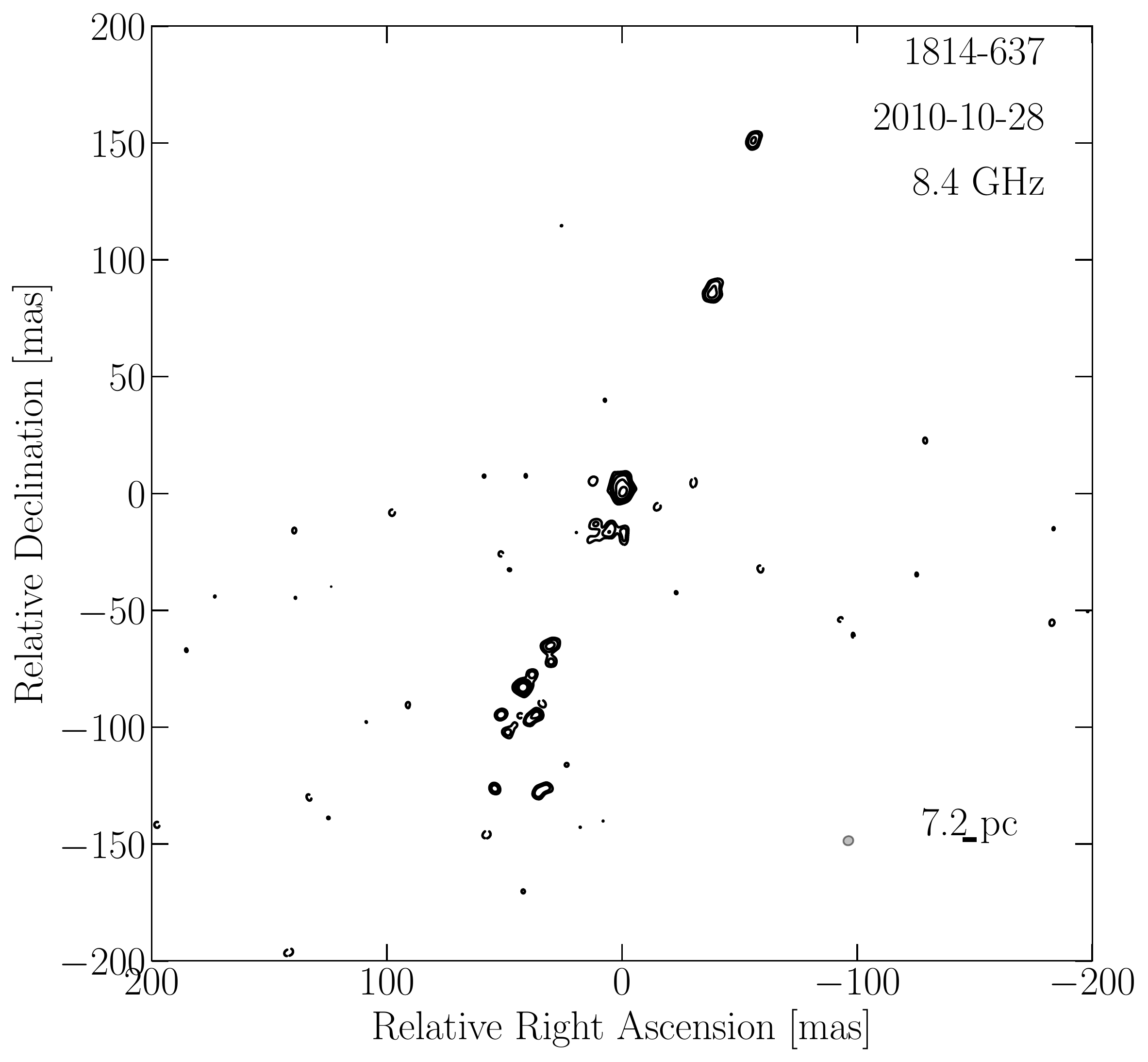}
\includegraphics[width=0.43\linewidth]{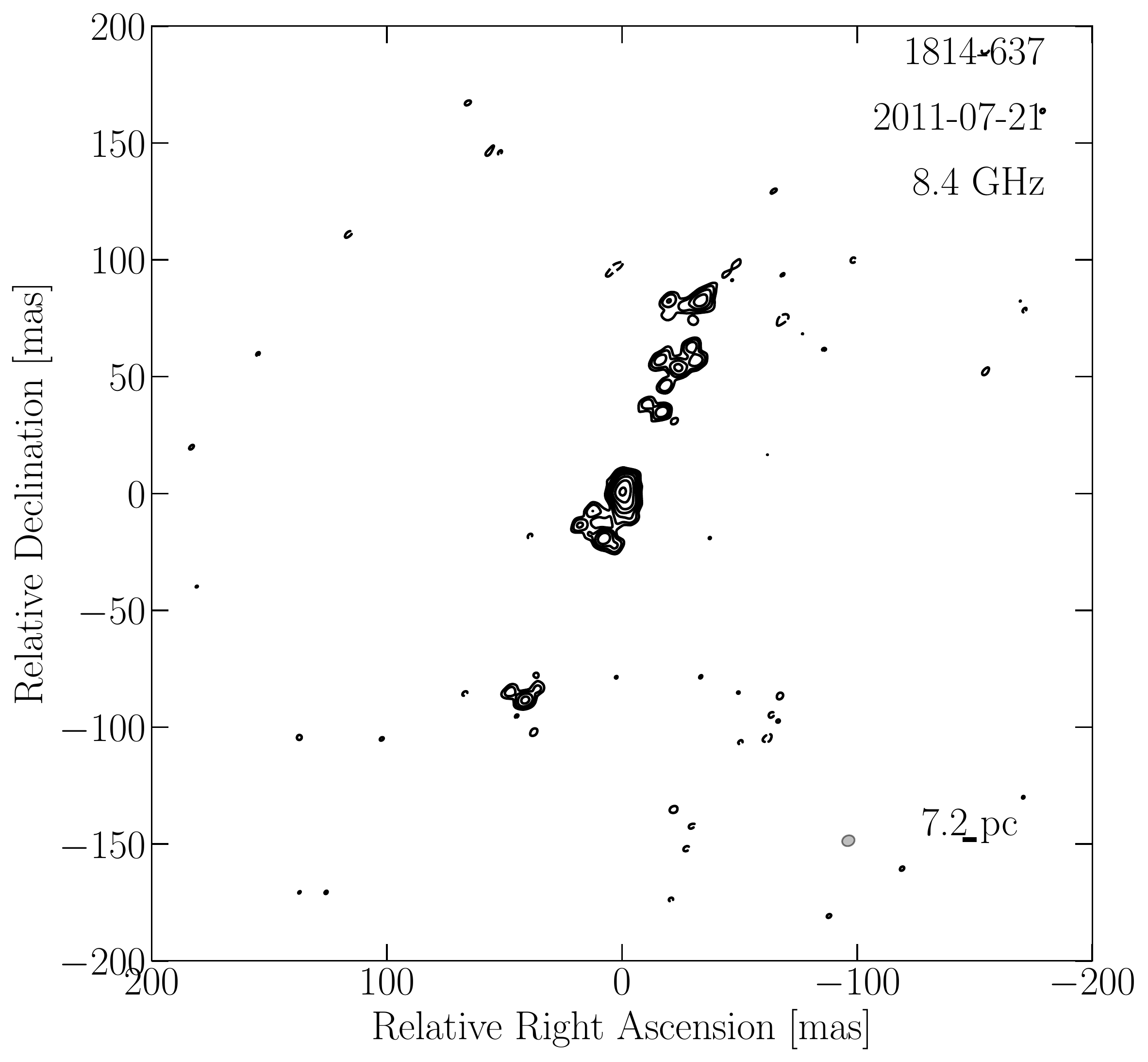}
\end{center}
\caption{Tapered images of PKS\,1814$-$63. The grey ellipse
  represents the beam size, while the black line indicates the linear
  scale at the source's redshift. Contours increase in steps of two starting from 5, 9, 1.5, 9, 9 times the noise level in each map, from top left to bottom, respectively.}
\label{1814_tap}
\end{figure*}
\begin{figure*}[!htbp]
\begin{center}
\includegraphics[width=0.43\linewidth]{Figures/Maps_new/2027-308_2008-06-09.pdf}
\includegraphics[width=0.43\linewidth]{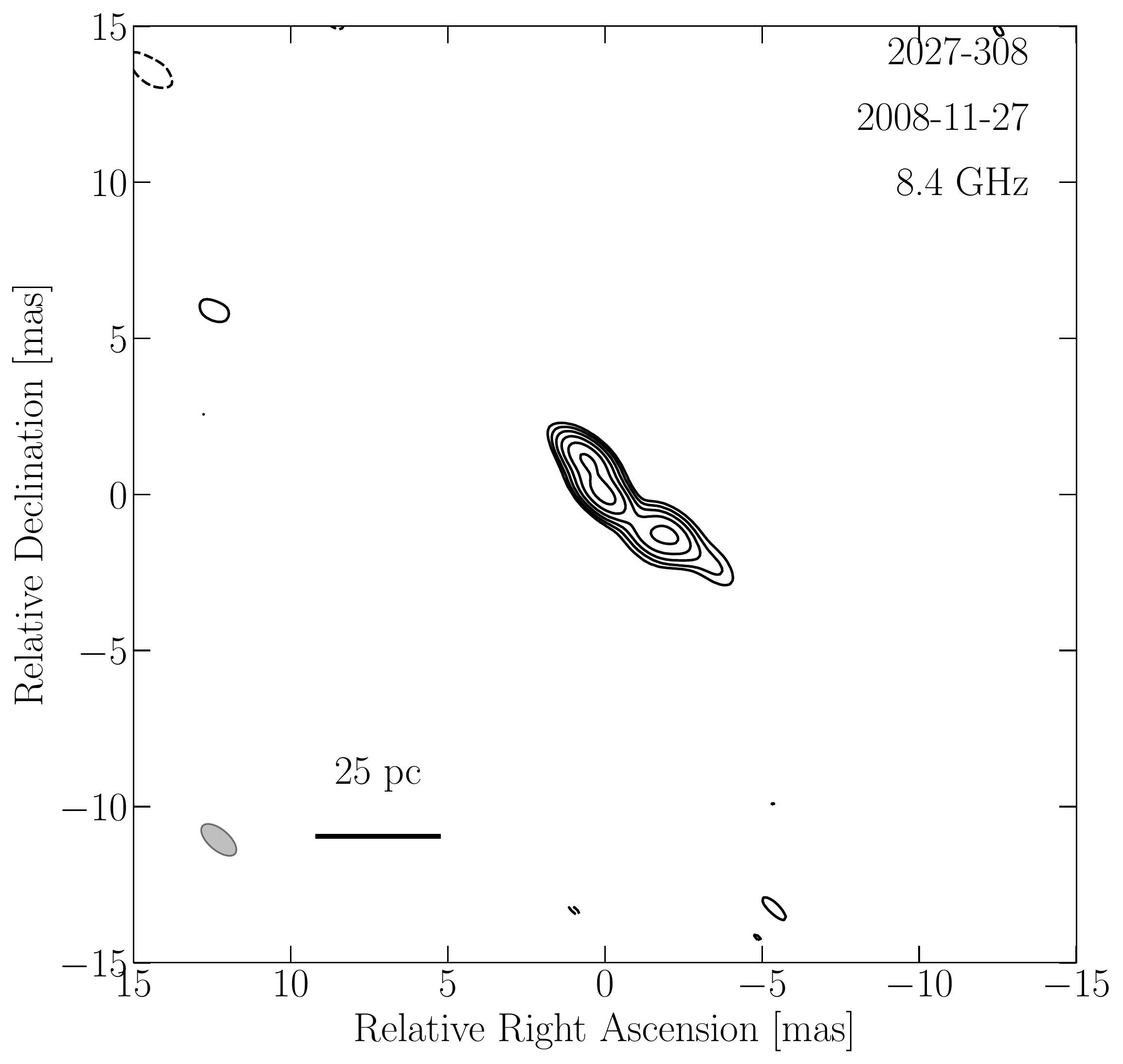}
\includegraphics[width=0.43\linewidth]{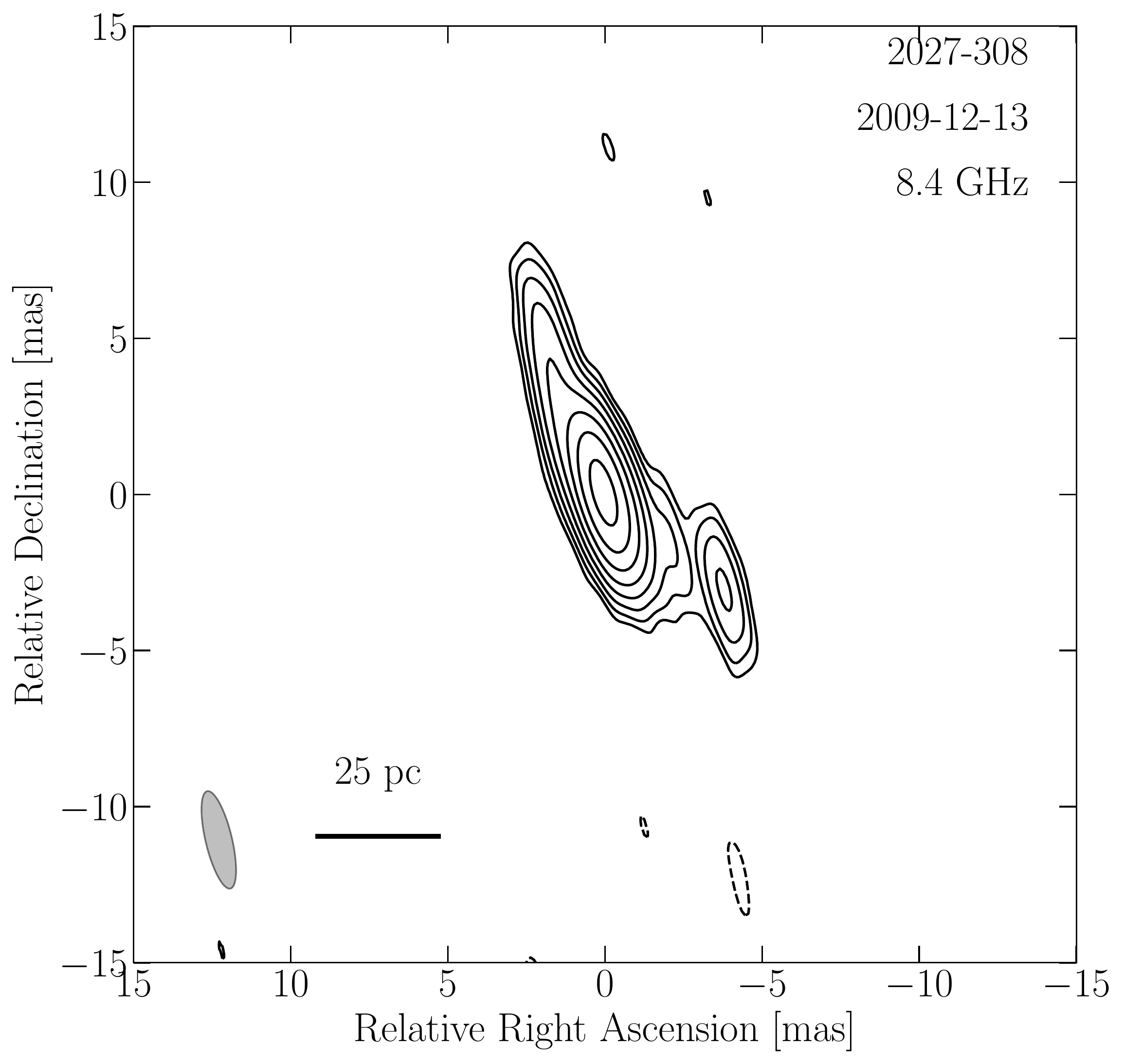}
\includegraphics[width=0.43\linewidth]{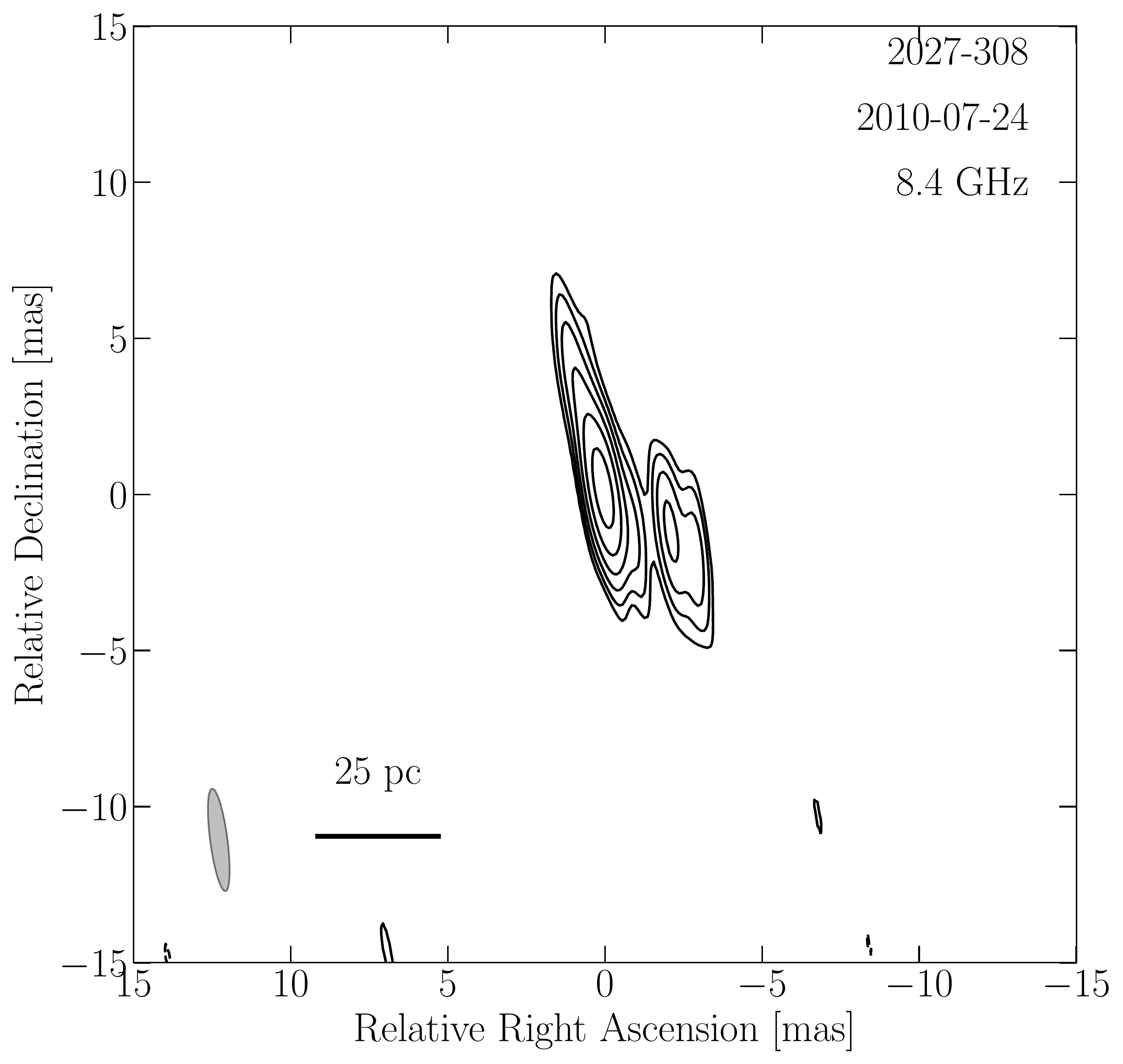}
\end{center}
\caption{Full-resolution images of PKS\,2027$-$308. The grey ellipse
  represents the beam size, while the black line indicates the linear
  scale at the source's redshift. Contours increase in steps of two starting from 0.8, 0.3, 2, 1 times the noise level in each map, from top left to bottom, respectively.}
\label{2027_fulla}
\end{figure*}
\begin{figure*}[!htbp]
\begin{center}
\includegraphics[width=0.43\linewidth]{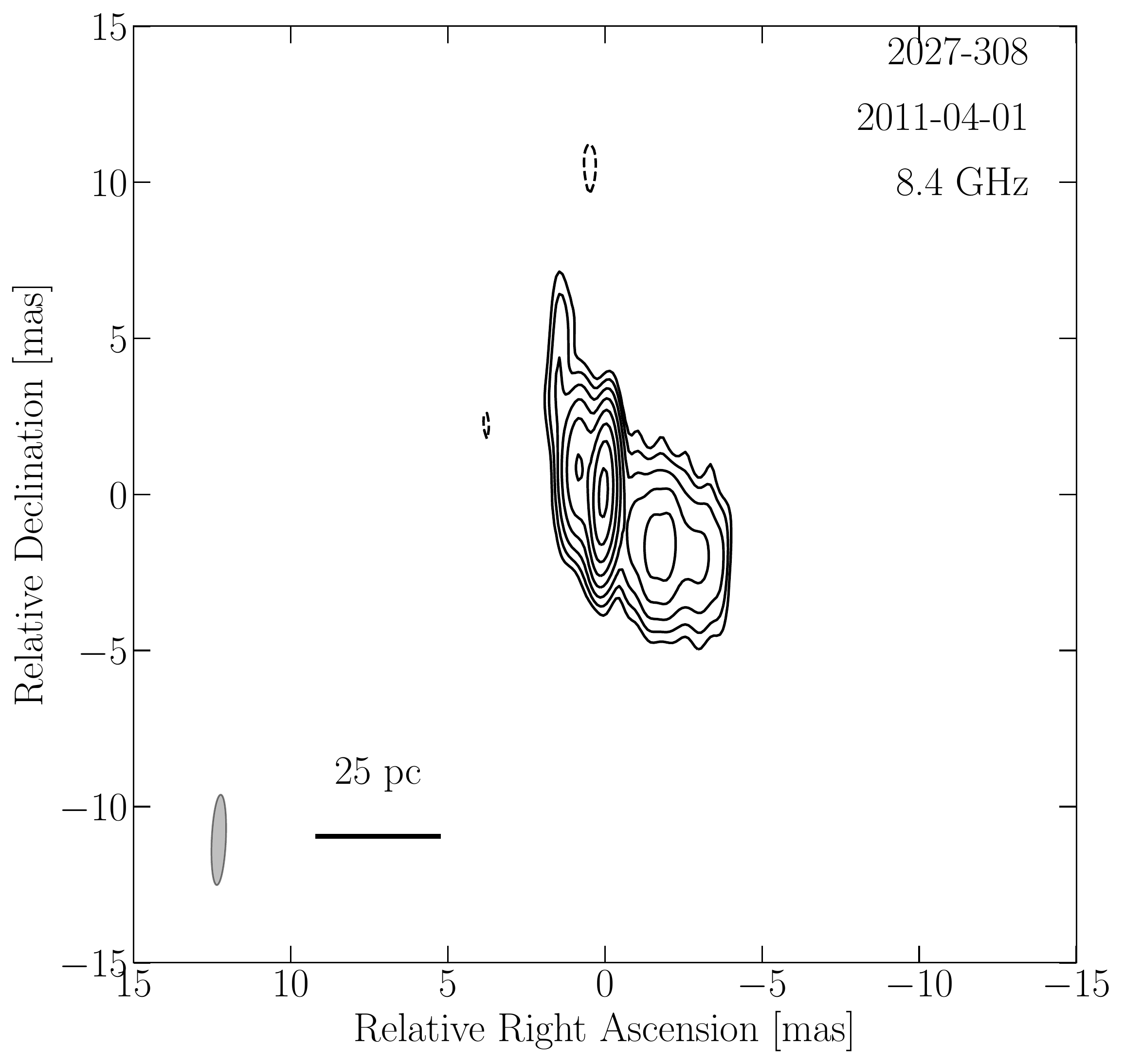}
\includegraphics[width=0.43\linewidth]{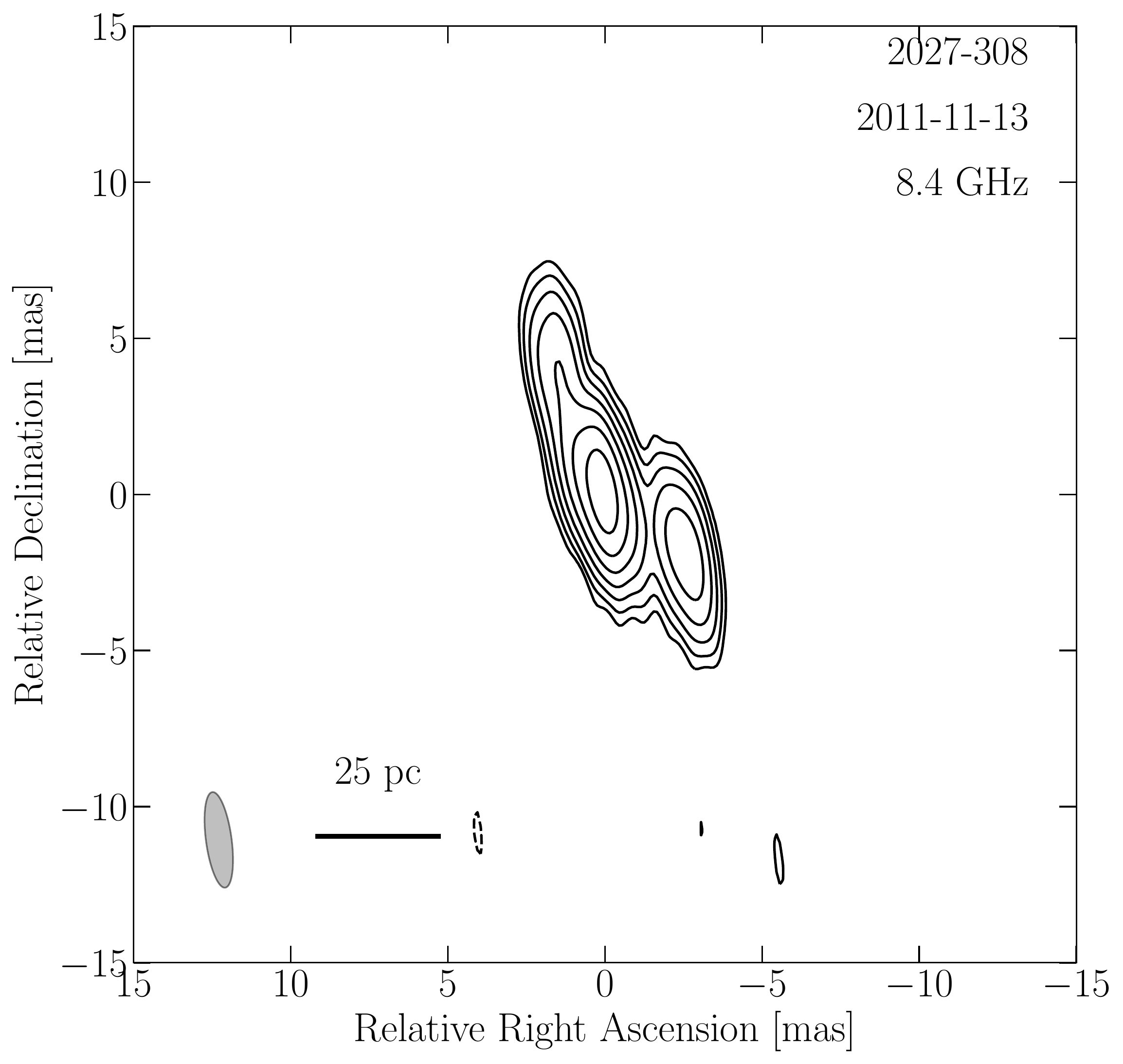}
\includegraphics[width=0.43\linewidth]{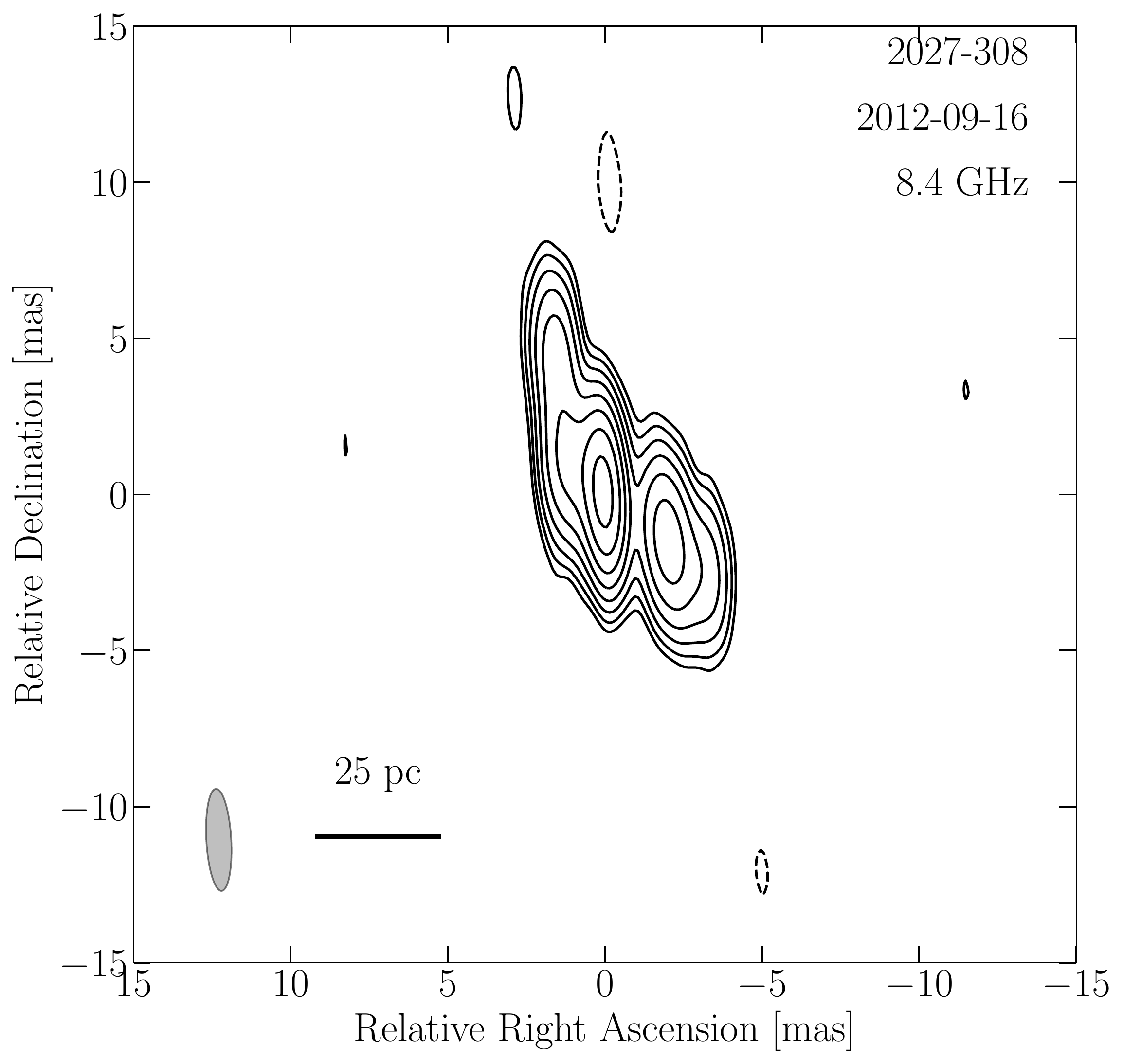}
\end{center}
\caption{Full-resolution images of PKS\,2027$-$308. The grey ellipse
  represents the beam size, while the black line indicates the linear
  scale at the source's redshift. Contours increase in steps of two starting from 1, 1.8, 2.2 times the noise level in each map, from top left to bottom, respectively.}
\label{2027_fullb}
\end{figure*}
\begin{figure*}[!htbp]
\begin{center}
\includegraphics[width=0.43\linewidth]{Figures/Maps_new/2152-699_2008-02-07.pdf}
\includegraphics[width=0.43\linewidth]{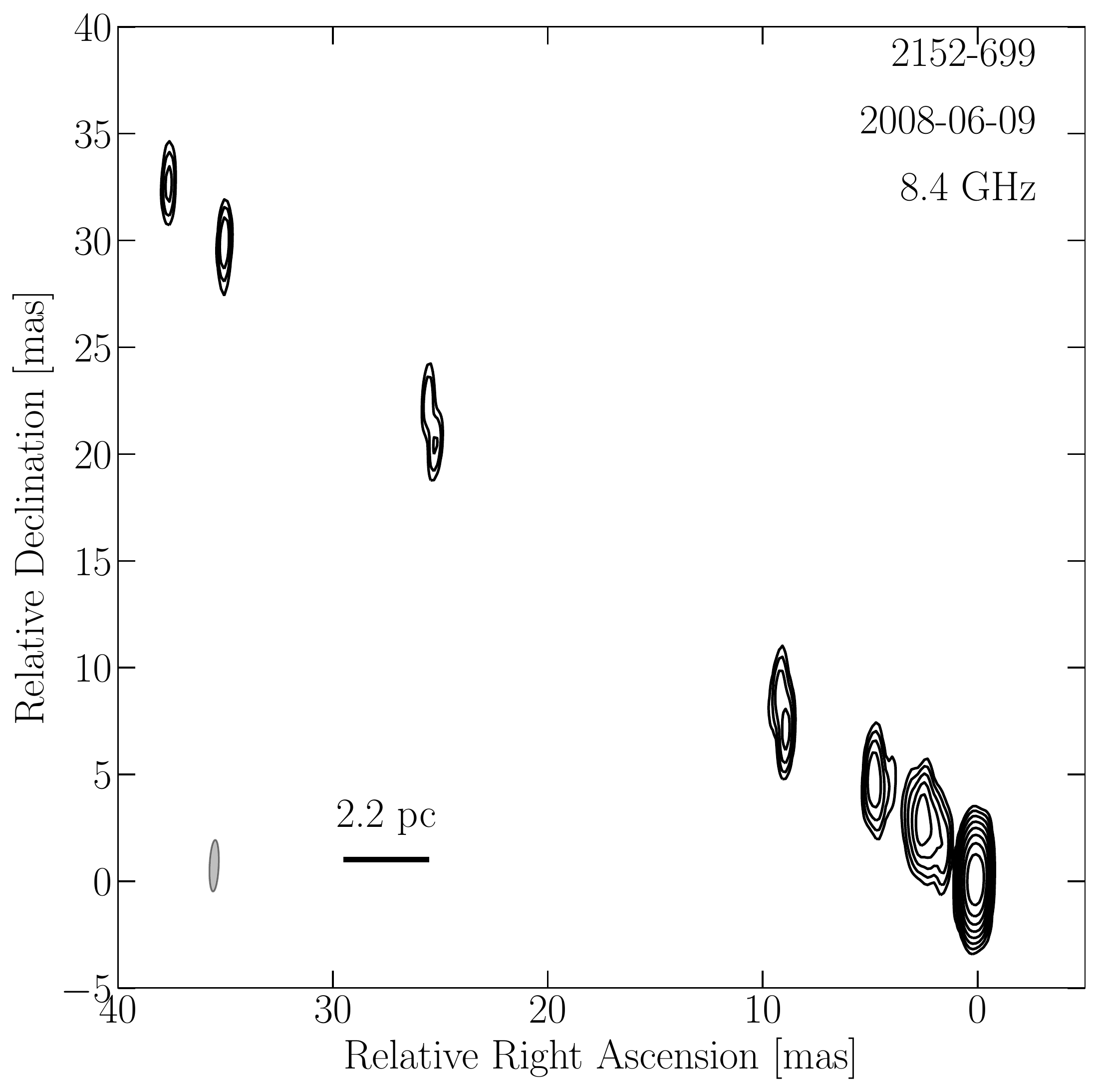}
\includegraphics[width=0.43\linewidth]{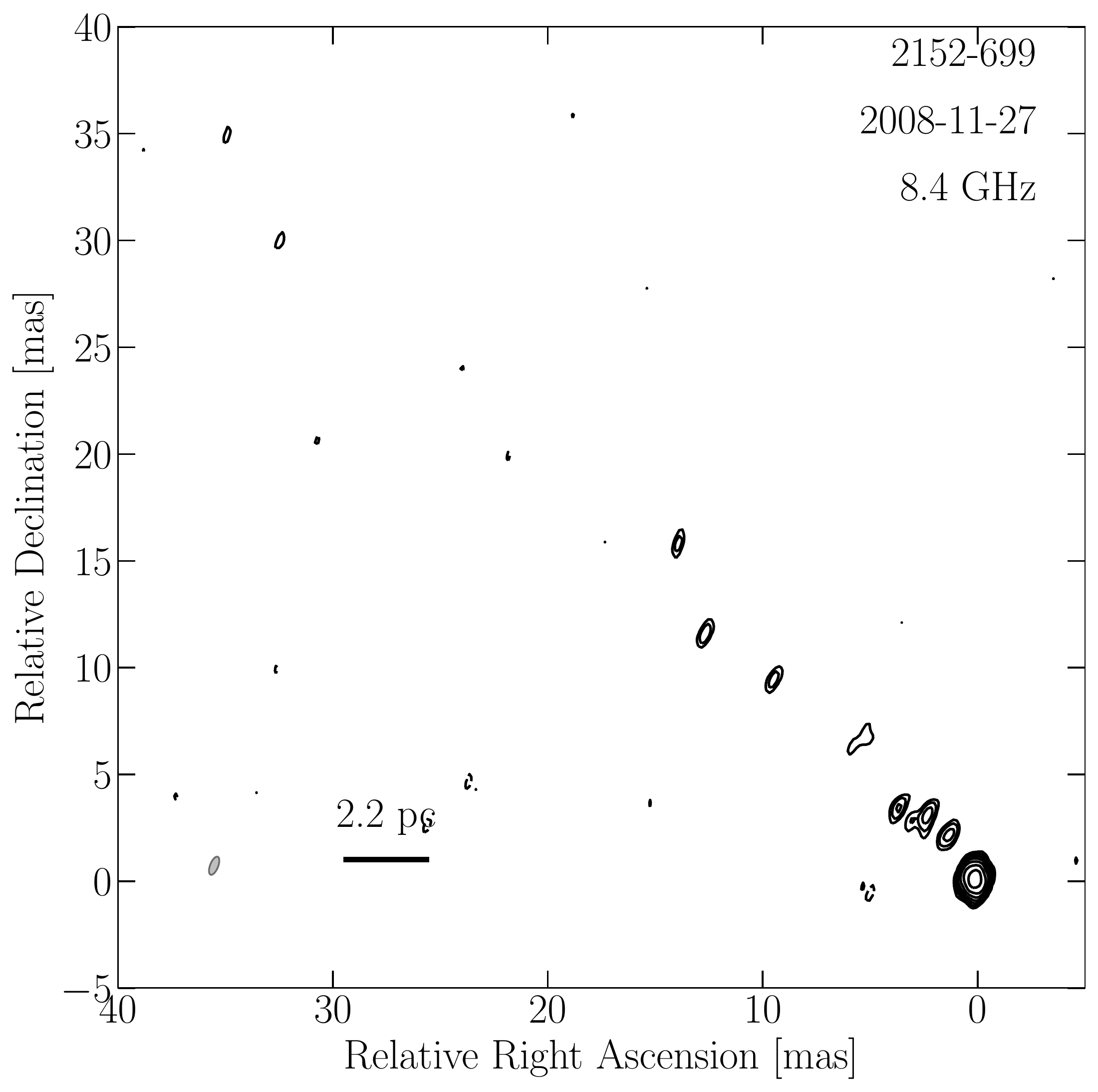}
\includegraphics[width=0.43\linewidth]{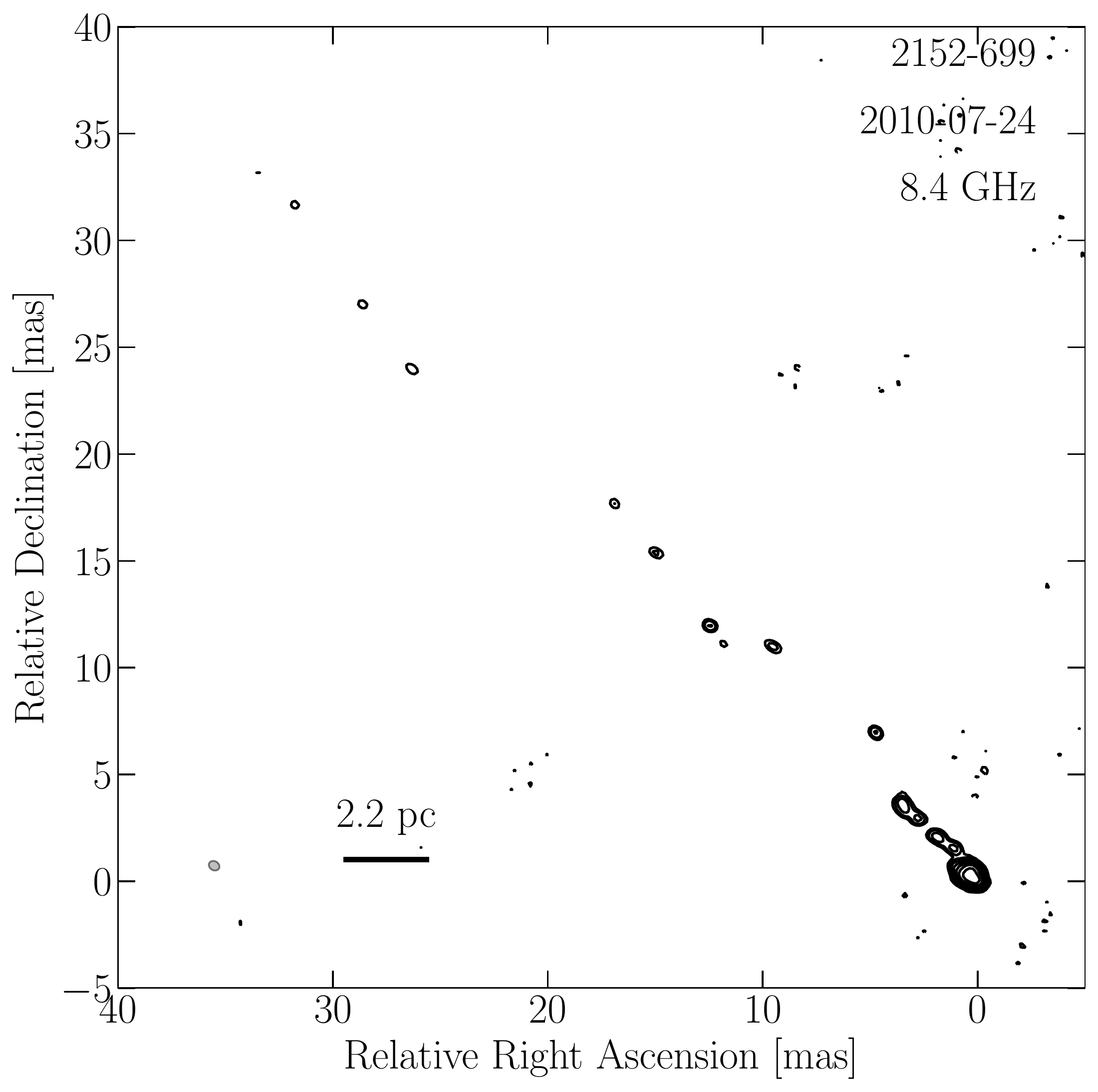}
\end{center}
\caption{Full-resolution images of PKS\,2153$-$69. The grey ellipse
  represents the beam size, while the black line indicates the linear
  scale at the source's redshift. Contours increase in steps of two starting from 0.6, 0.6, 1.3, 1.3 times the noise level in each map, from top left to bottom, respectively.}
\label{2152_fulla}
\end{figure*}
\begin{figure*}[!htbp]
\begin{center}
\includegraphics[width=0.43\linewidth]{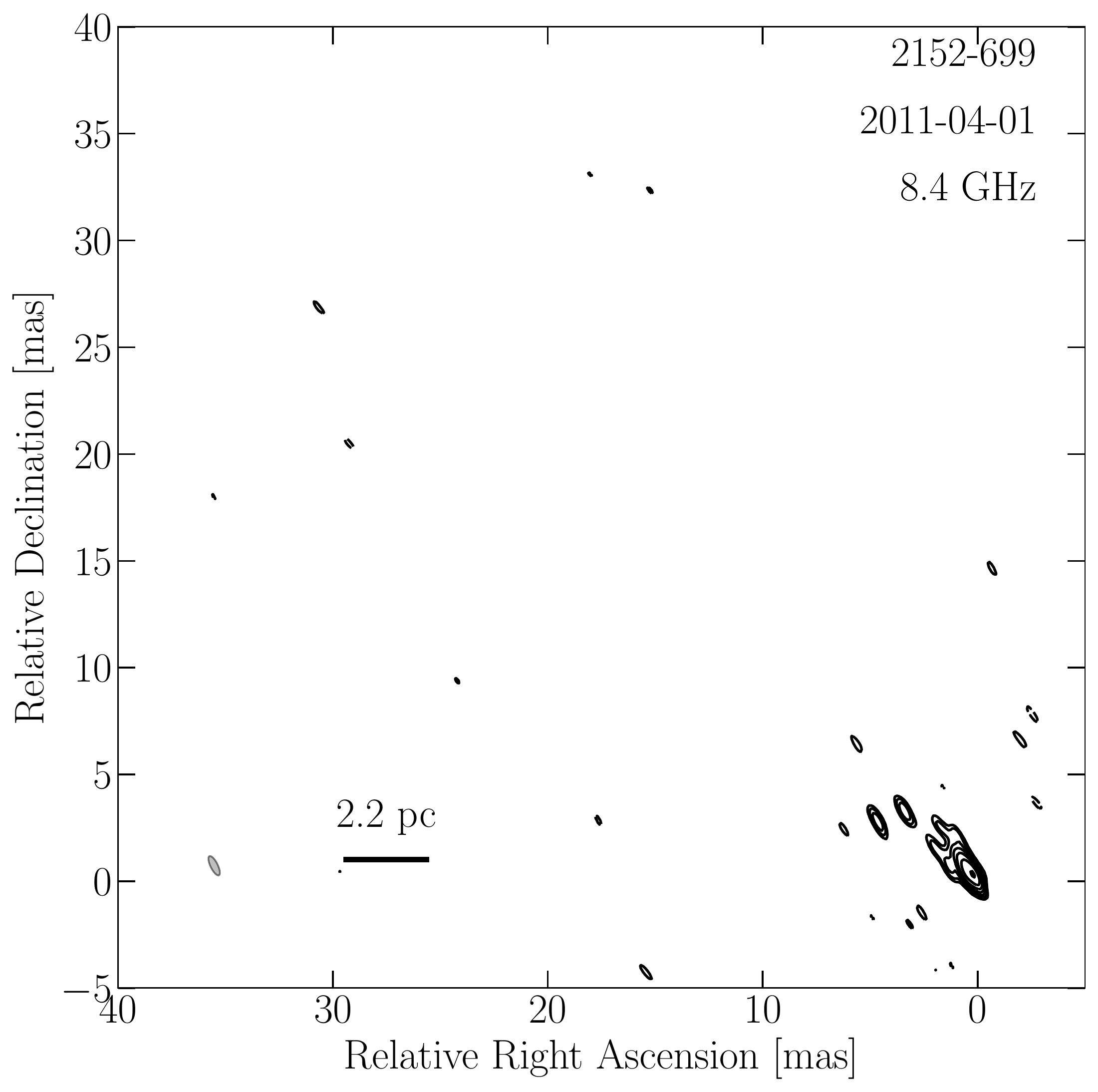}
\includegraphics[width=0.43\linewidth]{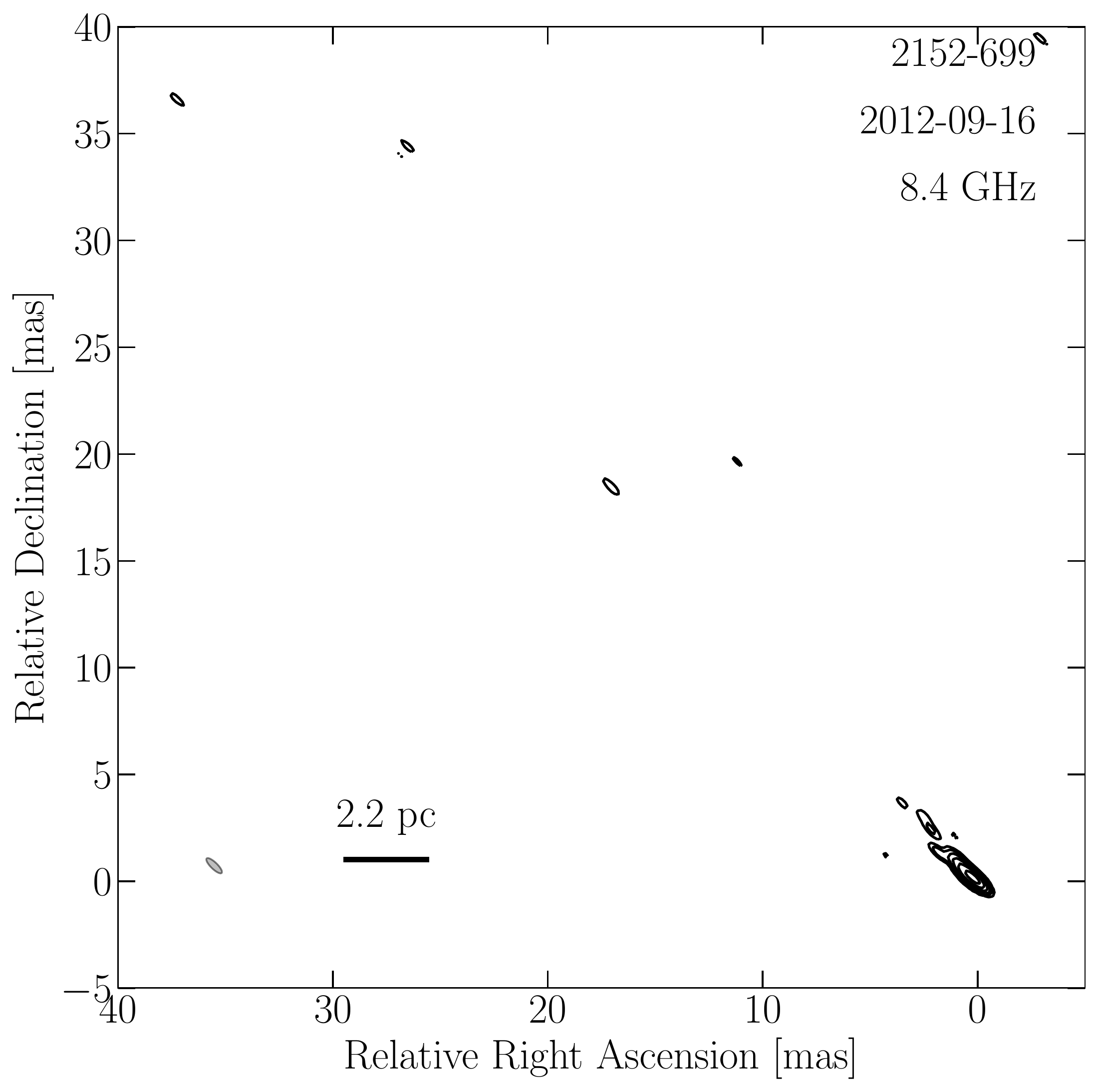}
\includegraphics[width=0.43\linewidth]{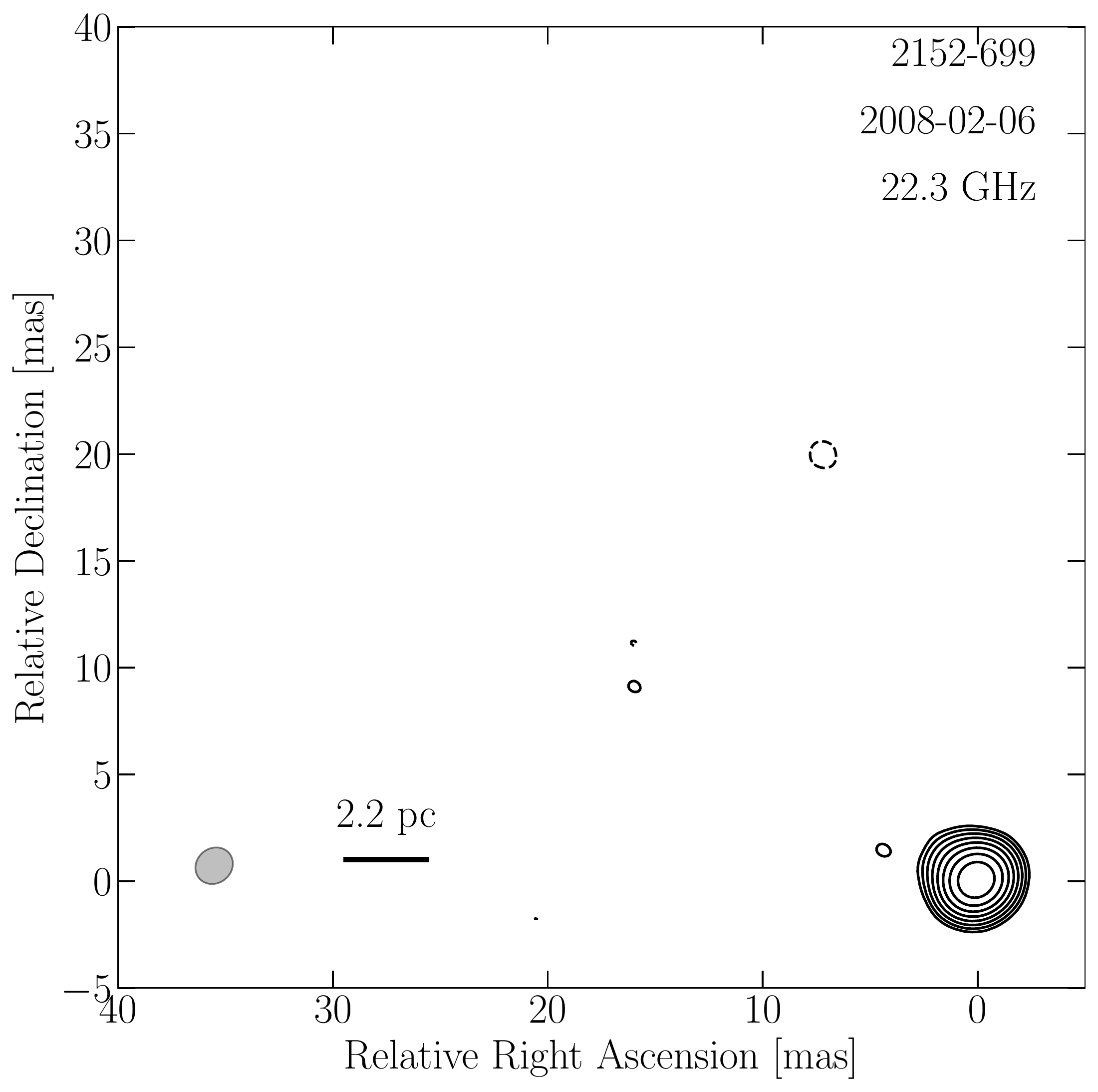}
\includegraphics[width=0.43\linewidth]{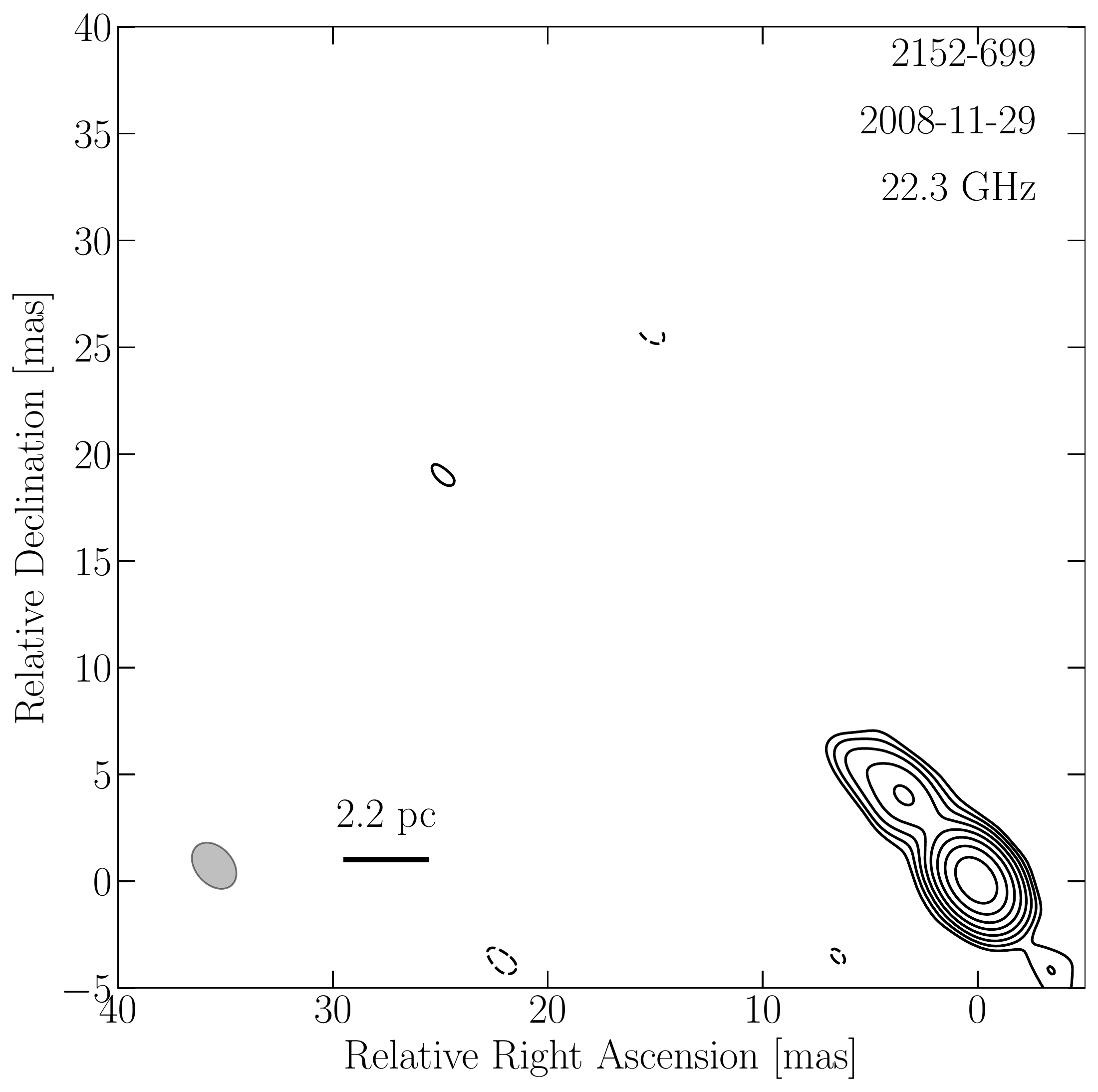}
\end{center}
\caption{Full-resolution images of PKS\,2153$-$69. The grey ellipse
  represents the beam size, while the black line indicates the linear
  scale at the source's redshift. Contours increase in steps of two starting from 1.3, 2, 1.3, 1.5 times the noise level in each map, from top left to bottom, respectively.}
\label{2152_fullb}
\end{figure*}
\clearpage

\begin{table*}

\caption{Details of the 8.4\,GHz TANAMI observations of PKS\,1258$-$321}             
\label{1258_tab}  
\begin{center}    

\begin{tabular}{c c c c c c c c }    
\hline\hline 
Obs.~date & Array configuration$^a$ & $S_\mathrm{total}$$^b$ & $S_\mathrm{peak}$$^b$ & RMS$^b$ & $b_\mathrm{maj}$$^c$ & $b_\mathrm{min}$$^c$ & P.A.$^c$ \\
(yyyy-mm-dd)  &                  & (Jy) & (Jy beam$^{-1}$) & (mJy beam$^{-1}$) & (mas) & (mas) & ($^\circ$) \\
\hline
2009-12-13 & AT-MP-HO-CD-TC & 0.13 & 0.12 & 0.09 & 6.89 & 2.87 & $-$89.3\\
2010-05-08 & AT-MP-HO-CD-PKS-TC-DSS43 & 0.10 & 0.09 & 0.08 & 3.94 & 3.55 & $-$75.9\\
2011-11-13 & AT-MP-HO-HH-CD-PKS-WW-DSS43-DSS45 & 0.13 & 0.12 & 0.14 & 4.57 & 3.04 & 84.0\\
2012-04-27 & AT-MP-HO-CD-PKS & 0.13 & 0.12 & 0.18 & 3.73 & 3.17 & 46.0\\
2013-03-14 & AT-HO-CD-PKS-KE-WW-DSS34-DSS45 & 0.15 & 0.14 & 0.15 & 4.11 & 3.98 & 38.0\\
\hline  
\end{tabular}
\end{center}

$^a$ AT: Australia Telescope Compact Array, CD: Ceduna, HH:
Hartebeesthoek, HO: Hobart, MP: Mopra, OH: GARS/O'Higgins, PKS:
Parkes, TC: TIGO, DSS43, DSS34 \& DSS45: NASA's Deep Space Network Tidbinbilla
(70\,m, 34\,m \& 34\,m), WW: Warkworth (12m), YG: Yarragadee, KE: Katherine, AK : ASKAP (single dish)\\
$^b$ Total flux density, peak flux density and RMS noise level in the
\texttt{CLEAN}-image. An error of 15\% is assumed.\\
$^c$ Major and minor axes and position angle of restoring beam.\\

\end{table*}
\begin{table*}

\caption{Details of the 8.4\,GHz TANAMI observations of IC\,4296}    
\label{1333_tab}  
\begin{center}  

\begin{tabular}{c c c c c c c c }    
\hline\hline 
Obs.~date & Array configuration$^a$ & $S_\mathrm{total}$$^b$ & $S_\mathrm{peak}$$^b$ & RMS$^b$ & $b_\mathrm{maj}$$^c$ & $b_\mathrm{min}$$^c$ & P.A.$^c$ \\
(yyyy-mm-dd)  &                  & (Jy) & (Jy beam$^{-1}$) & (mJy beam$^{-1}$) & (mas) & (mas) & ($^\circ$) \\
\hline
  2008-02-07 & AT-MP-HO-HH-CD-PKS & 0.21 & 0.15 & 0.044 & 3.81 & 1.06 & $-$7.8\\
  2008-06-09 & AT-MP-HO-CD-PKS & 0.22 & 0.19 & 0.045 & 3.94 & 3.26 & $-$82.9\\
  2008-11-27 & OH-AT-MP-HO-CD-PKS-DSS43 & 0.23 & 0.20 & 0.12 & 4.08 & 1.96 & $-$61.5\\
  2010-07-24 & AT-MP-HO-CD-PKS & 0.23 & 0.20 & 0.051 & 4.98 & 3.52 & 76.6\\
  2011-11-13 & AT-MP-HO-HH-CD-PKS-WW-TC & 0.12 & 0.037 & 0.30 & 1.72 & 0.45 & 3.15\\
\hline  
\end{tabular}
\end{center}

$^a$ See Table~\ref{1258_tab} for the antenna codes.\\
$^b$ Total flux density, peak flux density and RMS noise level in the
\texttt{CLEAN}-image. An error of 15\% is assumed.\\
$^c$ Major and minor axes and position angle of restoring beam.\\
\end{table*}
\begin{table*}

\caption{Details of the 8.4\,GHz TANAMI observations of PKS\,1549$-$790.}             
\label{1549_tab}  
\begin{center}    

\begin{tabular}{c c c c c c c c }    
\hline\hline 
Obs.~date & Array configuration$^a$ & $S_\mathrm{total}$$^b$ & $S_\mathrm{peak}$$^b$ & RMS$^b$ & $b_\mathrm{maj}$$^c$ & $b_\mathrm{min}$$^c$ & P.A.$^c$ \\
(yyyy-mm-dd)  &                  & (Jy) & (Jy beam$^{-1}$) & (mJy beam$^{-1}$) & (mas) & (mas) & ($^\circ$) \\
\hline
2008-02-07 & AT-MP-HO-CD-PKS & 1.75 & 0.15 & 0.90 & 1.07 & 0.38 & 5.55\\
2008-06-09 & AT-MP-HO-HH-CD-PKS & 1.00 & 0.32 &	1.13 & 3.06	& 0.65 & $-$20.2\\
2008-11-27 & TC-OH-AT-MP-HO-CD-PKS-DSS43 & 0.69 & 0.12 & 1.15 & 0.67 & 0.38 &	85.1\\
2009-12-13 & AT-MP-HO-CD-TC & 0.80 & 0.12 &  2.35 & 0.57 & 0.34 & 31.6\\
2012-09-17 & AT-HO-HH-CD-PKS-DSS34-DSS45-KE & 0.60 & 0.08 & 0.87 & 1.01	& 0.43 & 6.35\\
\hline  
\end{tabular}
\end{center}

$^a$ See Table~\ref{1258_tab} for the antenna codes.\\
$^b$ Total flux density, peak flux density and RMS noise level in the
\texttt{CLEAN}-image. An error of 15\% is assumed..\\
$^c$ Major and minor axes and position angle of restoring beam.\\

\end{table*}
\begin{table*}

\caption{Details of the 8.4\,GHz TANAMI observations of PKS\,1733$-$565}             
\label{1733_tab}  
\begin{center}    

\begin{tabular}{c c c c c c c c }    
\hline\hline 
Obs.~date & Array configuration$^a$ & $S_\mathrm{total}$$^b$ & $S_\mathrm{peak}$$^b$ & RMS$^b$ & $b_\mathrm{maj}$$^c$ & $b_\mathrm{min}$$^c$ & P.A.$^c$ \\
(yyyy-mm-dd)  &                  & (Jy) & (Jy beam$^{-1}$) & (mJy beam$^{-1}$) & (mas) & (mas) & ($^\circ$) \\
\hline
  2008-02-07 & AT-MP-HO-CD-PKS & 0.18 & 0.16 & 0.06 & 2.62 & 0.77 & $-$4.32\\
  2008-03-28 & AT-MP-HO-HH-CD-PKS-DSS43 & 0.19 & 0.16 & 0.04 & 2.95 & 0.76 & 1.06\\
  2008-08-08 & AT-MP-HO-HH-CD-PKS-DSS45 & 0.19 & 0.17 & 0.08 & 3.64 & 1.19 & 13.3\\
  2009-02-23 & AT-MP-HO-CD-PKS-TC-OH & 0.16 & 0.12 & 0.12 & 2.70 & 1.05 & 46.8\\
  2010-03-12 & AT-MP-HO-CD-PKS-DSS43 & 0.10 & 0.10 & 0.07 & 4.28 & 3.46 & 88.5\\
  2010-10-29 & AT-MP-HO-CD-PKS-DSS34-DSS45-TC-HH & 0.15 & 0.12 & 0.05 & 2.49 & 0.62 & $-$2.42\\
  2011-07-22 & AT-MP-HO-HH-CD-PKS-DSS43-DSS34 & 0.12 & 0.10 & 0.03 & 2.71 & 0.72 & 2.54\\
  2012-04-27 & AT-MP-HO-CD-PKS & 0.07 & 0.07 & 0.06 & 2.29 & 0.62 & 1.23\\

\hline  
\end{tabular}
\end{center}

$^a$ See Table~\ref{1258_tab} for the antenna codes.\\
$^b$ Total flux density, peak flux density and RMS noise level in the
\texttt{CLEAN}-image. An error of 15\% is assumed..\\
$^c$ Major and minor axes and position angle of restoring beam.\\

\end{table*}
\begin{table*}

\caption{Details of the 8.4\,GHz TANAMI observations of PKS\,1814$-$63}             
\label{1814_tab}  
\begin{center}    

\begin{tabular}{c c c c c c c c }    
\hline\hline 
Obs.~date & Array configuration$^a$ & $S_\mathrm{total}$$^b$ & $S_\mathrm{peak}$$^b$ & RMS$^b$ & $b_\mathrm{maj}$$^c$ & $b_\mathrm{min}$$^c$ & P.A.$^c$ \\
(yyyy-mm-dd)  &                  & (Jy) & (Jy beam$^{-1}$) & (mJy beam$^{-1}$) & (mas) & (mas) & ($^\circ$) \\
\hline
  2008-02-07 & AT-MP-HO-CD-PKS & 0.60 & 0.41 & 2.07 & 1.84 & 0.58 & 2.78\\
  2008-06-09 & AT-MP-HO-HH-CD-PKS & 0.63 & 0.52 & 5.00 & 4.14 & 3.18 & 71.4\\
  2008-11-27 & TC-OH-AT-MP-HO-CD-PKS-DSS43 & 0.77 & 0.25 & 1.85 & 1.44 & 0.63 & $-$55.2\\
  2010-10-29 & AT-MP-HO-CD-PKS-DSS34-DSS45-TC-HH & 0.44 & 0.10 & 0.97 & 2.09 & 0.65 & 0.23\\
  2011-07-22 & AT-MP-HO-HH-CD-PKS-DSS43-DSS34 & 0.39 & 0.12 & 4.92 & 1.99 & 1.58 & 78.9\\

\hline  
\end{tabular}
\end{center}

$^a$ See Table~\ref{1258_tab} for the antenna codes.\\
$^b$ Total flux density, peak flux density and RMS noise level in the
\texttt{CLEAN}-image. An error of 15\% is assumed..\\
$^c$ Major and minor axes and position angle of restoring beam.\\

\end{table*}
\begin{table*}

\caption{Details of the 8.4\,GHz TANAMI observations of PKS\,2027$-$308}             
\label{2027_tab}  
\begin{center}    

\begin{tabular}{c c c c c c c c }    
\hline\hline 
Obs.~date & Array configuration$^a$ & $S_\mathrm{total}$$^b$ & $S_\mathrm{peak}$$^b$ & RMS$^b$ & $b_\mathrm{maj}$$^c$ & $b_\mathrm{min}$$^c$ & P.A.$^c$ \\
(yyyy-mm-dd)  &                  & (Jy) & (Jy beam$^{-1}$) & (mJy beam$^{-1}$) & (mas) & (mas) & ($^\circ$) \\
\hline
  2008-06-09 & AT-MP-HO-HH-CD-PKS & 0.12 & 0.08 & 0.15 & 3.03 & 0.69 & $-$6.61\\
  2008-11-28 & TC-OH-AT-MP-HO-CD-PKS-DSS43 & 0.10 & 0.04 & 0.26 & 1.37 & 0.67 & 49.2\\
  2009-12-14 & AT-MP-HO-CD-TC & 0.11 & 0.07 & 0.08 & 3.20 & 0.84 & 13.3\\
  2010-07-24 & TC-AT-MP-HO-CD-PKS & 0.10 & 0.05 & 0.23 & 3.30 & 0.55 & 7.4\\
  2011-04-01 & AT-MP-HO-HH-CD-PKS-DSS43-WW & 0.12 & 0.05 & 0.09 & 2.9 & 0.45 & $-$2.53\\
  2011-11-14 & AT-MP-HO-HH-CD-PKS-WW-TC & 0.09 & 0.05 & 0.09 & 3.09 & 0.82 & 7.62\\
  2012-09-17 & AT-HO-HH-CD-PKS-DSS34-DSS45-KE & 0.10 & 0.05 & 0.05 & 3.27 & 0.79 & 3.14\\

\hline  
\end{tabular}
\end{center}

$^a$ See Table~\ref{1258_tab} for the antenna codes.\\
$^b$ Total flux density, peak flux density and RMS noise level in the
\texttt{CLEAN}-image. An error of 15\% is assumed..\\
$^c$ Major and minor axes and position angle of restoring beam.\\

\end{table*}
\begin{table*}

\caption{Details of the 8.4\,GHz TANAMI observations of PKS\,2153$-$69}             
\label{2152_tab}  
\begin{center}    

\begin{tabular}{c c c c c c c c }    
\hline\hline 
Obs.~date & Array configuration$^a$ & $S_\mathrm{total}$$^b$ & $S_\mathrm{peak}$$^b$ & RMS$^b$ & $b_\mathrm{maj}$$^c$ & $b_\mathrm{min}$$^c$ & P.A.$^c$ \\
(yyyy-mm-dd)  &                  & (Jy) & (Jy beam$^{-1}$) & (mJy beam$^{-1}$) & (mas) & (mas) & ($^\circ$) \\
\hline
  2008-02-08 & AT-MP-HO-CD-PKS & 0.48 & 0.21 & 0.25 & 1.67 & 0.31 & $-$5.83\\
  2008-06-09 & AT-MP-HO-HH-CD-PKS & 0.64 & 0.27 & 0.25 & 2.42 & 0.43 & $-$2.53\\
  2008-11-27 & TC-OH-AT-MP-HO-CD-PKS-DSS43 & 0.42 & 0.20 & 0.51 & 0.91 & 0.38 & $-$21.5\\
  2010-07-24 & TC-AT-MP-HO-CD-PKS & 0.32 & 0.20 & 0.63 & 0.52 & 0.42 & 59.0\\
  2011-04-01 & AT-MP-HO-HH-CD-PKS-DSS43-WW & 0.54 & 0.24 & 0.94 & 0.99 & 0.33 & 26.3\\
  2012-09-16 & AT-HO-HH-CD-PKS-DSS34-DSS45-KE & 0.52 & 0.25 & 1.20 & 0.97 & 0.31 & 45.7\\

\hline  
\end{tabular}
\end{center}

$^a$ See Table~\ref{1258_tab} for the antenna codes.\\
$^b$ Total flux density, peak flux density and RMS noise level in the
\texttt{CLEAN}-image. An error of 15\% is assumed..\\
$^c$ Major and minor axes and position angle of restoring beam.\\

\end{table*}
\clearpage
\section{Extended kinematic analysis information}
\label{app:kin}
\begin{figure}[!!htbp]
\begin{center}
\includegraphics[width=\linewidth]{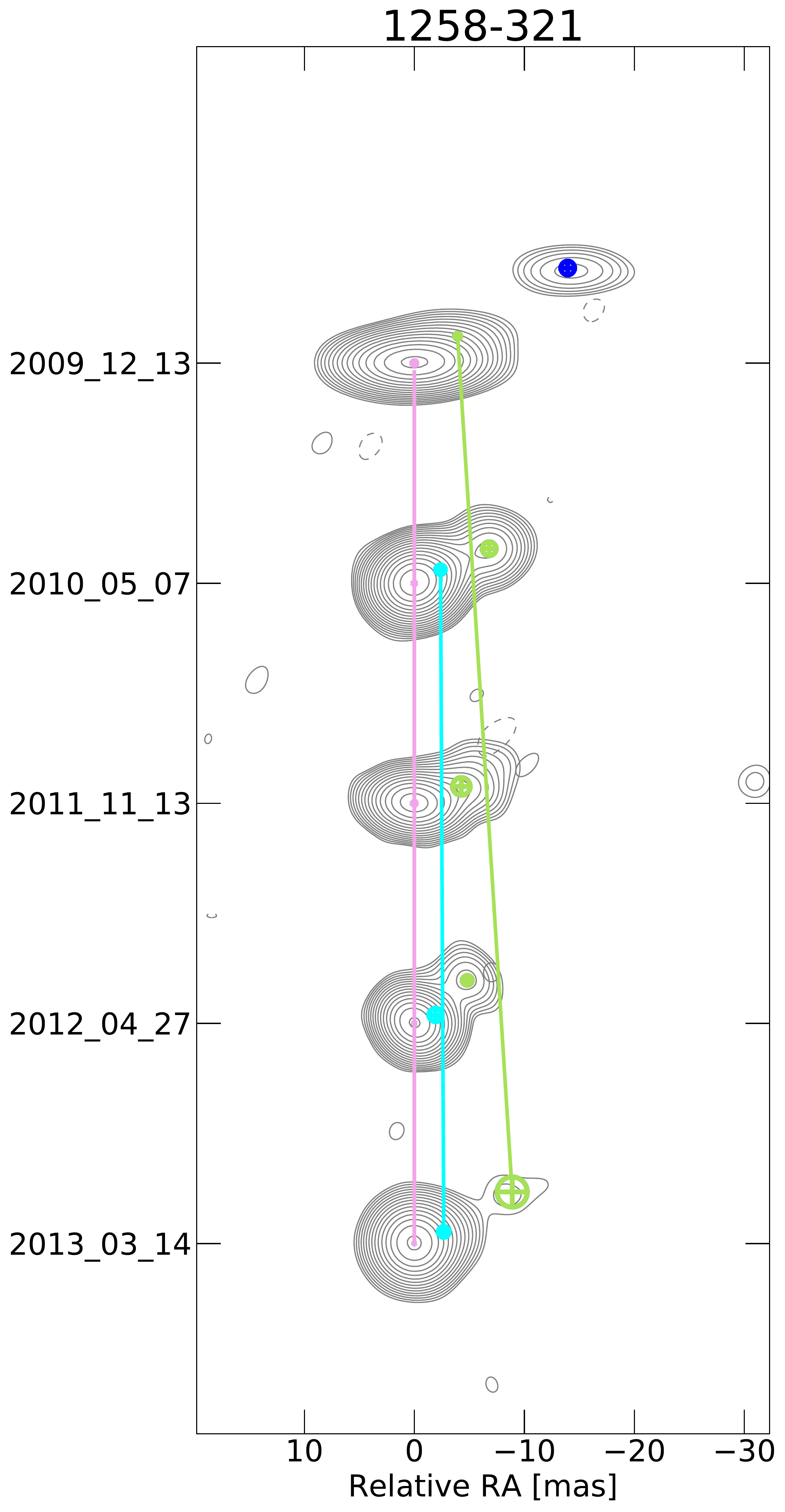}
\end{center}
\caption{Multi-epoch tapered images of 1258$-$321.}
\label{kina_1258}
\end{figure}

\begin{figure}[!!htbp]
\begin{center}
\includegraphics[width=\linewidth]{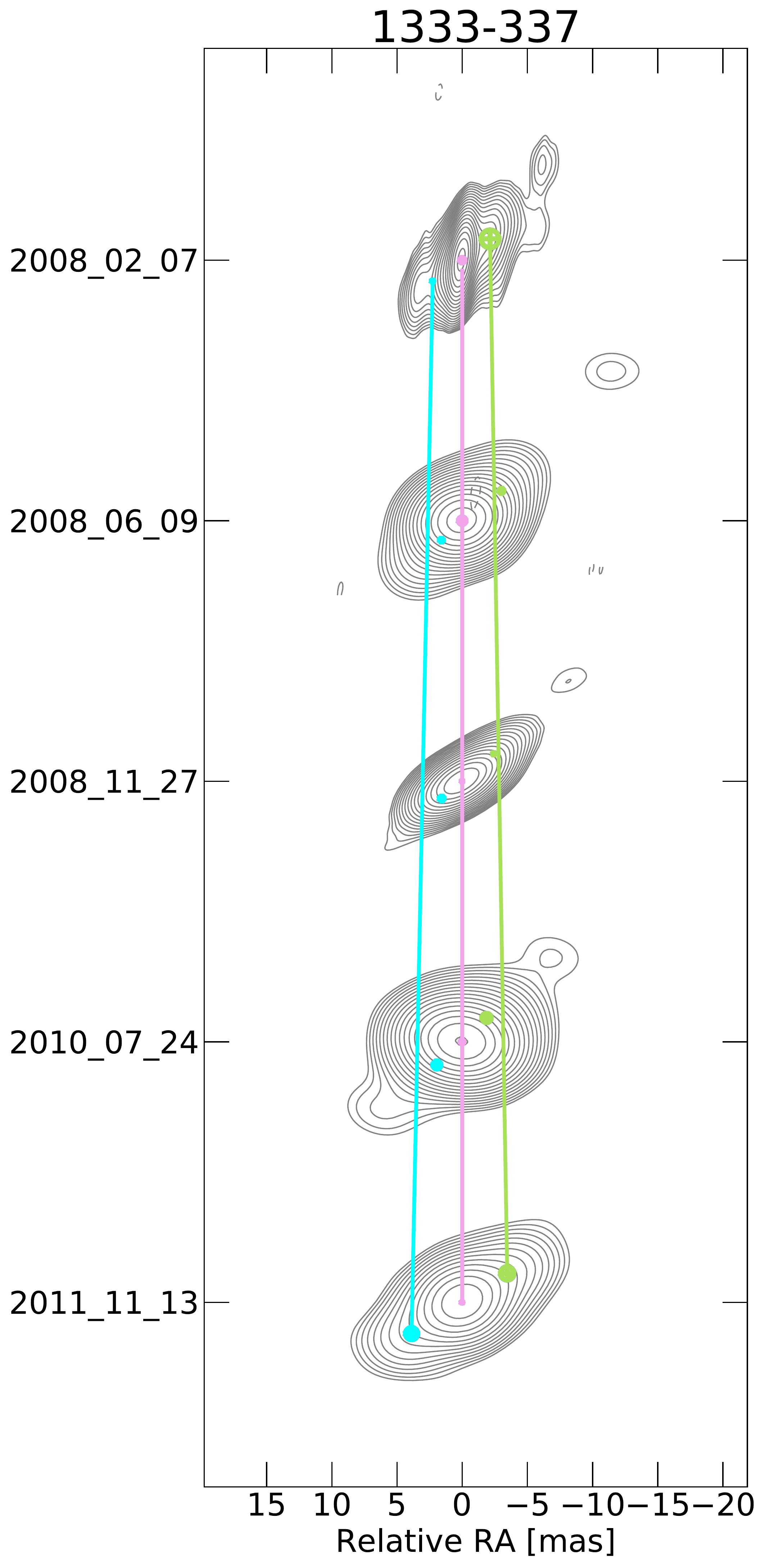}
\end{center}
\caption{Multi-epoch tapered images of 1333$-$337.}
\label{kina_1333}
\end{figure}

\begin{figure}[!!htbp]
\begin{center}
\includegraphics[width=\linewidth]{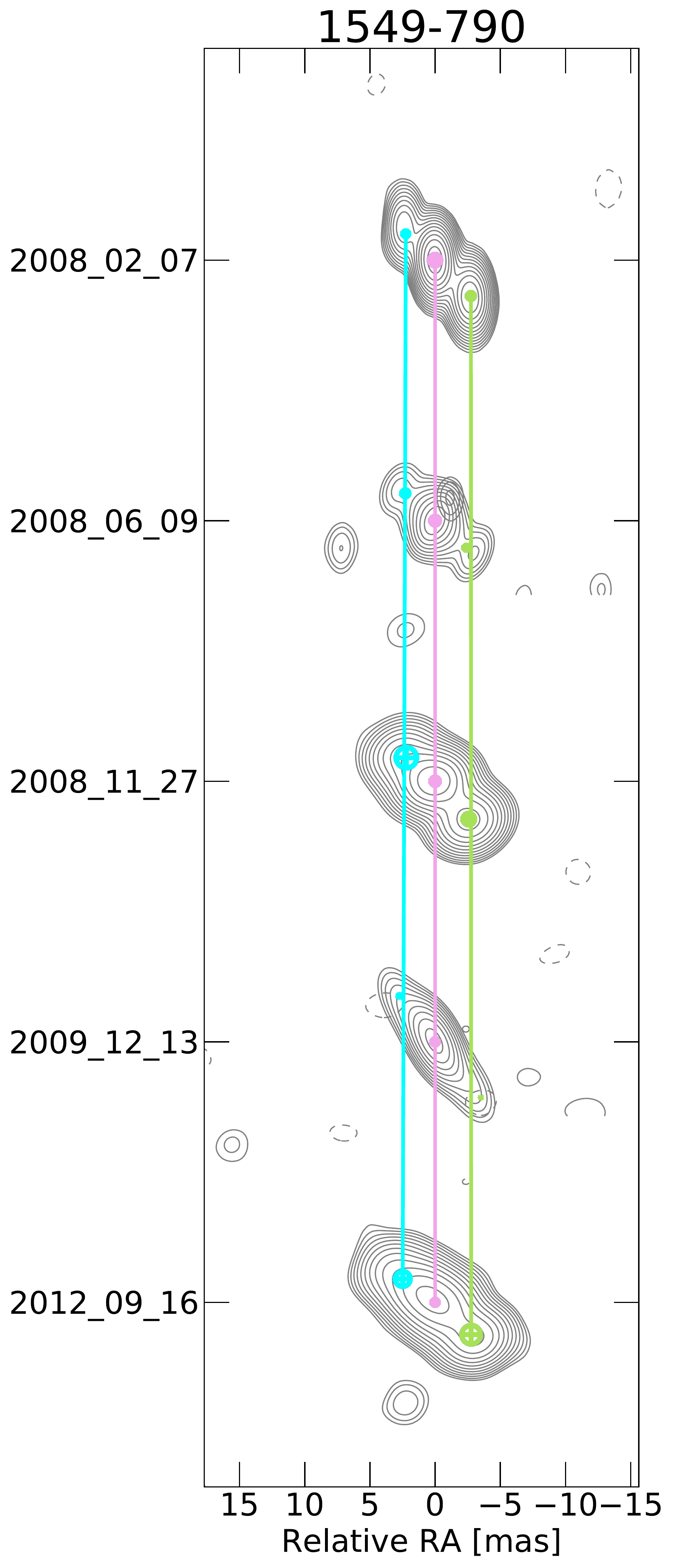}
\end{center}
\caption{Multi-epoch tapered images of 1549$-$790.}
\label{kina_1549}
\end{figure}

\begin{figure}[!!htbp]
\begin{center}
\includegraphics[width=0.7\linewidth]{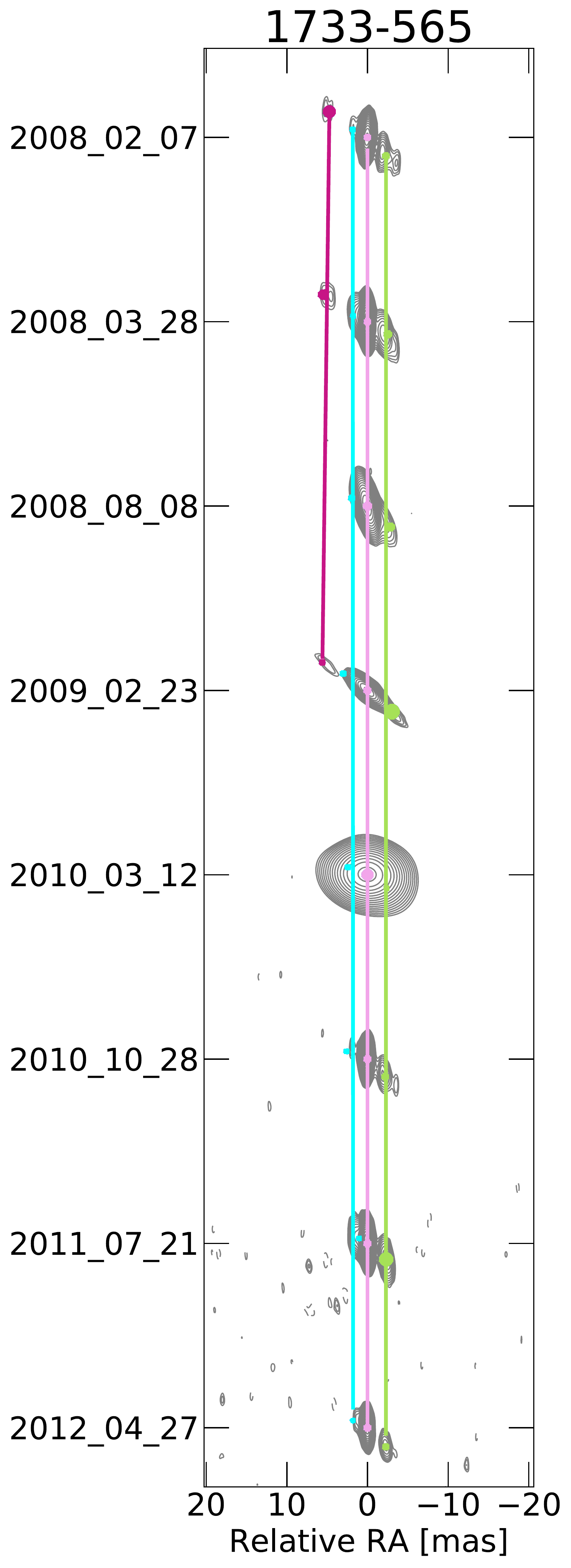}
\end{center}
\caption{Multi-epoch tapered images of 1733$-$565.}
\label{kina_1733}
\end{figure}

\begin{figure}[!!htbp]
\begin{center}
\includegraphics[width=0.8\linewidth]{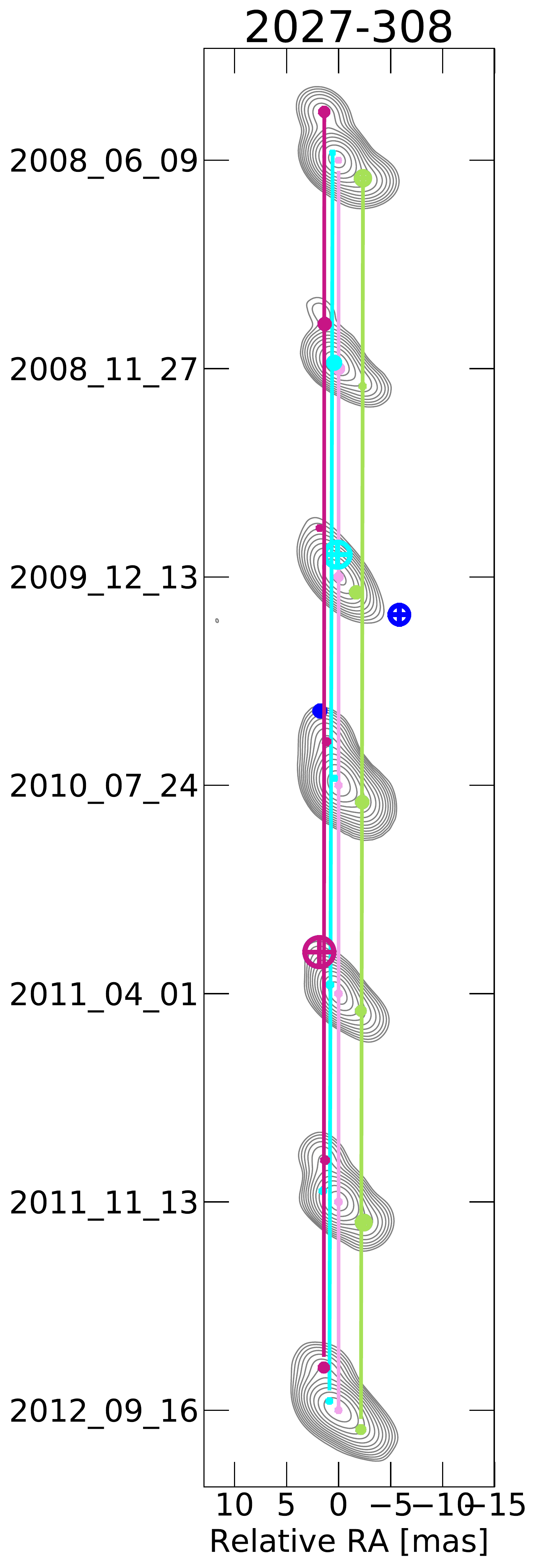}
\end{center}
\caption{Multi-epoch tapered images of 2027$-$308.}
\label{kina_2027}
\end{figure}

\begin{figure}[!!htbp]
\begin{center}
\includegraphics[width=\linewidth]{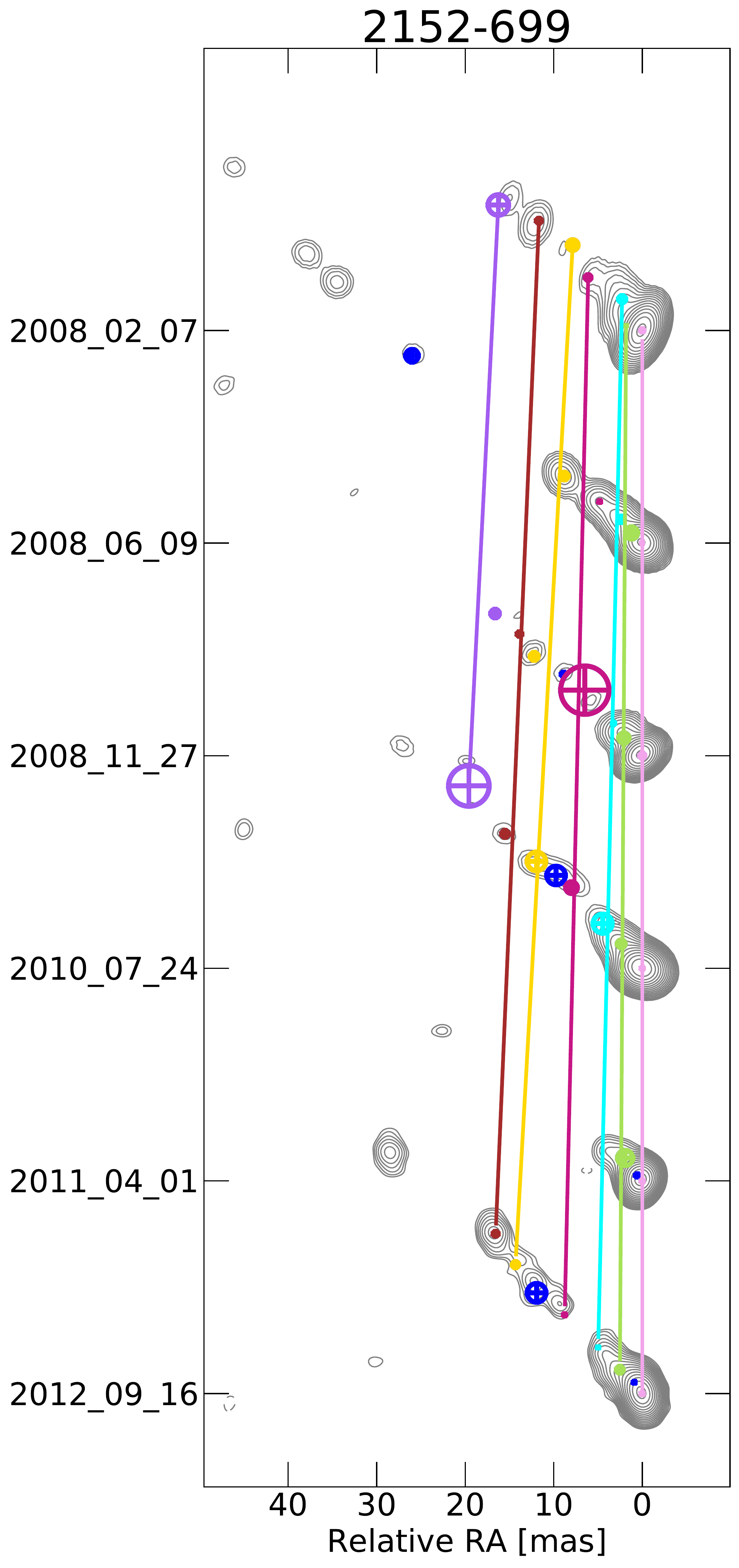}
\end{center}
\caption{Multi-epoch tapered images of 2152$-$699.}
\label{kina_2152}
\end{figure}

Here we include additional information illustrating the results of our kinematic analysis of the multi-epoch TANAMI data. Figures~\ref{kina_1258} through~\ref{kina_2152} show the multi-epoch images of TANAMI radio galaxies, with the corresponding component identification and tracking. Note that, for ease of representation, the distance between the images at different epochs is always constant, and therefore does not represent the time difference between each image pair. Moreover, the colored lines are not fits to the displayed component positions, but simple interpolations meant to clarify the identification and tracking of the different components.

Tables~\ref{1258_mods} through~\ref{2152_mods} list the flux density, radial distance, position angle and size for each circular Gaussian component identified in each source during the kinematic analysis. Note that all components have been shifted so that the core position is always at $(0,0)$ coordinates in all epochs. The position angle is given in the range $(\pi,-\pi)$, with the zero in the N-S direction (in image coordinates) and positive values in the counter-clockwise direction. Components that were not identified (blue crossed circles in Figures~\ref{kina_1258} through~\ref{kina_2152}) are not listed. Note that an apparent speed was fitted only for components detected in at least five epochs. 


\begin{figure*}[!htbp]
\begin{center}
\includegraphics[width=0.49\linewidth]{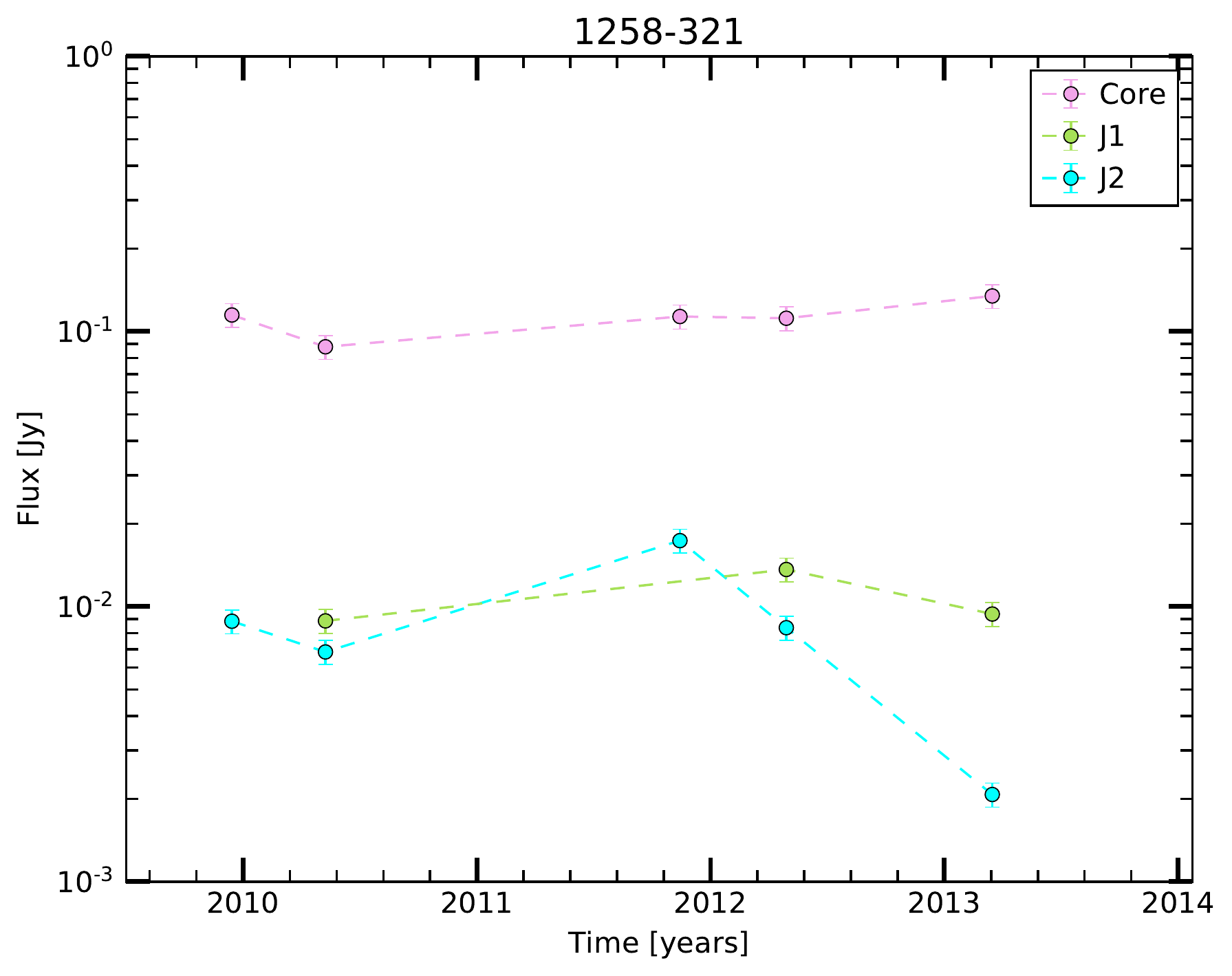}
\includegraphics[width=0.49\linewidth]{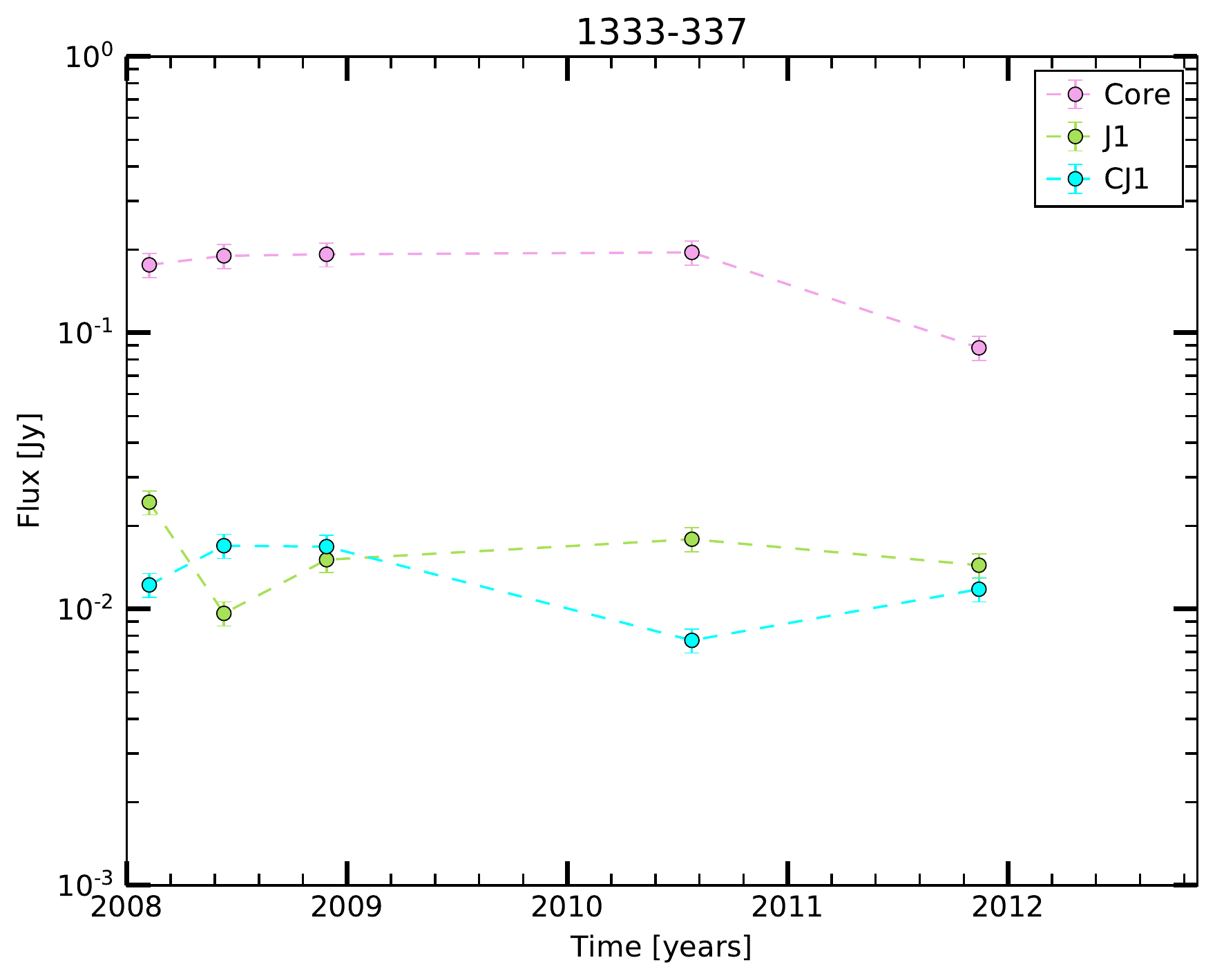}
\includegraphics[width=0.49\linewidth]{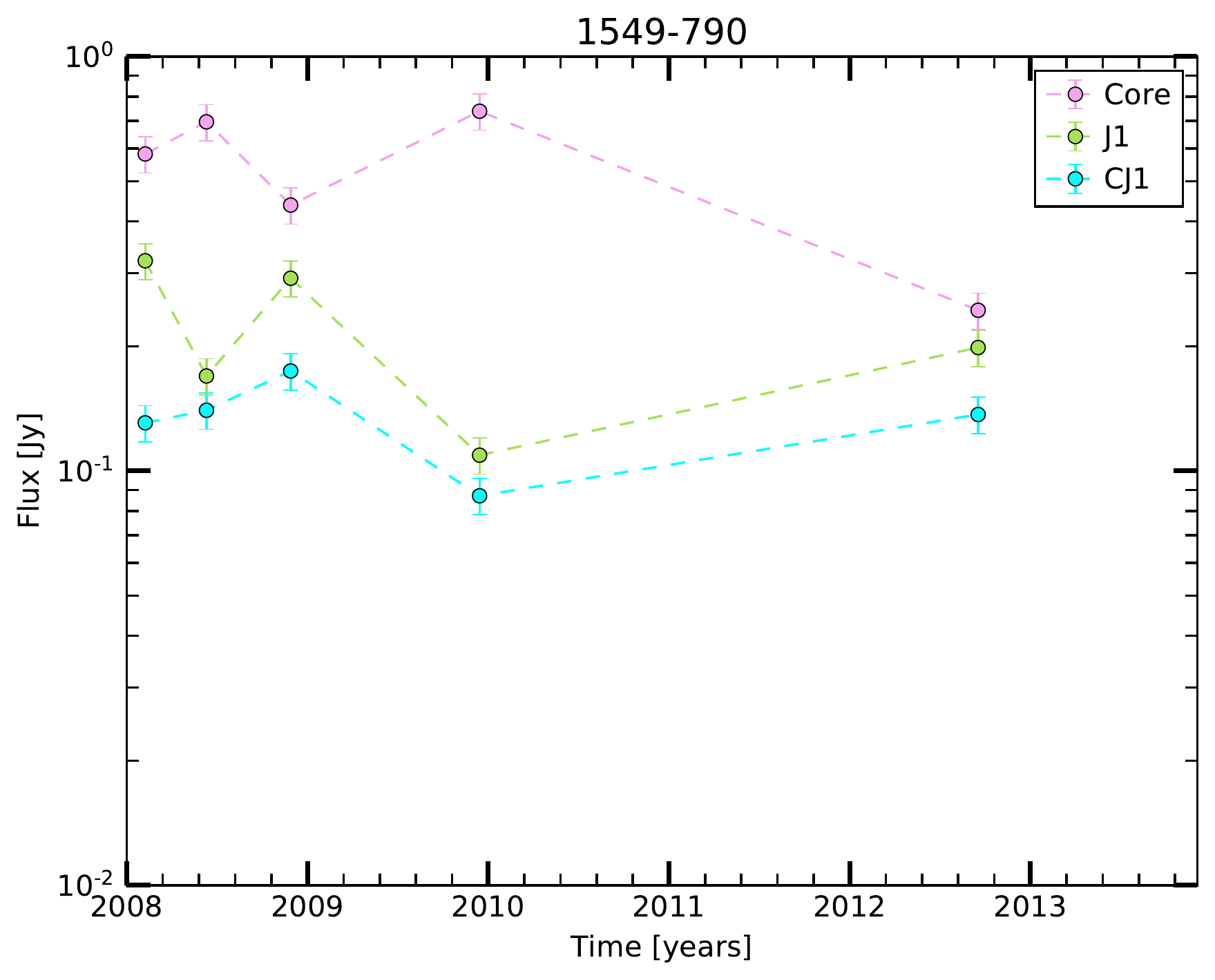}
\includegraphics[width=0.49\linewidth]{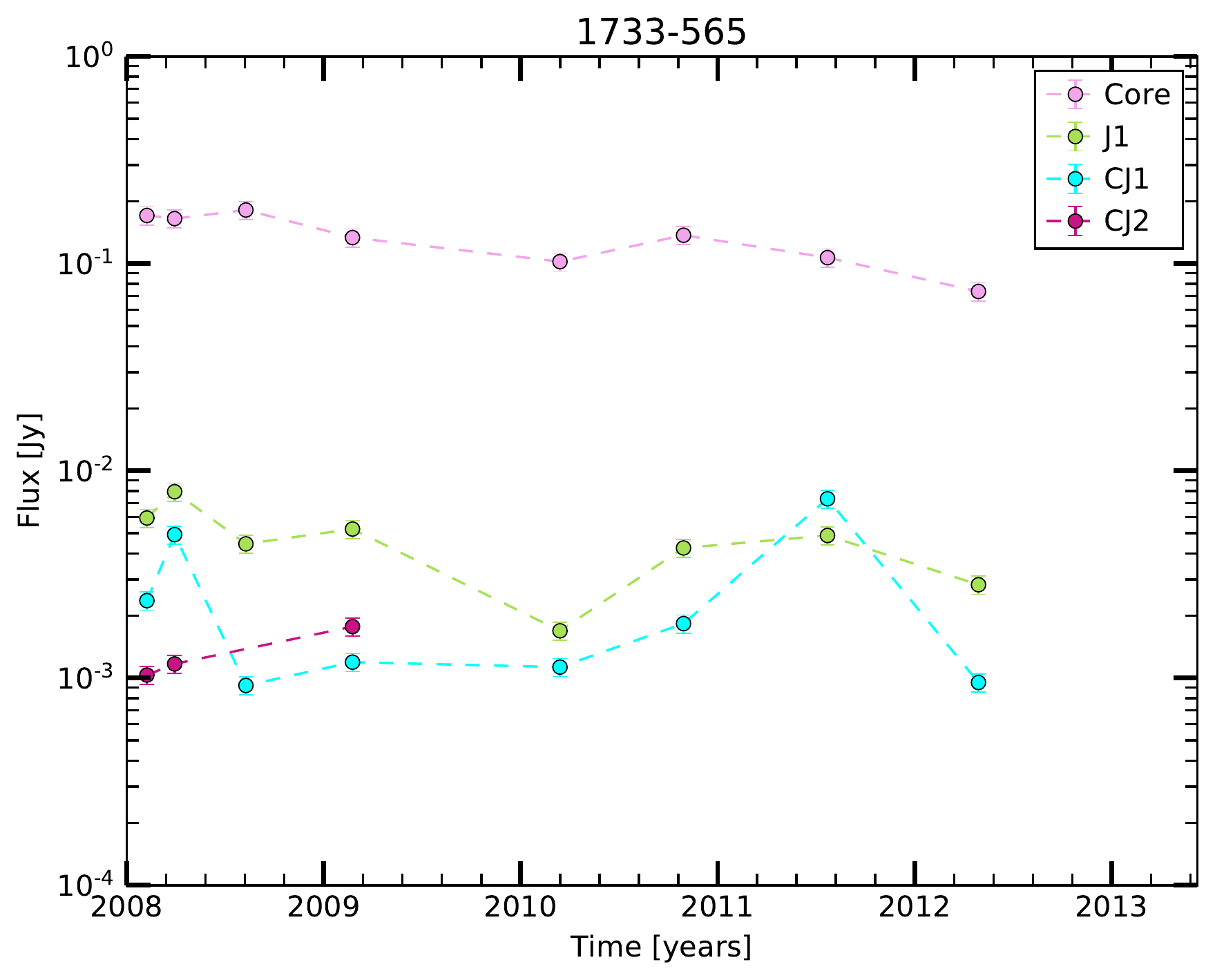}
\includegraphics[width=0.49\linewidth]{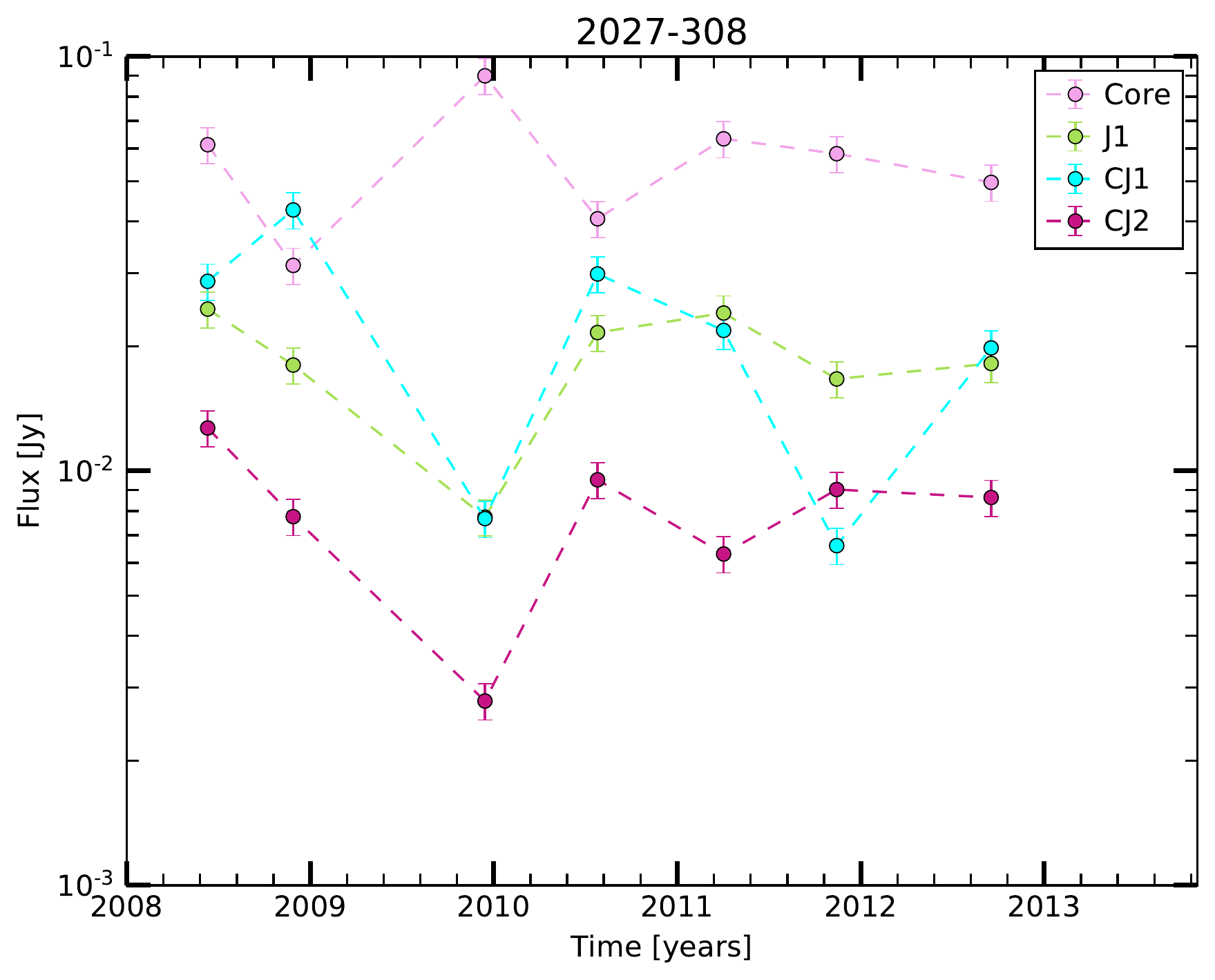}
\includegraphics[width=0.49\linewidth]{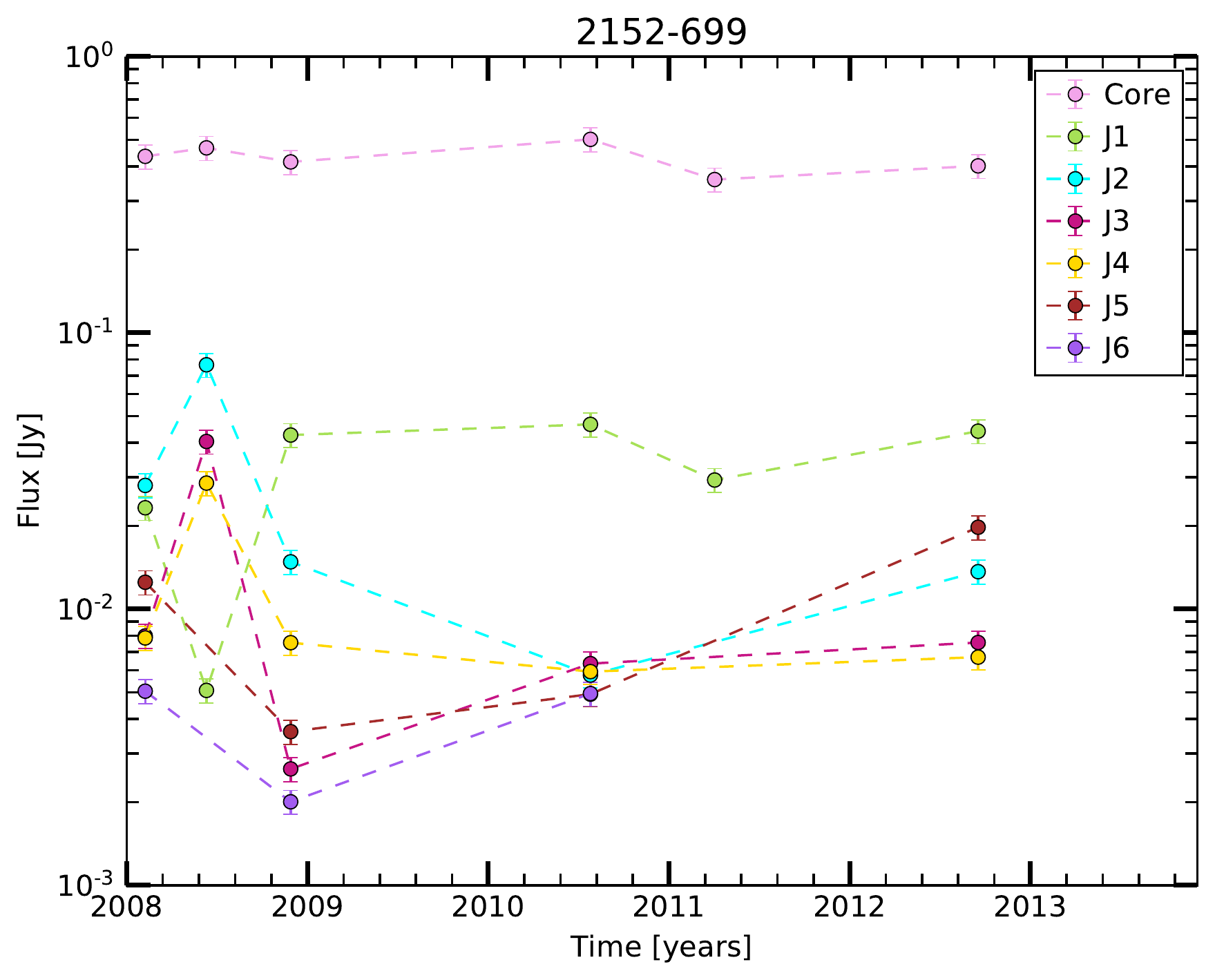}
\end{center}
\caption{Plots of flux of modeled Gaussian components versus time.}
\label{kin_flux}
\end{figure*}
\clearpage
\begin{table}
\caption{Difmap \texttt{Modelfit} parameters for the Gaussian components model of the TANAMI 8.4~GHz images of PKS\,1258$-$321.}    
\label{1258_mods}  
\small
\begin{center}  
\begin{tabular}{llcccc}    
\hline\hline 
Epoch & ID & $S^a$ (Jy) & $d^b$ (mas) & $\phi^c$ (deg) & Size (mas)\\
\hline
2009-12-14 & Core & 0.115 & 0.00 & -166.72 & 0.22\\
 & J2 & 0.009 & 4.63 & -58.07 & 0.28\\
2010-05-09 & Core & 0.088 & 0.00 & 156.29 & 0.10\\
 & J1 & 0.009 & 2.68 & -62.67 & 0.44\\
 & J2 & 0.007 & 7.49 & -65.24 & 0.64\\
2011-11-14 & Core & 0.113 & 0.00 & -41.25 & 0.16\\
 & J2 & 0.017 & 4.54 & -70.20 & 0.80\\
2012-04-28 & Core & 0.112 & 0.00 & -62.62 & 0.03\\
 & J1 & 0.014 & 2.08 & -68.12 & 0.61\\
 & J2 & 0.008 & 6.20 & -50.87 & 0.45\\
2013-03-16 & Core & 0.135 & 0.00 & -152.52 & 0.01\\
 & J1 & 0.009 & 2.87 & -68.31 & 0.50\\
 & J2 & 0.002 & 10.05 & -62.22 & 1.37\\
\hline  
\end{tabular}
\end{center}
$^a$ Flux density.\\
$^b$ Radial distance from the core.\\
$^c$ Position angle.\\
\end{table}

\begin{table}
\caption{Difmap \texttt{Modelfit} parameters for the Gaussian components model of the TANAMI 8.4~GHz images of IC\,4296.}    
\label{1333_mods}  
\small
\begin{center}  
\begin{tabular}{llcccc}    
\hline\hline 
Epoch & ID & $S^a$ (Jy) & $d^b$ (mas) & $\phi^c$ (deg) & Size (mas)\\
\hline
2008-02-07 & Core & 0.176 & 0.00 & -39.24 & 0.20\\
 & J1 & 0.024 & 2.69 & -52.59 & 0.69\\
 & CJ1 & 0.012 & 2.77 & 125.33 & 0.08\\
2008-06-10 & Core & 0.190 & 0.00 & -81.58 & 0.29\\
 & CJ1 & 0.017 & 2.18 & 133.03 & 0.13\\
 & J1 & 0.010 & 3.78 & -52.72 & 0.19\\
2008-11-28 & Core & 0.192 & 0.00 & 36.26 & 0.03\\
 & CJ1 & 0.017 & 2.05 & 130.03 & 0.17\\
 & J1 & 0.015 & 3.22 & -48.98 & 0.08\\
2010-07-26 & Core & 0.195 & 0.00 & 2.22 & 0.14\\
 & J1 & 0.018 & 2.62 & -45.52 & 0.35\\
 & CJ1 & 0.008 & 2.63 & 132.57 & 0.30\\
2011-11-14 & Core & 0.088 & 0.00 & -140.43 & 0.06\\
 & J1 & 0.014 & 4.11 & -57.20 & 0.52\\
 & CJ1 & 0.012 & 4.56 & 121.68 & 0.47\\
\hline  
\end{tabular}
\end{center}
$^a$ Flux density.\\
$^b$ Radial distance from the core.\\
$^c$ Position angle.\\
\end{table}

\begin{table}
\caption{Difmap \texttt{Modelfit} parameters for the Gaussian components model of the TANAMI 8.4~GHz images of PKS\,1549$-$79.}    
\label{1549_mods}  
\small
\begin{center}  
\begin{tabular}{llcccc}    
\hline\hline 
Epoch & ID & $S^a$ (Jy) & $d^b$ (mas) & $\phi^c$ (deg) & Size (mas)\\
\hline
2008-02-07 & Core & 0.582 & 0.00 & -56.41 & 0.45\\
 & CJ1 & 0.131 & 3.02 & 48.06 & 0.23\\
 & J1 & 0.322 & 3.89 & -135.08 & 0.28\\
2008-06-10 & Core & 0.696 & 0.00 & 26.97 & 0.34\\
 & CJ1 & 0.140 & 3.10 & 47.35 & 0.28\\
 & J1 & 0.170 & 3.18 & -130.89 & 0.19\\
2008-11-28 & Core & 0.438 & 0.00 & -28.46 & 0.33\\
 & CJ1 & 0.174 & 2.85 & 50.63 & 0.79\\
 & J1 & 0.292 & 3.89 & -138.32 & 0.45\\
2009-12-14 & Core & 0.738 & 0.00 & 64.25 & 0.25\\
 & CJ1 & 0.087 & 4.46 & 37.68 & 0.10\\
 & J1 & 0.109 & 5.54 & -140.60 & 0.01\\
2012-09-17 & Core & 0.244 & 0.00 & -24.03 & 0.24\\
 & CJ1 & 0.137 & 3.07 & 54.07 & 0.59\\
 & J1 & 0.199 & 3.71 & -131.91 & 0.72\\
\hline  
\end{tabular}
\end{center}
$^a$ Flux density.\\
$^b$ Radial distance from the core.\\
$^c$ Position angle.\\
\end{table}
\begin{table}
\caption{Difmap \texttt{Modelfit} parameters for the Gaussian components model of the TANAMI 8.4~GHz images of PKS\,1733$-$565.}    
\label{1733_mods}  
\small
\begin{center}  
\begin{tabular}{llcccc}    
\hline\hline 
Epoch & ID & $S^a$ (Jy) & $d^b$ (mas) & $\phi^c$ (deg) & Size (mas)\\
\hline
2008-02-07 & Core & 0.171 & 0.00 & 48.16 & 0.15\\
 & CJ1 & 0.002 & 2.08 & 61.63 & 0.05\\
 & J1 & 0.006 & 3.24 & -135.13 & 0.17\\
 & CJ2 & 0.001 & 5.71 & 56.00 & 0.49\\
2008-03-30 & Core & 0.165 & 0.00 & 128.94 & 0.14\\
 & CJ1 & 0.005 & 1.92 & 68.29 & 0.03\\
 & J1 & 0.008 & 2.98 & -121.55 & 0.22\\
 & CJ2 & 0.001 & 6.45 & 58.60 & 0.33\\
2008-08-09 & Core & 0.182 & 0.00 & 95.43 & 0.26\\
 & CJ1 & 0.001 & 2.18 & 63.11 & 0.13\\
 & J1 & 0.004 & 3.85 & -132.29 & 0.26\\
2009-02-23 & Core & 0.134 & 0.00 & 154.40 & 0.19\\
 & CJ1 & 0.001 & 3.67 & 55.31 & 0.10\\
 & J1 & 0.005 & 4.01 & -130.93 & 0.66\\
 & CJ2 & 0.002 & 6.58 & 58.26 & 0.08\\
2010-03-14 & Core & 0.102 & 0.00 & 47.57 & 0.43\\
 & CJ1 & 0.001 & 2.61 & 68.87 & 0.08\\
 & J1 & 0.002 & 2.90 & -125.16 & 0.04\\
2010-10-29 & Core & 0.137 & 0.00 & -29.71 & 0.18\\
 & CJ1 & 0.002 & 2.76 & 69.52 & 0.04\\
 & J1 & 0.004 & 3.10 & -134.81 & 0.17\\
2011-07-23 & Core & 0.107 & 0.00 & 178.79 & 0.19\\
 & CJ1 & 0.007 & 1.21 & 58.89 & 0.03\\
 & J1 & 0.005 & 3.03 & -130.44 & 0.59\\
2012-04-28 & Core & 0.073 & 0.00 & 91.32 & 0.17\\
 & CJ1 & 0.001 & 2.01 & 63.43 & 0.01\\
 & J1 & 0.003 & 3.27 & -135.75 & 0.14\\
\hline  
\end{tabular}
\end{center}
$^a$ Flux density.\\
$^b$ Radial distance from the core.\\
$^c$ Position angle.\\
\end{table}

\begin{table}
\caption{Difmap \texttt{Modelfit} parameters for the Gaussian components model of the TANAMI 8.4~GHz images of PKS\,2027$-$308.}    
\label{2027_mods}  
\small
\begin{center}  
\begin{tabular}{llcccc}    
\hline\hline 
Epoch & ID & $S^a$ (Jy) & $d^b$ (mas) & $\phi^c$ (deg) & Size (mas)\\
\hline
2008-06-10 & Core & 0.061 & 0.00 & -114.51 & 0.08\\
 & CJ1 & 0.029 & 0.93 & 38.51 & 0.04\\
 & J1 & 0.025 & 2.90 & -126.30 & 0.64\\
 & CJ2 & 0.013 & 4.85 & 16.47 & 0.34\\
2008-11-28 & Core & 0.031 & 0.00 & -42.65 & 0.38\\
 & CJ1 & 0.043 & 0.71 & 37.81 & 0.54\\
 & J1 & 0.018 & 2.85 & -126.22 & 0.14\\
 & CJ2 & 0.008 & 4.48 & 17.34 & 0.44\\
2009-12-14 & Core & 0.090 & 0.00 & -84.95 & 0.25\\
 & CJ1 & 0.008 & 2.14 & 2.65 & 1.21\\
 & J1 & 0.008 & 2.24 & -131.26 & 0.45\\
 & CJ2 & 0.003 & 5.03 & 21.30 & 0.12\\
2010-07-26 & Core & 0.041 & 0.00 & -40.67 & 0.13\\
 & CJ1 & 0.030 & 0.78 & 30.47 & 0.09\\
 & J1 & 0.022 & 2.79 & -125.51 & 0.45\\
 & CJ2 & 0.009 & 4.34 & 15.32 & 0.20\\
2011-04-03 & Core & 0.063 & 0.00 & 56.16 & 0.15\\
 & CJ1 & 0.022 & 1.20 & 44.99 & 0.18\\
 & J1 & 0.024 & 2.70 & -127.68 & 0.32\\
 & CJ2 & 0.006 & 4.39 & 24.93 & 1.39\\
2011-11-14 & Core & 0.058 & 0.00 & -44.63 & 0.15\\
 & CJ1 & 0.007 & 1.84 & 54.85 & 0.17\\
 & J1 & 0.017 & 3.12 & -129.34 & 0.62\\
 & CJ2 & 0.009 & 4.21 & 17.89 & 0.22\\
2012-09-17 & Core & 0.050 & 0.00 & -27.36 & 0.12\\
 & CJ1 & 0.020 & 1.25 & 45.02 & 0.12\\
 & J1 & 0.018 & 2.84 & -130.86 & 0.28\\
 & CJ2 & 0.009 & 4.34 & 19.06 & 0.34\\
\hline  
\end{tabular}
\end{center}
$^a$ Flux density.\\
$^b$ Radial distance from the core.\\
$^c$ Position angle.\\
\end{table}

\begin{table}
\caption{Difmap \texttt{Modelfit} parameters for the Gaussian components model of the TANAMI 8.4~GHz images of PKS\,2153$-$69.}    
\label{2152_mods}  
\small
\begin{center}  
\begin{tabular}{llcccc}    
\hline\hline 
Epoch & ID & $S^a$ (Jy) & $d^b$ (mas) & $\phi^c$ (deg) & Size (mas)\\
\hline
2008-02-07 & Core & 0.435 & 0.00 & 35.39 & 0.18\\
 & J1 & 0.023 & 1.91 & 73.49 & 0.04\\
 & J2 & 0.028 & 4.21 & 32.82 & 0.40\\
 & J3 & 0.008 & 8.55 & 45.65 & 0.32\\
 & J4 & 0.008 & 12.44 & 39.21 & 0.60\\
 & J5 & 0.013 & 17.02 & 43.30 & 0.28\\
 & J6 & 0.005 & 21.56 & 48.95 & 1.16\\
2008-06-10 & Core & 0.467 & 0.00 & -143.96 & 0.04\\
 & J1 & 0.005 & 1.67 & 46.58 & 0.69\\
 & J2 & 0.077 & 3.59 & 41.63 & 0.37\\
 & J3 & 0.040 & 6.71 & 45.80 & 0.10\\
 & J4 & 0.029 & 11.62 & 49.38 & 0.44\\
2008-11-28 & Core & 0.416 & 0.00 & -104.97 & 0.29\\
 & J1 & 0.043 & 2.88 & 46.22 & 0.59\\
 & J2 & 0.015 & 4.87 & 41.53 & 0.12\\
 & J3 & 0.003 & 9.84 & 41.26 & 2.73\\
 & J4 & 0.008 & 16.57 & 47.45 & 0.45\\
 & J5 & 0.004 & 19.51 & 45.30 & 0.26\\
 & J6 & 0.002 & 23.10 & 46.03 & 0.48\\
2010-07-26 & Core & 0.501 & 0.00 & 78.52 & 0.11\\
 & J1 & 0.047 & 3.64 & 40.67 & 0.45\\
 & J2 & 0.006 & 6.75 & 41.61 & 1.07\\
 & J3 & 0.006 & 12.12 & 41.28 & 0.66\\
 & J4 & 0.006 & 16.99 & 44.73 & 1.05\\
 & J5 & 0.005 & 21.70 & 45.62 & 0.38\\
 & J6 & 0.005 & 28.42 & 43.58 & 2.29\\
2011-04-03 & Core & 0.359 & 0.00 & 63.90 & 0.13\\
 & J1 & 0.029 & 3.23 & 36.72 & 0.87\\
2012-09-17 & Core & 0.402 & 0.00 & 42.75 & 0.11\\
 & J1 & 0.044 & 3.69 & 43.33 & 0.38\\
 & J2 & 0.014 & 7.21 & 43.79 & 0.06\\
 & J3 & 0.008 & 12.49 & 44.61 & 0.11\\
 & J4 & 0.007 & 20.42 & 44.56 & 0.35\\
 & J5 & 0.020 & 24.49 & 42.57 & 0.28\\
\hline  
\end{tabular}
\end{center}
$^a$ Flux density.\\
$^b$ Radial distance from the core.\\
$^c$ Position angle.\\
\end{table}

\end{appendix}

\end{document}